%% file: thesis.tex
\documentclass[12pt,lot, lof]{puthesis}
\newcommand{\proquestmode}{}

\title{Addressing Security and Privacy Challenges in Internet of Things}

\submitted{January 2017}  
\copyrightyear{2017}  
\author{Arsalan Mosenia}
\adviser{Niraj K. Jha}  
\department{Electrical Engineering}

\usepackage{cite}
\usepackage{graphicx}
\usepackage{amsmath}
\usepackage{gensymb}
\usepackage{caption}
\usepackage{subcaption}
\usepackage{multirow}
\usepackage{MnSymbol}%
\usepackage{wasysym}%
\usepackage{hhline}
\usepackage{tikz}
\usepackage{xcolor,colortbl}
\usepackage{color}

\usepackage{verbatim}

\usepackage{multirow}
\usepackage{longtable}

\usepackage{booktabs}

\ifdefined\printmode

\usepackage{url}

\else

\ifdefined\proquestmode

\usepackage{hyperref}
\hypersetup{bookmarksnumbered}

\makeatletter
\hypersetup{pdftitle=\@title,pdfauthor=\@author}
\makeatother

\else

\usepackage{hyperref}
\hypersetup{colorlinks,bookmarksnumbered}

\makeatletter
\hypersetup{pdftitle=\@title,pdfauthor=\@author}
\makeatother

\fi 
\fi 

\ifodd 0
\renewcommand{\maketitlepage}{}
\renewcommand*{\makecopyrightpage}{}
\renewcommand*{\makeabstract}{}

\else

\abstract{
\input{abstract}
}

\acknowledgements{
\input{acknowledgements}
}

\dedication{To my beloved parents and family who have supported me throughout my education.}

\fi  

\begin{document}

\makefrontmatter

\include{ch-intro/chapter-intro}
\include{ch-relatedwork/chapter-relatedwork}

\include{ch-hmonitoring/chapter-hmonitoring}

\include{ch-OpSecure/chapter-OpSecure}
\include{ch-physio/chapter-physio}

\include{ch-DISASTER/chapter-DISASTER}
\include{ch-CABA/chapter-CABA}

\include{ch-conclusion/chapter-conclusion}
\singlespacing

\bibliographystyle{IEEEtran}
\cleardoublepage
\ifdefined\phantomsection
  \phantomsection  
\else
\fi
\addcontentsline{toc}{chapter}{Bibliography}

\bibliography{thesis_b_file}

\end{document}

%% file: abstract.tex
Internet of Things (IoT), also referred to as the Internet of Objects, is envisioned as a holistic and transformative approach for providing numerous services. The rapid development of various communication protocols and miniaturization of transceivers along with recent advances in sensing technologies offer the opportunity to transform isolated devices into communicating smart things. Smart things, that can sense, store, and even process electrical, thermal, optical, chemical, and other signals to extract user-/environment-related information, have enabled services only limited by human imagination.

Despite picturesque promises of IoT-enabled systems, the integration of smart things into the standard Internet introduces several security challenges because the majority of Internet technologies, communication protocols, and sensors were not designed to support IoT. Several recent research studies have demonstrated that launching security/privacy attacks against IoT-enabled systems, \textit{in particular wearable medical sensor (WMS)-based systems}, may lead to catastrophic situations and life-threatening conditions. Therefore, security threats and privacy concerns in the IoT domain need to be proactively studied and aggressively addressed. In this thesis, we tackle several domain-specific security/privacy challenges associated with IoT-enabled systems. 

We first target health monitoring systems that are one of the most widely-used types of IoT-enabled systems. We discuss and evaluate several energy-efficient schemes and algorithms, which significantly reduce total energy consumption of different implantable and wearable medical devices (IWMDs). The proposed schemes make continuous long-term health monitoring feasible while providing spare energy needed for data encryption. 

Furthermore, we present two energy-efficient protocols for implantable medical devices (IMDs), which are essential for data encryption: (i) a secure wakeup protocol that is resilient against battery draining attacks, along with (ii) a low-power key exchange protocol that shares the encryption key between the IMD and the external device while ensuring confidentiality of the key. 

Moreover, we introduce a new class of attacks against the privacy of a patient who is carrying IWMDs. We describe how an attacker can infer private information about the patient by exploiting physiological information leakage, i.e., signals that continuously emanate from the human body due to the normal functioning of organs or IWMDs attached to (or implanted in) the body.

Further, we propose a new generic class of security attacks, called dedicated intelligent security attacks against sensor-triggered emergency responses (DISASTER), that is applicable to a variety of sensor-based systems. DISASTER exploits design flaws and security weaknesses of safety mechanisms deployed in cyber-physical systems (CPSs) to trigger emergency responses even in the absence of a real emergency. In addition to introducing DISASTER, we comprehensively describe its serious consequences and demonstrate the possibility of launching such attacks against the two most widely-used CPSs: residential and industrial automation/monitoring systems.

Finally, we present a continuous authentication system based on BioAura, i.e., information that is already gathered by WMSs for diagnostic and therapeutic purposes. We extensively examine the proposed authentication system and demonstrate that it offers promising advantages over one-time knowledge-based authentication systems, e.g., password-/pattern-based systems, and may potentially be used to protect personal computing devices and servers, software applications, and restricted physical spaces.

%% file: acknowledgements.tex
This thesis would not have been possible without the help of so many people in so many ways. I would like to express my sincere appreciation to all those who supported me to complete the Ph.D. program at Princeton University.

I have had the honor and privilege to work under the supervision of Prof. Niraj K. Jha. who helped me take the first steps as a researcher. His guidance and support have helped me with academic research and writing from the first day of the Ph.D. program and have been very beneficial in shaping my Ph.D work. Apart from his immense wisdom and insight, his enthusiasm to explore new research directions and novel ideas has been a constant source of inspiration to me. His patience, kindness, understanding, and empathy towards his students are truly admirable. He has been extremely caring during my days of struggle. Although I have immensely benefited from his technical expertise, I still have so much to learn from his superior discipline and kind disposition.

I also received much advice and guidance from Profs. Anand Raghunathan, Susmita Sur-Kolay, and Mehran Mozaffari-Kermani. Discussion with them were always an exceptional source of new ideas. Without their continuous support, encouragement, and inspiration, preparing this thesis would not have been feasible. 

I am deeply grateful for the guidance of Prof. Prateek Mittal, who inspired me in many ways and offered me the opportunity to become a teaching assistant in ELE 432 that he taught in Spring 2016. I have also been fortunate to collaborate with him on a research study during my last year of Ph.D. work.

I would like to thank my thesis readers, Profs. Niraj K. Jha, Susmita Sur-Kolay, and Prateek Mittal, for the extensive efforts in polishing and revising this thesis. I am also thankful to Profs. Niraj K. Jha, Anand Raghunathan, and David Wentzlaff for agreeing to be on my final public oral committee. 

Profs. Niraj K. Jha, Prateek Mittal, and Sharad Malik kindly offered me teaching opportunities in their courses. Those were extremely valuable and enjoyable experiences from which I realized how much devotion makes perfection. I greatly appreciate their help, guidance, and trust in me. 

The faculty in the Electrical Engineering and Computer Science Departments have been a phenomenal source of knowledge and guidance. I would like to thank all of them. I am also thankful to have group-mates like Mehran Mozaffari-Kermani, Mohammad Shoaib, Sourindra Choudhuri, Abdullah Guler, Fisher Wu, Xiaoliang Dai, Debajit Bhattacharya, Bochao Wang, Hongxu Yin, Ozge Akmandor, and Ajay Bhoj for their companionship. I would like to express my sincere gratitude towards Colleen Conrad, Lori Bailey, Heather L. Evans, and Stacey Weber as well; they have been extremely helpful throughout my time in the Department of Electrical Engineering.

I also appreciate the help I received from my friends who have made this journey unforgettable. I have had so many great moments with Behsan Behzadi, Pouya Sharifi, Amir Mazaheri, Fariborz Mirzaei, Siamak Garabaglu, Hooman Mohajeri, Mohammad Shahrad, Hossein Valavi, Hasan Nosratabadi, Moein Malekaghlagh, Mehdi Yazdi, Hesam Aklaghpour, Vahid Askari, Hooman Kayyatzadeh, and Ali Jamal. I have been fortunate to have them in my life. 

Finally, my heartfelt thanks to my beloved parents, grandparents, and my wonderful brother Arman. I would not been able to accomplish anything without their unwavering love, support, and understanding.

This work was supported by NSF under Grant no.~CNS-1219570 and CNS-1617628 and Project X Fund from Princeton University. 

%% file: ch-intro/chapter-intro.tex
\chapter{Introduction\label{ch:intro}}
Internet of Things (IoT) does not have a unique definition. However, a broad interpretation of IoT is that it provides any service over the traditional Internet by enabling human-to-thing, thing-to-thing, or thing-to-things communications \cite{IoT1}. IoT represents the interconnection of heterogeneous entities, where the term entity refers to a human, sensor, or potentially anything that may request/provide a service \cite{IOTHET}. 

The emergence of the IoT paradigm is one of the most spectacular phenomena of 
the last decade. The development of various communication protocols, along 
with the miniaturization of transceivers, provides the opportunity to transform 
an isolated device into a communicating thing. Moreover, computing power, 
energy capacity, and storage capabilities of small computing or sensing devices 
have significantly improved while their sizes have decreased
drastically. Boosted by the rapid development of IoT-enabled systems in recent years, Internet-connected wearable medical sensors (WMSs) are garnering an ever-increasing attention in both academic and industrial research. Although WMSs were initially developed to enable low-cost solutions for continuous health monitoring, the applications of WMS-based systems now range far beyond health care. As discussed later in this chapter, several research efforts have proposed the use of such systems in diverse application domains, e.g., education, human-computer interaction, and security. 

As a side effect of rapid advances in the design and development of IoT-enabled systems, the number of potential threats and possible attacks against security of such systems, \textit{in particular WMS-based systems that rely on resource-constrained sensors}, and privacy of their users has grown drastically. Thus, security threats and common privacy concerns need to be studied and addressed in depth. This would greatly simplify the development of secure smart devices that enable a plethora of services for human beings, ranging from building automation to health monitoring, in which very different things, e.g., temperature sensor, light sensor, and medical sensors, might interact with each other or with a human carrying a smart computing device, e.g., a smartphone, tablet, or laptop. This dissertation aims to explore and address different security/privacy issues associated with IoT-enabled systems with a special focus on WMS-based systems. 

In the rest of this chapter, we first describe the IoT paradigm, and then discuss WMS-based systems.

\section{The IoT paradigm}
In this section, we first discuss different IoT reference models described in the literature. Then, we describe the scope of IoT applications. Thereafter, we explain what security means in the scope of IoT. Finally, we discuss who the attackers that target the IoT might be, and what motivations they might have.

\subsection{IoT reference models}
\begin{figure*}
    \centering
    \begin{subfigure}[b]{0.4\textwidth}
        \includegraphics[width=170pt,height=150pt]{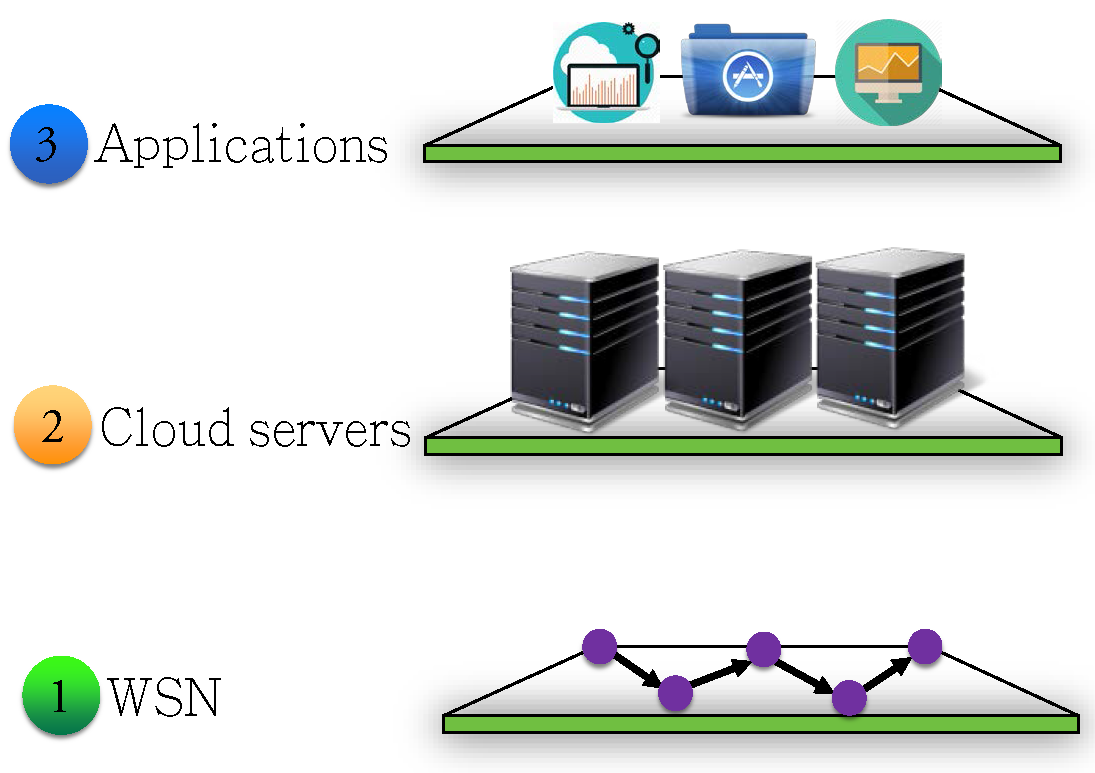}
        \caption{Three-level model\cite{LVL3}}
        \label{fig:gull}
    \end{subfigure}
    \begin{subfigure}[b]{0.4\textwidth}
        \includegraphics[width=170pt,height=150pt]{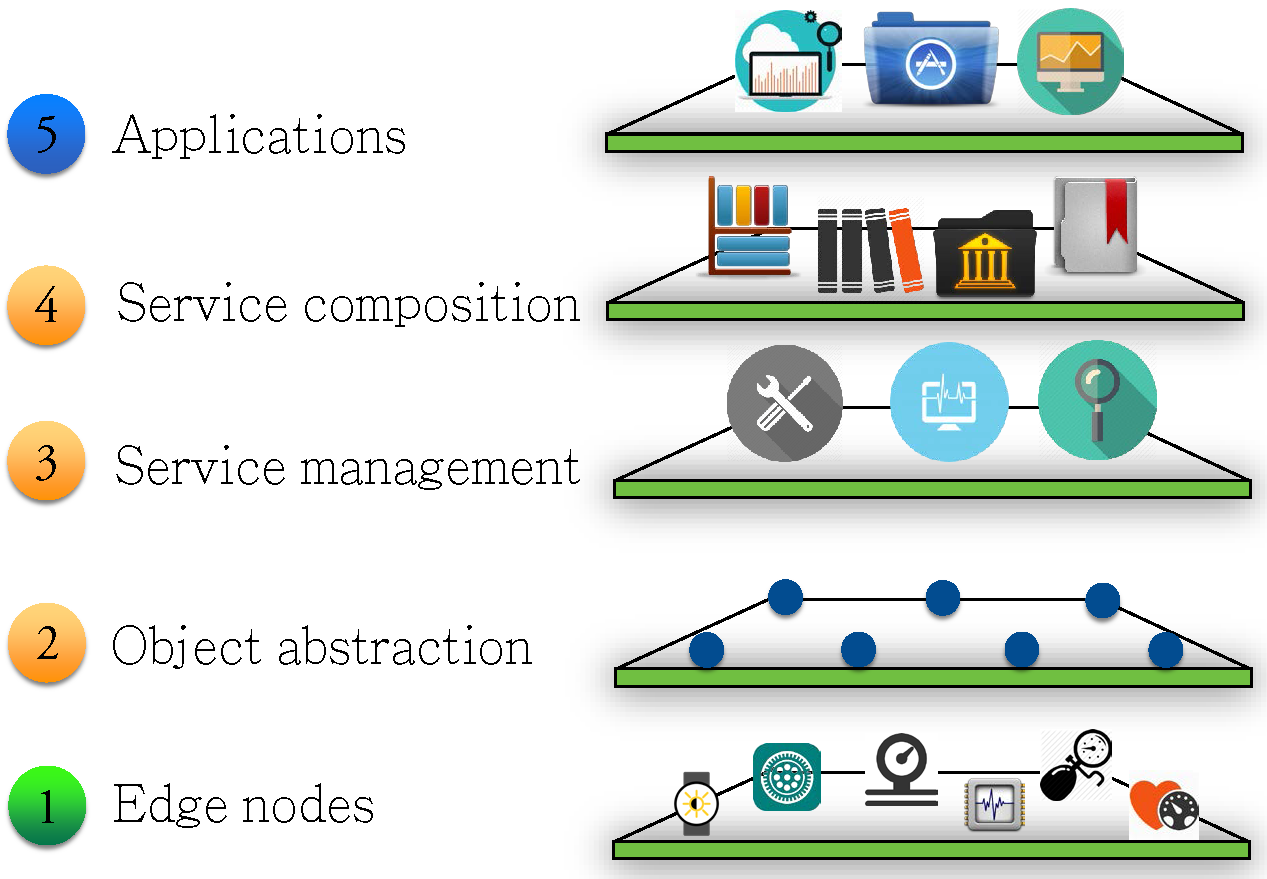}
        \caption{Five-level model \cite{IOTHET}}
        \label{fig:tiger}
    \end{subfigure}
\par\bigskip
    \begin{subfigure}[b]{0.8\textwidth}
        \includegraphics [width=340pt,height=200pt]{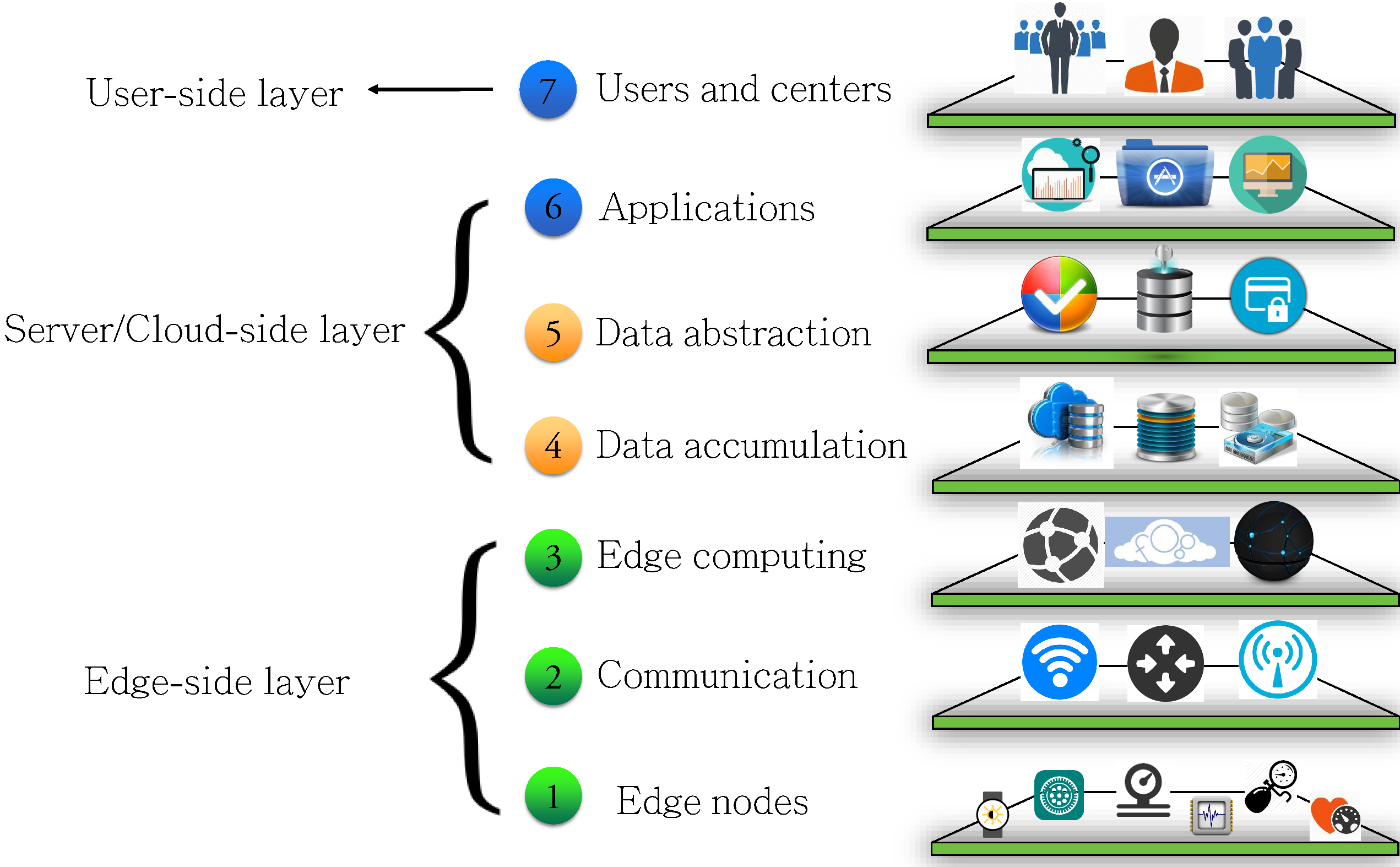}
        \caption{CISCO's seven-level model \cite{CISCO}}
        \label{fig:mouse}
    \end{subfigure}
    \caption{Three IoT reference models}\label{fig:reference}
\end{figure*}

Three IoT reference models have been widely discussed in academic and 
industrial publications. Fig.~\ref{fig:reference} shows these models and their 
different levels. The three-level model \cite{LVL3} is among the first 
reference models proposed for IoT. It depicts IoT as an extended version of 
wireless sensor networks (WSNs). In fact, it models IoT as a combination of 
WSNs and cloud servers, which offer different services to the user. The 
five-level model \cite{IOTHET} is an alternative that has been proposed to 
facilitate interactions among different sections of an enterprise by 
decomposing complex systems into simplified applications consisting of an 
ecosystem of simpler and well-defined components \cite{IOTHET}. In 2014, CISCO 
suggested a comprehensive extension to the traditional three-level and 
five-level models. CISCO's seven-level model has the potential to be 
standardized and thus create a widely-accepted reference model for 
IoT \cite{CISCO}. In this model, data flow is usually bidirectional. However, 
the dominant direction of data flow depends on the application. For example, 
in a control system, data and commands travel from the top of the model 
(applications level) to the bottom (edge node level), whereas, in a monitoring 
scenario, the flow is from bottom to top. 

In order to summarize IoT security attacks and their countermeasures in a level-by-level fashion, we use the CISCO reference model in Chapter \ref{ch:relatedwork}. Next, we briefly describe each level of this model. 

\noindent \textit{\textbf{Level 1-Edge devices}}: The first level of this
reference model typically consists of computing nodes, e.g., smart
controllers, sensors, RFID readers, etc., and different versions of RFID
tags. Data confidentiality and integrity must be taken into account from
this level upwards.

\noindent  \textit{\textbf{Level 2-Communication}}: The communication level consists of all the components that enable transmission of information or commands: (i) communication between devices in the first level, (ii) communication between the components in the second level, and (iii) transmission of information between the first and third levels (edge computing level). 

\noindent \textit{\textbf{Level 3-Edge computing}}: Edge computing, also called fog computing, is the third level of the model in which simple data processing is initiated. This is essential for reducing the computation load in the higher level as well as providing a fast response. Most real-time applications need to perform computations as close to the edge of the network as possible. The amount of processing in this level depends on the computing power of the service providers, servers, and computing nodes. Typically, simple signal processing and learning algorithms are utilized here.

\noindent \textit{\textbf{Level 4-Data accumulation}}: Most of the
applications may not need instant data processing. This level enables
conversion of data in motion to data at rest, i.e., it allows us to
store the data for future analysis or to share with high-level computing
servers. The main tasks of this level are converting the format from network 
packets to database tables, reducing data through filtering and selective 
storing, and determining whether the data are of interest to higher levels.

\noindent \textit{\textbf{Level 5-Data abstraction}}: This level provides the opportunity to render and store data such that further processing becomes simpler or more efficient. The common tasks of entities at this level include normalization, de-normalization, indexing and consolidating data into one place, and providing access to multiple data stores.

\noindent  \textit{\textbf{Level 6-Applications}}: The application level provides information interpretation, where software cooperates with data accumulation and data abstraction levels. The applications of IoT are numerous and may vary significantly across markets and industrial needs.

\noindent  \textit{\textbf{Level 7-Users and centers}}: The highest level of the IoT is where the users are. Users make use of the applications and their analytical data.

\subsection{Scope of applications}

In this section, we first discuss the scope of 
IoT applications.

Smart homes and buildings, electronic health aids, and smarter vehicles are 
just some of the IoT instances \cite{MY_NEW_3, MY_NEW_4, MY_NEW_2}. Each smart device may provide several services 
to enable a more intuitive environment. However, we are not even close to 
exhausting the possible uses of IoT. IoT provides an opportunity to combine sensing, communication, networking, 
authentication, identification, and computing, and enables numerous services 
upon request such that access to the information of any smart thing is possible 
at any time. Fig.~\ref{fig:app} demonstrates various applications of the
IoT, which we describe next:

\begin{figure*}
\centering
\includegraphics[trim = 0mm 0mm 0mm 0mm, clip, width=430pt,height=280pt]{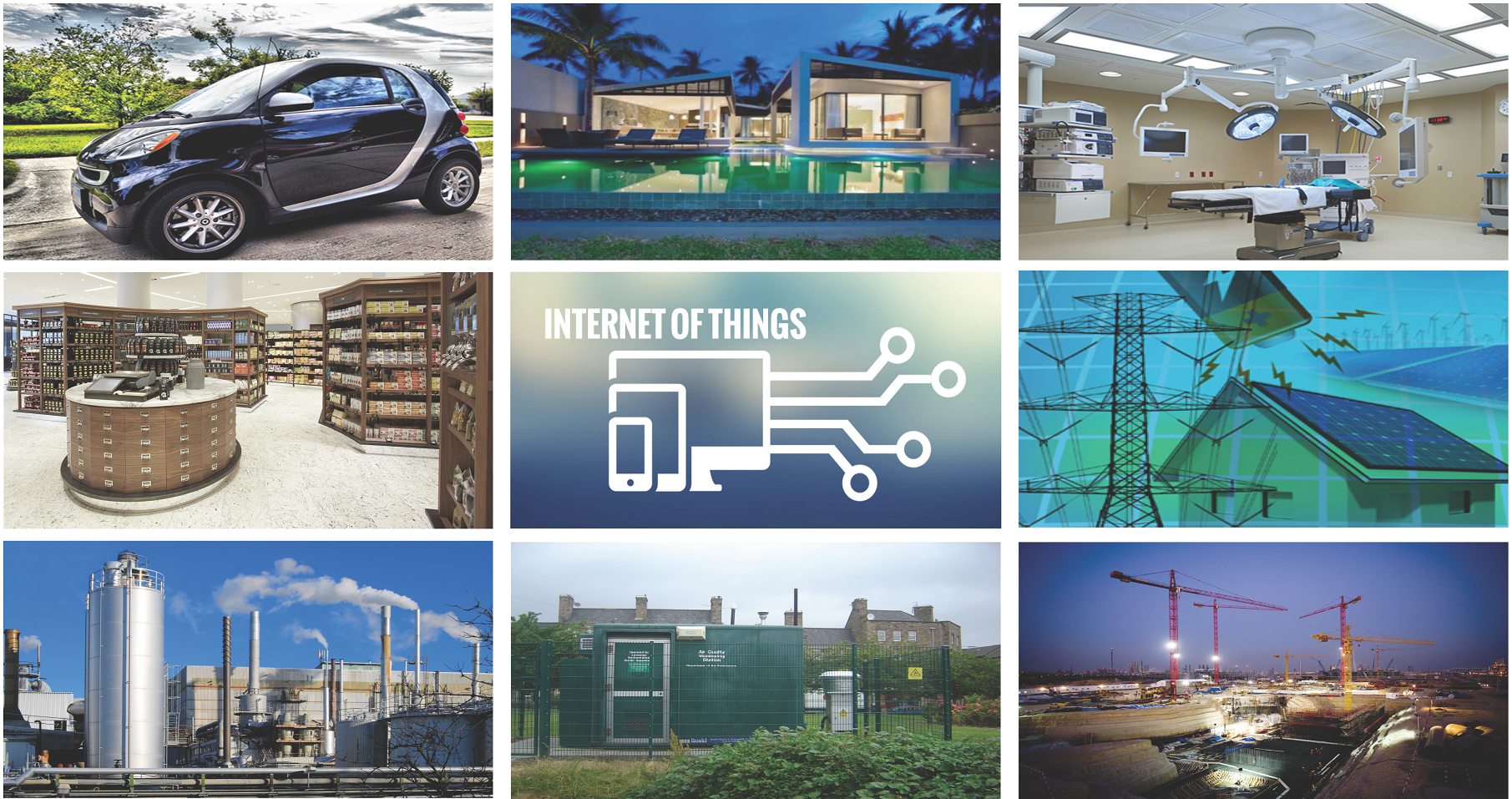}
\caption{Different applications of IoT \cite{IOT_SURVEY}}
\label{fig:app}
\end{figure*}

\noindent \textit{\textbf{1. Smart vehicles}}: Smart vehicles have started to revolutionize 
traditional transportation.  Small IoT-based systems can enable remote 
locking/unlocking of cars, download of roadmaps, and access to traffic 
information. Moreover, Internet-connected cars provide significant protection 
against theft. \\
\noindent \textit{\textbf{2. Smart buildings}}: Smart homes and buildings enable
effective energy management. For example, smart thermostats, which have 
embedded sensors and data analysis algorithms, can control air conditioners 
based on user preferences and habits. Moreover, smart controllers can adjust 
lighting based on user's usage. Several household items, e.g., 
refrigerators, televisions, and security systems, could have their own 
processing units, and provide over-the-Internet services. These smart devices 
greatly enhance users' convenience. Remotely-controllable devices receive 
commands from users to perform actions that have an effect on the surrounding 
environment. Thus, attacks on these devices may lead to physical 
consequences \cite{HOMES}. \\
\noindent \textit{\textbf{3. Health monitoring}}: Recent advances in biomedical sensing and 
signal processing, low-power devices, and wireless communication have 
revolutionized health care. IoT-based long-term personal health monitoring and 
drug delivery systems, in which various physiological signals are captured, 
analyzed, and stored for future use, provide a fundamentally new approach to 
health care \cite{CNIA}. Smart medical devices are already in use in fitness, diet, and 
health monitoring systems \cite{ALINIA}. The future of IoT-based health care systems 
lies in designing personal health monitors that enable early detection of 
illnesses. \\
\noindent \textit{\textbf{4. Energy management}}: Use of smart IoT-based systems, which 
integrate embedded sensors and actuation components, enables a proactive 
approach to optimizing energy consumption. In particular, power outlets, lamps, 
fridges, and smart televisions, which can be controlled remotely, are expected 
to share information with energy supply companies to optimize the energy 
consumption in smart homes. Moreover, such things allow the users to remotely 
control or manage them, and enable scheduling that can lead to a 
significant reduction in energy consumption.\\
\noindent \textit{\textbf{5. Construction management}}: Monitoring and management of modern 
infrastructure, e.g., bridges, traffic lights, railway tracks, and buildings,
are one of the key IoT applications \cite{CONS}. IoT can be used for 
monitoring any sudden changes in structural conditions that can lead to 
safety and security risks. It can also enable construction and 
maintenance companies to share information about their plans. For example, a 
construction company can let GPS companies know its maintenance plans for the 
roads and, based on that, the smart GPS devices can choose an alternative 
route, which avoids the road under construction. \\
\noindent \textit{\textbf{6. Environmental monitoring}}: The use of smart things with 
embedded sensors enables environmental monitoring as well as detection of 
emergency situations, e.g., a flood, which require a fast response. In 
addition, the quality of air and water can be examined by IoT-based 
devices. Moreover, humidity and temperature can be easily monitored \cite{ENVI}.\\
\noindent \textit{\textbf{7. Production and assembly line management}}: IoT-based smart 
systems allow rapid manufacturing of new products and an interactive response 
to demands by enabling communication between sensors and 
controlling/monitoring systems \cite{PROD}. Moreover, intelligent management 
approaches that use real-time measurements can also enable energy 
optimization and safety management. \\
\noindent \textit{\textbf{8. Food supply chain}}: The food supply chain model is fundamentally distributed and sophisticated. IoT can provide valuable information for 
managers of this chain. Although IoT is already in use within the supply 
management systems, its current benefits are limited. One of the most obvious 
and significant advantages of IoT in supply management is that it ensures 
security and safety of the products by utilizing IoT-based tracking 
\cite{FTRACK}. These devices can raise a warning in case of a security 
breach at any unauthorized level of the supply management system.

\subsection{Definition of security in the scope of IoT}
Next, we define two of the most commonly-used terms in the scope of IoT: a 
secure thing and a security attack. When defining what a secure thing is, it 
is important to understand the characteristics that define security. 
Traditionally, security requirements are broken down into three main 
categories: (i) confidentiality, (ii) integrity, and (iii) availability,
referred to as the CIA-triad. Confidentiality entails applying a set of rules 
to limit unauthorized access to certain information. It is crucial for IoT 
devices because they might handle critical personal information, e.g., medical 
records and prescription. For instance, an unauthorized access to personal 
health devices may reveal personal health information or even lead to 
life-threatening situations \cite{SUG_2}. Integrity is also necessary
for providing a reliable service. The device must ensure that the received 
commands and collected information are legitimate. An integrity compromise may 
lead to serious adverse consequences. For example, integrity attacks against 
medical devices, e.g., an insulin pump \cite{PrincetonInsulin} or a pacemaker 
\cite{Zero-Power}, may have life-threatening outcomes. IoT availability is 
essential for providing a fully-functioning Internet-connected environment. 
It ensures that devices are available for collecting data and prevents
service interruptions.
\begin{table*}[t] 
\caption{Security requirements} 
\centering 
\begin{tabular}{|l| l| c|} 
\hline\hline 
Requirement & Definition & Abbreviations\\ [0.5ex]
\hline 
Confidentiality & Ensuring that only authorized users access the & C \\[0.2ex]
& information  & \\[1ex]
\hline 
Integrity & Ensuring completeness, accuracy, and absence & I\\[0.2ex]
& of unauthorized data manipulation & \\[1ex]
\hline 
Availability & Ensuring that all system services are available, & A\\[0.2ex]
& when requested by an authorized user & \\[1ex]
\hline 
Accountability & An ability of a system to hold users responsible & AC\\[0.2ex] %
& for their actions & \\[1ex]
\hline 
Auditability  & An ability of a system to conduct persistent & AU\\[0.2ex]
& monitoring of all actions  & \\[1ex]
\hline 
Trustworthiness & An ability of a system to verify identity and  & TW\\[0.2ex]
& establish trust in  a third party & \\[1ex]
\hline 
Non-repudiation & An ability of a system to confirm occurrence/ & NR \\[0.2ex]
& non-occurrence of an action & \\[1ex]
\hline 
Privacy & Ensuring that the system obeys privacy policies  & P \\[0.2ex]
& and enabling users to control their data & \\[1ex]
\hline 
\end{tabular} 
\label{table:requirements}
\end{table*} 

The insufficiency of the CIA-triad in the context of security has been 
addressed before \cite{TRIAD, INCIA2,INCIA3}.  Cherdantseva et al.~\cite{TRIAD} show 
that the CIA-triad does not address new threats that emerge in a collaborative 
security environment. They provide a comprehensive list of 
security requirements by analyzing and examining a variety of information, 
assurance, and security literature. This list is called the IAS-octave and 
is proposed as an extension to CIA-triad. 
Table~\ref{table:requirements} summarizes the security requirements in the IAS-octave, and provides their definitions and abbreviations. We define:
\begin{itemize}
\item \textit{Secure thing}: A thing that meets all of the above-mentioned security requirements. 
\item \textit{Security attack}: An attack that threatens at least one of the above-mentioned security requirements. 
\end{itemize}

\subsection{Potential attackers and their motivations}
Next, we briefly discuss who the attackers that target the IoT might be, and 
what motivations they may have.

IoT-based systems may manage a huge amount of 
information and be used for services ranging from industrial management to 
health monitoring. This has made the IoT paradigm an interesting target for 
a multitude of attackers and adversaries, such as occasional hackers, 
cybercriminals, hacktivists, government, etc.

Potential attackers might be interested in stealing sensitive information, 
e.g., credit card numbers, location data, financial accounts' passwords, and 
health-related information, by hacking IoT devices. Moreover, they might try to 
compromise IoT components, e.g., edge nodes, to launch attacks against a 
third-party entity. Consider an intelligence agency that infects millions of 
IoT-based systems, e.g., remote monitoring systems, and smart devices, e.g., 
smart televisions. It can exploit the infected systems and devices to spy on 
a person of interest or to conduct an attack on a large scale. Also, 
hacktivists or those in opposition might be interested in attacking smart 
devices to launch protests against an organization. 

\section{WMS-based systems}
Aging population and rapidly-rising costs of health care have triggered 
a lot of interest in WMSs. Traditionally, 
in-hospital monitoring devices, such as electrocardiogram (ECG) and 
electroencephalogram (EEG) monitors, have been used to sense and store raw 
medical data, with processing being performed later on another machine, 
e.g., an external computer \cite{CNIA, MY_NEW_1}. Several trends in communication, 
signal processing, machine learning, and biomedical sensing have converged to 
advance continuous health monitoring from a distant vision to the verge of 
reality. Foremost among these trends is the development of Internet-connected 
WMSs, which can non-invasively sense, collect, and even process different 
types of body-related data, e.g., electrical, thermal, and optical signals 
generated by the human body. 

WMSs enable proactive prevention and remote detection of health issues, 
thus with the potential to significantly reduce health care costs 
\cite{AGING_1, AC_3}. Since the introduction of the first wearable heart 
monitor in 1981 \cite{FIRST_HR}, numerous types of WMS-based systems have 
been proposed, ranging from simple accelerometer-based activity monitors 
\cite{AC_1,AC_3} to complex sweat sensors \cite{SW_2}. 
WMS-based systems have also been developed for continuous long-term health 
monitoring \cite{CNIA,CONT2}. 

In the last decade, with the pervasive use of Internet-connected WMSs, the 
scope of applications of WMS-based systems has extended far beyond health 
care. For example, such systems have targeted application domains 
as diverse as education, information security, and human-computer interaction 
(HCI). Park et al.~\cite{SMART_KG} introduced a WMS-based teaching assistant 
system, called SmartKG. It collects, manages, and fuses data gathered by 
several wearable badges to prepare valuable feedback to the teacher. Barreto et al.~\cite{EMG_HEAD} designed and 
implemented a human-computer interface, which uses EEG and electromyogram (EMG) signals gathered from the subject's head to control computer cursor movements.

Despite the emergence of numerous WMS-based systems in recent years, potential 
challenges associated with their design, development, and implementation are 
neither well-studied nor well-recognized. The rest of this section: 
\begin{itemize}
\item provides a brief history of wearable computing devices and WMSs and 
discusses how their market is growing,
\item explains in depth the scope of applications of WMS-based systems,
\item describes the architecture of a typical WMS-based system and discusses
constituent components, and the limitations of these components,
\item suggests an inclusive list of desirable design goals and requirements 
that WMS-based systems should satisfy.
\end{itemize}

\subsection{History and market growth}
\label{HISTORY}
Wearable devices have a history that goes back longer than most people expect. 
The first truly wearable computer appeared in 1961, when Edward O. Thorpe and 
Claude Shannon created Roulette Predictor \cite{FIRST_WEAR}, a wearable 
computer that could be concealed in a shoe and accurately predict where the 
ball would land on a roulette circle. Integration of wearable sensors in 
wearable computing devices was done in 1981, when Polar Electro Company 
introduced the first wearable heart rate monitors for professional athletes 
\cite{FIRST_HR}. After that, several companies were founded to offer various 
services based on WMSs. However, due to the immaturity of the sensing 
technology, implementation complexities, e.g., heat management, limitations of 
wearable sensors, e.g., small local storage, and security/privacy concerns, 
the majority of such companies experienced a difficult time commercializing 
their products, and as a result, went through bankruptcy \cite{BANK_RUP}.

Rapid advances in communication protocols and the miniaturization of 
transceivers in 1990s, along with the emergence of different WMSs in the early 
2000s, transformed the market of wearable technologies. In the last decade, 
the rapidly-falling prices of WMSs and components used to implement WMS-based 
systems have changed the application landscape \cite{FALL_1,FALL_2,
GROWTH_0}. In addition, the rising market of personal smart devices, e.g., 
smartphones and tablets, that are powerful, ubiquitous, and less 
resource-limited relative to wearable sensors, has enabled a plethora of 
services, ranging from continuous health monitoring \cite{CNIA} to
continuous user authentication \cite{CABA}. The introduction of Apple's App 
Store for the iPhone and iPod Touch in July 2008, Google's Android Market 
(now called Google Play Store), and RIM's BlackBerry App World in 2009, 
enabled easy distribution of third-party applications, further boosting the 
WMS industry \cite{APPLE,GOOGLE}.

Global Wearable Sensors Market \cite{GWSMI} recently published a report that 
includes information from 2011 to 2016. This report indicates that the 
introduction of smart watches from companies like Samsung, Sony, and Nike 
has given a significant boost to the wearable technology market.  As of 2016, 
North America has the highest penetration of wearable sensors since
Americans tend to be early adopters of new technologies. However, 
Asia is expected to show the highest growth rate in a few years 
due to the presence of developing countries like India and China \cite{PENT}. 
A recently-published report provided by IHS Technology \cite{IHS} forecasts 
that the number of WMSs will rise by 7$\times$ from 67 million units in 2013 to 
466 million units in 2019. Another recent report by Business Insider 
\cite{CABA_INT} claims that 33 million wearable devices were sold in 2015 only 
for health monitoring. It forecasts that this number will reach 148 million by 
2019. In the years after that, such usage is expected to explode further.

\subsection{Scopes of applications}
\label{SCOPE}
\begin{figure*}
\centering
\includegraphics[trim = 5mm 130mm 7mm 0mm ,clip, width=430pt,height=200pt] {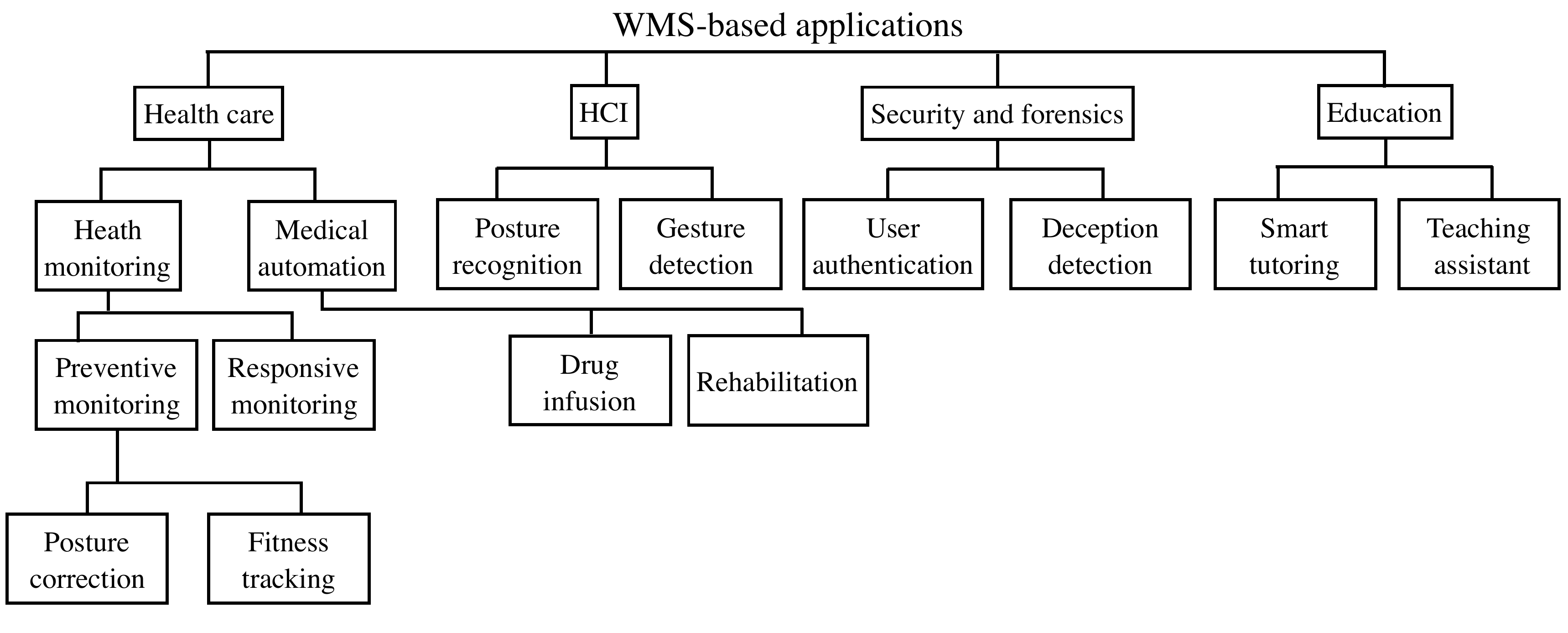}
\caption{The scope of applications of WMS-based systems}
\label{fig:APPS}
\end{figure*}

In this section, we describe various applications of WMS-based systems 
(a summary is shown in Fig.~\ref{fig:APPS}).

\subsubsection{Health care}
Rapid advances in WMS-based systems are transforming and revolutionizing 
health care. Medical WMS-based systems are of two main types: (i) health 
monitoring systems that monitor the patient to prevent the occurrence of a 
medical condition or detect a disease at an early stage, and (ii) medical 
automation systems, which offer continuous treatment or rehabilitation 
services. Next, we describe each type.

\noindent \textbf{\textit{1. Health monitoring systems}}: Prevention and early detection of medical conditions are essential for promoting wellness. Unfortunately, conventional clinical diagnostic 
practices commonly fail to detect health conditions in the early stages since 
diagnosis is typically performed after the emergence of major health
symptoms, and previous medical data on the patient are often very sketchy.
Furthermore, clinical practices are difficult to carry out in out-of-hospital 
environments.

In order to address the above-mentioned drawbacks of traditional clinical 
practices, several research studies have targeted WMS-based health monitoring 
systems. Such systems can be divided into two categories based on their 
main task: (i) preventive systems that aim to provide an approach to prevent 
diseases before the emergence of their symptoms, and (ii) responsive systems 
that attempt to detect health conditions at an early stage and provide health 
reports to the patient or the physician. Next, we describe each category.

\noindent \textbf {Preventive systems:} Preventive health monitoring systems 
provide real-time feedback to the user in an attempt to correct 
behaviors that might lead to adverse health conditions in the future. 
They promote healthy behaviors and lower the probability of serious illnesses 
by automatically detecting/predicting unhealthy activities and warning the 
user about them \cite{PREVENT_HEALTH}. Posture correctors and fitness trackers 
are two of the most widely-known types of preventive health monitoring systems.

\indent \textit {Posture corrector:} A poor posture results in muscle 
tightening, shortening, or weakening, causing several health conditions, 
e.g., back pain and spinal deformity \cite{BAD_POST}. Posture correctors
\cite{POST_1,POST_2,POST_3} monitor the user's movements and habits and offer 
real-time feedback upon the detection of any posture abnormality, e.g., 
slouching when sitting in front of a computer display. In fact, they help the 
user maintain a healthy posture while performing daily activities. \\
\indent \textit {Fitness tracker:} Such trackers are in widespread use and 
their market is rapidly growing. Although they may use different sensing 
technologies, they all have a common characteristic: they non-invasively 
measure some types of fitness-related parameters, e.g., calories burned, heart 
rate, number of steps taken \cite{FIT_APP}, and even sleep patterns 
\cite{SLEEP}. State-of-the-art fitness trackers play a significant role in the IoT paradigm by enabling object-to-object 
communication, transmission of user's data to the Cloud, and remote monitoring of user's activities \cite{FIT_IOT1}. For example, a fitness tracker, which can communicate with other objects, may be able to gather data from gymnasium equipment to support aspects of fitness progress awareness, such as shopping suggestions to support the user's fitness regime \cite{FIT_IOT2}. 

\noindent \textbf {Responsive systems:} Responsive health monitoring systems 
aim to detect medical conditions at an early stage by monitoring and analyzing 
various biomedical signals, e.g., heart rate, blood glucose, blood sugar, EEG, 
and ECG, over a long time period. For example, the CodeBlue project 
\cite{CODE_BLUE} examined the feasibility of using interconnected sensors for 
transmitting vital health signs to health care providers. Nia et al.~\cite{CNIA} 
proposed an extremely energy-efficient personal health monitoring system based 
on eight biomedical sensors: (1) heart rate, (2) blood pressure, (3) oxygen 
saturation, (4) body temperature, (5) blood glucose, (6) accelerometer, 
(7) ECG, and (8) EEG. MobiHealth \cite{MOBI_H} offered an 
end-to-end mobile health platform for continuous health care monitoring. 

\noindent \textbf{\textit{2. Medical automation systems}}: Unlike health monitoring systems, medical automation systems enhance 
the user's quality of life after/during the emergence of health issues. They mitigate health issues or minimize disease symptoms 
by actively providing essential therapy. Based on their functionality, medical 
automation systems can be divided into two main categories: drug infusion 
and rehabilitating systems. 

\noindent \textbf {Drug infusion systems:} Drug infusion systems enable safe 
injection of pharmaceutical compounds, e.g., nutrients and medications, into 
the body to achieve desired therapeutic effects. Automatic drug infusion 
systems control the drug release profile, absorption, and distribution to 
enhance the treatment efficacy and safety as well as patient convenience and 
compliance \cite{DRUG_INF}. Insulin delivery systems are one of the most 
commonly-used types of drug infusion systems. They continuously monitor the 
patient's blood glucose level using wearable glucose sensing patches and 
inject a prescribed amount of insulin into the blood stream when necessary.

\noindent \textbf {Rehabilitation systems:} Such systems have attracted
a lot of attention in the past two decades. They are currently used by 
patients after a major operation, sensory loss, stroke, severe accident, or 
brain injury \cite{REHAB_16}. They are also used to help patients who suffer 
from serious neurological conditions, e.g., Parkinson's disease or post-stroke 
condition \cite{GAIT_2}. Gait and/or motor abilities analysis is often
used in rehabilitation in hospitals and health care centers \cite{SURVEY_GAIT}.

An example of WMS-based rehabilitation system is Valedo \cite{VALEDO}, 
which is a medical back-training device developed by Hocoma AG to enhance 
patient compliance. It gathers trunk movements using two WMSs, transfers them 
to a game environment, and guides the patient through exercises targeted at
low back pain therapy. Another example is Stroke Rehabilitation Exerciser 
\cite{PHIL} developed by Philips Research, which coaches the patient through 
a sequence of exercises for motor retraining. Salarian et al.~\cite{SALARIAN} 
proposed a method for enhancing the gait of a patient with Parkinson's 
disease. Hester et al.~\cite{HESTER} proposed a WMS-based system to facilitate 
post-stroke rehabilitation. 

\subsubsection{HCI}
In our daily conversations, the existence of common contexts, i.e., implicit 
information that characterizes the situation of a person or place that is 
relevant to the conversation, helps us convey ideas to each other and react 
appropriately. Unfortunately, the ability to share context-dependent ideas does 
not transfer well to humans interacting with machines. The design of WMS-based 
human-computer interfaces has notably improved the richness of communications 
in HCI \cite{RICH}. In particular, various WMS-based gesture detection and 
emotion recognition systems have been proposed in the literature to enhance 
HCI. 

\noindent \textbf{\textit{1. Gesture detection systems}}: Several applications, such as sign-language recognition and remote control 
of electronic devices, need to respond to simple gestures made by humans. 
In the last decade, many WMS-based gesture recognition mechanisms have been 
developed to process sensory data collected by WMSs, e.g., magnetometer 
\cite{MAG_1,MAG_2}, accelerometer \cite{AC_1,AC_3}, and gyroscopes 
\cite{PINGU}, to recognize user's gesture and enable gesture-aware HCI. 

Although gestures from any part of the body can be used for interacting with 
a computing device, previous experimental research efforts \cite{RING_13} 
have demonstrated that finger-based gesture detection mechanisms are more 
successful in practice since their information entropy is much larger than 
that of interactions based on other human body parts. As a result, several 
research studies \cite{HAND1,HAND2,HAND3,PINGU} have focused on developing 
algorithms to detect hand gestures in real-time. A promising example of 
WMS-based gesture detection systems is Pingu \cite{PINGU}, a smart wearable 
ring that is capable of recognizing simple and tiny gestures from user's ring 
finger. 

\noindent \textbf{\textit{2. Emotion recognition systems}}:
Wearable technology was first used to detect emotions/feelings by Picard et 
al.~\cite{AFFECTIVE}. Since then, several researchers have used different sets 
of WMSs to detect different emotions/feelings, e.g., stress 
\cite{STRESS_1,STRESS_2}, depression \cite{DEPRESSION}, and happiness 
\cite{HAPPY}). However, we humans still cannot agree on how we define certain 
emotions, even though we are extremely good at expressing them. This fact has 
made emotion recognition a technically challenging field. However, it is becoming increasingly important in HCI studies as its 
advantages become more apparent. 

\subsubsection{Information Security and Forensics}
Next, we discuss two well-known types of WMS-based systems developed in the 
domain of information security and forensics for deception detection and 
authentication. 

\noindent \textbf{\textit{1. Deception detection systems}}:
The examination of the truthfulness of statements made by victims, suspects, 
and witnesses is of paramount importance in legal settings. Real-time 
WMS-based deception detection systems attempt to facilitate security screening 
and criminal investigation, and also augment human judgment \cite{MESER}. They
process sensory data collected by various types of WMSs (commonly heart rate, 
blood pressure, and accelerometers) to detect suspicious changes in the 
individual's mental state, e.g., a rapid increase in stress level,
behavior, e.g., involuntary facial movements, and physiological signals, e.g., 
an increase in the heart rate. For example, PokerMetrics \cite{POKER_METRICS} 
is a lie detection system that processes heart rate, skin conductance, 
temperature, and body movements to find out when the user is bluffing during 
a poker tournament. FNIRS-based polygraph \cite{FNIRS} is another 
fairly accurate lie detection system that processes data collected by a 
wearable near-infrared spectroscope.

\noindent \textbf{\textit{2. Authentication systems}}:
Authentication refers to the process of verifying a user's identity based on 
certain credentials \cite{AUTH_WMS}. A rapidly-growing body of literature on the 
usage of behaviometrics, i.e., measurable behavior such as frequency of 
keystrokes, and biometrics, i.e., strongly-reliable biological traits such as 
EEG signals, for authentication has emerged in the last two decades 
\cite{SU_1_0,SU_1_1,SU_1_2}.

Design of WMS-based authentication is an emerging research domain that is 
attracting increasing attention. Several research efforts have investigated 
the feasibility of using the data collected by WMSs as behaviometrics or 
biometrics. In particular, various research studies \cite{SMART_W1,SMART_W3} 
have demonstrated that the data collected by smart watches, e.g., 
acceleration, orientation, and magnetic field, can be used to distinguish 
users from each other. Furthermore, the use of EEG \cite{EEG_5} and 
ECG \cite{ECG_3} signals, as biomedical traits with high 
discriminatory power for authentication, has received widespread 
attention. Although EEG/ECG-based authentication systems have shown promising 
results, they have been unable to offer a convenient method for 
\textit{continuous user authentication} for two reasons. First, the 
size/position requirements of the sensors that capture EEG/ECG signals 
significantly limit their applicability \cite{ECG_3,EEG_CAP}. Second, 
processing of EEG/ECG signals for authentication is resource-hungry 
\cite{EEG_PRO}. A recently-proposed WMS-based authentication system, called 
CABA \cite{CABA}, has attempted to effectively address these drawbacks by 
using an ensemble of biomedical signals that can be continuously and 
non-invasively collected by WMSs.

\subsubsection{Education}
Next, we describe how technological advances in WMSs are transforming 
education by opening up new opportunities for employing smart tutoring and 
teaching assistant systems. 

\noindent  \textbf{\textit{1. Smart tutoring}}:
With the rapid development of online tutoring and exponential increase in 
the number of massive open online course websites, many research 
projects have been conducted on computer-based tutoring systems, which aim to 
select suitable instructional strategies based on the learner's reactions, 
mental conditions, emotional states, and feedback (see \cite{SU_LEARN} for a 
survey). Moreover, there is a strong motivation in the military community for 
designing adaptive computer-based tutoring systems to provide effective 
training in environments where human tutors are unavailable \cite{MIL2,MIL1}. 
WMS-based tutoring systems can recognize the user's emotional condition, level 
of understanding, physical state, and stress level by collecting and 
processing sensory data, e.g., user's heart rate and blood pressure. They can 
also predict learning outcomes, e.g., performance and skill acquisition, and 
continuously adapt their teaching/training approaches to optimize learning 
efficiency \cite{SU_LEARN}.

\noindent  \textbf{\textit{2. Teaching assistant}}: WMS-based teaching assistant systems can 
continuously collect and process various forms of biomedical signals from 
students, and analyze their voices, movements, and behaviors 
in order to reach a conclusion about the lecturer's quality of presentation 
and listeners' level of satisfaction. They can facilitate the teaching process 
by continuously assisting the lecturer in delivering and subsequently making 
the learning process shorter, more efficient, more pleasant, and even 
entertaining. For example, Grosshauser et al.~\cite{DANCE} have designed a 
WMS-based teaching assistant system that monitors movements of dancers and 
provides feedback to their teacher. Park et al.~\cite{SMART_KG} have designed 
SmartKG that relies on several wearable badges to provide valuable information 
about kindergarten students to their teacher.

\begin{figure*}
\centering
\includegraphics[trim = 5mm 100mm 70mm 0mm ,clip, width=430pt,height=230pt] {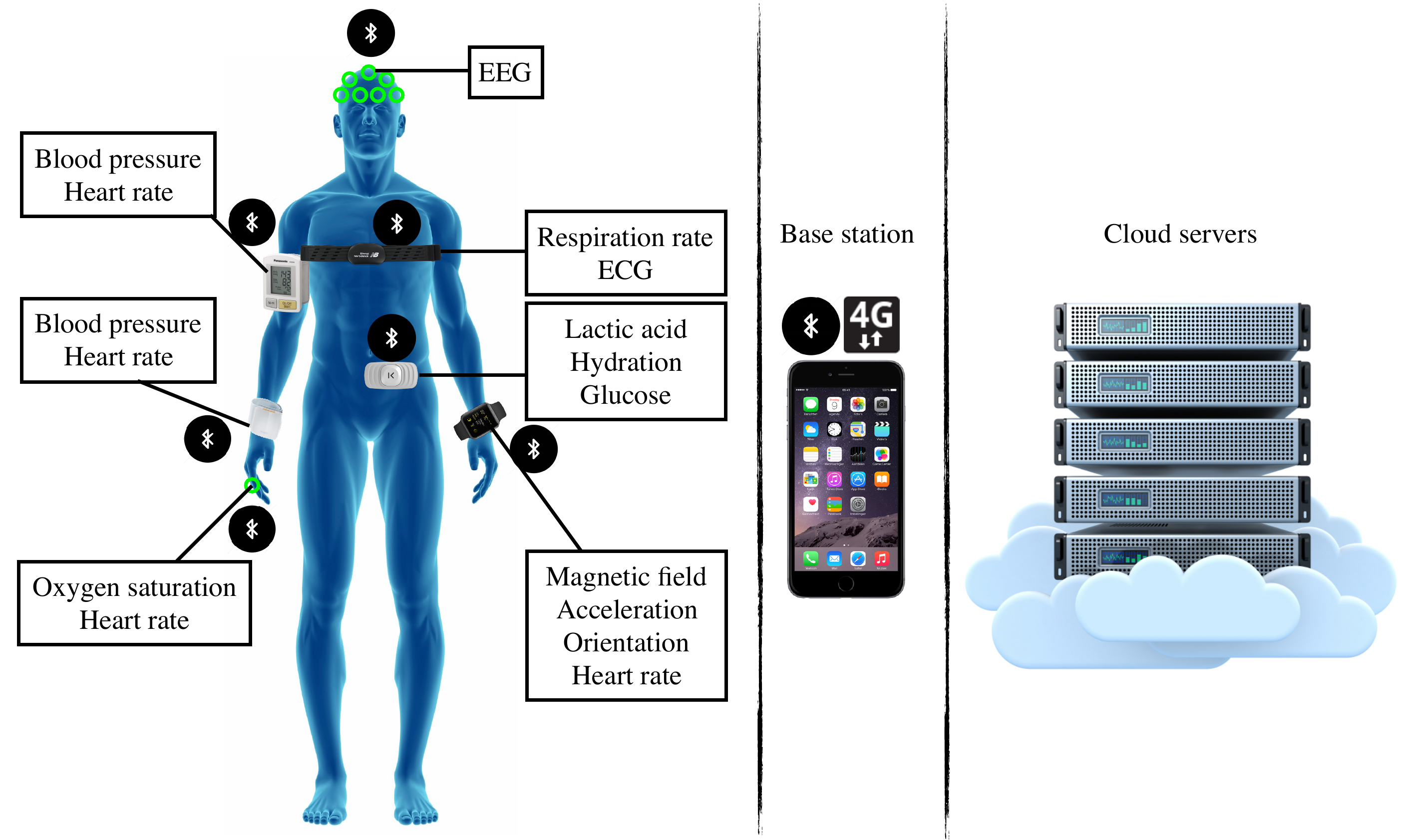}
\caption{The three main components of WMS-based systems: WMSs, the base station, and Cloud servers \cite{WMSSURVEYOURS}.}
\label{fig:ARCH}
\end{figure*}

\subsection{Main components of WMS-based systems}
\label{COMPONENTS}
In this section, we describe the components that constitute a typical 
WMS-based system, and their limitations. As shown in Fig.~\ref{fig:ARCH}, a 
typical WMS-based system consists of three main components: WMSs, the 
base station, and Cloud servers. Next, we describe each.

\subsubsection{WMSs}
With continuing performance and efficiency improvements in computing and 
real-time signal processing, the number and variety of WMSs have increased 
significantly, ranging from simple pedometers to sophisticated heart-rate 
monitors. WMSs sense electrical, thermal, chemical, and other signals from the user's body. The majority of these sensors, e.g., EEG and ECG, directly sense and collect biomedical 
signals. However, a few sensors, e.g., accelerometers, gather raw data that can be used to extract health-related information. Table \ref{table:sensorspec} lists various commonly-used WMSs in an alphabetical order, along with a short description for each sensor. 

\begin{table*}[ht] 
\caption{Common WMSs} 
\centering 
\begin{tabular}{|l| l|} 
\hline\hline 
Sensor & Description\\ [0.5ex]
\hline 
Accelerometer & measures changes in the acceleration of the device \\ [0.2ex]
& caused by user's movements\\ 
\hline
Blood pressure sensor & measures systolic and diastolic blood pressures\\
\hline
ECG sensor & measures the electrical activity of the heart\\
\hline
EEG sensor & measures the electrical activity of the brain\\
\hline
EMG sensor & records electrical activity produced by muscles\\
\hline
Glucometer & measures approximate blood glucose concentration\\
\hline
GSR sensor & measures continuous variation in the electrical\\
& characteristics of the skin\\
\hline
Gyroscope & measures changes in device orientation caused by  \\
& user's movements\\
\hline
Heart rate sensor & counts the number of heart contractions per minute\\ 
\hline
Magnetometer & specifies user's direction by examining the changes  \\
& in the earth's magnetic field around the user\\
\hline
Microphone & records acoustic sounds generated by the body \\
& (used for respiration analysis or emotion detection)\\
\hline
Near-infrared spectroscope & provides neuroimaging technology to examine an \\
& aspect of brain function\\
\hline
Oximeter & measures the ratio of oxygen-saturated hemoglobin \\
& to the total hemoglobin count in the blood \\ 
\hline
Pedometer & counts each step a person takes by detecting the \\
&  motion of the person's hands or hips\\
\hline
Respiration rate sensor & counts how many times the chest rises in a minute\\
\hline
Strain sensor & measures strain on different body parts \\
& (used to detect when the user is slouching)\\
\hline
Thermometer & measures an individual's body temperature \\ 
\hline 
\end{tabular} 
\label{table:sensorspec} 
\end{table*}

Despite the variety of WMSs available, they share two common limitations that 
must be considered while designing a WMS-based system: small storage 
capacity and limited energy. 

\noindent \textit{\textbf{1. Small storage}}: Storing a large amount of data in a WMS is 
not feasible for two reasons. First, adding a large storage to a WMS 
dramatically increases its energy consumption, and as a result, significantly 
decreases its battery lifetime \cite{CNIA}. Second, the size constraints of 
WMSs impose specific storage constraints. The WMS size needs to be kept small 
to ensure user convenience.

\noindent \textit{\textbf{2. Limited energy}}: The small on-sensor battery with limited 
energy capacity is one of the most significant factors that limits the volume 
of data sampled and transmitted by WMSs.  It is still feasible to wirelessly 
transmit all raw data without performing any on-sensor processing if devices 
are charged regularly, e.g., on an hourly basis. However, forcing the user to 
frequently recharge the WMSs would impose severe inconvenience. As described 
later in Section \ref{ON_SENSOR}, on-sensor processing may significantly 
preserve battery lifetime by extracting salient information from the data and transmitting it.

The above-mentioned limitations of WMS-based systems have three direct 
consequences. First, the data generated by WMSs cannot be stored on them for a 
long period of time and should be transmitted to other devices/servers. 
Second, only extremely resource-efficient algorithms can be implemented on 
WMSs. Third, WMSs cannot usually support traditional cryptographic mechanisms, 
e.g., encryption, and are vulnerable to several security attacks, e.g., 
eavesdropping. 

\subsubsection{Base station}
Due to limited on-sensor resources (small storage and limited energy), the 
sensory data are frequently sent to an external device with more computation 
power. This device is referred to as the \textit{base station}. It may range 
from smartphones to specialized computing devices, known as central 
hubs \cite{CNIA}. They commonly have large data storage, and powerful network 
connectivity through cellular, IEEE 802.11 wireless, and Bluetooth interfaces, 
and powerful processors \cite{RESOR}. Smartphones have become the dominant 
form of base stations since they are ubiquitous and powerful and provide all 
the technologies needed for numerous applications \cite{SMART_P1}. Moreover, 
smartphones support highly-secure encrypted transmission, which deters
several potential attacks against the system \cite{CABA}.

The base station has its own resource constraints, though much less severe, in 
terms of storage and battery lifetime.  Continuous processing along with 
wireless transmission to the Cloud may drain the base station's battery 
within a few hours, and as a result, cause user inconvenience. Base stations 
typically perform lightweight signal processing on the raw data and 
re-transmit a fraction or a compressed form of data to Cloud servers for 
further analysis and long-term storage.

\subsubsection {Cloud servers}
Since both WMSs and base stations are resource-constrained, sensory data are 
commonly sent to Cloud servers for resource-hungry processing and long-term 
storage. Depending on the wireless technology used, the data can be sent 
either directly or indirectly (through a base station, such as a smartphone) 
to the Cloud. In addition to the huge storage capacity and high computational 
power that Cloud servers can provide for WMS-based applications, 
they facilitate access to shared resources in a pervasive 
manner, offering an ever-increasing number of online on-demand services. 
Furthermore, Cloud-based systems support remote update of software, without 
requiring that the patient install any software on the personal devices, thus 
making system maintenance quick and cost-effective. This makes Cloud-based 
systems a promising vehicle for bringing health care services to rural 
areas \cite{RURAL}.

Despite the promise of Cloud servers in this context, utilizing them in 
WMS-based systems has two drawbacks. First, Cloud-based systems are highly 
dependent on the reliability of the Internet connection. Outage of Internet 
service may have serious consequences. For example, unavailability of a 
seizure prediction system (that tries to detect the occurrence of a seizure a 
few seconds before the patient's body starts shaking) may lead to a
life-threatening situation. Second, the use of Cloud servers increases 
the response time (the time required to collect sensory data, process 
them, and provide a response or decision). As a result, there may be a 
significant deterioration of the quality of service in real-time applications.

\subsection {Design goals}
\label{GOALS}
\begin{figure}[ht]
\centering
\includegraphics[trim = 88mm 48mm 83mm 42mm ,clip, width=250pt,height=240pt]{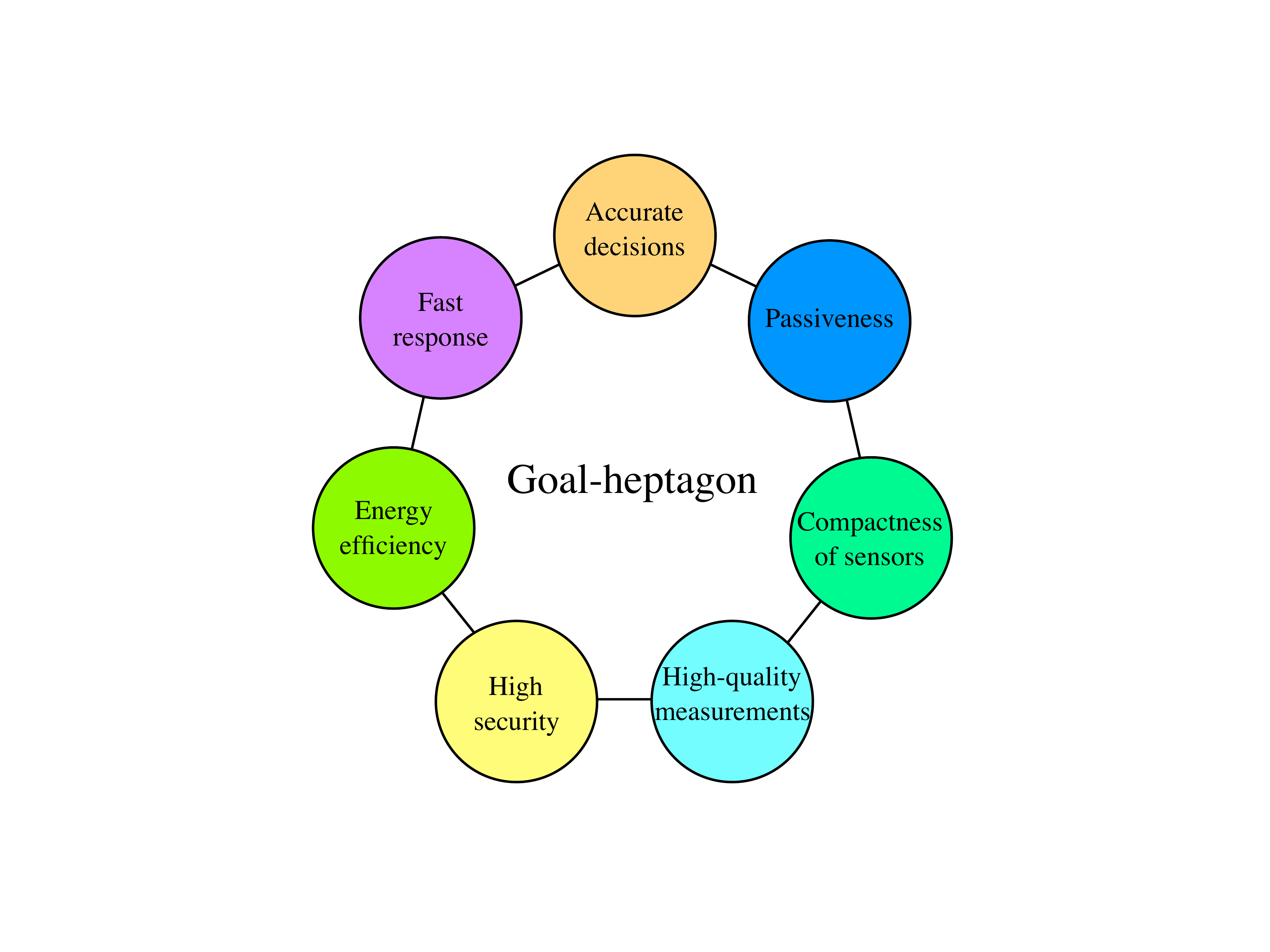}
\caption{Goal-heptagon: Desiderata for WMS-based systems \cite{WMSSURVEYOURS}.} 
\label{fig:DESIGN}
\end{figure}

Although the scope of applications of WMS-based systems is quite wide, they 
share several common design goals. We reviewed many recent research studies on 
the design and development of different types of WMS-based systems and 
realized that, unfortunately, there is no standard inclusive list of desirable 
goals in the literature. In this section, we suggest such a list. 
Fig.~\ref{fig:DESIGN} summarizes seven general design goals that should be 
considered in designing WMS-based systems. Next, we present the
rationale behind each goal.\\
\noindent \textit{\textbf{1. Accurate decisions}}: WMS-based systems process the input 
data, e.g., an EEG signal, and return decisions as output, e.g., whether a 
seizure is occurring or not. The quality of the service provided by a 
WMS-based system depends on the accuracy of decisions made by the system. For 
instance, a WMS-based authentication system must confidently determine if the 
user is authorized to use restricted resources, or a posture corrector must 
accurately decide whether the user's posture is healthy.\\
\noindent \textit{\textbf{2. Fast response}}: A short response time is a desirable 
design goal for the majority of systems. In order to ensure user convenience, 
it is obviously desirable for the system to quickly respond to user requests. 
Moreover, a short response time is essential for an authentication system, in 
which the system must quickly authenticate a legitimate user and reject an 
impostor \cite{CABA}. Furthermore, a long response time may endanger user 
safety in some scenarios. For example, if an insulin pump fails to immediately detect 
an emergency, e.g., hyperglycemia or hypoglycemia, and provide a response when it is necessary, the patient might suffer from 
life-threatening conditions \cite{INSU_DELAY}.\\
\noindent \textit{\textbf{3. Energy efficiency}}: The battery used in a WMS
is typically the greatest contributor to both size and weight. As a result, WMSs
typically have very limited on-sensor energy \cite{B_SIZE}. Rapid depletion of
battery charge, necessitating frequent, e.g., on a hourly basis, battery
replacement/recharge would deter wide adoption of the device \cite{PREVENT}.
Thus, all components embedded in WMSs and the signal processing algorithms
implemented on them must be energy-efficient.\\
\noindent \textit{\textbf{4. High security}}: The emergence of the IoT paradigm has
magnified the negative impact of security attacks on sensor-based systems.
Furthermore, the demonstration of several attacks in recent research efforts
(see \cite{IOT_SURVEY} for a survey) has led to serious security concerns and
highlighted the importance of considering security requirements. To ensure
system security, different security requirements must be proactively addressed. As mentioned earlier, security requirements are often broken down into three main categories: (i) confidentiality, (ii) integrity, and (iii) availability, referred to as the CIA-triad \cite{TRIAD}. \\
\noindent \textit{\textbf{5. High-quality measurements}}: Undoubtedly, the quality of 
the decisions offered by a WMS-based system depends on the quality of 
sensory measurements provided by WMSs. It has been shown that user's activities may negatively impact the quality of data obtained by the WMSs, e.g., running significantly deteriorates the quality of the signal 
collected by EEG sensors \cite{RUN}. Hence, WMSs should be designed 
to provide accurate and noise-robust measurements during different 
daily activities, especially intensely physical ones. \\
\noindent \textit{\textbf{6. Compactness of sensors}}: To ensure user convenience, WMSs
must be kept lightweight and as small as possible. Moreover, in many scenarios,
the presence of specific WMSs, e.g., blood glucose monitor, may reveal the
presence of an illness along with the current level of the illness, leading to
serious privacy concerns \cite{STIGMA}. Thus, WMSs should be designed
to be compact so that the user can easily hide them.\\
\noindent \textit{\textbf{7. Passiveness}}: Passiveness, i.e., minimal user
involvement, is a key consideration in designing a WMS-based system. It is 
very desirable that WMSs be calibrated transparently to the user and sensory 
data be measured independently of user activities \cite{CALIB_INV}. Obviously, 
if a wearable device, e.g., a smart watch, asks the user to calibrate internal 
sensors, e.g., accelerometers and magnetometers, manually, it may be quite 
annoying to the user \cite{ACCEL}.

\input{ch-intro/intro_contributions}

%% file: ch-intro/intro_contributions.tex
\section {Contributions of the thesis}
To mitigate the security/privacy issues in the IoT domain while considering domain-specific limitations (e.g., limited energy and small storage capacity) of various components utilized in IoT-based systems, this thesis provides low-energy solutions that make data encryption practical for resource-constrained sensors (e.g., WMSs). Furthermore, in order to shed light on new domain-specific security/privacy issues associated with IoT-based systems, two novel security attacks are introduced in this thesis. Moreover, a novel IoT-enabled continuous authentication system is presented, which aims to address the security issues and weaknesses of previously-proposed authentication systems. Our main contributions can be summarized as follows:

\begin{enumerate}
\item The thesis first targets health monitoring systems, one of the most widely-known types of IoT-based systems that are envisioned as key to enabling a holistic approach to health care. It describes different solutions for reducing the total energy consumption of different implantable and wearable medical devices (IWMDs) utilized in continuous health monitoring systems. The proposed solutions can significantly increase the battery lifetimes of different IWMDs while offering spare energy for encrypting medical data before transmitting them. 

\item The thesis introduces OpSecure, an optical secure communication channel between an implantable medical device (IMD) and an external device, e.g., a smartphone. OpSecure enables an intrinsically user-perceptible unidirectional data transmission, suitable for physically-secure communication with minimal size and energy overheads. Based on OpSecure, we design and implement two protocols: (i) a low-power wakeup protocol that is resilient against remote battery draining attacks, and (ii) a secure key exchange protocol to share the encryption key between the IMD and the external device. The proposed protocols complement lightweight symmetric encryption mechanisms, so that common security attacks against insecure communication channels can be prevented.

\item The thesis shows how security/privacy attacks against health monitoring systems extend far beyond wireless communication to/from IWMDs, and explains why encryption cannot always provide a comprehensive solution for eliminating security/privacy attacks on IWMDs. In particular, it describes the possibility of privacy attacks that target physiological information leakage, i.e., signals that continuously emanate from the human body due to the normal functioning of its organs. Furthermore, it discusses several novel attacks on privacy by leveraging information leaked from them during their normal operation.

\item The thesis then introduces a generic security attack that is applicable to a variety of cyber-physical systems (CPSs), i.e., systems with integrated physical and processing capabilities that can interact with humans and the environment. These attacks are called dedicated intelligent security attacks against sensor-triggered emergency responses (DISASTER). DISASTER targets safety mechanisms deployed in automation/monitoring CPSs and exploits design flaws and security weaknesses of such mechanisms to trigger emergency responses even in the absence of a real emergency. In addition to introducing DISASTER, it describes the serious consequences of such attacks, and demonstrates the feasibility of launching DISASTER against the two most widely-used sensor-based systems: residential and industrial automation/monitoring systems. Moreover, it provides several countermeasures that can potentially prevent DISASTER and discusses their advantages and drawbacks.

\item Finally, the thesis presents continuous authentication based on biological aura (CABA), a novel user-transparent system for continuous authentication based on information that is already gathered by WMSs for diagnostic and therapeutic purposes. The presented continuous authentication system can offer a promising alternative to one-time knowledge-based authentication systems (e.g., password-/pattern-based authentication systems) and potentially be used to protect personal computing devices and servers, software applications, and restricted physical spaces.
\end{enumerate}

\section{Thesis outline}
The rest of this thesis is organized as follows. Chapter \ref{ch:relatedwork} discusses related work. Chapter \ref{ch:hmonitoring} quantifies the energy and storage requirements of continuous personal health monitoring systems and presents several schemes to reduce the overheads of wirelessly transmitting, storing, and encrypting/authenticating the data. Chapter \ref{ch:OpSecure} discusses OpSecure and two protocols that can be used in conjunction with symmetric encryption to protect the wireless channel between the IMD and an external device from different security attacks. Chapter \ref{ch:physio} describes the concept of physiological information leakage and how such leakage can be exploited by attackers even if the communication channels are encrypted. Chapter \ref{ch:DISASTER} introduces DISASTER and describes its consequences. Moreover, it suggests several countermeasures to mitigate such attacks. Chapter \ref{ch:CABA} presents CABA, describes its various applications, and discusses how it can be extended to user identification and adaptive access control authorization. Chapter \ref{ch:conclusion} concludes the thesis and presents ideas for future research. 

%% file: ch-relatedwork/chapter-relatedwork.tex
\chapter{Related Work\label{ch:relatedwork}}
In this chapter, we discuss related work and provide background for several key concepts used in this thesis. In Section \ref{IOT_COUNT_CH2}, we summarize different attacks and threats on the edge-side layer of IoT along with their countermeasures. In Section \ref{TOPICS}, we describe several research directions related to the domain of WMSs.

\input{ch-relatedwork/IOT_SURVEY_PART2}

\input{ch-relatedwork/topic1}

%% file: ch-relatedwork/IOT_SURVEY_PART2.tex
\section{Vulnerabilities of IoT and their countermeasures}
\label{IOT_COUNT_CH2}
In this section, we discuss different attacks and threats on the edge-side layer of IoT and describe possible countermeasures against them.
\subsection{Known vulnerabilities}
\begin{figure*}
\centering
\includegraphics[scale=0.85, trim = 80mm 140mm 80mm 10mm] {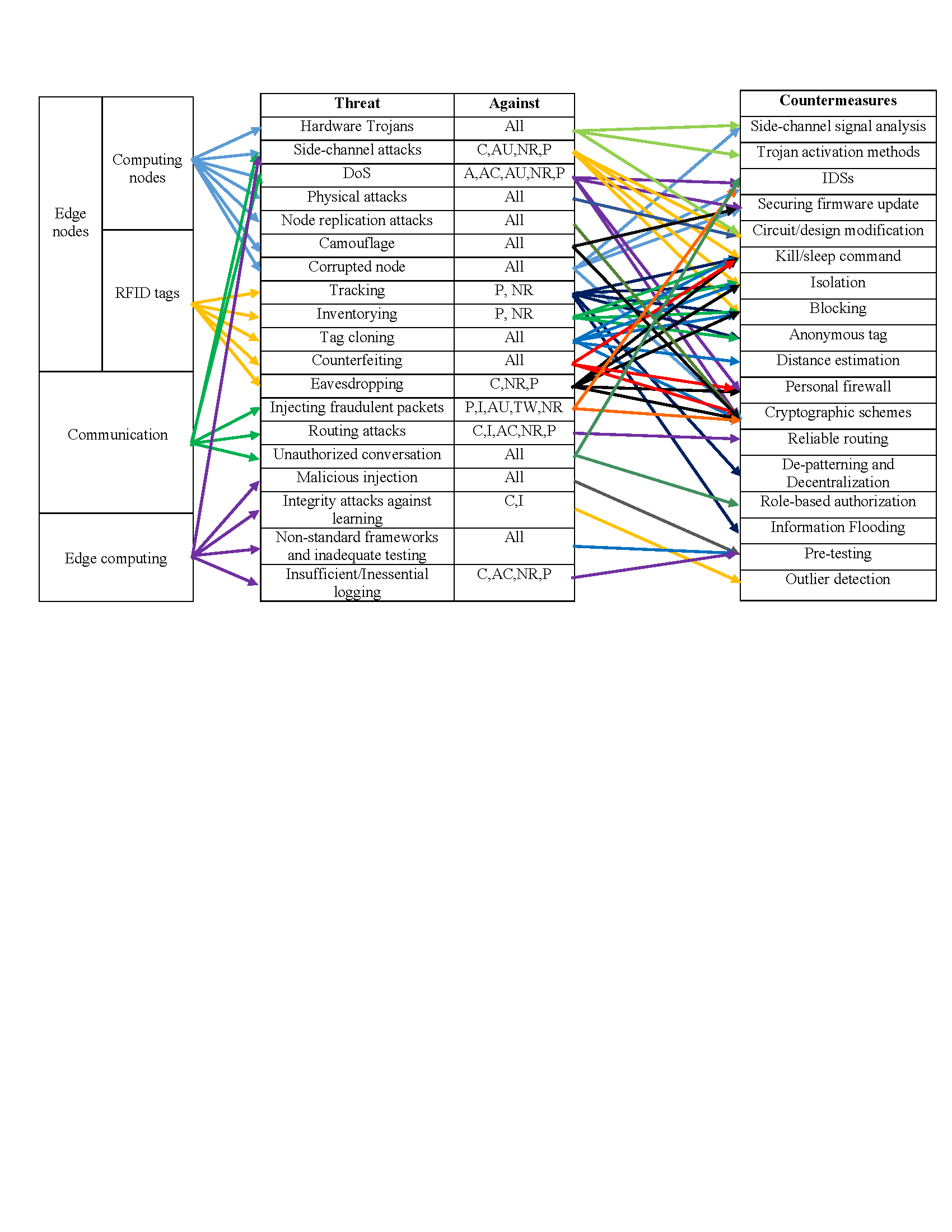}
\caption{Summary of attacks and countermeasures \cite{IOT_SURVEY}}
\label{fig:SUM}
\end{figure*}

Next, we provide an in-depth analysis of possible attacks and vulnerabilities at each level of the edge-side layer (edge nodes, communication, and edge computing). Fig.~\ref{fig:SUM} summarizes several attacks and their countermeasures that are discussed in this section. We describe the left side of this figure (attacks) first. The security requirements and their abbreviations that are used in
Fig.~\ref{fig:SUM} are given in Table \ref{table:requirements}.

\subsubsection{Edge nodes}
Next, we discuss various attacks against the first level of the 
reference model that includes computing nodes and RFID tags. 

\noindent \textit{\textbf{1. Computing nodes}}:
We begin with attacks against the edge computing nodes, e.g., RFID 
readers, sensor nodes, and compact controlling nodes.\\
\noindent \textbf{Hardware Trojan:} Hardware Trojans have emerged as a
major security concern for integrated circuits
\cite{HT1,HT2,HT3,HT4,HT5}. Hardware Trojan is a malicious modification
of an integrated circuit, which enables the attacker to use the circuit
or to exploit its functionality to obtain access to data or software
running on the integrated circuits (ICs) \cite{KOOSHANFAR}. In order to insert a hardware Trojan in the original circuitry, the attacker maliciously alters the design before/during fabrication and specifies a triggering mechanism that activates the malicious behavior of the Trojan \cite{HT1}. Trojans are generally divided into two categories based on their triggering mechanisms \cite{KOOSHANFAR,HT5}: (i) externally-activated Trojans, which can be triggered by an antenna or a sensor that can interact with the outside world, and (ii) internally-activated Trojans that are activated after a certain condition is met inside the integrated circuit, e.g., a Trojan that wakes up after a specific timespan when it receives a triggering signal from a countdown circuitry added by the attacker. \\
\noindent \textbf{Non-network side-channel attacks}: Each node may reveal critical 
information under normal operation, even when not using any wireless 
communication to transmit data. For example, the electromagnetic (EM) 
signature, i.e., the EM 
waves emitted by the node, can provide valuable information about the status 
of the device. The declassification of TEMPEST documents \cite{NSA} in 2007, 
and the recent publications of some EM-based attacks \cite{EM1,EM2,PHYSIO} have 
started to develop the idea of non-network side-channel threats. For example, 
in a recent work, researchers were able to demonstrate 
how the acoustic/EM signals leaked from a medical device can provide valuable 
information about the patient or the device \cite{PHYSIO}. As mentioned in 
that work, detection of the existence of known signals or protocols may 
endanger the safety of the user, e.g., if the user has a device that is very 
expensive. Moreover, this type of attack may lead to a serious privacy issue 
in medical systems. For example, consider a subject who wears a medical device 
indicating a certain medical condition that carries a social stigma. Detecting 
the presence of this device can embarrass the patient. In addition, specific 
side-channel information from the devices may provide significant information 
about the individual's health condition, e.g., glucose level, blood
pressure, etc.\\
\noindent \textbf{Denial of Service (DoS) attacks:} There are three well-known types of DoS attacks against edge computing nodes: battery draining, sleep deprivation, and outage attacks.\\
\indent \textit{1. Battery draining}: Due to size constraints, nodes usually 
have to carry small batteries with very limited energy capacity. This has made 
battery-draining attacks a very powerful attack that may indirectly 
lead to serious consequences, such as a node outage or a failure to report
an emergency. For example, if an attacker can find a way to deplete the 
battery of a smoke detector, he will be able to disable the fire 
detection system \cite{BD5}. Such attacks could destroy a network if 
recharging the nodes is difficult \cite{BD6}. An example of a battery-draining 
attack is when an attacker sends tons of random packets to a node and
forces the node to run its checking mechanisms, e.g., authentication 
mechanism. Several battery-draining attacks have been 
discussed in the literature \cite{BD1,BD2,BD3,BD4}. \\
\indent \textit{2. Sleep deprivation}: Sleep deprivation is a specific type of 
DoS attack in which the victim is a battery-powered node with a limited energy 
capacity. In this type of attack, the attacker attempts to send an undesired 
set of requests that seem to be legitimate. Therefore, detection of this type 
of attack is much harder than that of a simple battery-draining attack. The 
idea of sleep deprivation was first described by Stajano \cite{SD1}. The 
research effort by Martin et al. is one of the first publications to closely 
examine the impact of sleep deprivation attacks on energy-constrained 
devices \cite{BD1}.\\
\indent \textit{3. Outage attacks}: Edge node outage occurs 
when an edge device stops performing its normal operation \cite{MY_NEW_5}. In some cases,
a set of devices or an administrator device may stop functioning. Outage may 
be a result of an unintended error in the manufacturing 
process, battery draining, sleep deprivation, code injection or unauthorized 
physical access to the node. One of the most famous examples of outage attacks 
is injecting Stuxnet \cite{ST1} into Iran's nuclear process control program. 
Stuxnet manipulates the industrial process control sensor signals such that 
the infected system loses its ability to detect abnormal behavior. Therefore, 
the system does not shut down even in an emergency situation \cite{ST1,ST2}. \\
\noindent \textbf{Physical attacks/tampering}: Edge devices operate in hostile 
environments in which physical access to the devices may be possible, thus
making them highly vulnerable to hardware/software 
attacks. The attacker, with a physical access to the device, may extract 
valuable cryptographic information, tamper with the circuit, modify 
programming, or change the operating system \cite{PHY1,PHY2,PHY3,PHY4, NEST}. Physical attacks against the edge nodes may 
cause permanent destruction. Therefore, their main purpose is to extract 
information for future use, e.g., find the fixed shared key. Such a 
well-known recent attack was on the Nest thermostat \cite{NEST}, in which the 
attacker replaces the default firmware with a malicious one. This 
attack enables the attacker to control the thermostat, even when he no longer 
has physical access to the device.\\
\noindent \textbf{Node replication attacks}: In such an attack, the attacker 
adds a new node, e.g., a malicious one, to an existing set of nodes by 
replicating one node's identification number. This attack can lead to a 
significant reduction in network performance. Moreover, the attacker can 
easily corrupt or misdirect packets that arrive at the replica \cite{PARNO}. 
This attack usually causes severe damage to the system by enabling the attacker 
to obtain required access to extract cryptographic shared keys \cite{NSURV2}.  
Moreover, node replicas may revoke authorized nodes by executing 
node-revocation protocols \cite {PARNO, CHAN}.\\
\noindent \textbf{Camouflage}: In this type of attack, the attacker inserts a 
counterfeit edge node or attacks an authorized node in order to hide at the 
edge level. Afterwards, the modified/counterfeit node can operate as a normal 
node to obtain, process, send, or redirect packets \cite{NSURV1,NSURV2}. 
Moreover, such a node can function in a passive mode in which it only conducts 
traffic analysis.

\noindent \textbf{Corrupted/malicious node}: The main goal of 
corrupting nodes is to gain unauthorized access to the network they
belong to. Malicious nodes injected into a network can obtain access to other 
nodes, possibly controlling the network on behalf of the attacker 
\cite{NSURV2}. A malicious node can also be used by the attacker to inject 
false data into the system or prevent delivery of true messages \cite{NSURV1}.

\noindent \textbf{\textit{2. RFID tags}}: Next, we discuss the attacks against RFID tags.\\ 
\noindent \textbf{Tracking}: Covert reading of RFID tags is a significant threat. 
Unfortunately, almost all such tags provide a unique identifier. As a result, 
a nearby unauthorized reader can easily and effectively read a tag that is 
attached to a product or an individual. Such a reading provides very strong 
tracking information \cite{BLOCKER}. In the simplest form of the attack, an 
attacker uses a large number of RFID readers to read these fixed identifiers. 
The threat grows and becomes more important when a tag identifier is combined 
with personal information, e.g., credit/loyalty card number and personal 
profile \cite{WEIS}.

\noindent \textbf{Inventorying}: There are certain types of tags that carry 
valuable information about the products they are attached to. In 
particular, electronic product code (EPC) tags have two custom fields: a 
manufacturer code and a product code. As a result, an individual who has
an EPC tag is subject to inventorying \cite{RFIDSURVEY}, i.e., a tag reader can examine what products the individual has. This threat leads to 
serious privacy concerns. For example, the attacker might recognize what types 
of medical device, e.g., an insulin pump, a patient is wearing and, 
therefore, what illnesses, e.g., diabetes, he suffers from. 

\noindent \textbf{Physical attacks/tampering}: This type of attack can
be launched when the attacker has full physical access to a tag. In this 
attack, the tags can be physically manipulated and modified in a laboratory 
setup \cite{RFIDSUR}. There are several known physical attacks against RFIDs. 
Among them are probe attacks, material removal, circuit manipulation, and clock 
glitching \cite{PHYSICAL}. These attacks are used for extracting information from the tag, or modifying the tag for counterfeiting. 

\noindent \textbf{Tag cloning}: Tag cloning (also referred to as spoofing) and 
impersonation of RFID tags could be very profitable to hackers, and 
extremely dangerous for the company's reputation. Potential damage 
can be amplified through a high level of automation \cite{CLONING}. An 
attacker may use tag cloning to access restricted areas, bank accounts, or 
sensitive information.

\noindent \textbf{Counterfeiting}: In counterfeiting, the attacker modifies the 
identity of an item, typically by means of tag manipulation. Generally, the 
attacker needs less information to launch counterfeiting attacks
relative to spoofing attacks. In these attacks, a tag is partially 
manipulated. Westhues describes how an RF tape-recorder can be
constructed to read commercial proximity cards and partially 
simulate their signals to bypass building security systems \cite{WEST}.

\noindent \textbf{DoS attacks}: In DoS attacks, the RF 
channels are jammed such that the tags cannot be read by the tag readers and, 
as a result, the intended services based on the RFID tags become unavailable. 
For example, an attacker can lock down a whole building by jamming all the 
RFID-based doors.  Additional vulnerabilities of RFID authentication protocols 
to DoS attack have been discussed in \cite{DOSRFID}.

\noindent \textbf{Eavesdropping}: In this attack, the main goal of the 
attacker is to intercept, read, and save messages for future analysis. The 
intercepted data can be used as an input to other attacks, such as tag cloning. 
The concept of eavesdropping attacks against RFIDs is not new and is 
frequently mentioned in the literature. Recent reports by the National 
Institute of Standards and Technology \cite{NIST} and the Department of 
Homeland Security \cite{DHS}, in addition to several published surveys, 
e.g. \cite{RFIDSURVEY,RFIDSUR,EAV4}, all mention risks of eavesdropping in the RFID environment. In particular, several practical attack scenarios and their experimental setups have been discussed in \cite{EAV4}.

\noindent \textbf{Side-channel attacks}: Such attacks 
use state-of-the-art tools to intercept and process communications in order to extract information from various patterns, even when the messages are encrypted. For example, if an attacker 
reads the tags at the entrance of a building, he can guess 
the number of individuals in the building at any moment by counting the 
number of communications. Over-the-air timing attacks against RFID tags and 
their efficacy are open research problems \cite{RFIDSURVEY}. Carluccio et al. have 
described the use of EM emanations to launch
a power-analysis attack against RFIDs \cite{SIDE1}.

\subsubsection{Communication}
Next, attacks against the communication level of the IoT reference model are 
discussed. \\
\noindent \textbf{Eavesdropping}: At the communication level, eavesdropping 
(also called sniffing) refers to intentionally listening to private 
conversations over the communication links \cite{EVE_1}. It can provide invaluable 
information to the attacker when the data are unencrypted. In 
this situation, usernames and/or passwords are often easy to extract. When 
packets also carry access control information, such as node configuration, 
shared network password, and node identifiers, eavesdropping can provide 
critical information. The attacker can use and process this captured 
information to design other tailored attacks. For example, if an attacker can 
successfully extract the information that is required to add a new node to the 
set of authorized nodes, he will easily be able to add a malicious node to the 
system.

\noindent \textbf{Side-channel attacks}: Although side-channel attacks are not
easy-to-implement, they are powerful attacks against encryption. They
pose a serious threat to the security and reliability of cryptographic 
implementations. As mentioned earlier, side-channel attacks can also be 
launched at the edge node level. In contrast to the attack at the edge
node level, 
the side-channel attacks at the communication level are usually non-invasive. 
They only extract information that is often unintentionally leaked. 
Typical examples of unintentional information leakage are time between two 
consecutive packets, frequency band of communications, and communication 
modulation. An important characteristic of non-invasive attacks is that they 
are undetectable, and as a result, there is no easy defense against them 
except to minimize leakage or else add noise to the leaked information.

\noindent \textbf{DoS attacks}: The most common and well-known DoS attack at
the communication level is a standard attack that jams the transmission of 
radio signals. Two types of active jamming attacks have been defined in 
the literature \cite{JAM1,JAM2}: (i) continuous jamming that involves complete 
jamming of all transmissions, and (ii) intermittent (also called 
non-continuous) jamming in which jamming is periodic and, as a result, 
the nodes can send/receive packets periodically. While the goal of constant 
jamming is to block all transmissions, with intermittent jamming, 
the attacker intends to lower the performance of time-sensitive systems. 
Consider a fire detection system that can detect unusual changes in the level 
of gases in the environment and calls the fire department in case of an 
emergency. An attacker can easily make the system unreliable by
intermittently jamming node-to-node and node-to-base transmissions. In this 
scenario, the system will become out-of-service if the attacker uses constant 
jamming. Several research efforts have examined the possibility and 
effectiveness of launching DoS attacks against various transmission protocols, 
including Bluetooth \cite{BLE_Jam}, ZigBee \cite{ZIGBEEJ}, and 6LowPan 
\cite{LOWPAN}.  In addition to active jamming attacks, the attacker can launch 
DoS against communication using malicious nodes or routers. The attacker may 
insert a node/router that intentionally violates the communication protocol in 
order to generate collisions or jam the communications \cite{NSURV2}. A 
malicious router/node may also refuse to route messages or attempt to 
misdirect them. This could be done intermittently or constantly. Constant DoS 
attacks are usually easy to detect, whereas detection of intermittent ones 
requires accurate and efficient monitors. 

\noindent \textbf{Injecting fraudulent packets}: An attacker can 
inject fraudulent packets into communication links using three different attack 
methods: (i) insertion, (ii) manipulation, and (iii) replication (also called 
replay) \cite{NSURV2}. In insertion scenarios, the attacker inserts new 
packets in network communication. In other words, an insertion attack
has the ability to generate and send malicious packets that seem legitimate. 
Manipulation attacks involve capturing the packet, and then modifying, 
e.g., updating header information, checksum, and data, and sending the 
manipulated packet. In replication attacks, the 
attacker captures the packets that have been previously exchanged between two 
things in order to replay the same packets. Generally, a stateless system, 
which does not keep track of previous packets or previous state of the system, 
is quite vulnerable to replay attacks.

\noindent \textbf{Routing attacks}: Attacks that affect how messages are routed 
are called routing attacks. An attacker may use such attacks to spoof, 
redirect, misdirect, or drop the packets at the communication level. The 
simplest type of routing attack is an altering attack in which the attacker 
changes the routing information, e.g., by generating routing loops or false 
error messages. In addition to altering attacks, several other serious 
attacks have been proposed, e.g., Black Hole \cite{BHOLE,GHOLE}, Gray Hole 
\cite{GHOLE}, Worm Hole \cite{WHOLE}, Hello Flood \cite{HFLOOD,HFLOOD2}, and 
Sybil \cite{SYBIL}. We briefly describe them next.

\noindent \textit{1. Black Hole}: A Black Hole attack is launched by using a  malicious node, which attracts all the traffic in the network by advertising that it has the shortest path to the destination in the network. As a result, all packets are sent to the malicious node, and the attacker can process the packets or simply drop them. 

\noindent \textit{2. Gray Hole}: A Gray Hole attack is a variation of Black Hole attack in which the nodes selectively drop some packets.

\noindent \textit{3. Worm Hole}: A Worm Hole attack is a severe attack that can be launched 
even when authenticity and confidentiality are guaranteed in all
communications. In this attack, an attacker first records packets at one 
location in the network and then tunnels them to a different location.

\noindent \textit{4. Hello Flood}: A Hello Flood attack is based on the fact that a node must 
broadcast ``HELLO PACKETS" to show its presence to neighbors. The receiving 
nodes may assume that they are within the communication range of the sender. 
In this attack, an attacker uses a malicious node with high transmission power 
to send ``HELLO PACKETS" to every other node in the  network and claim
to be their neighbor. 

\noindent \textit{5. Sybil}: In a Sybil attack, the attacker adds/uses Sybil nodes, which are 
nodes with fake identities. Sybil nodes can out-vote honest nodes in the 
system.

\noindent \textbf{Unauthorized conversation}: Every edge node needs to 
communicate with other nodes in order to share data or access their data. 
However, each node should only talk to a subset of nodes that need its data. 
This is an essential requirement for every IoT system, in particular, ones
consisting of both insecure and secure nodes. For example, in 
a smart home scenario, the thermostat requires the smoke detector's data in 
order to shut down the heating system in an emergency situation. However, if 
the insecure smoke detector can share (get) information with (from) every 
other node, an attacker might be able to control the whole home automation 
system by hacking the smoker detector.

\subsubsection{Edge computing level}
Edge (fog) computing is an emerging technology. Thus, its
vulnerabilities have not yet been adequately explored. 
The few research efforts that address attacks on edge 
computing mainly focus on possible threats to sensor networks 
\cite{FOGS1,FOGS2, MY_NEW_6}. Next, we discuss and suggest some attack 
scenarios against an edge computing based scheme. Although some of these 
attacks were designed to target conventional systems and 
networks, they are also applicable to the edge computing based systems. 

\noindent \textbf{Malicious injection}: Insufficient validation of the input 
may enable malicious input injection. An attacker could inject a malicious 
input that causes the service providers to perform operations on behalf of the 
attacker. For example, an attacker may add an unauthorized component to one of 
the levels below (communication or edge node levels) that is capable of 
injecting malicious inputs into the servers. Afterwards, the attacker might be 
able to steal data, compromise database integrity, or bypass authentication. 
Standard database error messages returned by a database may also assist the 
attacker.  In situations where the attacker has no knowledge of the database's 
tables, forcing an exception may reveal more details about each table and the 
names of its fields \cite{SQL}.

\noindent \textbf{Integrity attacks against machine learning}: Two types of 
attacks can be launched against machine learning methods that are used
in IoT systems: causative and 
exploratory \cite{POS0}. In causative attacks, the attacker changes the 
training process by manipulating the training dataset, whereas in exploratory 
attacks, he exploits vulnerabilities without altering the training process. 
Recent research has introduced a new type of causative attack, called the 
poisoning attack \cite{PODEF,POS2,POS3}. In a poisoning attack, the attacker 
adds precisely-selected invalid data points to the training dataset. In an edge 
computing based system, an attacker might be able to launch this attack 
against the learning algorithm by directly accessing the server or computing 
nodes, or he might be able to add malicious data to the dataset by adding 
a sufficient number of malicious nodes to lower levels of the IoT model. The 
main motivation is to cause the classification algorithm to deviate from 
learning a valid model by manipulating the dataset.

\noindent \textbf{Side-channel attacks}: Earlier, we mentioned several types 
of side-channel attacks against the components at the edge node 
and communication levels. In addition, an attacker might use 
the information leaked from additional components, e.g., service providers and 
servers, to launch side-channel attacks. For example, a service, which 
generates verbose fault warnings, provides a useful tool for designers and 
developers. However, the same warnings can provide extravagant information in 
operational environments.

\noindent \textbf{Non-standard frameworks and inadequate testing}: Non-standard 
coding flaws can give rise to serious privacy and security concerns. Moreover, 
since the nodes typically need to connect to intermediate servers, the 
consequences of a compromise might be amplified. The development of an edge 
computing based system is a sophisticated process because it requires combining heterogeneous resources and devices that are often made by different manufacturers \cite{MOBILE}. In addition, there is neither a 
generally-accepted framework for the implementation of edge computing based 
systems nor a standard set of policies. As a result, several privacy and 
security flaws of these systems may remain undetected.

\noindent \textbf{Insufficient/inessential logging}: Logging is a nice 
approach for detecting an intrusion or a hacking attempt. Developers should 
log events such as successful/unsuccessful authentication attempts, 
successful/unsuccessful authorization attempts, and application errors. 
The edge computing based systems may be damaged as a result of insufficient 
logging \cite{LOGGING}. It is also recommended that the log files be encrypted.

\subsection{Existing countermeasures}
Here, the right side of Fig.~\ref{fig:SUM} that consists of several 
countermeasures is discussed. Next, we describe each defense in a 
level-by-level fashion. 

\subsubsection{Solutions for security issues in edge nodes}

First, we describe countermeasures for addressing attacks against the edge 
nodes.

\noindent \textbf{\textit{1. Computing nodes}}:
We start with solutions for attacks against computing nodes. \\
\noindent \textbf{Side-channel analysis:}
Side-channel signal analysis provides an effective approach for the detection of both hardware Trojans and malicious firmware/software installed on a device.\\
\indent \textit{1. Trojan detection:} Side-channel signals, including
timing \cite{KOOSHANFAR,TIMING_1,TIMING_2}, power
\cite{POWER_1,POWER_2,POWER_3}, and spatial temperature
\cite{POWER_1,THERM_1} can be used for Trojan detection. The presence of
a Trojan in a circuit commonly affects power and/or delay
characteristics of wires and gates in the circuit, and alters heat
distribution on the silicon IC. In order to detect a hardware Trojan,
side-channel signal-based Trojan detection mechanisms compare physical
characteristics and/or the heat distribution map of a suspicious IC to the
ones of a Trojan-free reference IC. Power-based analyses offer an
activity monitoring method that can be utilized to detect suspicious
activities within the IC, enabling detection of Trojans. Timing-based
methods enable the detection of Trojans by testing the IC using
efficient delay tests, which are sensitive to small changes in the
circuit delay along the affected paths and can differentiate Trojans
from process variations. Spatial temperature-based mechanisms rely on
infrared imaging techniques, which provide thermal maps of ICs. Silicon is transparent in the infrared spectral region and this transparency enables us to obtain maps of thermal infrared emissions using infrared imaging techniques \cite{THERM_1}.\\
\indent \textit{2. Malicious firmware/software detection:} The
effectiveness of side-channel signal analysis in detecting malicious
firmware/software installed on a device has been shown by several
previous research efforts \cite{MAL_1,MAL_2,MAL_3, HOS_MAL_1, HOS_MAL_2}. As mentioned earlier
in Section IV, side-channel signals can reveal valuable information
about the device's operation. Similar to the Trojan detection mechanism,
malware detection methods can process side-channel signals to detect
abnormal behaviors of the device, e.g., a significant increase in its power 
consumption, which are the results of a malware installed on the device.  \\
\noindent \textbf{Trojan activation:} Trojan activation strategies aim
to partially/fully activate the Trojan circuitry to facilitate Trojan
detection. Several Trojan activation approaches have been proposed in the 
last decade \cite{KOOSHANFAR,TRJA_2,TRJA_1}. The common goal of such
strategies is to magnify and detect the disparity between the behavior,
outputs, or side-channel leakages of a Trojan-free circuit and the ones
of a Trojan-inserted circuit. For example, Chakraborty et al. proposed MERO \cite{MERO}, an efficient methodology to derive a compact set of test patterns (minimizing test time and cost), while maximizing the Trojan detection coverage. MERO can increase the detection sensitivity of many side-channel Trojan detection. The basic concept is to detect low probability conditions at the internal nodes, select candidate Trojans triggerable by a subset of these rare conditions, and then derive an optimal set of vectors that can trigger each of the selected low probability nodes.\\
\noindent \textbf{Policy-based mechanisms and intrusion detection systems
(IDSs)}: Policy-based approaches are promising mechanisms for solving security 
and privacy problems at this IoT level. Violation of essential policies can 
be detected continuously by introducing an IDS \cite{NSURV2}. An IDS ensures that general rules are not broken. It provides a reliable approach to defend against battery-draining and sleep deprivation attacks by detecting unusual requests to the node. Several recent and ongoing research efforts provide efficient IDS designs for monitoring the edge nodes and detecting 
potential threats \cite{POL1,POL2,POL3,POL4,POL5}. 

\noindent \textbf{Circuit modification:} Changing the circuit is one 
of the most effective defenses against physical, side-channel, and Trojan attacks. In the following, for each of these attacks, we briefly discuss how specific circuit changes and modifications may address/prevent the attack. \\
\indent \textit{1. Tamper proofing and self-destruction:}  Nodes may be
integrated with physical hardware that enhances protection against
physical attacks. For example, to protect against tampering of sensors,
several mechanical/electrical tamper-proofing methods for designing the
physical packages of the nodes have been proposed and have traditionally
been used in home automation sensors, e.g., smoke detectors. Moreover, using self-destruction mechanisms provides an alternative approach to defend against physical attacks \cite{PROOF}. \\
\indent \textit{2. Minimizing information leakage:} There are also some well-known approaches for 
addressing side-channel attacks including, but not limited to, adding 
randomized delay \cite{SIDE1} or intentionally-generated noise 
\cite{SIDE_FIN}, balancing Hamming weights \cite {SIDE2}, using constant 
execution path code \cite{SIDE2}, improving the cache architecture 
\cite{SIDE3}, and shielding \cite{PHYSIO}.\\
\indent \textit{3. Integrating physically unclonable function (PUF) into the 
circuity:} A PUF is a noisy function embedded into an integrated circuit 
\cite{PUFS0}. When queried with a challenge $x$, a PUF generates a response 
$y$ that depends on both $x$ and the unique intrinsic physical properties of 
the device \cite{PUFS1,PUFS2}. PUFs are assumed to be physically unclonable, 
unpredictable, and tamper-evident. PUFs enable unique device identification 
and authentication \cite{PUFS1,PUFS3}, and offer Trojan detection mechanisms 
\cite{KOOSHANFAR}. Any unintended modification of the circuit physical layout 
changes the circuit parasitic parameters that can be detected by Trojan 
detection methods.\\
\noindent \textbf{Securing firmware update}: Each firmware update can be 
launched remotely or directly. In the case of remote firmware update, the base or server broadcasts a command (CMD) to announce that there 
is a new version of firmware available. Then, a node with the new firmware 
broadcasts an advertisement (ADV) to neighboring nodes. The nodes that are 
willing to update their firmware and have also received ADV compare the new 
version with their existing version, and send requests (REQ) if they need an 
update. Eventually, the advertiser starts sending data to the requesters. 
Providing a secure method for remotely updating the firmware requires 
authentication of CMD, ADV, REQ, and data packets. Moreover, risks 
posed by DoS attacks during each step of the protocol should be considered 
\cite{CMD}. In addition to remote firmware updates, some nodes support direct 
updates of the firmware, e.g., using a USB cable. In this case, the integrity 
of the firmware should be checked, and the user, who tries to update the 
firmware, should be authenticated, because a lack of sufficient integrity 
check mechanisms may enable an attacker to replace legitimate device firmware 
with a malicious one \cite{NEST}.

\noindent \textbf{\textit{2. RFID tags}}:
Next, we describe solutions and suggestions for addressing attacks against RFID tags. \\
\noindent \textbf{Kill/sleep command}: A kill scenario is built into the 
manufacturing process of RFID tags. An RFID tag has a unique PIN, e.g., a 
32-bit password. Upon receiving the correct PIN from the RFID reader, the tag 
can be killed, i.e., the tag will not be able to transmit any further 
information after receiving this command \cite{RFIDSUR}. There is an alternative 
approach called a sleep command that puts the tags to sleep, i.e., makes them 
inactive for a period of time \cite{RFIDSURVEY}. Although these ideas seem simple at 
first glance, designing and implementing secure and effective PIN
management schemes need sophisticated techniques.\\
\noindent \textbf{Isolation}: A very effective way of protecting the privacy of 
tags is to isolate them from all EM waves. One way is to build and use 
isolation rooms. However, building such rooms is usually very expensive. An 
alternative approach is to use an isolation container that is usually made of 
a metal mesh \cite{RFIDSURVEY}. This container, which can block EM waves of certain 
frequencies, is called a Faraday cage \cite{FARA}. Another approach to 
is to jam all nearby radio channels using an active RF 
jammer which continuously interrupts specific RF channels.  \\
\noindent \textbf{Blocking}: Juels et al. proposed a protection scheme called 
blocking \cite{BLOCKER}. It adds a modifiable bit to the tag that is called 
a privacy bit. A `0'  privacy bit indicates that public scanning is allowed 
for the tag, whereas a `1'  bit marks the tag as private. This scheme requires 
a certain type of tag (called blocker tag), which is a special RFID tag that 
prevents unintended scanning. However, the idea of using blocker tags has two 
main limitations: (i) it requires the use of a modified version of RFID tags, 
and (ii) unreliable transmission of the tags may easily lead to privacy 
failure even when the blocking scheme is implemented. Another blocking 
approach, called soft blocking, has been proposed in \cite{SOFT}. 
It relies on auditing of reader configurations to enforce a set of policies 
that is defined in software. This set guarantees that readers can only read 
public tags. Then, a monitoring device can passively examine if a reader is 
violating tag policies.\\
\noindent \textbf{Anonymous tag}: A novel idea based on look-up table mapping 
has been proposed by Kinoshita \cite{KINO}. The key contribution of his work 
is a scheme to store a mapping between an anonymous ID and a real ID of each 
tag in such a way that a attacker cannot find the mapping algorithm 
to recover the real ID from the anonymous one. The mapping may represent a key 
encryption algorithm or a random value mapped to the real ID. Note that 
although the anonymous ID emitted by an RFID tag has no intrinsic valuable 
information, it can still enable tracking as long as the ID is fixed over time 
\cite{RFIDSUR}. In order to address the tracking problem, the anonymous ID should 
be re-issued frequently. \\
\noindent \textbf{Distance estimation}: Use of signal-to-noise ratio 
as a metric to determine the distance between a reader and a tag is proposed 
in \cite{RFIDSURVEY}. For the first time, Fishkin et al. claim that it is 
possible to derive a metric to estimate the distance of a reader that tries to 
read the tag information. This enables the tag to only provide distance-based 
information. For example, the tag might release general information, e.g., the 
product type, when scanned at 10 meters distance, but release its unique 
identifier at less than 1 meter distance.\\
\noindent \textbf{Personal firewall}: A personal RFID firewall \cite{FIRE} 
examines all readers' requests to tags. The firewall can be assumed to be 
implemented in a device that supports high computation needs and provides 
enough storage capacity, e.g., a cellphone. The firewall enables the setting 
of sophisticated policies. For example, ``my tag should not release my 
personal information when I am not within 50 meters of my work place".\\
\noindent \textbf{Crypthographic schemes}: Three types of cryptographic schemes are widely discussed in the previous literature to address the security attacks against RFID tags:\\
\indent \textit{1. Encryption:} Full encryption usually requires significant
hardware. Therefore, its implementation in RFID tags has not been feasible 
due to the need for the tags to be low-cost (a few cents). Feldhofer 
\cite{FELDHOFER} proposed an authentication mechanism based on the Advanced 
Encryption Standard (AES). However, for a standard implementation of AES, 
20-30K gates are typically needed \cite{PERIS}, whereas RFID tags can only 
store hundreds of bits and support 5-10K logic gates. The limitations 
arising from gate count and cost suggest that the tag can only devote 
250-3500 gates to the security mechanism. The traditional implementation of 
AES was not appropriate until Jung et al. proposed a novel implementation of 
AES that requires only 3595 logical gates \cite{AESRFIDFELD}. However, no 
fully-developed version of AES has been implemented in any RFID tag.\\
\indent \textit{2. Hash-based schemes:} Such schemes are 
widely used for addressing security issues in the RFID technology.
Recent research on hash functions can be found in \cite{HASH1,HASH2,HASH3,HASH4,HASH5}. A simple security mechanism based on hash functions is proposed in \cite{WEIS}. In this work, two states are defined for each tag: (i) locked state in which a tag responds to all 
queries with its hashed key, and (ii) unlocked state in which the tag
carries out its normal operation. To unlock a tag, the reader sends a request, 
including the hashed key, to a back-end database and waits to get the key. 
After getting the key, the reader sends the key to the locked tag. Then,
the tag changes its state to unlocked. Although this significantly improves 
RFID security, the problem of tracking still remains. To address this issue, 
Weis et al. \cite{WEIS} propose a more sophisticated scheme, in which the 
hashed key is changed in a manner that is unpredictable.\\
\indent \textit{3. Lightweight cryptographic protocols:} In order to address 
the security and privacy issues of RFID tags by taking into account their cost 
requirements, several lightweight cryptographic protocols have been suggested. 
For example, Peris et al. propose a minimalist lightweight mutual 
authentication protocol for low-cost RFID tags \cite{PERIS}. They claim that 
their method provides an adequate security level for certain applications, and 
can be implemented with only slightly more than 300 gates, which is 
quite acceptable even for the most limited RFID tags. Moreover, a simple 
scheme for mutual authentication between tags and readers is proposed by 
Molnar et al. \cite{MOLNAR}. Their protocol uses a shared secret and a 
pseudorandom function to protect the messages exchanged between the tag and 
the reader. Another example is extremely-lightweight challenge-response 
authentication protocols described in \cite{AUTH1}. These protocols can be 
used in authenticating tags, but can be broken by a powerful adversary 
\cite{RFIDSUR}.\\
\noindent \textbf{Circuit modification:} In addition to the
previously-mentioned applications of PUF (device
identification/authentication and hardware Trojan detection), several
research efforts have proposed different anti-counterfeiting mechanisms
to prevent RFID tag cloning by integrating PUFs into RFID tags
\cite{PUFTAG1,PUFTAG2,PUFTAG3,PUFTAG4}. For example, consider an authentication 
mechanism that aims to identify the user based on his RFID tag. It can 
generate a set of challenge-response pairs for each tag during the enrollment 
phase and store it in a database. At a later point in time, during 
verification, it can compare the response provided by the user's PUF-based 
RFID tag for a chosen challenge from the database with the corresponding 
response in the database \cite{PUFTAG4,PUFTAG3}.

\subsubsection{Solutions for security issues in communication}
Next, we discuss solutions for addressing the security issues that 
exist at the communication level of the reference model. 

\noindent \textbf{Reliable routing}: An essential characteristic of IoT 
networks that complicates implementation of secure routing protocols is that intermediate 
nodes or servers might require direct access to message content before 
forwarding it. As mentioned earlier, several valid attacks against routing 
have been proposed in the literature. Karlof et al. have addressed most 
major attack scenarios \cite{RO0}.  They provide the first detailed security 
analysis of major routing protocols and practical attacks against them, along 
with countermeasures. Various other research efforts have also tried to 
address security and privacy concerns in routing \cite{RO1,RO2,RO3,RO4}.

\noindent \textbf{IDS}: IDS is essentially needed at the communication level as 
a second line of defense to monitor network operations and communication links, 
and raise an alert in case of any anomaly, e.g., when a pre-defined policy is 
ignored. Traditional IDS approaches \cite{TIDS1,TIDS2,POL4} are usually 
customized for WSNs or for the traditional Internet. 
However, few recent IDS proposals address the security and privacy concerns 
of IoT directly. SVELTE \cite{SVELTE} is one of the 
first IDSs designed to meet the requirements of the IPv6-connected 
nodes of IoT. It is capable of detecting routing attacks, such as spoofed or 
altered information, and Black Hole attack. Another intrusion detection method 
for the IoT has been proposed in \cite{IDS2}.

\noindent \textbf{Cryptographic schemes}: Using cryptographic schemes, e.g., 
strong encryption, to secure communication protocols is one of the most effective defenses against a variety of attacks, including eavesdropping and simple routing attacks, at the communication level. Several 
encryption methods have been proposed to address security issues in 
communication \cite{AESBOOK,ENC2}. The encryption-decryption techniques, 
developed for traditional wired networks, are not directly applicable to most 
IoT components, in particular, to small battery-powered edge nodes. Edge nodes 
are usually tiny sensors that have limited battery capacity, processing
power, and memory. Using encryption increases memory usage, energy 
consumption, delay, and packet loss \cite{LIGHT}. Variants of AES have 
yielded promising results for providing secure communication in IoT.  
Moreover, different lightweight encryption methods have been proposed,
e.g., CLEFIA \cite{CLEFIA} and PRESENT \cite{PRESENT}. 
Unfortunately, at this time, there are no promising public key encryption 
methods that provide enough security while meeting lightweight requirements 
\cite{LIGHT}.

\noindent \textbf{De-patterning and decentralization}: De-patterning and 
decentralization are two of the major methods proposed to 
provide anonymity and defense against side-channel attacks. There is always a 
trade-off between anonymity and the need to share information. De-patterning 
data transmissions can protect the system against side-channel attacks, e.g., 
traffic analysis, by inserting fake packets that can significantly alter the 
traffic pattern, when required. An alternative method for ensuring anonymity 
is distribution of sensitive data through a spanning tree such that no node 
has a complete view of the original data. This method is called 
decentralization \cite{DEP}. 

\noindent \textbf{Role-based authorization}: In order to prevent a
respone to requests by intruders or malicious nodes in the system, a 
role-based authorization system verifies if a component, e.g., edge node, 
service provider, or router, can access, share, or modify the information. 
Moreover, for every communication, the authorization system should check 
whether the two parties involved in the action have been validated and have 
required authority \cite{RBAUTH}.

\noindent \textbf{Information flooding}: Ozturk et al. propose flooding based 
anti-traffic analysis mechanisms to prevent an external attacker from tracking 
the location of a data source, since that information may release the location 
of things \cite{OZTURK}. They have proposed three different approaches to 
flooding: (i) baseline, (ii) probabilistic, and (iii) phantom. In baseline 
flooding, every node in the network forwards a packet once and only once. In 
probabilistic flooding, only a subset of nodes within the entire network 
contributes to data forwarding and the others discard the messages they 
receive. In phantom flooding, when the source sends a message, the message 
unicasts in a random fashion (referred to as a random walk phase). Then, the 
message is flooded using the baseline flooding technique (referred to as 
the flooding phase).

\subsubsection{Solutions for security issues at the edge computing level}
Next, we describe countermeasures and solutions for addressing the 
security attacks and issues at the edge computing level. 

\noindent \textbf{Pre-testing}: Testing of updates and design implementations
is important before they can be used in a critical system 
\cite{MICROSOFT}. The behavior of the whole system and its components, 
e.g., routers, edge nodes, servers, etc., should be closely examined by 
feeding different inputs to the system and monitoring the outputs. In 
particular, pre-testing attempts to identify the set of possible attack 
scenarios and simulate these scenarios to see how the system responds 
\cite{SAT}. It also specifies what information should be logged and what 
information is too sensitive to be stored. In addition, the input files should 
be closely examined to prevent the danger of malicious injection. For example, 
the attacker should not be able to execute any command by injecting it into 
the input files.

\noindent \textbf{Outlier detection}: The common goal of almost all defenses 
against integrity attacks on machine learning methods is to reduce the 
influence of adding invalid data points to the result. 
These invalid data points are deemed outliers in the training set. 
Rubinstein et al. have designed a defense framework against poisoning attacks 
based on robust statistics to alleviate the effect of poisoning \cite{RUBIN}. 
In addition, a bagging defense against such integrity attacks has been 
proposed by Biggio et al. \cite{BAGGING}. They examine the effectiveness of 
using bagging, i.e., a machine learning method that generates multiple 
versions of a predictor and utilizes them to get an aggregated predictor by 
getting averages over the versions or using a plurality vote \cite{BAGDEF},
in reducing the influence of outlying observations on training 
data. Mozaffari-Kermani et al. have presented several countermeasures against 
poisoning attacks in the area of healthcare \cite{MEHRAN}. They have evaluated 
the effectiveness of their schemes and identified the machine learning 
algorithms that are easiest to defend.

\noindent \textbf{IDS}: IDSs can detect the existence 
of a malicious node that tries to inject invalid information
into the system or violate the policies. Several recent research efforts have 
proposed IDS based methods to address the injection issue 
\cite{DIG,IDSINJ1,IDSINJ2,IDSINJ3}. For example, Son et al. describe the 
design and implementation of DIGLOSSIA \cite{DIG}, a new tool that precisely 
and efficiently detects code injection attacks on servers.

%% file: ch-relatedwork/topic1.tex
\section{Emerging research directions in the domain of WMS-based systems}
\label{TOPICS}
As described in Chapter \ref{ch:intro}, with the pervasive use of Internet-connected WMSs, the scope of applications of WMS-based systems has extended far beyond what has been
traditionally imagined. In this section, we describe several research directions that are closely related to the domain of WMSs and discuss how previous research studies have 
attempted to facilitate the design and development of WMS-based systems. In particular, we briefly discuss the related research work present at the intersection of WMS and the following research areas: (i) design of low-power sensors, (ii) minimally-invasive capture methods, (iii) security and privacy, (iv) calibration and noise cancellation, and (v) big data.

\subsection{Design of low-power sensors}
The on-sensor energy has three major consumers: (i) sampling, 
(ii) transmission, and (iii) on-sensor computation \cite{CNIA}. Thus, for each 
WMS, the energy consumption of one or a combination of these consumers should 
be reduced to enhance its battery lifetime. Next, we summarize what solutions 
previous studies have proposed to reduce the energy consumption of each of 
these energy consumers.

\subsubsection{Sampling}
The sampling energy is mainly the energy consumed by the analog-to-digital 
converter (ADC). The total energy consumption of an ADC can be divided into: 
(i) I/O energy, (ii) reference energy, (iii) sample-and-hold energy, (iv) ADC 
core energy, and (v) input energy \cite{ADCOR}. However, separate 
calculations of these values is difficult. Hence, the total on-chip ADC 
energy consumption per sample (EADC) is commonly reported in the literature. 
In order to enhance ADC energy efficiency, several architectures have been 
proposed in the last two decades, including but not limited to, 
asynchronous \cite{ASY}, cyclic \cite{CYC}, and delta-sigma \cite{DELSIG} 
(see \cite{ADCOR} for a survey). As extensively discussed in \cite{CNIA}, with 
recent advances in the design and development of ADCs, the sampling energy 
consumption of a WMS has become negligible in comparison to its total energy 
consumption. 

\subsubsection{Transmission protocols}
A key consideration in the design of a WMS is the communication technology 
(radio and protocol) used to connect the WMSs with the base station. Several 
transmission protocols have been implemented on low-power wireless chipsets 
to enable energy-efficient data transmission. These protocols include, but are 
not limited to, ANT/ANT+ \cite{ANTBASICS}, ZigBee \cite{ZIGBEE}, Bluetooth Low Energy (BLE) \cite{BLE}, and Nike+ \cite{NIKE}. 
Among them, three protocols have become dominant in the market: ANT, ZigBee, 
and BLE. Dementyev et al.~\cite{COMPARISON} analyzed the power consumption of 
these protocols. They found that BLE typically achieves the lowest power 
consumption, followed by ZigBee and ANT. BLE has become a promising solution for short-range 
transmissions between WMSs and the base station since it benefits from the 
widespread use of Bluetooth circuitry integrated in smartphones. In addition 
to energy-efficient transmission, new protocols commonly offer 
lightweight strong encryption, e.g., a modified form of AES \cite{AESBOOK}, to provide confidentiality as well as per-packet 
authentication and integrity. This prevents several security attacks, e.g., 
eavesdropping and integrity attacks, against WMS-based systems.

\subsubsection{On-sensor computation}
\label{ON_SENSOR}
The required signal processing varies significantly from one application to 
another. In most applications, sensors perform lightweight signal processing 
on the data, e.g., compression, using on-sensor resources and then transmit 
the processed data to the base station for further processing, e.g., indexing and machine learning. Due to limited on-sensor resources, on-sensor computation, with attendant energy overhead, can
be avoided for applications in which the sampling rate of biomedical signals 
is low, e.g., monitoring the patient's body temperature \cite{CNIA}. However, 
in some applications, on-sensor computation is beneficial and preferred over 
off-sensor computation due to one of the following reasons \cite{BEHNAM_3,BEHNAM_4}. First, on-sensor 
computation may significantly reduce the transmission energy (and as a result 
the total energy consumption of the device) even though it imposes extra 
energy consumption for computation. For example, if an EEG sensor can 
detect abnormal changes in the data, it only needs to transmit a small fraction of 
the data that includes those changes. Second, for some applications, in 
particular mission-critical applications, the communication delay or the 
possibility of unavailability of the Cloud or Internet may not be tolerable. 

Next, we briefly discuss three commonly-used types of on-sensor algorithms.

\noindent \textit {\textbf{1. Aggregation}}: In practice, a WMS does not usually need to 
transmit data as fast as it collects them. Hence, it can first aggregate
multiple sensory measurements in one packet and only then transmit the packet. 
In this scenario, the total number of bits transmitted remains the same. 
However, the average number of transmitted packets over a fixed time period is 
reduced due to the aggregation. This can significantly reduce the 
transmission energy of WMSs \cite{AGG_1}. The number of samples that can be 
aggregated in a single packet varies from one device to another based on its 
resolution, and is specified based on what is a tolerable response time 
\cite{CNIA}.

\noindent \textit {\textbf{2. Compression}}: Compression algorithms reduce the 
number of bits needed to represent data. On-sensor compression is commonly 
utilized to decrease transmission energy by reducing the total number of 
transmitted bits \cite{COMPRESS_1}. It can also save on-sensor storage by 
dropping non-essential information from the raw data \cite{COMPRESS_2}. For 
example, compressive sensing \cite{COMP_SENSING} offers a promising signal 
compression technique for acquiring and reconstructing a continuous signal by 
exploiting the sparsity of the signal to recover it from far fewer samples 
than required by the Shannon-Nyquist sampling theorem. Compressive sensing is 
recommended as a promising on-sensor compression technique in several recent 
research studies on WMSs \cite{shoaib,ENCOMP}. It significantly reduces 
the number of bits required to represent a signal and, at the same time, 
enables energy-efficient feature extraction and classification in the 
compressed domain \cite{shoaib}. 


\noindent \textit {\textbf{3. Lightweight classification}}: Classification is defined 
as the problem of identifying to which category from a given set a new 
observation belongs. In a typical classification problem, a feature extraction 
procedure first extracts a set of features from the raw data. Then, a learning 
algorithm (also called classifier) trains a model based on a training set of 
data containing observations whose category membership is known. After 
training, the classifier infers the category of new observations 
using the trained model and the features extracted from the new data samples. 
As a resource-limited device, a WMS may consume a considerable percentage of 
its energy to extract features and classify data samples. In order to reduce 
the energy required for inference, both feature extraction and classification 
must be energy-efficient. Compressive sensing-based feature extraction and 
classification \cite{shoaib} can significantly reduce the number of features 
that need to be processed, while maintaining high accuracy. Simple 
classifiers, e.g., decision trees \cite{DT} and perceptrons
\cite{PRCEPT}, enable lightweight classification when the use of traditional 
classification methods, e.g., support vector machine \cite{SVM}, which need 
more computational power and storage, is not tolerable. A few recent research 
studies have proposed application-specific classification algorithms, e.g., 
seizure detection based on EEG signals \cite{SEZ_1}, arrhythmia 
detection based on ECG signals \cite{ARH_1}, and physical activity 
classification based on acceleration data \cite{AC_1,AC_MA2}.

\subsection{Minimally-invasive capture methods}
As mentioned in Section \ref{GOALS}, passiveness is one of the key design 
goals of a WMS-based system. In order to ensure passiveness, WMSs should 
exploit minimally-invasive capture methods. Prior to the emergence of WMSs, 
such methods were developed to enhance user convenience 
for in-hospital settings. For example, EEG capture was invented by Berger in 
1924 \cite{BERGER}. The emergence of WMSs has magnified the need for
such methods. As a result, several novel sensing approaches, e.g., 
for glucose sensing \cite{GLU_1,GLU_2} and sweat analysis 
\cite{SW_2,SW_3}, have been developed for wearable watches and 
patches. They can non-invasively analyze on-skin chemical substances and 
minimize/eliminate the need for incisions or surgery. For example, Gao et 
al.~\cite{SW_2} have proposed a wearable sweat-analyzing patch that 
selectively measures sweat metabolites, e.g., glucose and lactate, and 
electrolytes, e.g., sodium and potassium ions, from on-skin liquids. Designing 
novel minimally-invasive methods for gathering biomedical data using wearable 
technology is an ongoing research direction that has attracted significant 
attention in recent years.

\subsection{Security and privacy}
The pervasive use of WMSs along with the emergence of the IoT paradigm during 
the last decade have led to several threats and attacks against the security 
of WMS-based systems and the privacy of individuals \cite{IOT_SURVEY}. Next, 
we describe such threats/attacks along with their most well-known 
countermeasures.

\subsubsection{Security threats and attacks}
Unfortunately, the security threats against WMS-based systems are not 
well-recognized. This has made WMSs targets for a multitude of adversaries, 
such as cybercriminals, occasional hackers, hacktivists, government, and 
anyone interested in accessing the sensitive information gathered, stored, or 
handled by WMS-based systems, e.g., health conditions or details of a 
prescribed therapy. 

As described in Section \ref{COMPONENTS}, a WMS-based system commonly consists 
of three main components: WMSs, base station, and Cloud servers. Possible 
security attacks against each of these components should be recognized and 
studied in depth in order to address attacks/threats against the whole system. Many well-known types of security attacks against the components/objects commonly used in IoT-based systems are summarized in \cite{IOT_SURVEY}. Since the majority 
of WMSs are connected to the Internet (either directly or through a 
smartphone), almost all such attacks are also applicable to WMS-based systems. 
Subashini et al.~\cite{SUBASHINI} describe various security attacks against 
the Cloud. These attacks/challenges include, but are not limited to, web 
application vulnerabilities such as Structured Query Language (SQL) injection, 
authorization/access control, integrity attacks, and eavesdropping. Moreover, 
many survey articles \cite{WSN_1,WSN_2,WSN_4} summarize security attacks 
against wireless sensor networks (WSNs) that are also applicable to WMS-based 
systems. 

Among previously-proposed attacks against WMS-based systems, the most 
well-known ones are: (i) eavesdropping on the communication channel to record 
unencrypted packets (an attack against confidentiality), and (ii) injection of 
illegitimate packets into the communication channel by reverse engineering 
the communication protocol (an attack against integrity). Encryption is the 
most effective approach for preventing these attacks. However, traditional 
encryption mechanisms are not suitable for WMSs due to on-sensor resource 
constraints. In order to reduce the resource overheads of encryption, several 
lightweight encryption mechanisms \cite{LW1,LW2} have been proposed in 
recent studies. Unfortunately, finding a practical low-power key exchange 
mechanism to securely share the encryption key is still a challenge, but
with some solutions on the horizon \cite{SECURE_VIBE}.     

\subsubsection{Privacy concerns}
With the exponential increase in the number of WMS-based systems, ensuring 
user privacy is becoming a significant challenge. Smart wearable
devices, e.g., smart watches, are equipped with many compact built-in
WMSs, e.g., accelerometers and heart rate sensors, and powerful communication 
capabilities in order to offer a large number of services. They collect, 
process, and store several types of private user-related data. 

Several recent research efforts have demonstrated how WMS-based 
systems may intentionally/unintentionally reveal the personal or corporate 
secrets of the user \cite{SEC_1,SEC_2,SEC_3}. For example, Wang et 
al.~\cite{SEC_1} demonstrate the feasibility of extracting the user's password 
by processing data gathered by the smart watch. The use of encryption may 
reduce leakage of private information by protecting the communication channel. 

\subsection{Calibration and noise cancellation}
The negative impact of various disturbances on the data collected by WMSs 
has been extensively discussed in recent research. In particular, it has been 
shown that environmental noise \cite{AMBI,AMBI_2}, user movement 
\cite{SALEH,MOVE_2}, and changes in sensor locations 
\cite{ALINIA,DISP_1,DISP_2} can impact sensory measurements
significantly. For example, Salehizadeh et al.~\cite{SALEH} discuss how sudden 
user movements can negatively impact pulse oximetry measurements, thus
leading to inaccurate readings and even loss of signal. Alinia et 
al.~\cite{ALINIA} demonstrate that a change in the location of an accelerometer 
can impact the quality of sensory readings. 

The above examples demonstrate that the various sources of noise should be 
taken into account while designing WMSs, and each sensor should be calibrated 
to ensure reliability and validity of measurements \cite{ALINIA}. Several noise 
cancellation and filtering techniques, e.g., for ECG \cite{CAN_10} and
EEG \cite{CAN_2}, have been proposed to mitigate the impact of noise. 
Furthermore, various user-independent and user-oriented calibration algorithms 
have been developed to calibrate different sensors, e.g., accelerometer 
\cite{ACCEL,AUTO_CALIB0}, magnetometer \cite{MAGNET_CALIB_1}, 
and gyroscope \cite{GYRO_CALIB_1}. Prior to a measurement, a 
user-independent (user-oriented) calibration algorithm calibrates the sensor 
without (with) the user's involvement based on the data gathered by the sensor 
itself and other sensors embedded in the system.
However, there is still a significant gap between the quality of measurements 
provided by wearable sensors and that of in-hospital monitoring devices. 
Unlike the in-hospital environment in which the user remains almost 
stationary, the user's position frequently changes during various daily 
activities. This makes the design of high-precision WMSs, which can provide 
high-quality measurements comparable to in-hospital equipment, a very complex 
task.

\subsection{Big data}
WMSs have the potential to generate big datasets over a short period of time. 
For example, a typical wearable EEG sensor generates over 120 MB of data per 
day \cite{CNIA}. With improvements in battery, sensor, and storage technologies, 
even more data might be generated by WMSs. Processing such large
datasets is a complex task due to the following reasons.

\begin{enumerate}
\item Data heterogeneity: Different WMSs collect different types of 
signals \cite{BIG_H1}. Moreover, due to on-sensor resource constraints, the 
data may not necessarily be acquired continuously or even at a fixed sampling 
rate, adding to the heterogeneity of data \cite{BIG_D1}. 

\item Noisy measurements: As mentioned earlier, there are numerous sources of 
noise and disturbances that can corrupt the raw data or deteriorate their 
quality. In addition, the dataset might have several hours of data
missing, when the user is not wearing one or multiple WMSs \cite{MISS}.

\item Inconsistency in data representation: Two devices containing the same 
sensors may offer very different types of raw data, e.g., older activity 
monitors generate a proprietary measure called an activity count, i.e., how 
often the acceleration magnitude exceeds some preset threshold, whereas newer 
ones commonly provide raw acceleration data \cite{BIG_D1}. Often, researchers 
and the industry use their own (often proprietary) data types and standards to 
report raw data.
\end{enumerate}

Previous WMS-related research on big data has mainly focused on extracting 
valuable information from big datasets generated by WMSs. Three of the most 
well-known research areas that aim to address the above challenges in 
processing large datasets are: (i) big data analytics, (ii) standardization, 
and (iii) data cleaning. 

\subsubsection{Big data analytics} This entails developing new methods and 
technologies to analyze big datasets, enabling a variety of services.  The aim 
is to provide fast and efficient algorithms to extract valuable information 
and trends embedded in the large datasets generated by WMSs. Sensory data 
encode aspects of human movement, but simultaneously, and at a higher level 
of abstraction, sleep patterns, physical strength, and mobility, and even 
complex aspects of physical/mental health \cite{LIFE_LOG}. The complex 
relationships and correlations present in big datasets might facilitate noise 
detection, and as a result, enable the negative impact of the noise artifacts 
to be mitigated \cite{BIG_D1}. 

\subsubsection{Standardization} Standardizing how wearable sensor datasets 
are stored and transferred can enable simultaneous processing of datasets from 
multiple locations around the globe. A naive solution may be as follows: all 
devices should record and transmit raw signal data in a standard format
and sample rate without any on-board preprocessing \cite{BIG_D1}. However, 
this is currently not a universally-practicable solution, mainly because 
battery restrictions for some applications, e.g., arrhythmia detection, 
necessitate preprocessing of data on the device \cite{CNIA, ONS1}. Furthermore, 
different applications inherently require different sample rates due to the 
nature of signals being monitored \cite{CNIA}. Thus, if standards mandate 
sample rates and filter settings, the rules should be application-specific.

\subsubsection{Cleaning} One of the first steps in data processing is 
data cleaning. This is the process of identifying and fixing data errors 
\cite{CLEAN_2}. Errors can be discovered in datasets by: (i) detecting 
violations of predefined integrity rules, (ii) finding inconsistent patterns in 
data, (iii) locating data duplicates, and (iv) searching for outlier values 
(see \cite{CLEAN_1} for a survey). A few innovative systems, e.g., 
NADEEF \cite{NADEEF} and Bigdansing \cite{BIGDANSING}, provide end-to-end 
solutions to data cleaning. However, there is still a lack of end-to-end 
off-the-shelf efficient systems for data cleaning \cite{CLEAN_2}. 

\subsection{Cloud computing}
As discussed in Section \ref{COMPONENTS}, a large number of WMS-based systems 
rely on Cloud servers. Despite the promise of the Cloud in this context 
(access to shared resources in a pervasive manner, large storage capacity, and 
high computational power), there are several challenges 
that need to be addressed for on-Cloud WMS-based services (see 
\cite{CLOUD_C_1,CLOUD_C_2,CLOUD_C_3,CLOUD_C_4} for survey articles). We 
briefly summarize them next.

\subsubsection{Availability/reliability} Many researchers have investigated 
the negative consequences of Cloud failure \cite{CERIN, SHAH_65}. Frequent 
failures of Cloud servers have serious consequences, e.g., increased energy 
consumption \cite{SHAH_31}, propagated service disruptions \cite{SHAH_66}, and, 
more importantly, adverse impact on the reputation of the provider 
\cite{SHAH_33}. In order to offer a smooth and continuous service, Cloud 
providers use redundancy techniques \cite{REDUN} (that back up data and 
store them in multiple data centers geographically spread across the world). 
As a result, the average system demand is several times smaller than server
capacity, imposing significant costs on the provider. To alleviate this
burden, an availability-tuning mechanism \cite{SHAH} has been suggested. It 
allows the customers to express their true availability needs and be charged 
accordingly. 

\subsubsection{Access control} The Cloud environment introduces new challenges 
on access control due to large scale, multi-tenancy (a software architecture 
in which a single instance of an application runs on a server and serves 
multiple groups of users), and host variability within the Cloud 
\cite{MUL_ACC}. In particular, multi-tenancy imposes new requirements on 
access control as intra-Cloud communication (provider-user and user-user) 
becomes popular \cite{POLICE}. Recent research efforts have been targeted 
at new access control techniques \cite{POLICE,AU1,AU2}, specifically 
designed for the Cloud. Masood et al.~summarize and compare the majority of 
newly-proposed Cloud-specific access control methods in \cite{AU_SURVEY}.

\subsubsection{Standardization and portability} Standardization of an 
efficient user interface is essential for ensuring user convenience. Web 
interfaces enable the user to access and analyze data on personal
devices, e.g., a smartphone.  Unfortunately, such web interfaces commonly 
impose a significant overhead because they are not specifically designed for 
smartphones or mobile devices \cite{STANDARD_INTER}. In addition to 
standardizing the user interface, standardization of data formats is
also essential to enable user-friendly services. If a Cloud provider stores 
data in its own proprietary format, users cannot easily move their data to 
other vendors \cite{PETCO}. 

\subsubsection{Bandwidth limitation} This is one of the fundamental challenges 
that needs to be handled in on-Cloud WMS-based systems when the number of 
users increases drastically, in particular for applications that need frequent 
data upload/download. Managing bandwidth allocation in a gigantic 
infrastructure, such as the Cloud that consists of several heterogeneous 
entities and millions of users, is very difficult. Some recent research efforts 
\cite{ALOC1,ALOC3} propose efficient bandwidth allocation methods for 
Cloud infrastructures. For example, Wei et al.~propose an allocation based on 
game theory \cite{ALOC3}.

%% file: ch-hmonitoring/chapter-hmonitoring.tex
\chapter{Energy-Efficient Long-term Continuous Personal Health Monitoring \label{ch:hmonitoring}}

Continuous health monitoring using wireless body-area networks of implantable 
and wearable medical devices (IWMDs) is envisioned as a transformative
approach to health care. Rapid advances in biomedical sensors,
low-power electronics, and wireless communications have brought this vision
to the verge of reality. However, key challenges still remain to be addressed.
The constrained sizes of IWMDs imply that they are designed with very limited 
processing, storage, and battery capacities. Therefore, there is a very strong 
need for efficiency in data collection, analysis, storage, and 
communication.

In this chapter, we first quantify the energy and storage requirements
of a continuous personal health monitoring system that uses eight
biomedical sensors: (1) heart rate, 
(2) blood pressure, (3) oxygen saturation, (4) body temperature, (5) blood 
glucose, (6) accelerometer, (7) electrocardiogram (ECG), and (8)
electroencephalogram (EEG). Our analysis suggests that there exists
a significant gap between the energy and storage requirements for long-term 
continuous monitoring and the capabilities of current devices. 

To enable energy-efficient continuous health
monitoring, we propose schemes for sample aggregation, anomaly-driven
transmission, and compressive sensing to reduce the overheads
of wirelessly transmitting, storing, and encrypting/authenticating
the data. We evaluate these techniques and demonstrate that they
result in two to three orders-of-magnitude improvements in energy and storage
requirements, and can help realize the potential of long-term continuous health 
monitoring \cite{CNIA}.

\section{Introduction}
Rapid technological advances in biomedical sensing and
signal processing, low-power electronics, and wireless networking are
transforming and revolutionizing health care. Prevention and early detection
of disease are increasingly viewed as critical to promoting wellness rather
than just treating illness. In particular, continuous long-term
health monitoring, where various physiological signals are captured,
analyzed, and stored for future use, is envisioned as key
to enabling a proactive and holistic approach to health care.

Several trends in computing and communications technology have converged
to advance continuous health monitoring from a distant vision to the
verge of practical feasibility. Foremost among these
is the evolution of IWMDs.
Traditionally, medical monitoring systems,
such as ECG and EEG monitors, have been used to simply gather raw data,
with signal processing and data analysis being performed offline.
However, with the continuing performance and energy efficiency
improvements in computing, real-time signal processing
has become possible. In the last decade, the
number and variety of IWMDs have increased significantly, ranging from
simple wearable activity and heart-rate monitors to sophisticated
implantable sensors. Moreover, advances in low-power wireless
communications enable radios to be integrated into even the most energy- 
and size-constrained devices.
This has led to the possibility of composing IWMDs into wireless 
body-area networks (WBANs) \cite{20Y,X1}. 

WBANs are opening up new opportunities for continuous health monitoring and
proactive health care \cite{21Y}. A typical WBAN for health monitoring
consists of (i) implantable and wearable sensors, which are
attached to the body or even implanted under the skin 
to measure vital signs and body signals, e.g., body temperature,
heartbeat, blood pressure, etc. and (ii) external devices (which
could be smartphones) that act as base stations to collect,
store, display, and analyze the data. 

Many recent and ongoing research efforts have addressed the design and deployment of WBANs. The CodeBlue project \cite{CODE_BLUE} focused on designing wireless sensor networks for medical applications. It included an ad-hoc network to transmit vital health signs to health care providers. Otto et al.~\cite{Y6} designed a system architecture to address various challenges posed by the need for reliable communication within the WBAN, and between 
the WBAN and a medical server. The MobiHealth project \cite{Y105} offered an end-to-end mobile health platform for health care monitoring. Different sensors, attached 
to a MobiHealth patient, enabled constant monitoring and transmission of vital signals. They considered security, reliability of communication resources, and quality of service (QoS) guarantees. 

Notwithstanding advances in IWMDs and WBANs,
some key technical challenges need to be addressed in order to enable
long-term continuous health monitoring. Due to size constraints
and the inconvenience or infeasibility of battery replacement,
IWMDs need to be highly energy-efficient. IWMDs as well as
the external devices that aggregate the monitored data have limited
storage capacity. 
Finally, health care applications also impose strict 
requirements for privacy, security, and reliability \cite{X1}. 

This chapter aims to address the challenging question of \textit{whether it is feasible to energy- and storage-efficiently provide long-term continuous health monitoring based on state-of-the-art technology}. In this chapter:
\begin{itemize}
\item We first discuss the traditionally used sense-and-transmit monitoring scheme to establish a baseline for our analyses. We evaluate a system that consists of eight biomedical sensors: (1) heart rate, (2) blood pressure, (3) oxygen 
saturation, (4) body temperature, (5) blood glucose, (6) accelerometer, (7) ECG, and (8) EEG.
\item We present analytical models that can be used to estimate the energy and storage requirements for these biomedical sensors. Our analysis suggests a significant gap between the energy and storage requirements for long-term continuous monitoring and the capabilities of current devices.

\item To address the aforementioned gaps in health monitoring, we propose and evaluate three schemes to reduce the overheads of sensing, storing, and 
wirelessly transmitting the data:
\begin{enumerate}
\item First, we explore a simple scheme based on aggregation of samples to 
amortize the communication protocol overheads and reduce the number of 
transmissions.
\item Second, we explore anomaly-driven transmission in which the sensors perform 
on-sensor signal processing to identify time intervals of interest, and only 
transmit/store data from these intervals.
\item Finally, we explore the concept of compressive sensing
(CS) \cite{INTROFORCS}, together with a newly developed approach for
computation on compressively-sensed data \cite{shoaib, shoaib2}, to drastically
reduce energy and storage.
\end{enumerate}

\item We demonstrate that the proposed schemes can potentially result in two to three orders-of-magnitude reduction in energy and storage requirements, and therefore may be instrumental in 
enabling continuous long-term health monitoring.

\item We  compare  all  proposed  schemes  and discuss how a continuous 
long-term health monitoring system should be configured based on
patients' needs and physicians' recommendations.
\end{itemize}

The rest of the chapter is organized as follows. Section \ref{BACK_SEC0} describes different components, which form a WBAN and the communication protocols that can be used to connect them together. Section \ref{BASELINE_SEC0} describes the baseline continuous health monitoring scheme. Section \ref{MODELS_SEC0} presents our analytical models and an analysis of the energy and storage requirements for the baseline WBAN using these models. Section \ref{PROPOSED_SEC0} describes the proposed schemes that 
include sample aggregation, anomaly-driven sampling, and CS-based
computation, and evaluates their energy impact.
Section \ref{STORAGE_SEC0} evaluates the impact of the proposed schemes on storage requirements. Section \ref{COMPARE_SEC0} compares different schemes and summarizes the medical considerations in configuration and optimization of different sensors. Finally, Section \ref{CONC_SEC0} concludes the chapter. 

\section{Different components of a general-purpose health monitoring system}
\label{BACK_SEC0}
In this section, we first describe two fundamental components that form a medical WBAN, namely biomedical sensors and the base station. Second, we discuss the communication protocols, which can used to connect them together.

\subsection{Health monitoring with networked wireless biomedical sensors}
Biomedical sensors have been used for health monitoring for a long time 
\cite{R2}. They sense electrical, thermal, optical, chemical, and other 
signals to extract information that are indicative of a patient's health condition. Examples of such sensors include oxygen saturation, glucose, blood pressure, heart rate, ECG, EEG, and several forms of imaging. 

\begin{figure}[ht]
\centering
\includegraphics [trim = 5mm 75mm 130mm 25mm ,clip, width=250pt,height=170pt]{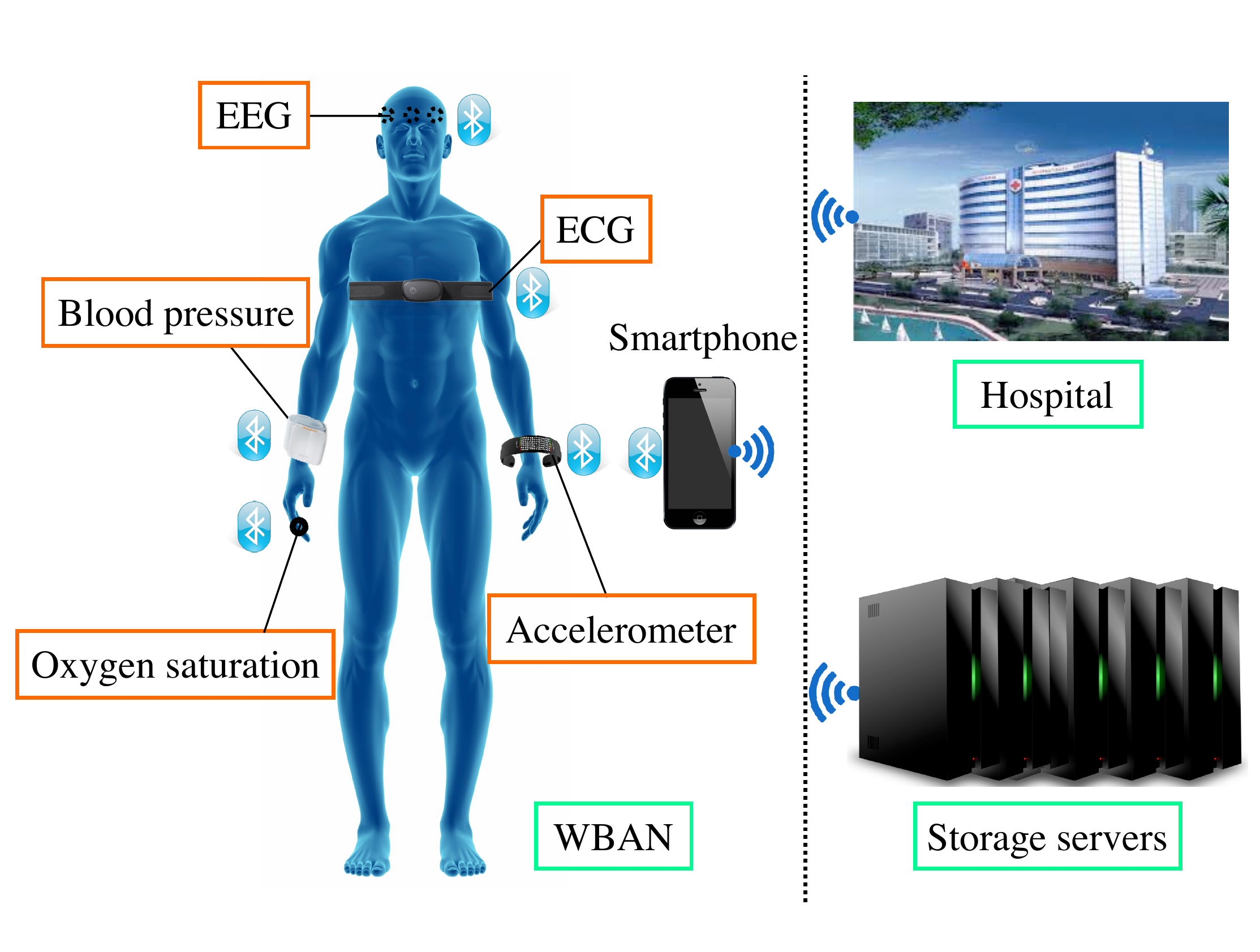}
\caption{A personal health care system.}
\label{fig:WSN}
\end{figure}

In addition to the biomedical sensors, an important component of a WBAN,
as shown in Fig.~\ref{fig:WSN},
is the base station or hub, a more capable device that aggregates data
from the biomedical sensors, visualizes health data for the patient, performs simple analytics, and communicates the health data to remote health providers or health databases. The base station, which could be a customized device or a commodity mobile device such as a smartphone, contains a more capable processor, data storage, and one or more wide-area network interfaces.

\subsection{Communication protocol}
A key consideration in the design of a WBAN is the
communication technology (radio and protocol) used to connect the medical 
sensors with the base station. Energy efficiency, security, and 
interoperability are some of the key factors that must be considered in this 
context.

Dementyev et al.~analyzed the power consumption characteristics of three 
popular emerging standards -- ANT, ZigBee, and BLE -- in a duty-cycled sensor 
node scenario \cite{COMPARISON}.  They found that BLE
achieves the lowest power consumption, followed by ZigBee and ANT. 
Most of the power consumption differences can be attributed to the time taken 
for a sensor to reconnect to the base station after waking up and the 
efficiency of the sleep mode used between transmissions of successive packets. 
In addition to low power consumption, BLE provides several other advantages for continuous health monitoring:
\begin{enumerate}
\item Smartphones have become dominant over other forms of base stations for 
potential use in the health monitoring system. BLE benefits from the 
widespread use of Bluetooth technology since BLE can be easily integrated into classical Bluetooth circuitry, and almost all new smartphones 
support BLE.
\item BLE is optimized for use in devices that need to communicate small packets wirelessly.
\item BLE is optimized to provide a low-rate ($<$ 270 $kb/s$) wireless
data transfer. As shown later, the maximum transmission rate of all sensors 
is much less than 270 $kb/s$.
\item BLE provides a long transmission range (more than 100 $m$) that 
enhances user convenience. 
\item Due to the privacy and safety concerns in medical systems, security is a key consideration in WBAN design. BLE supports strong encryption (Advanced Encryption Standard) to provide confidentiality as well as per-packet authentication and integrity.
\end{enumerate}

Thus, in our work, we use BLE for short-range transmissions between medical sensors and the base station.
\section{Baseline continuous health monitoring system}
\label{BASELINE_SEC0}
In this section, we first describe our baseline WBAN targeted at long-term 
continuous health monitoring that consists of eight sensors. Then, we discuss 
its energy and storage requirements.
\subsection{Baseline WBAN}
As mentioned earlier, we use eight biomedical sensors in the WBAN. In the 
baseline WBAN, each sensor node gathers raw data at a specific sampling 
frequency related to its application. Then, the node generates a BLE packet 
using a single sample and sends the raw data to the base station for further 
analysis. In this scheme, each sensor transmits the sample as soon as it is 
gathered, and the base station is responsible for processing. In order to 
implement the WBAN, first, it is required to specify the sampling rate for 
each sensor. This rate must be chosen in such a way that the requirements of 
different applications are met. The rates vary significantly from one sensor 
to another. Moreover, the same sensor may need to have different sampling 
rates in different applications \cite{19Y}. We have investigated the range of 
possible sampling rates for each sensor by reviewing the medical literature 
published between 1997 and 2014. Next, we provide these ranges for various 
sensors.

\begin{itemize}
\item \textbf{Heart rate}: The heart rate is commonly sampled at 6-8 $Hz$ 
frequency. For example, this sampling rate is currently used in fetal heart 
rate monitors \cite{FHR}. While the typical human heart rate is 65-82 beats
per minute ($bpm$), the rate can sometimes exceed 180 $bpm$. These considerations
suggest a sampling rate of 2-8 $Hz$ \cite{Hedge}.     
\item \textbf{Blood pressure}: During a typical ambulatory blood pressure 
monitoring session, the blood pressure is commonly measured every 15 to 30 
minutes over a 24-hour period \cite{BPS}. In some cases (e.g., occurrence of a 
hemorrhage), the blood pressure should be sampled at a much higher frequency. 
For example, Adibuzzaman et al.~have investigated the use of a blood pressure 
waveform sampled at 100 $Hz$ to monitor physiological system variations during a 
hemorrhage \cite{BP100}.
\item \textbf{Oxygen saturation}: The sampling rate of continuously-monitored 
oxygen saturation is suggested to be in the 0.001 $Hz$ to 2.00 $Hz$ range 
\cite{EVAN, KAML, 19Y}.  For example, Evans et al.~use 
measurements at 5-min intervals (sampling rate of 0.003 $Hz$) to monitor 
critically ill, mechanically ventilated adult patients during intrahospital 
transport \cite{EVAN}.
\item \textbf{Temperature}: The body temperature normally fluctuates over the 
day. Continuous monitoring of these small fluctuations is suggested by 
different researchers for a variety of applications \cite{SIMON, 19Y}. For 
example, Simon et al.~suggest measurements at 10-min intervals to determine 
the influence of circadian rhythmicity and sleep on 24-hour leptin variations 
\cite{SIMON}. However, some applications require a higher sampling rate 
(e.g., 1 $Hz$) \cite{19Y}. Thus, we assume the sampling rate of the body 
temperature sensor to be in the 0.001 $Hz$ to 1 $Hz$ range.
\item \textbf{Blood sugar}: Blood sugar measurements every 5 to 15 minutes are used in a variety of medical applications \cite{19Y, Sugar4}. However, some applications, such as continuous glucose monitoring to detect a sudden rise or drop in the glucose level of
diabetics, require a higher sampling rate ($\sim$100 $Hz$)~\cite{19Y}. 
\item \textbf{Accelerometer}: An accelerometer is widely used for physical activity detection. Its sampling rate typically lies in the 30 $Hz$ to 400 $Hz$ range. However, a lower sampling rate (e.g., down to 2 $Hz$) might be 
enough for some applications \cite{HE_ACC1, ACC2, 19Y, SMART}.
\item \textbf{ECG}: Determining the frequency content of an ECG signal by 
investigating its frequency spectrum is usually difficult because it is hard 
to distinguish between frequency components of signal and noise. Berson 
et al.~record over-sampled ECG signals and then apply different low-pass 
filters to them \cite{BERSON}. They describe the effect of filtering on 
amplitude variations, concluding that at least a sampling frequency of 
50-100 $Hz$ is necessary to prevent amplitude errors. Moreover, Simon et al.~demonstrated that a 1000 $Hz$ sampling rate is enough for the majority of ECG-based applications \cite{FSIMON}. We consider ECG sampling rates
in the 100-1000 $Hz$ range. 
\item \textbf{EEG}: Traditionally, the range of 
EEG frequencies that was accepted to be clinically relevant was in/below the 
gamma band (40-100 $Hz$). However, filtering of the EEG signal at around 70 $Hz$ 
and using at least a 200 $Hz$ sampling rate are commonly suggested by medical 
literature \cite{HIGH2}. Moreover, recent studies have shown that EEG signals 
may also have physiological relevance in high-frequency bands 
(e.g., 100-500 $Hz$) \cite{HIGH1, HIGH2}. Based on the above discussion, we
consider EEG sampling rates in the 100-1000 $Hz$ range.
\end{itemize}

Next, we consider the sampling resolution of each sensor, where resolution is 
defined as the number of bits required for representing a sample. 
We reviewed several recent publications in the biomedical literature to obtain
these resolutions.

\begin{itemize}
\item \textbf{Heart rate}: An accurate and compact low-power heart rate sensor 
for home-based health care monitoring is described and implemented 
in \cite{HRR}. It shows that a resolution of 10 bits is appropriate for 
providing an accurate measurement of the heart rate. 
\item \textbf{Blood pressure}: We consider 16 bits of resolution for blood
pressure samples, which is commonly used in  
commercial blood pressure monitoring devices \cite{19Y}.
\item\textbf{Oxygen saturation}: We consider 8 bits of resolution for
oxygen saturation based on the data reported in~\cite{OX1,OX}. 
\item\textbf{Temperature}: The body temperature varies within the 35 to 
$42 \celsius$ range. An 8-bit resolution is sufficient for body temperature 
sampling.
\item\textbf{Blood sugar}:  Measurements of blood sugar are based on color 
reflectance. The meter quantifies the color change and generates a 
numerical value that represents the concentration of glucose. A 16-bit 
resolution has been shown to be adequate for blood sugar monitoring devices 
\cite{SUGARSITE}. 
\item\textbf{Accelerometer}: 
We consider 12-bit resolution, which has been used in a variety of 
wearable accelerometer applications and commercial devices 
\cite{19Y,HE_ACC1,FREESCALE}. 
\item\textbf{ECG}:
Ultra low-power ECG sensors, which are commonly used in long-term monitoring, 
support  8 or 12 bits of resolution \cite{ECG1, ECG2, MIT8}. A 
resolution of 8 bits may result in a small but noticeable quantization error. Researchers have shown that greater than 8 bits of resolution will meet ECG requirements \cite{MIT12}. Therefore, we assume a resolution of 12 bits.
\item\textbf{EEG}: 
Several low-power wearable EEG sensors \cite{EEGR1, EEGR2} use 10- or 
12-bit analog-to-digital converter (ADC) units. The recording should represent samples down to 0.5 $\mu V$ and up to plus/minus several millivolts. We consider a 12-bit resolution. 
\end{itemize}

Table \ref{table:sensorspec_HE} summarizes information on sensors, their 
resolution and sampling rate, and the maximum wireless data transmission rate. 

\begin{table}[ht] 
\caption{Resolution, sampling rate, and maximum transmission rate} 
\centering 
\begin{tabular}{c c c c} 
\hline\hline 
Sensor & Resolution & Sampling & Maximum transmission \\ [0.5ex]
       & (bits/sample) & rate (Hz) & rate (bits/s)\\ [0.5ex]
\hline 
Heart rate & 10 & 2-8 & 80 \\ 
Blood pressure & 16 & 0.001-100 & 1600\\ 
Oxygen saturation & 8 & 0.001-2 & 16 \\ 
Temperature & 8 & 0.001-1 & 8 \\ 
Blood sugar & 16 & 0.001-100 & 1600\\
Accelerometer & 12 & 2-400 & 4800\\
ECG & 12 & 100-1000 & 12000\\
EEG & 12 & 100-1000 & 12000 \\[1ex] 
\hline 
\end{tabular} 
\label{table:sensorspec_HE} 
\end{table} 

\subsection{Energy and storage requirements}
Next, we discuss energy and storage requirements for a continuous health 
monitoring system.

Energy consumption can be divided into three categories: sampling, 
data transmission, and data analysis \cite{C10}. Wireless data transmission is 
usually the major energy-consumer. The available energy in each sensor node is 
often quite limited. The battery used in the node is typically the largest 
contributor in terms of both size and weight. Battery lifetime is 
a very important consideration in biomedical sensors. In particular, battery 
replacement of implanted sensors may require surgery and, hence, impose cost 
and health penalties \cite{21Y}. Thus, biomedical sensors often need to
maintain their functionality for months or even years without the need
for a battery change. For instance, an implanted pacemaker requires a 
battery lifetime of at least five years. Furthermore, during communication, 
biomedical sensors generate heat that may be absorbed in nearby tissue,
with possible harmful effects. Hence, the energy consumption should 
also be minimized from this perspective \cite{21Y}. 

Moreover, a WBAN imposes specific storage requirements. Although 
WBANs facilitate health monitoring and early detection of health problems, 
physicians usually want access to raw data so that they can independently 
verify the accuracy of on-sensor processing.  Thus, it is important to enable 
medical personnel to access all or at least important chunks of raw data. 
However, storing the raw data in the sensor nodes is not 
feasible for two main reasons. First, IWMD sizes need to be kept small
to facilitate patient mobility.  Second, adding a large storage to a sensor 
increases its energy consumption drastically, and as a result, battery lifetime 
decreases dramatically. Therefore, we may think of storing the data in the 
base station.  However, the base station (e.g., a smartphone) may have
its own resource constraints, though much less severe, in terms of
storage and battery lifetime.  In addition, in order to provide data 
backup, we may want to periodically send stored data from the base station to 
storage servers. Therefore, the costs of long-term storage using reliable 
storage services (e.g., Amazon S3 \cite{SMART_ANDROID_3}) should also be considered. Thus, 
it is important to minimize storage requirements for long-term 
health monitoring while maintaining adequate information for future reference.

\section{Analytical models for the evaluation of WBAN's energy and storage 
requirements}
\label{MODELS_SEC0}
In this section, we first describe the analytical models that we use to 
abstract the essential characteristics of the continuous health monitoring 
system. Then, we use the model to evaluate the baseline IWMDs.

\subsection{Analytical models}

Analytical models can be used to predict system requirements. 
They are much more efficient than performing simulation. Next, we describe 
the models used to quantify the energy consumption and storage
requirements of the continuous health monitoring system.  Table 
\ref{table:model} provides the list of variables used in our models.

\begin{table}[ht] 
\caption{Variables, unit, and description} 
\centering 
\begin{tabular}{l l l} 
\hline\hline 
Variable & Unit & Description \\ [0.5ex]
\hline 
$E_{total}$ & $J/day$ & Total energy consumption of a biomedical sensor\\
$E_s$ & $J/day$ & Energy consumption of sampling\\
$E_t$ & $J/day$ & Energy consumption of transmission \\
$E_c$ & $J/day$ & Energy consumption of computation\\
$E_{ADC}$ & $J/sample$ & Energy consumption of sampling per sample\\
$f_t$ & $Hz$ & Transmission frequency \\
$f_s$ & $Hz$ & Sampling frequency \\
$N$ & $-$ & Sampling resolution \\
$S$ & $1/day$ & $\#$samples per day \\
$C$ & $1/day$ & $\#$transmissions per day \\
$P_{send}$ & $W$ & Average power consumption in the sending mode \\
$P_{standby}$ & $W$ & Average power consumption in the standby mode \\
$I_{send}$ & $A$ & Average drained current in the sending mode \\
$I_{standby}$ & $A$ & Average drained current in the standby mode \\
$T_{send}$ & $s$ & Sending time \\
$T_{standby}$ & $s$ & Standby time \\
$V_{supply}$ & $V$ & Supply voltage\\
$SR$ & $B/year$ & Required amount of storage in a year\\[1ex] 
\hline 
\end{tabular} 
\label{table:model} 
\end{table} 

\vspace*{2mm}
\subsubsection{Energy consumption}
As mentioned earlier in Section \ref{BASELINE_SEC0}, energy consumption of a sensor 
has three major components: sampling, transmission, and on-sensor computation. 
Therefore, we assume that total energy consumption of the sensor 
($E_{total}$) can be written as: 
\begin{equation}
E_{total}=E_s+E_t+E_c
\end{equation} 

\noindent \textit{\textbf{1. Sampling energy}} 

\begin{figure} [h]
\centering
\includegraphics [trim = 125mm 96mm 110mm 95mm ,clip, width=250pt,height=178pt]{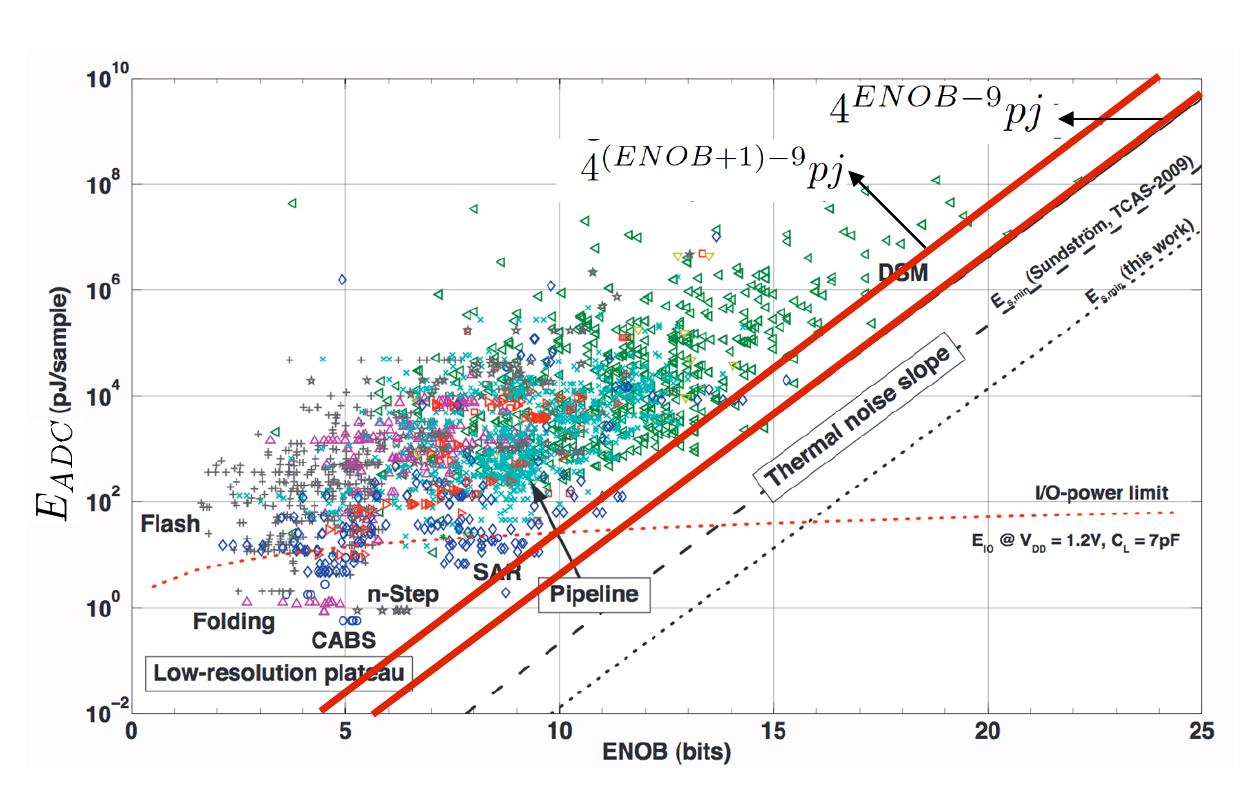}
\caption{Scatter plot of the reported $E_{ADC}$ vs. ENOB bits for different 
ADC architectures: asynchronous ($\circ$), cyclic ($\square$), delta-sigma 
($\triangleleft$), flash ($+$), folding ($\bigtriangleup$), pipeline 
($\times$), successive approximation ($\diamond$), subranging 
($\triangleright$), n-Slope ($*$), n-Step ($\star$), and other 
($\bigtriangledown$)} 
\label{figure:ENOB}
\end{figure}

Next, we discuss the sampling energy that is consumed by the ADC chip. 
The total energy consumption of an ADC chip can be divided 
into: (i) I/O energy, (ii) reference energy, (iii) sample-and-hold energy, 
(iv) ADC core energy, and (v) input energy \cite{ADCOR}.  However, 
separate calculation of these values is difficult. Thus, we use the actual 
values of the total on-chip ADC energy consumption per sample ($E_{ADC}$) 
reported in \cite{ADCOR}. It summarizes the experimental results from more 
than 1400 scientific papers published between 1974 and 2010. 
Fig.~\ref{figure:ENOB} shows the scatter plot of the reported $E_{ADC}$ 
in each of these papers vs. the effective number of bits (ENOB), where ENOB is 
defined as:\\
\begin{equation}
ENOB=\frac{SNR-1.76}{6.02},
SNR=\frac{P_{signal}}{P_{noise}}
\end{equation}

ENOB is always less than the resolution for all ADC 
chips. In particular, for medium-resolution ADCs ($8 \leq N \leq 16$) that are 
used in biomedical sensors, $ENOB \leq N-1$ provides a better boundary for the 
ENOB. For example, Verma et al.~presented a low-power 12-bit resolution ADC 
for WSNs \cite{ADCENOB12}. The ENOB of this ADC is reported to be 10.55 bits.

As shown in Fig.~\ref{figure:ENOB}, the $E_{ADC}$ of modern medium-resolution 
ADCs is within the $4^{ENOB-9}$ $pJ$ to $4^{(ENOB+1)-9}$ $pJ$ range. Therefore, 
the sampling energy consumption per day ($E_s$) can be upper-bounded as 
follows:\\
\begin{equation}
E_s = E_{ADC} * S
\end{equation}
\begin{equation}
E_{ADC} < 4^{(ENOB+1)-9}pJ \leq 4^{(N-9)} pJ,\\ 
\end{equation}
\begin{equation}
S = f_s (\frac{1}{s}) * 60 (\frac{s}{min})* 60 (\frac{min}{hr}) * 24 (\frac{hr}{day})\\
\end{equation}
\begin{equation}
\implies E_s < f_s (\frac{1}{s}) * 60 (\frac{s}{min})* 60
(\frac{min}{hr}) * 24 (\frac{hr}{day}) * 4^{(N-9)} pJ
\end{equation}

Table \ref{table:sam} shows the upper-bound values of $E_s$ for all the 
sensors. As discussed later, $E_s$ values for all sensors are negligible in 
comparison to their total energy consumption. Hence, we can safely assume that 
$E_{total} \approx E_t + E_c$. 

\begin{table}[ht] 
\caption{Upper-bound values of $E_s$} 
\centering 
\begin{tabular}{c c} 
\hline\hline 
Sensor & $E_s$ (J/day) \\
\hline 
Heart rate & 2 e-6\\ 
Blood pressure & 1 e-1 \\ 
Oxygen saturation & 4 e-8 \\ 
Temperature & 4 e-8 \\ 
Blood sugar & 1 e-1  \\
Accelerometer & 2 e-3 \\
ECG & 5 e-3 \\
EEG & 5 e-3 \\[1ex] 
\hline 
\end{tabular} 
\label{table:sam} 
\end{table} 

\vspace*{2mm}
\noindent \textit{\textbf{2. Transmission energy}}

In our experiments, we used the Texas Instruments CC2541 Development Kit as 
the BLE transmission device. To provide a quantitative comparison, we 
experimentally measured the energy consumption of the transmission chip in a 
cyclic scenario. In a cyclic transmission, the transmitter takes $T_{send}$ 
seconds to send the data to the base station and then enters a standby phase 
for $T_{standby}$ seconds. Hence, the average energy consumption of 
transmission can be calculated as follows: \\
\begin{equation}
E_t = (T_{send} * P_{send} + T_{standby} * P_{standby}) * C
\end{equation}
\begin{equation}
C=f_t (\frac{1}{s}) * 60(\frac{s}{min})* 60 (\frac{min}{hr}) * 24
(\frac{hr}{day})
\end{equation}
$T_{send}$ is a fixed value and measured as 2.6 milliseconds for 
a single packet transmission. $T_{standby}$ depends on the transmission 
frequency ($f_t$):\\
\begin{equation}
T_{standby} = \frac{1}{f_t} - T_{send}
\end{equation}
\noindent $P_{send}$ and $P_{standby}$ can be obtained by measuring the 
current drained from the battery with supply voltage $V_{supply}$. We 
calculated the average power consumption for a single packet transmission 
using a standard oscilloscope. $P_{send}$ and $P_{standby}$ were 
found to be 30.5 $mW$ and 2.5 $\mu W$, respectively, where $V_{supply}$ is set 
to 2.5 $V$. In order to measure 
the power consumption of a single packet transmission, we also considered 
different packet sizes (varying from 1 $B$ to 20 $B$). Our experimental results show 
that variations in transmission energy of a single packet are negligible when 
the packet size changes from 1 $B$ to 20 $B$. However, since 
$P_{send} \gg P_{standby}$, a higher transmission rate obviously leads to 
a higher energy consumption.

\noindent \textit{\textbf{3. Computation energy}}

Computation energy varies significantly from one biomedical application to 
another. In most applications, the computation energy can be divided into 
feature extraction energy and classification energy.  
Since a feature extraction function can be converted into matrix
form, the feature extraction energy can be estimated as the energy 
consumption of a matrix multiplication function. The classification energy 
can be estimated based on the reported values of classification energy per 
vector for various methods. However, obtaining a general model for computation 
energy is difficult because of its dependence on the application.
In this work, when we consider on-sensor computation energy, we use the values reported in \cite{shoaib,shoaib2}.

\subsubsection{Storage requirement}

Next, we provide an analytical model for estimating the amount of required 
storage for one-year storage of raw medical data. When there is no on-sensor 
computation, this only depends on the sampling frequency ($f_s$) and sampling 
resolution ($N$):\\

\begin{equation}
SR = f_s (\frac{1}{s})* N (bits)* (\frac {1 B}{8 bits}) *
31536000(\frac{s}{year})
\end{equation}

\noindent However, simple on-sensor computation can significantly decrease the 
amount of required storage. For example, if the computation method can 
efficiently detect points of interest from the raw data, we may only
need to store those specific points for further analysis. Moreover, on-sensor 
data compression, e.g., in CS-based applications, can also decrease the number 
of transmitted bits from the sensor to the base station by compressing the raw 
data before transmission.

\subsection{Evaluation of the baseline WBAN} 

Next, we evaluate the energy consumption and storage requirement for the 
baseline scheme, described in Section \ref{BASELINE_SEC0}, using the models 
described above.\\ 

\subsubsection{Evaluation of the energy consumption}

Since each sensor has its own sampling rate and resolution, its energy 
consumption differs from that of others. Table \ref{table:schemeone_energy} 
shows the minimum and maximum amounts of energy consumption for different 
devices in this baseline scenario.  They correspond to the minimum and maximum sampling rates, respectively. Table \ref{table:schemeone_battery} shows the 
battery lifetime of each sensor.  The minimum/maximum battery lifetimes are 
reported assuming that each sensor node uses a regular coin cell battery 
(CR2032).  A regular coin cell battery is commonly used in biomedical sensors.
Not surprisingly, ECG and EEG sensors are seen
to consume the most amount of energy. Thus, these sensors are the 
main obstacles to providing long-term health monitoring. 
\begin{table}[ht] 
\caption{Minimum and maximum values of total energy consumption} 
\centering 
\begin{tabular}{c c c} 
\hline\hline 
Sensor & Minimum (J/day) & Maximum (J/day)\\ [0.5ex] 
\hline 
Heart rate & 13.99 & 55.23\\ 
Blood pressure & 0.26 & 686.88\\ 
Oxygen saturation & 0.26  & 14.00 \\ 
Temperature & 0.26  & 7.13 \\ 
Blood sugar & 0.26  & 686.88\\
Accelerometer & 14.00 & 2747.52 \\
ECG & 686.88 & 6868.80 \\
EEG & 686.88 & 6868.80 \\[1ex] 
\hline 
\end{tabular} 
\label{table:schemeone_energy} 
\end{table} 

\begin{table}[ht] 
\caption{Minimum and maximum battery lifetimes of different sensors} 
\centering 
\begin{tabular}{c c c} 
\hline\hline 
Sensor & Minimum (days) & Maximum (days)\\ [0.5ex] 
\hline 
Heart rate & 48.8 & 192.90 \\ 
Blood pressure & 3.93 & 10125.69\\ 
Oxygen saturation & 192.86 & 10125.69 \\ 
Temperature & 378.68 & 10125.69 \\ 
Blood sugar & 3.93 & 10125.69\\
Accelerometer & 0.98 & 192.86 \\
ECG & 0.39 & 3.93 \\
EEG & 0.39 & 3.93 \\[1ex] 
\hline 
\end{tabular} 
\label{table:schemeone_battery} 
\end{table} 

\subsubsection{Evaluation of the storage requirement}

Next, we evaluate the baseline system from the storage perspective. We readily 
realize the baseline transmission scheme requires a significant amount of storage.  Table \ref{table:memory} shows the minimum and maximum amounts of storage required for long-term health 
monitoring in this system. The minimum (maximum) value corresponds to the 
minimum (maximum) sampling frequency.  Since EEG and ECG signals require the 
largest amount of storage, we mainly target these signals for storage reduction.

\begin{table}[ht] 
\caption{Minimum and maximum storage required for long-term storage} 
\centering 
\begin{tabular}{c c c} 
\hline\hline 
Sensor & Minimum (MB/yr) & Maximum (GB/yr)\\ [0.5ex] 
\hline 
Heart rate & 75.18 & 0.29\\ 
Blood pressure & 0.07 & 5.87\\ 
Oxygen saturation & 0.03  & 0.06 \\ 
Temperature & 0.03  & 0.03 \\ 
Blood sugar & 0.07  & 5.87\\
Accelerometer & 90.23 & 17.62 \\
ECG & 4511.26 & 44.06 \\
EEG & 4511.26 & 44.06 \\[1ex] 
\hline 
\end{tabular} 
\label{table:memory} 
\end{table} 

\section{Improving the energy efficiency of continuous health monitoring}
\label{PROPOSED_SEC0}
In this section, we first propose three schemes for signal processing and
transmission that can be used in a WBAN. Then, we evaluate and 
compare these schemes from the energy perspective. 
We divide the sensors into two different categories 
based on their transmission rate: low-sample-rate sensors (heart rate,
blood pressure, oxygen saturation, temperature, blood sugar,
accelerometer) and high-sample-rate sensors (EEG and ECG).  Then, we
use the following three schemes to reduce the energy consumption of each node.

\begin{itemize}
\item We accumulate multiple samples in one packet before transmitting the raw 
data in order to decrease the number of transmitted packets.  The base station 
is responsible for processing and storage of the raw data. This approach is 
applicable to both high-sample-rate and low-sample-rate sensors.

\item We process the data in high-sample-rate sensors (EEG and ECG) using 
traditional signal processing methods. Then, we transfer just a fraction of 
the raw data from the sensor node for storage in the base station based on the 
result of computation.

\item We suggest using CS-based computation in high-sample-rate sensor nodes 
before data transmission. Again, we just transfer a small fraction of the raw 
data from the sensor node for storage in the base station based on the result 
of on-sensor computation.
\end{itemize}

Although on-sensor computation leads to some extra computational energy 
consumption, it reduces transmission energy consumption significantly
due to the reduction in the amount of data transmitted. This is especially true 
when the transmission rate of a sensor is very high and important events (e.g., 
seizure, heart attack) are rare.  However, in the case of low-sample-rate 
sensors, the decrease in transmission energy does not offset the
increase in computational energy.  Therefore, we do not employ any on-sensor 
computation for low-sample-rate sensors.

Each scheme is discussed in the following subsections and compared against the 
baseline scheme.  We estimated the minimum/maximum energy consumption of each 
sensor in different scenarios, and based on that, we computed the 
minimum/maximum battery lifetime. 

\subsection{Sample aggregation}
In practice, we do not usually need to transmit the data as fast as we gather 
them. Thus, we could first accumulate multiple samples (up to 20 $B$) in one 
packet and only then transmit the packet.  The total number of bits transmitted 
remains the same.  However, the average number of transmitted packets per 
second is reduced due to the accumulation.  The number of samples that can be 
accumulated in a single packet varies from one device to another based on 
its resolution. In addition, the data processing algorithm in the base station 
might have been optimized with a specific number of required samples in
mind. Therefore, the number of samples per packet may need to be
varied between 1 and the maximum number.  For the devices being
evaluated, Table \ref{table:packetPerSample} shows the maximum number of 
samples that can be gathered into a single packet. 

\begin{table}[ht] 
\caption{Maximum number of samples in one packet} 
\centering 
\begin{tabular}{c c} 
\hline\hline 
Sensor & \#Samples\\ [0.5ex] 
\hline 
Heart rate & 16 \\ 
Blood pressure & 10 \\ 
Oxygen saturation & 20   \\ 
Temperature & 20  \\ 
Blood sugar & 10  \\
Accelerometer & 13 \\
ECG & 13 \\
EEG & 13 \\[1ex] 
\hline 
\end{tabular} 
\label{table:packetPerSample} 
\end{table} 

In order to calculate the total energy consumption of a sensor, we also
need to consider the storage energy required for storing multiple
packets before transmission. To store 20 $B$, which is the maximum number of 
bytes that can be sent in a single transmission, we consider the energy 
consumption of a 160-cell buffer. This storage energy remains fixed for the 
maximum and minimum transmission rates. However, the maximum (minimum) energy 
consumption is calculated as the energy consumption of transmission using 
the maximum (minimum) rate plus the energy consumed by the 160-cell buffer. 
Using the SRAM cell energy reported for the 90 $nm$ technology node in \cite{Y}, 
we calculate the minimum and maximum energy consumption of each device, as 
shown in Table \ref{table:secondSchemeEnergy}.  The minimum and maximum battery 
lifetimes of each sensor are shown in Table \ref{table:second_battery}. 
Relative to the baseline, this method provides up to 13.58$\times$ reduction 
in maximum energy consumption for low-sample-rate sensors. The maximum 
and minimum energy consumptions of high-sample-rate sensors are reduced by 
12.98$\times$ and 12.83$\times$, respectively.

\begin{table}[ht] 
\caption{Minimum and maximum values of total energy consumption while using 
the sample aggregation scheme} 
\centering 
\begin{tabular}{c c c} 
\hline\hline 
Sensor & Minimum (J/day) & Maximum (J/day)\\ [0.5ex]
\hline 
Heart rate & 1.50 & 4.07\\ 
Blood pressure & 0.65 & 69.38\\ 
Oxygen saturation & 0.65  & 1.33 \\ 
Temperature & 0.64  & 0.98 \\ 
Blood sugar & 0.65  & 69.38\\
Accelerometer & 1.70 & 212.13 \\
ECG & 53.52 & 529.36 \\
EEG & 53.52 & 529.36 \\[1ex] 
\hline 
\end{tabular} 
\label{table:secondSchemeEnergy} 
\end{table} 

\begin{table}[ht] 
\caption{Minimum and maximum battery lifetimes of different sensors while 
using sample aggregation scheme} 
\centering 
\begin{tabular}{c c c} 
\hline\hline 
Sensor & Minimum (days) & Maximum (days)\\ [0.5ex] 
\hline 
Heart rate & 663.39 & 1800 \\ 
Blood pressure & 38.92 & 4153.85 \\ 
Oxygen saturation & 2030.08 & 4153.85 \\ 
Temperature & 2715.10 & 4218.75 \\ 
Blood Sugar & 38.92 & 4153.85 \\
Accelerometer & 12.73 & 1588.24 \\
ECG & 5.10 & 50.45 \\
EEG & 5.10 & 50.45 \\[1ex] 
\hline 
\end{tabular} 
\label{table:second_battery} 
\end{table} 

\subsection{Anomaly-driven transmission}

Next, we evaluate a process-and-transmit scheme that is more appropriate for 
high-sample-rate sensors (ECG and EEG), which consume significant
amounts of energy.  If we first process raw data in the sensor nodes 
themselves and then just transmit some small chunks of data based on the 
processing results, we can reduce the transmission rate significantly. In 
this scenario, whenever we detect an abnormal activity, we are required 
to transmit the raw data corresponding to the abnormal event, in order
to facilitate offline evaluation of the data. The computational energy in each 
sensor node and data transmission rate directly depend on the intended 
application.  We evaluated seizure detection and arrhythmia detection as applications for EEG and ECG sensors, respectively. The traditional computation that we have considered for seizure/arrhythmia detection is as follows. First, we sample the signal at the Nyquist sampling rate. Second, we use a feature extraction algorithm (spectral energy analysis for EEG and Wavelet transform for ECG) to extract the important feature of the signal and build a feature vector. Third, we classify the feature vectors using a binary classifier \cite{trad1,trad2,trad3,shoaib,shoaib2}.

Let us consider an EEG sensor first. We assume signal processing in this 
sensor is based on a traditional algorithm for seizure detection, as 
described in \cite{shoaib,shoaib2}. The frequency of epileptic seizures 
varies from person to person. In some cases, seizures may even be
separated by years. On the other extreme, seizures might occur every day. 
Williamson et al.~\cite{will} studied 90 patients and reported the mean 
seizure frequency and mean duration to be 4.7 per month (range: 3 to 9 per 
month) and 3.8 minutes (range: 1 to 20 minutes), respectively. 
Based on their result, if the EEG sensor just transmits the small fraction of 
data corresponding to seizures, the sensor needs to transmit information 
over a duration of 17.8 minutes per month, on an average. 
Table \ref{table:totalEne_EEG_Trad} shows the average total energy 
consumption of the EEG sensor when we use the traditional signal processing 
method described in \cite{shoaib,shoaib2} and only transmit important chunks 
of data whenever an abnormality is detected. The minimum (maximum) value 
corresponds to the minimum (maximum) sampling frequency. In this scheme, the 
processing module consumes the major part of energy. Relative to the baseline, 
it provides up to 177$\times$ reduction in total energy consumption 
for the EEG sensor. Table \ref{table:Trad_Bat} shows the minimum and maximum 
battery lifetimes of the EEG sensor in this scheme. 

\begin{figure}[h]
\centering
\includegraphics [trim = 55mm 35mm 30mm 30mm ,clip, width=260pt,height=190pt]
{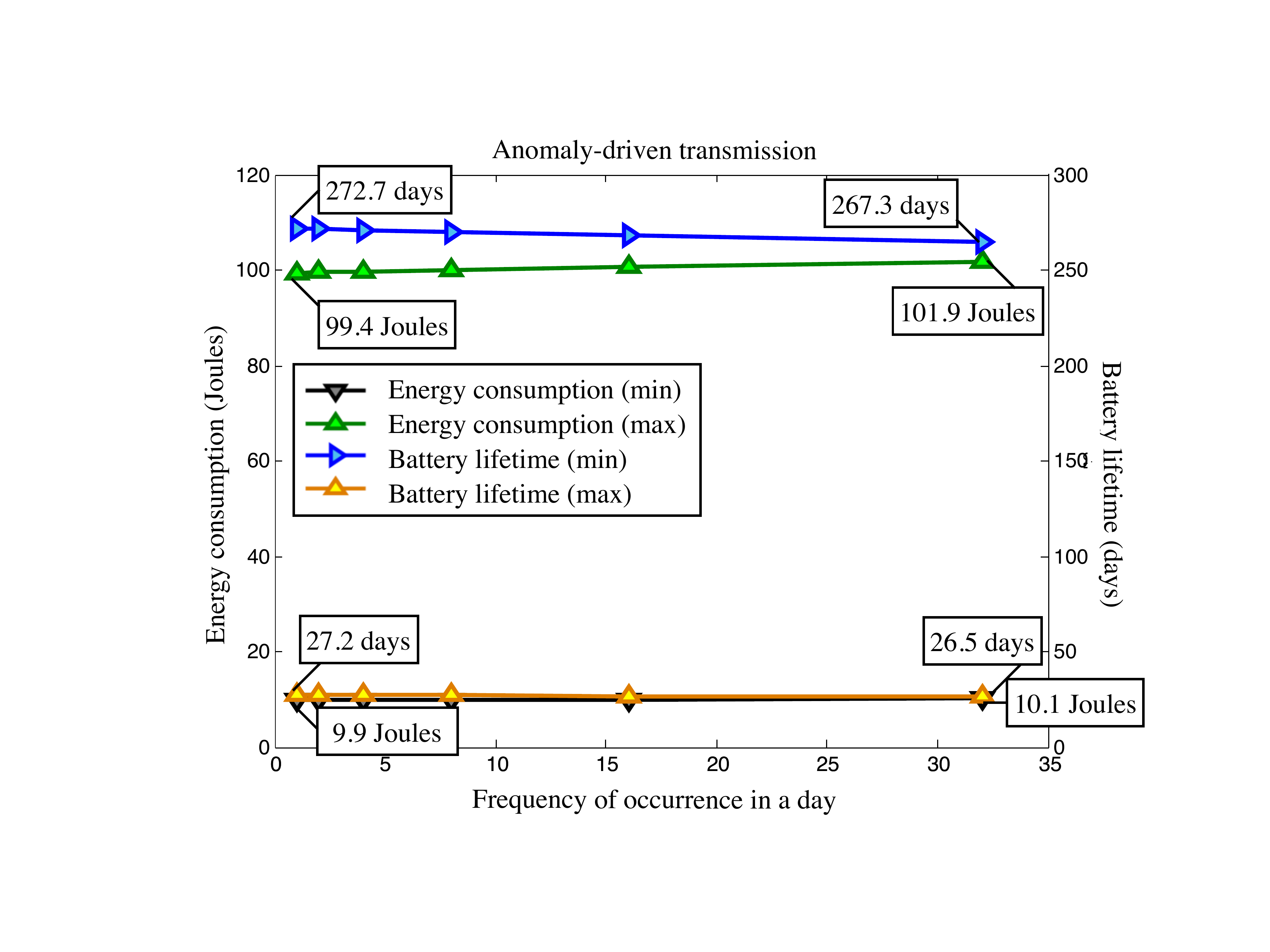}
\caption{Energy consumption and battery lifetime of the ECG sensor for the anomaly-driven method with respect to frequency of occurrence of arrhythmia in a day.} 

\label{figure:arrPlot_BAT}
\end{figure}

Next, we consider ECG sensors, and assume that the signal processing method is 
the traditional computation method for arrhythmia detection, as discussed in 
\cite{shoaib2}. Unlike seizure, the frequency of occurrence of arrhythmia 
varies significantly. There are different types of arrhythmia: each may lead to
intermittent or consistent symptoms. Therefore, it is difficult to 
predict the frequency of occurrence for arrhythmia. 
Fig.~\ref{figure:arrPlot_BAT} shows the total 
energy consumption and battery lifetime of the ECG sensor with respect to frequency of occurrence of arrhythmia in a day, respectively. We assume that 
after detecting an abnormal event, the sensor transmits the information of a 
standard one-minute ECG strip to the base station.

\begin{table}[ht] 
\caption{Average total energy consumption of the EEG 
sensor for the anomaly-driven method} 
\centering 
\begin{tabular}{c c c} 
\hline\hline 
Sensor & Minimum (J/day) & Maximum (J/day)\\ [0.5ex] 
\hline 
EEG & 36.27 & 38.83 \\[1ex] 
\hline 
\end{tabular} 
\label{table:totalEne_EEG_Trad} 
\end{table} 

\begin{table}[ht] 
\caption{Average battery lifetimes for the EEG sensor for 
the anomaly-driven method}
\centering 
\begin{tabular}{c c c} 
\hline\hline 
Sensor & Minimum (days) & Maximum (days)\\ [0.5ex] 
\hline 
EEG & 69.53 & 74.44 \\[1ex] 
\hline 
\end{tabular} 
\label{table:Trad_Bat} 
\end{table} 

\subsection{CS-based computation and transmission}

As the third scheme, we evaluate an approach for computation and data 
transmission that can reduce the energy consumption of EEG and ECG sensors 
significantly. As mentioned earlier, since the total energy consumption of EEG 
and ECG sensors is very high due to their high data transmission rates, if we 
can process the raw data in these sensors and transmit only small chunks of 
data upon the occurrence of an abnormal event, the transmission energy may be 
reduced significantly.  However, now the computation energy becomes the major 
energy bottleneck.  Hence, we try to reduce it through CS-based computation.
First, we briefly describe CS. 

CS (also called compressive sampling or sparse sampling) is a signal processing 
approach for efficiently sampling and reconstructing a signal \cite{INTROFORCS}. The common goal 
of various signal processing 
approaches is to reconstruct a signal from a finite number of measurements. Without any prior knowledge or assumptions about the signal, this task is not feasible due to the fact that there is no way to reconstruct an arbitrary signal in an interval in which it is not measured. However, under certain conditions and assumptions, the signal can be reconstructed using a finite 
number of samples. In the CS approach, a signal can be recovered from far 
fewer samples than required by Nyquist sampling. Recovering a signal using 
the CS approach relies on two fundamental principles: sparsity and 
incoherence. 

\begin{enumerate}
\item Sparsity: This requires that the signal be sparse in some domain 
(i.e., the signal's representation in some domain should have many 
coefficients close to or equal to zero) \cite{BEHNAM_1, BEHNAM_2}. CS can be used to compress 
an $N$-sample signal $X$ that is sparse in a secondary basis $\Psi$. 
Previous research has shown that ECG and EEG signals are sparse enough in the Wavelet transform space \cite{ECGSPA} and Gabor space \cite{R7X, R8X, R9X}, respectively.

\item Incoherence: This indicates that unlike the signal of interest, the 
sampling/sensing waveforms have an extremely dense representation in the 
transformed domain.
\end{enumerate}

The main limitation of the classical CS approach is as follows. Although the 
signal can be recovered using only a few samples, the traditional signal 
processing methods are not designed to process the compressed form of
the signal. Therefore, the signal needs to be reconstructed before processing 
by the traditional signal processing methods. Unfortunately, reconstruction 
of a signal from its compressed representation is an energy-intensive task 
and cannot be performed on sensors due to their energy constraints. In 
WBANs, it is often necessary to process the data sampled by the biomedical 
sensors, e.g., to detect anomalies or compute statistics of interest. 
In this work, we evaluate a modified version of the classical CS approach 
that enables ECG and EEG signals to be processed on the sensor without being 
reconstructed (Fig.~\ref{figure:COMPS}). The need for reconstruction can be 
circumvented by performing signal processing computations directly in the CS 
domain. Shoaib et al.~have developed precisely such a method 
\cite{shoaib, shoaib2}, and demonstrated applications to various biomedical 
signals. This method reduces the computation energy significantly because 
much fewer data samples need to be processed. Generally, this method consists 
of three steps:

\begin{figure} [h]
\centering
\includegraphics [trim = 20mm 110mm 10mm 45mm ,clip, width=380pt,height=148pt]{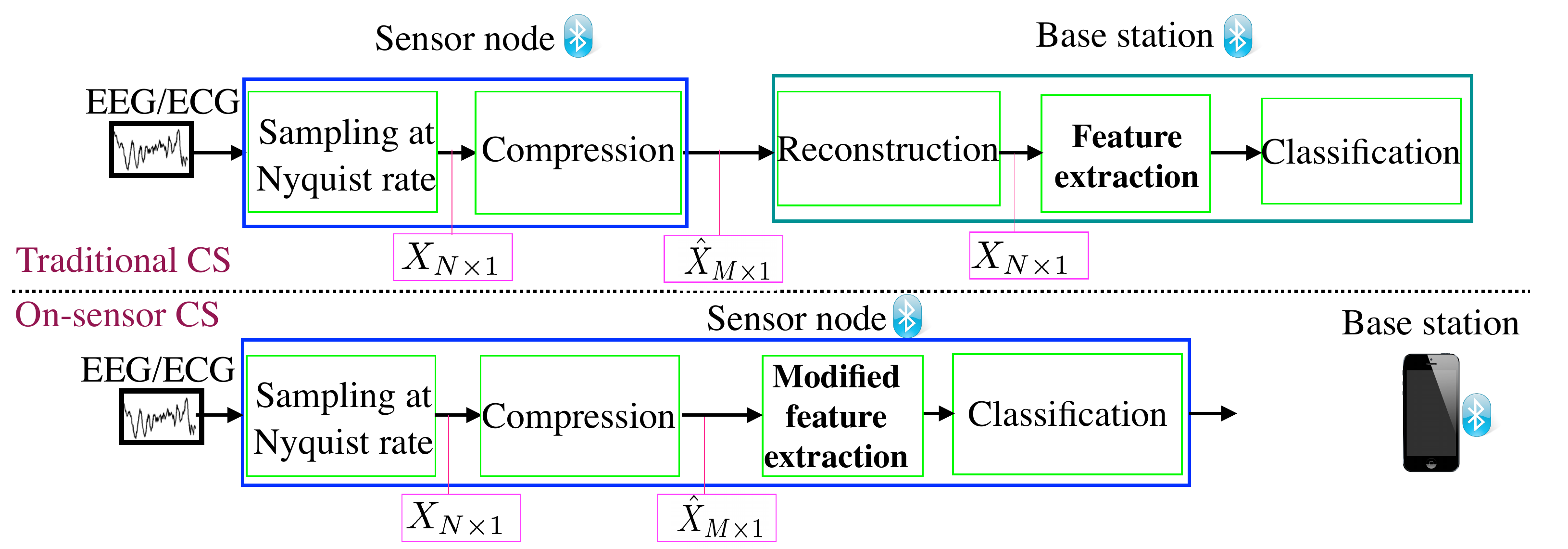}
\caption{Traditional CS vs. on-sensor CS-based computation.} 
\label{figure:COMPS}
\end{figure}

\begin{enumerate}
\item First, we compress the signal of interest using a low-rank random 
projection matrix. If we can represent the signal ($X$) as $\Psi * s$, where 
$s$ is a vector of K-sparse coefficients, a low-rank random matrix $\Phi$ 
can be found to transform $X$ to a set of $M$ samples where 
$O(Klog(N/K)) < M \ll N$. We can then use the following equation for obtaining 
the compressed samples (denoted by $\hat{X}$):

\begin{align}
\hat{X}_{M\times1} = \Phi_{M \times N} \times X_{N\times 1}
\end{align}

\item Second, we generate a feature extraction operation in the CS domain 
($\hat{H}$) from its equivalent in the Nyquist domain ($H$) by minimizing 
the error in the inner product between feature vectors. For any feature 
extraction method, which can be represented by matrix $H$, we can derive an 
equivalent $\hat{H}$ matrix in the CS domain \cite{shoaib, shoaib2}.
\item Third, we compute $\hat{Y}= \hat{H} \times \hat{X}$ and provide 
$\hat{Y}$ to the classification process.
\end{enumerate}

The compression ratio is given by $\alpha=N/M$.  It denotes the amount of 
compression obtained by the projection. Because CS leads to a drastic reduction
in the number of samples, it has the potential for reducing the energy
consumption of various sensors, including biomedical sensors. Direct 
computation on compressively-sensed data enables classification to be 
performed on the sensor node with one to two orders of magnitude energy
reduction. We exploit this method for long-term continuous health monitoring.

In order to choose a reasonable compression ratio ($\alpha$), we first need to 
compare the outcomes of the CS-based method for different compression ratios. 
Next, we discuss sensitivity (also called recall) and number of false alarms per hour (FA/h) for different compression ratios. Sensitivity represents the true positive rate. It measures the percentage of actual positives that are correctly identified, such as the percentage of seizure conditions that are correctly classified as seizure. FA/h is the number of false positive outcomes in an hour of detection. Such an outcome is an error in classification since a test result indicates the presence of a medical condition that is not actually present.

\begin{figure} [h]
\centering
\includegraphics [trim = 30mm 20mm 40mm 30mm ,clip, width=250pt,height=170pt]{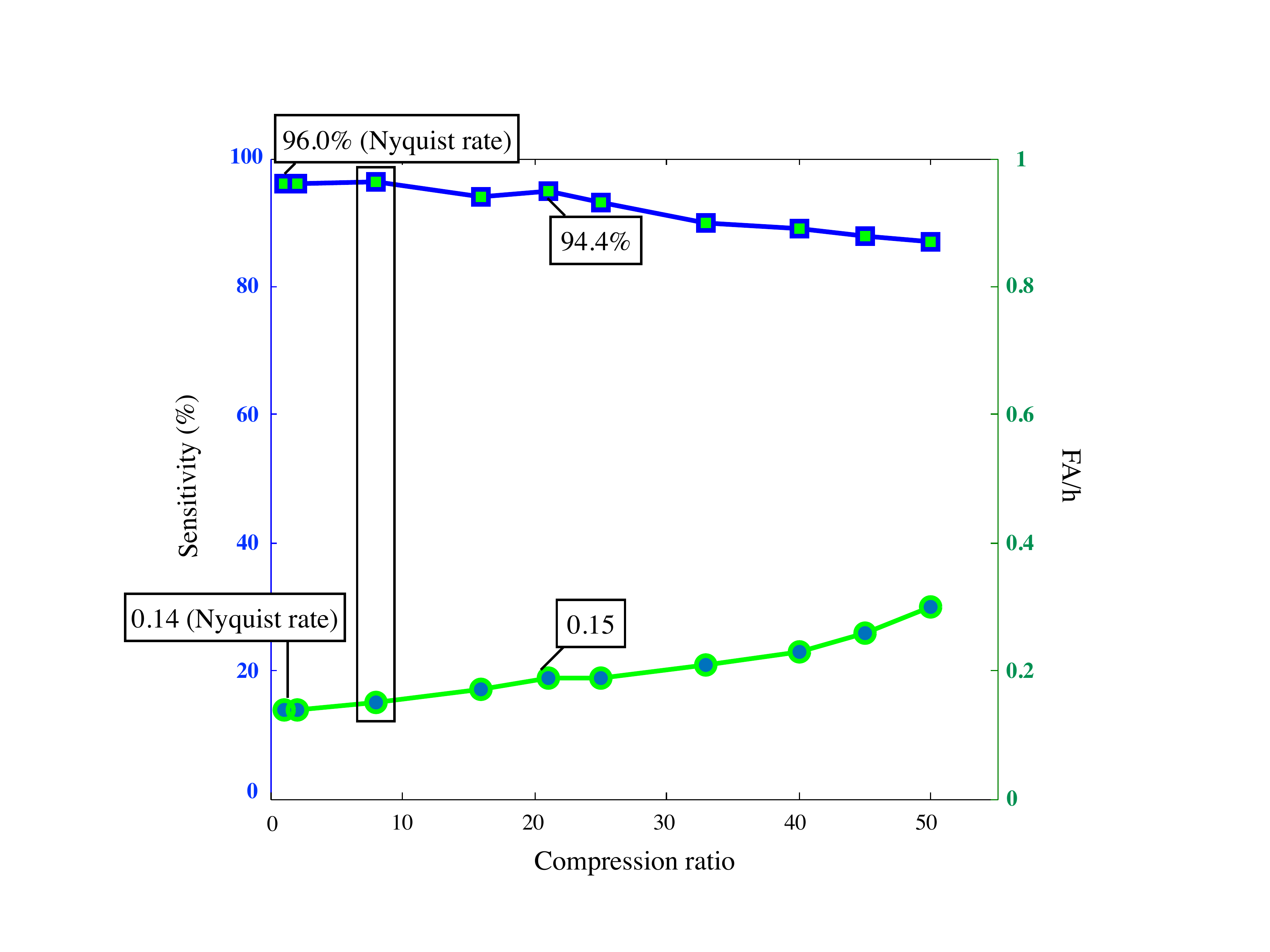}
\caption{Sensitivity and FA/h of seizure detection classification with
respect to compression ratio. Sensitivity and FA/h CS-based method using 
$\alpha=8\times$ are almost equal to the sensitivity and FA/h of the 
traditional method using Nyquist sampling ($\alpha=1\times$).} 
\label{figure:SN}
\end{figure}

Fig.~\ref{figure:SN} shows the sensitivity and FA/h for
seizure detection with respect to different compression ratios. A
compression ratio $\alpha$ of 8$\times$ is seen to maintain
sensitivity and FA/h for seizure classification. Moreover, an 8$\times$ 
compression ratio also exhibits similar results for arrhythmia detection 
\cite{shoaib, shoaib2}. Thus, we assume this ratio for deriving the next 
set of results.

Next, we examine the EEG sensor in the context of seizure detection.
Using the CS-based algorithm for seizure detection, the average value of 
total energy consumption of the EEG sensor (Table \ref{table:CS_EG}) is much 
less than that of the anomaly-driven signal processing method 
(Table \ref{table:totalEne_EEG_Trad}). Relative to the baseline, 
the total energy consumption of the EEG sensor is reduced by up to 724$\times$ 
in this scheme. Table \ref{table:CS_BAT} shows the battery lifetime of 
the EEG sensor, which improves by a similar ratio.

\begin{figure} [h]
\centering
\includegraphics [trim = 50mm 40mm 40mm 25mm ,clip, width=250pt,height=170pt]{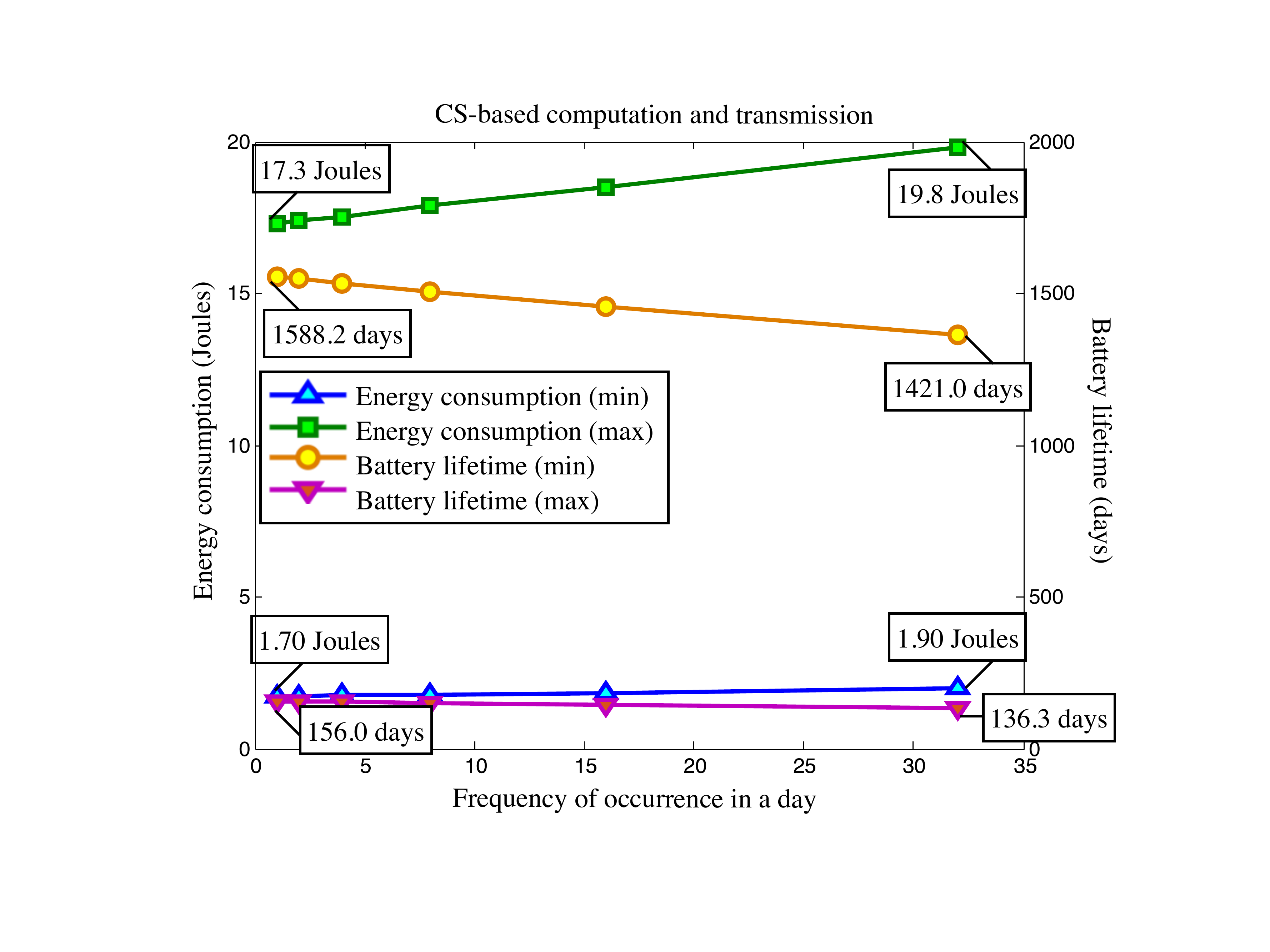}
\caption{Energy consumption and battery lifetime of the ECG sensor for the CS-based method with respect to frequency of occurrence of arrhythmia in a day.} 
\label{figure:arrCS_ENG}
\end{figure}

Next, we examine an ECG sensor in the context of arrhythmia detection. 
Fig.~\ref{figure:arrCS_ENG} shows the total energy consumption and battery lifetime of the ECG sensor with respect to 
the frequency of occurrence of arrhythmia in a day. Similar to the previous scheme, we assumed that 
after detecting an arrhythmia, the ECG sensor transmits the information of 
a standard one-minute ECG strip to the base station.

\begin{table}[ht] 
\caption{Average total energy consumption of the EEG sensor for 
CS-based computation} 
\centering 
\begin{tabular}{c c c} 
\hline\hline 
Sensor & Minimum (J/day) & Maximum (J/day)\\ [0.5ex] 
\hline
EEG & 6.93 & 9.50 \\[1ex] 
\hline 
\end{tabular} 
\label{table:CS_EG} 
\end{table} 

\begin{table}[ht] 
\caption{Average battery lifetimes of the EEG sensor for 
CS-based computation} 
\centering 
\begin{tabular}{c c c} 
\hline\hline
Sensor & Minimum (days) & Maximum (days)\\ [0.5ex] 
\hline 
EEG & 284.43 & 389.45 \\[1ex] 
\hline 
\end{tabular} 
\label{table:CS_BAT} 
\end{table} 

\subsection{Summary of proposed schemes}

Next, we summarize the results.

\begin{figure}[h]
\centering
\includegraphics [trim = 85mm 110mm 80mm 20mm ,clip, width=250pt,height=160pt]
{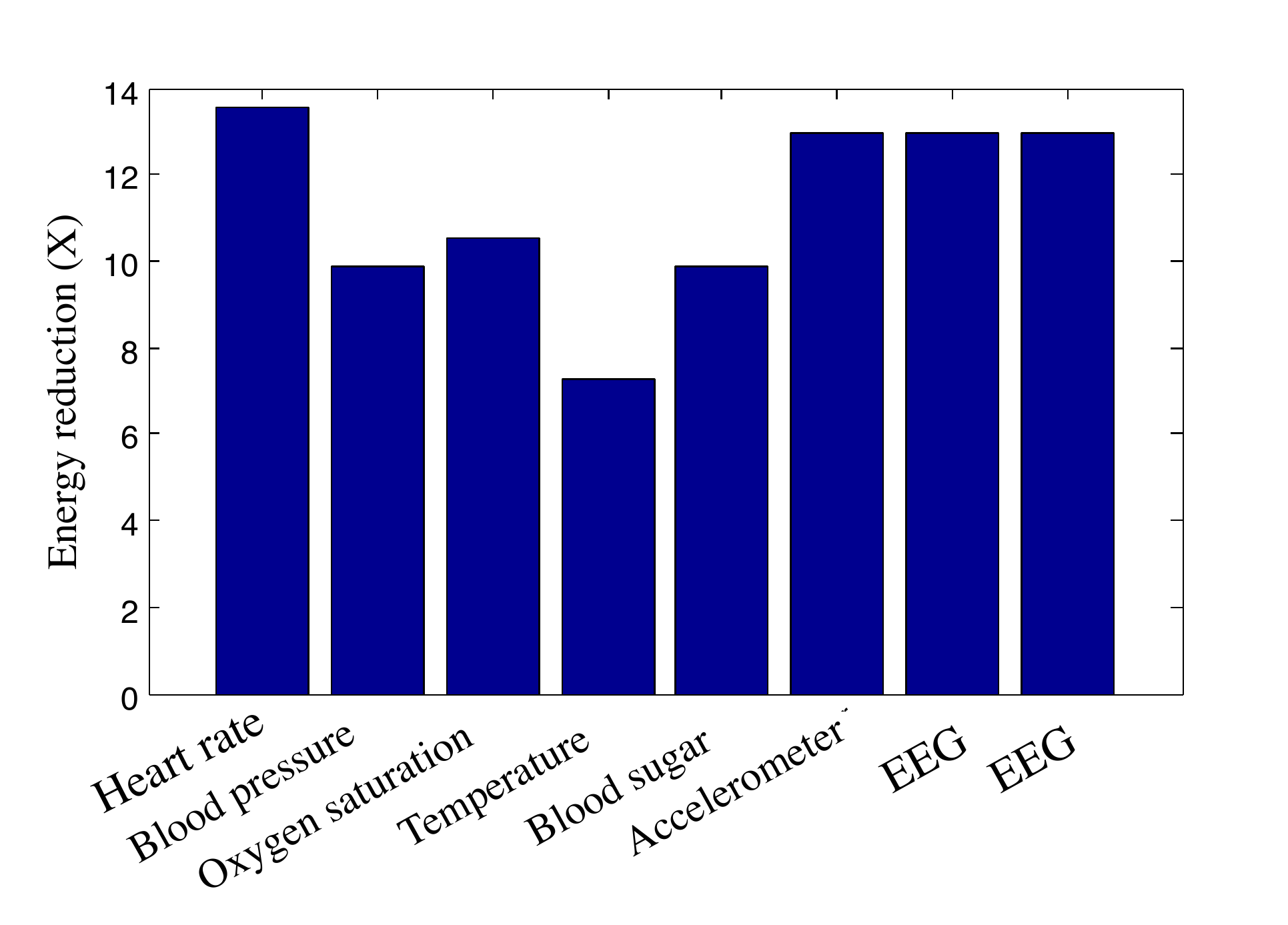}
\caption{Energy reduction in each sensor when the sensor accumulates 
multiple samples in one packet. Raw data are assumed to be gathered at the 
maximum frequency.}
\label{figure:bargraph_all}
\end{figure}

Fig.~\ref{figure:bargraph_all} shows the energy reduction in each sensor for 
the sample aggregation scheme.  The energy reduction is an order of
magnitude relative to the baseline.

\begin{figure}[h]
\centering
\includegraphics [trim = 65mm 118mm 60mm 118mm ,clip, width=250pt,height=120pt]{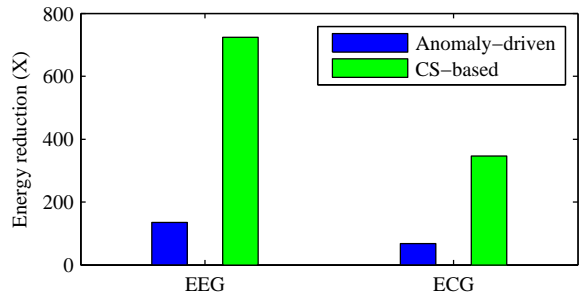}
\caption{Energy reduction in EEG and ECG sensors. The number of arrhythmia 
events in a day is assumed to be 32, and raw data are assumed to be gathered 
at the maximum frequency.}
\label{figure:bargraph_high}
\end{figure}

Fig.~\ref{figure:bargraph_high} shows the energy reduction in EEG and ECG 
sensors when the maximum sampling frequency is employed. The CS-based approach
can be seen to result in two to three orders of magnitude energy reduction
relative to the baseline.
 
\section{Storage requirements}
\label{STORAGE_SEC0}
We have described three schemes for decreasing the energy consumption
of sensors: (i) sample aggregation, (ii) anomaly-driven, and 
(iii) CS-based computation in the node. The first scheme cannot reduce the 
amount of required storage because we just accumulate multiple packets in 
order to reduce the number of transmissions, but we still transmit all the 
data. However, if we can process the raw data in the sensor nodes and just 
transmit a chunk of raw data that is essential for future analysis, we would 
be able to reduce the amount of required storage significantly.

When anomaly-driven or CS-based signal processing is
done on the sensor node, the sensor node samples, processes, and then 
transmits the data based on the result of processing. However, in the
case of CS-based computation, the data can be transmitted in compressed
form and reconstructed in the base station or server for
further processing if needed.  

Let us consider EEG sensors first. Based on the results in \cite{will}, we 
assume the mean seizure frequency and mean seizure duration to be 4.7 per 
month and 3.8 minutes, respectively. Therefore, as mentioned earlier, the EEG 
sensor needs to transmit information for a duration of 17.8 minutes per
month, on an average.  Table \ref{table:memory_siz} shows the average amount 
of storage required for storing the raw data in the two schemes for seizure 
detection based on EEG signal analysis. In this table, the minimum (maximum) 
value corresponds to the minimum (maximum) sampling frequency. The 
anomaly-driven scheme can be seen to reduce the amount of storage required for 
storing these signals by 2418$\times$. The CS-based scheme provides another 
8$\times$ reduction on top of this.

\begin{figure} [h]
\centering
\includegraphics[trim = 60mm 113mm 70mm 115mm ,clip, width=250pt,height=150pt] {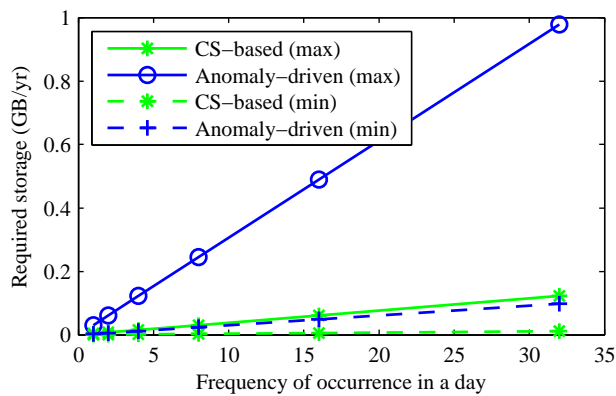}
\caption{The amount of storage required for storing important chunks of ECG 
signals based on the results of computation.}
\label{figure:memoryplot}
\end{figure}

As mentioned earlier, unlike seizures, the frequency with which arrhythmia 
occurs may vary significantly.  In order to provide a quantitative analysis 
for storage requirements in the case of arrhythmia detection, we assume that, 
after each detection, the sensor transmits the information of a standard 
one-minute ECG strip to the base station. Fig.~\ref{figure:memoryplot} shows 
the amount of required storage for the anomaly-driven and CS-based
schemes with respect to the frequency of occurrence.  Again, we observe the
significant advantage of the CS-based scheme. 
    
\begin{table}[ht] 
\caption{Average storage required for long-term storage of processed data} 
\centering 
\begin{tabular}{l c c} 
\hline\hline 
Sensor & Minimum (MB/yr) & Maximum (MB/yr)\\ [0.5ex] 
\hline 
EEG (Anomaly) & 1.87 & 18.65 \\
EEG (Compressed) & 0.23 & 2.33 \\[1ex] 

\hline
\end{tabular} 
\label{table:memory_siz} 
\end{table} 

\section{Choosing the appropriate scheme and hardware platform}
\label{COMPARE_SEC0}
In this section, we first compare the different schemes we presented, and 
discuss how the appropriate scheme can be chosen for each sensor. Second, we 
discuss two different types of hardware platforms: application-specific 
integrated circuit (ASIC) and general-purpose commercial products. We describe 
the potential benefits of using ASIC hardware.

\subsection{The appropriate scheme for each sensor}
Each scheme has its own advantages and disadvantages. For example, sample 
aggregation decreases energy consumption at the cost of increased latency. 
Schemes that use on-sensor computation can significantly increase battery 
lifetime, however, provide less raw data to the physicians. 

Choosing an appropriate scheme for each sensor depends on medical
considerations such as tolerable latency and patient's condition.
Next, we discuss different considerations that should be taken into account 
by designers, in addition to the battery lifetime and storage requirement. 

\noindent \textit{\textbf{Latency}}: Latency is the time interval between the occurrence of an 
anomaly and the response that is provided by medical devices, physicians 
or medical personnel. Tolerable latency depends on the patient's condition. 
\begin{itemize}
\item \textit{Example 1:} Consider a continuous health monitoring system that 
is used to monitor a healthy subject who does not have any history of a 
serious illness. The system can be configured for this subject to provide 
routine medical check by collecting and sending medical information to 
physicians or hospitals at long intervals (e.g., once a day). In fact, 
latency is not an important factor in this case, and the sensor can be 
configured to minimize the energy and storage requirements. For example, we 
can use the CS-based computation method for both EEG and ECG sensors and use 
the aggregation method for other low-frequency sensors (e.g., temperature) to 
maximize the battery lifetime of all sensors.

\item \textit{Example 2:} Consider a continuous health monitoring system
that is used to monitor a subject who has previously been diagnosed with high 
blood glucose. As a result, any rapid rise in blood glucose should be detected 
and addressed immediately. In such a scenario, the latency that might be added 
by using sample aggregation for blood glucose levels may not be acceptable.
\end{itemize}

Among all the discussed schemes, sample aggregation is the only one that may 
lead to a non-negligible increase in latency. Therefore, the number of samples 
that can be aggregated in one packet before transmission can be limited by 
the latency that can be tolerated.

\noindent \textit{\textbf{2. Amount of raw medical data transmitted}}: Physicians may want to examine 
raw medical data over a specific time period to verify on-sensor computation. 
The amount of raw information that needs to be transmitted and stored for 
further analysis varies from one device to another. It also depends on 
the medical condition of the patient.

Schemes that use on-sensor computation (anomaly-driven transmission and 
CS-based computation and transmission) only transfer a small portion of 
raw data containing important information about the occurrence of the anomaly. 
However, if more medical information is required to be transferred to the base 
station, the designers should use the other schemes or send more raw 
data (e.g., over an hour of measurements) after detecting an anomaly. 

\noindent \textit{\textbf{3. Extensibility}}: This is a design consideration where the
implementation takes future modifications of the algorithms into
consideration. High extensibility implies that applications of a biomedical 
sensor can be extended in the future with a minimum level of effort. 
Generally, schemes that rely on on-sensor computation are less extensible in 
comparison to schemes that transfer raw data to the base station due to the 
fact that they are designed to minimize the energy consumption and the amount 
of required storage in certain applications (e.g., arrhythmia detection). 
Therefore, if a physician wants to change the computation algorithm of the 
medical device, another device should be designed and used, or at least the 
device's firmware should be updated each time.

Table \ref{table:ADVAN} compares various schemes. 

\begin{table*}[t] 
\caption{Comparison of different schemes} 
\centering 
\begin{tabular}{l c c c c c c} 
\hline\hline 
Scheme & Latency & Amount of raw
data transmitted & Extensibility  \\ [0.5ex]
\hline 
Baseline  & Low & All raw data & High\\ 
Sample aggregation & Varies & All raw data & High\\ 
Anomaly-driven & Low & A portion of collected data & Low \\ 
CS-based & Low & A portion of compressed data & Low \\ 
\hline 
\end{tabular} 
\label{table:ADVAN} 
\end{table*} 

Potentially, different schemes can be used in the health monitoring system 
for different sensors.  Since the sensors are located on different parts
of the body, they cannot share on-sensor resources (e.g., the battery).
Thus, their battery lifetimes are independent.

We can also use a combination of schemes even in just one sensor. For example, 
we can combine one of the schemes that uses on-sensor computation 
(anomaly-driven or CS-based) 
with the sample aggregation scheme to reduce total energy consumption even 
more. However, since in anomaly-driven and CS-based schemes, the computation 
energy is dominant and the transmission energy is only a small fraction of 
total energy consumption, the addition of the sample aggregation scheme 
will not provide a significant energy reduction.

\subsection{The hardware platform}
An appropriate hardware platform can be chosen from various
general-purpose commercial products or else designed as ASIC hardware.
General-purpose commercial products enable the designers to implement an 
algorithm or prototype of a biomedical sensor quickly. However, they are not 
optimized for the specific application. Anomaly-driven and CS-based schemes 
use some algorithms to process the raw data on the EEG or ECG sensor nodes 
before transmission. An ASIC could be designed for these algorithms. In 
particular, in our computation schemes, the on-sensor computation algorithm 
uses a support vector machine as a classifier to detect anomalies 
(arrhythmia and seizure). Specialized processors and architectures that enable efficient handling of data structures used by the classifier could reduce computation energy even further \cite{Amirali_1, Amirali_2, Amirali_3, Amirali_4, HOS_ENER_1, HOS_ENER_2, HOS_ENER_3}. Further energy reduction can be achieved 
through supply voltage scaling. The total energy is determined primarily by 
the sum of dynamic (active-switching) energy and the static (leakage) energy. 
However, reduction in active-switching energy due to supply voltage scaling 
is opposed by an increase in leakage energy. Therefore, there is an optimal 
supply voltage at which the circuit attains its minimum energy consumption 
and still work reliably. This could be addressed in an ASIC. However, such an 
ASIC may not be desirable from a cost perspective and does not improve 
transmission energy. 

\section{Chapter summary}
\label{CONC_SEC0}
In this chapter, we discussed a secure energy-efficient system for long-term continuous health monitoring.  We discussed and evaluated various schemes with the help of eight biomedical sensors that would typically be part of a WBAN.  We also evaluated the storage requirements for long-term analysis and storage. 

Among the four schemes we evaluated (including the baseline scheme), we showed 
that the CS-based scheme provides the most computational energy savings 
(e.g., up to 724$\times$ for ECG sensors) because it needs to process much 
fewer signal samples.  For low-sample-rate sensors, we can achieve significant 
energy savings by simply accumulating the raw data before transmitting them to 
the base station. 

In addition, the CS-based scheme also allows us to reduce the 
storage requirements significantly.  For example, for an EEG sensor based 
seizure detection application, we achieve total storage savings of up to 
19344$\times$. The results indicate that long-term continuous health 
monitoring is indeed feasible from both energy and storage points of view.

Finally, we compared all proposed schemes and discussed how a continuous 
long-term health monitoring system should be configured based on patients' 
needs and physicians' recommendations.

%% file: ch-OpSecure/chapter-OpSecure.tex
\chapter{OpSecure: A Secure Optical Communication Channel for Implantable Medical Devices \label{ch:OpSecure}}

Implantable medical devices (IMDs) are opening up new opportunities for 
holistic health care by enabling continuous monitoring and treatment of various 
medical conditions, leading to an ever-improving quality of life for patients. 
Integration of radio frequency (RF) modules in IMDs has provided wireless 
connectivity and facilitated access to on-device data and post-deployment 
tuning of essential therapy. However, this has also made IMDs susceptible to 
various security attacks. Several lightweight encryption mechanisms have been 
developed to prevent well-known attacks, e.g., integrity attacks that send 
malicious commands to the device, on IMDs. However, lack of a secure key 
exchange protocol (that enables the exchange of the encryption key while 
maintaining its confidentiality) and the immaturity of already-in-use wakeup 
protocols (that are used to turn on the RF module before an authorized data 
transmission) are two fundamental challenges that must be addressed to ensure 
the security of wireless-enabled IMDs.

In this chapter, we introduce OpSecure, an optical secure communication 
channel between an IMD and an external device, e.g., a smartphone. OpSecure 
enables an intrinsically user-perceptible unidirectional data transmission, 
suitable for physically-secure communication with minimal size and 
energy overheads. Based on OpSecure, we design and implement two protocols: 
(i) a low-power wakeup protocol that is resilient against remote battery 
draining attacks, and (ii) a secure key exchange protocol to share the 
encryption key between the IMD and the external device. We investigate the two 
protocols using a human body model \cite{OPSECURE_PAPER}.

\section{Introduction}
IMDs are revolutionizing health care by offering 
continuous monitoring, diagnosis, and essential therapies for a variety of 
medical conditions. IMDs can capture, process, and store various types of 
physiological signals, and are envisioned as the key to enabling a holistic 
approach to health care \cite{CNIA}. Rapid technological advances in wireless 
communication, sensing, signal processing, and low-power electronics are 
transforming the design and development of IMDs. State-of-the-art IMDs, e.g., 
pacemakers and implantable drug infusion systems, commonly support short-range 
wireless connectivity, which enables remote diagnosis and/or monitoring of 
chronic disorders and post-deployment therapy adjustment \cite{Zero-Power}. 
Moreover, wireless connectivity allows health care professionals to 
non-intrusively monitor the device status, e.g., physicians can gauge the 
device battery level without performing any surgery. 

Despite the numerous services that wireless connectivity offers, it may make 
an IMD susceptible to various security attacks. Previous research efforts 
\cite{Zero-Power,GOL_MIT,PHYSIO,SECURE_VIBE,PrincetonInsulin,MEDMON} have demonstrated how wireless 
connectivity may be a security loophole that can be exploited by an attacker. 
For example, Halperin et al.~\cite{Zero-Power} show how an attacker can exploit 
the security susceptibilities of the wireless protocol utilized in an 
implantable cardioverter defibrillator (ICD) to perform a battery draining 
attack against the device. This is an attack that aims to deplete the device 
battery by frequently activating/using the RF module. Moreover, they
show that it is feasible to exploit these susceptibilities to change on-device 
data or the current operation of the device. Gollakota et al.~\cite{GOL_MIT} 
explain how an adversary can eavesdrop on an insecure communication channel 
between an IMD and its associated external device to extract sensitive 
information about the patient.

To prevent battery draining attacks, an attack-resilient wakeup protocol, 
which activates the RF module before every authorized communication, must be 
used. Today's IMDs often employ a magnetic switch, which turns on their RF 
module in the presence of an external magnet. Unfortunately, it has been shown 
that magnetic switches cannot prevent battery draining attacks since they can 
be easily activated by an attacker (without the presence of a nearby magnet) if 
a magnetic field of sufficient strength is applied \cite{MAG_VUL,Zero-Power}. 

In order to secure the RF wireless channel between the IMD and the external 
device and avert the risk of eavesdropping on the channel, the use of 
cryptographic techniques, e.g., data encryption, has been suggested 
\cite{SUG_1,SUG_2}. However, traditional cryptographic techniques are not 
suitable for IMDs due to limited on-device resources, e.g., limited
storage and battery energy. For example, asymmetric encryption mechanisms are 
not applicable to resource-constrained IMDs since they would significantly 
decrease the IMD battery lifetime \cite{SECURE_VIBE_4,SUG_1}. Several 
lightweight symmetric encryption mechanisms have been proposed in the last 
decade to ensure the security of communication protocols utilized in 
IMDs (see \cite{L_SURV} for a survey). While symmetric cryptography may offer 
a secure lightweight solution, it is greatly dependent on a secure key 
exchange protocol. Such a protocol enables sharing of the encryption key 
between the IMD and the external device. As extensively described later in 
Section \ref{RELATED}, previously-proposed key exchange protocols have
various shortcomings since \textit{they either add significant overheads to 
the IMD or are susceptible to remote eavesdropping.}

In this chapter, we present practical key exchange and wakeup protocols, which 
complement lightweight symmetric encryption mechanisms, to thwart common 
security attacks against insecure communication channels. 
We introduce a secure optical communication channel, which we call 
OpSecure. We discuss the design and implementation of a low-power wakeup 
protocol and a secure key exchange protocol based on OpSecure.
Our main contributions can be summarized as follows:
\begin{enumerate}
\item We introduce OpSecure, an optical secure unidirectional (from the 
external device to the IMD) communication channel.
\item We present an attack-resilient low-power wakeup protocol for IMDs based on OpSecure.
\item We propose a secure key exchange protocol, which enables sharing
of the encryption key between IMDs and their associated external devices.  
\item We discuss the design and implementation of a prototype IMD platform that 
supports the proposed protocols and present evaluation results for the
prototype.  
\end{enumerate}

The remainder of this chapter is organized as follows. In Section 
\ref{PROBLEM}, we explain why wakeup and key exchange protocols are essential 
for IMDs and briefly discuss the shortcomings of previously-proposed protocols. 
We present OpSecure and summarize its advantages in Section \ref{PROPOSED}. 
We also propose a wakeup protocol and a key exchange protocol based on 
OpSecure. In Section \ref{EXP}, we describe our prototype and experimental 
setup. We evaluate the prototype implementation that supports both the proposed 
protocols (wakeup and key exchange) in Section \ref{EXAM}. Finally, we conclude 
in Section \ref{CONC}.

\section{Problem definition}
\label{PROBLEM}
In this section, we first explain the role of wakeup and key exchange protocols 
in providing secure communication for IMDs. Then, we present a brief 
overview of prior related work on these protocols and summarize their
shortcomings.  

\subsection{Wakeup and key exchange protocols}
\begin{figure}[ht]
\centering
\includegraphics[trim = 52mm 73mm 66mm 80mm ,clip, width=240pt,height=140pt]{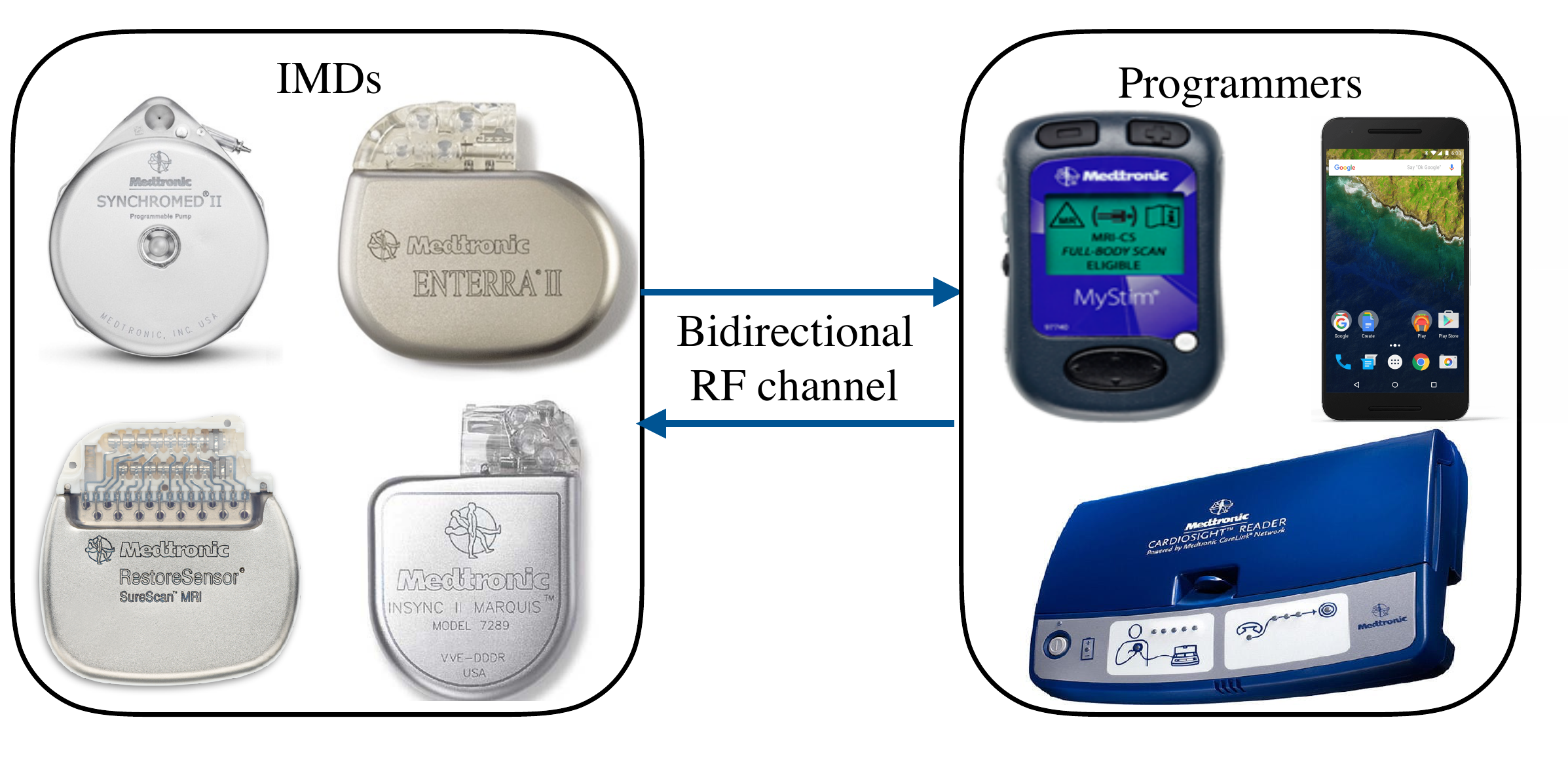}
\caption{Overall system architecture: IMD and external device have a 
bidirectional RF channel that supports symmetric encryption, e.g., Bluetooth 
Low Energy.} 
\label{fig:ARCH_OPSECURE}
\end{figure}

As mentioned earlier, the IMD and its associated external device commonly have 
an RF channel that is used for bidirectional data communication. 
We assume that both devices are capable of using symmetric encryption for 
protecting the data sent over the RF channel. The overall system architecture 
that we target is illustrated in Fig.~\ref{fig:ARCH_OPSECURE}. 

Due to severe on-sensor energy constraints, the RF module must be enabled 
only when absolutely needed, e.g., when an authorized physician wants to 
access on-device data. Thus, prior to each data transmission, the RF module 
should be activated using a pre-defined wakeup protocol. This protocol must 
satisfy two main design requirements. First, it must be resilient against 
battery draining attacks so that an attacker cannot activate the RF module. 
Second, it should add negligible size and energy overheads to the device. 

After enabling the RF module by the wakeup protocol, data can be transmitted 
over the bidirectional communication channel that supports symmetric 
encryption. Since symmetric encryption is based on an encryption key, an 
exchange protocol must be used to securely exchange the encryption key between 
the IMD and the external device. Every practical key exchange protocol must 
satisfy the following design requirements. First, it must guarantee the 
confidentiality of the encryption key and be resilient to remote eavesdropping. 
Second, its size and energy overheads must be minimal. Third, it must ensure 
that health care professionals can access and control the IMD without a notable 
delay in an emergency situation in which the patient needs immediate medical 
assistance.   

\subsection{Related work}
\label{RELATED}
Next, we summarize previous research efforts on both wakeup and key exchange 
protocols and highlight their shortcomings.  

\subsubsection{Wakeup protocols}
As mentioned earlier, a magnetic switch is commonly integrated into today's 
IMDs to turn on the RF module when needed. However, magnetic switches are 
vulnerable to battery draining attacks since they can be remotely activated 
\cite{Zero-Power}. A few wakeup protocols have recently been presented in the 
academic literature. For example, the wakeup protocol presented by Halperin 
et al.~\cite{Zero-Power} relies on an authentication technique in which the IMD 
harvests the RF energy supplied by the external device itself. The RF module 
is powered by the battery only after the external device is authenticated. 
However, the RF energy harvesting subsystem needs an antenna, which imposes a 
significant size overhead on the IMD. Kim et al.~\cite{SECURE_VIBE} suggest a 
wakeup scheme in which the IMD activates the RF module when it detects the 
vibration generated by an external electrical motor. Their scheme adds minimal 
size and energy overheads to the IMD since it only needs the addition of a 
low-power accelerometer to the IMD. However, in practice, the patient's 
regular activities, e.g., running, may unintentionally and frequently turn on 
the RF module, and as a result, deplete the device battery.

\subsubsection{Key exchange protocols}
The use of a pre-defined password, which is stored on the device and known 
to the user, is a longstanding tradition in the security community. However, 
a key exchange approach that needs active user involvement, e.g., asking the 
user to remember a password and give it to authorized physicians upon
request, is not suitable for IMDs since the user may not be able to cooperate 
with health care professionals in an emergency, e.g., when the patient is 
unconscious. In order to minimize user involvement, previous research studies 
have proposed several user-independent key exchange protocols. Next, we 
summarize them and discuss their shortcomings.

\noindent \textit{\textbf{1. Ultraviolet tattoos:}} Schechter \cite{UV_TATOO} presented a 
scheme in which a fixed user-selected human-readable key is tattooed directly 
on the patient's body using ultraviolet ink. In this protocol, 
all devices that need to communicate with the IMD must be equipped with a 
small, reliable, and inexpensive ultraviolet light-emitting diode and an input 
mechanism for key entry. This tattoo-based approach has two limitations. 
First, the design requires the patient to agree to acquire a tattoo, which
significantly limits its applicability \cite{PASS_TATOO}. Second, if the 
password becomes compromised, access by the attacker cannot be prevented easily 
since the password cannot be changed in a user-convenient manner. 

\noindent \textit{\textbf{2. Physiological signal-based key generation:}} A few 
physiological signal-based key generation protocols have been proposed 
\cite{SECURE_VIBE_13, SECURE_VIBE_14, SECURE_VIBE_15}, which can be used to 
generate a shared key for the IMD and the external device from synchronized 
readings of physiological signals, such as an electrocardiogram (ECG). 
Unfortunately, the robustness and security properties of keys generated using 
such techniques have not been well-established \cite{SECURE_VIBE}.

\noindent \textit{\textbf{3. Using an acoustic side channel:}} Halperin et al.~presented a key 
exchange protocol based on acoustic side channels in \cite{Zero-Power}. 
Unfortunately, their protocol is susceptible to remote acoustic eavesdropping 
attacks \cite{REM_ACO} and, as a result, does not offer a secure key exchange 
protocol. Moreover, it is not reliable in noisy environments since they 
utilized a carrier frequency within the audible range. Furthermore, it imposes 
a significant size overhead \cite{SECURE_VIBE}.

\noindent \textit{\textbf{4. Using a vibration side channel:}} Kim et al.~\cite{SECURE_VIBE} 
proposed a key exchange protocol that relies on a vibration side channel, 
i.e., a channel in which the transmitter is a vibration motor, and the receiver 
is an accelerometer embedded in the IMD. This protocol requires negligible 
size and energy overheads. However, it has two shortcomings. First, since 
electrical motors generate capturable electromagnetic and acoustic waves 
during their normal operation \cite{PHYSIO}, an adversary might be able to 
extract the key from signals leaked from the vibration motor. Second, since 
the method uses an accelerometer to detect vibrations, regular physical 
activities, e.g., running, may be interpreted as key transmission. This can 
reduce the battery lifetime of the IMD since the device needs to listen to 
the communication channel even when there is no actual transmission.

In this chapter, we aim to address the above-mentioned shortcomings of 
previously-proposed protocols, \textit{in particular size/energy overheads 
and vulnerability to eavesdropping}, through a simple low-power, yet secure 
key exchange protocol using \textit{visible light}. 

\section{The proposed channel and protocols}
\label{PROPOSED}
In this section, we first describe OpSecure and highlight its advantages. 
Then, we discuss the two proposed protocols that are based on OpSecure.

\subsection{OpSecure: The proposed channel}
Optical data transmission (also called light-based wireless communication) is 
a well-known communication type that has attracted increasing attention in 
recent years due to its potential to offer high-speed wireless communication 
(as a complement or an alternative to WiFi) for a variety of portable devices, 
e.g., smartphones and laptops. Previous research studies \cite{LIGHT_1,LIGHT_2} 
demonstrate that optical communication channels can enable high-rate data 
transmission (the transmission rate can vary from several hundred $Mb/s$ to a 
few $Gb/s$). In an optical channel, data packets flow from a light source 
(transmitter) to a light sensor (receiver). Therefore, to establish a 
bidirectional communication channel between two devices, both devices must 
have a light source and a light sensor.  

There is a basic domain-specific challenge that must be addressed when 
developing an optical communication scheme for IMDs: integration of light 
sources into IMDs imposes significant size and energy overheads on these 
resource-constrained devices. Hence, it is not feasible to transmit data from an IMD to an external device via an optical channel even though such a 
channel can potentially enable two-way communication. Unlike light sources, 
state-of-the-art already-in-market light sensors are sufficiently compact and 
energy-efficient to be embedded in IMDs. Therefore, a one-way communication 
channel, which transmits data from the external device to the IMD, can be 
implemented with minimal size and energy overheads. We implement such a 
channel and call it OpSecure.

OpSecure is intrinsically secure due to its close proximity requirement and 
high user perceptibility. Visible light attenuates fast in the body and, hence, 
can only be captured within a very close range. As demonstrated later in 
Section \ref{EXAM}, if the light source is in contact with the body and 
directed at the IMD, it can penetrate deep enough into the body to reach the 
IMD. However, a passive adversary cannot eavesdrop on OpSecure without an 
eavesdropping device attached to the body, which is very likely to be noticed 
by the patient. 

\begin{figure}[ht]
\centering
\includegraphics[trim = 52mm 43mm 66mm 80mm ,clip, width=240pt,height=180pt]{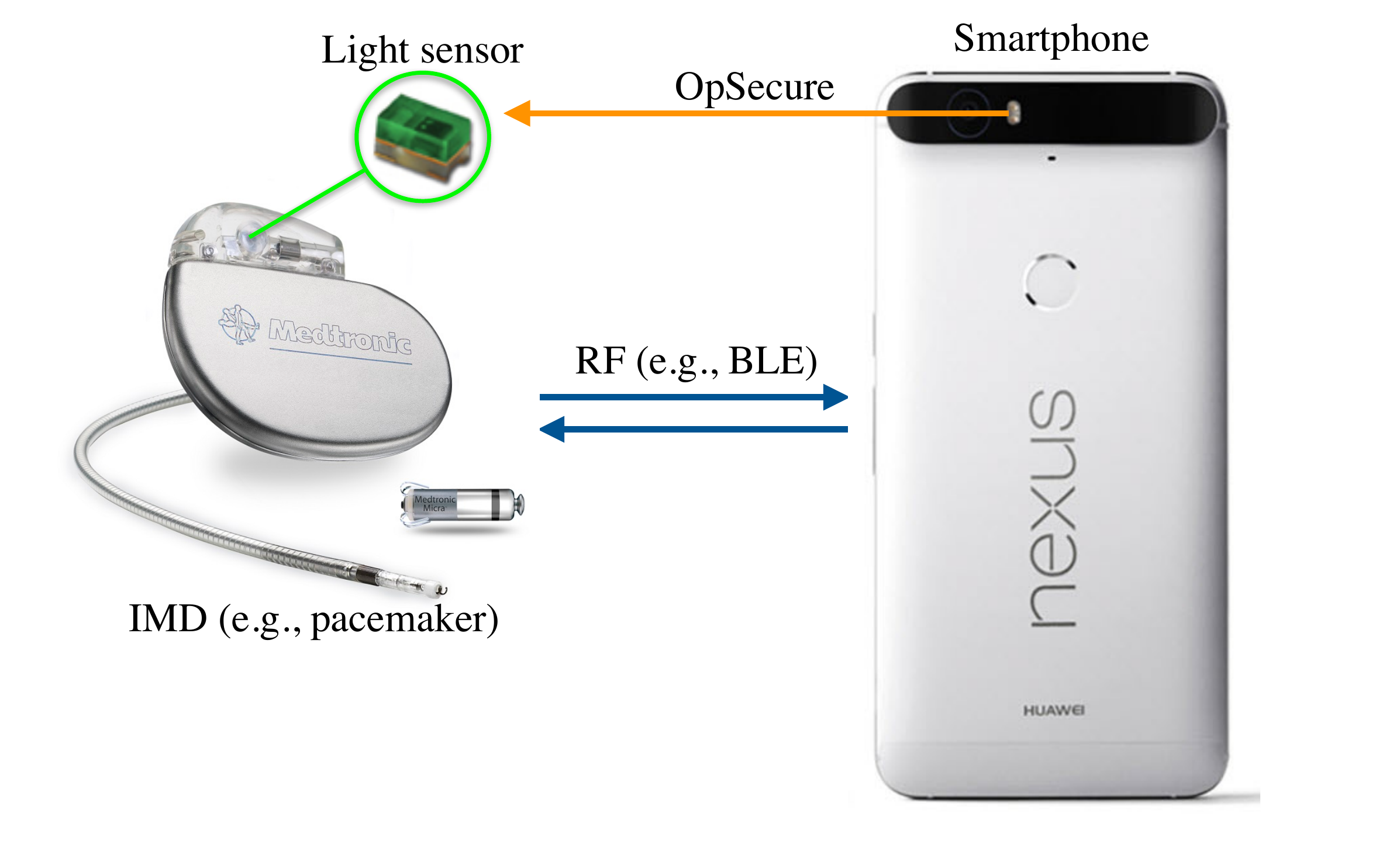}
\caption{The IMD (pacemaker) has an embedded light sensor, and the 
smartphone flashlight acts as a light source.} 
\label{fig:SM_PC}
\end{figure}

As illustrated in Fig.~\ref{fig:ARCH_OPSECURE}, the external device may vary from 
specialized IMD programmers, i.e., external devices that are specifically 
designed to query the IMD data or send commands to the IMD, to general-purpose 
portable devices, e.g., smartphones. As in the case of vibration-based
key exchange \cite{SECURE_VIBE}, we implement our prototype using a smartphone 
used as the external device (see Fig.~\ref{fig:SM_PC}). This has three 
advantages. First, the component 
that we need in the external device for establishing OpSecure is already 
present in smartphones (the flashlight can be used as a light source). Second, 
smartphones have become the dominant form of base stations for a large number 
of medical devices since they are ubiquitous and powerful, and provide 
various technologies needed for numerous applications \cite{SMART_P1}. As a 
result, they can be used as a base station for collecting and processing 
several types of physiological data (including data collected by IMDs) 
\cite{CNIA}. Third, smartphones can easily support highly-secure encrypted 
transmission, which deters several potential attacks against the 
IMD \cite{CABA}.

However, note that OpSecure can also be implemented on other devices that
are used to communicate with the IMD with minimal overheads if they can be 
equipped with a small light source and an input mechanism for key entry.


\subsection{The proposed protocols}
Next, we describe both the wakeup and key exchange protocols that we have 
developed based on OpSecure.  

\subsubsection{Wakeup protocol} 
As mentioned earlier, when the light source is in contact with the human body, 
visible light can penetrate deep enough into the body to reach the IMD.
However, it attenuates very fast in the body. We exploit this fundamental 
characteristic of visible light to develop a wakeup protocol that works as 
follows. The smartphone fully turns on its flashlight and the IMD wakes up 
periodically to check if a light source is on the body, i.e., it checks if the 
intensity of the light received by the IMD is above a pre-defined threshold. 
\textit{The presence of an on-body light source} that points to the IMD is 
interpreted as the presence of a trusted external device.

As shown later in Section \ref{EXAM}, the proposed wakeup protocol adds 
minimal size and energy overheads to the device. Unlike a majority of 
previously-proposed protocols, it also provides immunity against battery 
draining attacks. In fact, an attacker, who wants to wake up the RF module, 
needs to attach a light source to the patient's body at a location close to 
the IMD. Such an action can be easily detected by the patient.

\subsubsection{Key exchange protocol} 
Assuming the IMD and the external device use a bidirectional RF communication 
protocol that supports symmetric encryption, our protocol can be used to 
transmit a randomly-generated key from the smartphone to the IMD. For each key 
exchange:

\noindent \textit{\textbf{Step 1:}} The smartphone first generates a random key 
$K \in \{{0,1}\}^N = k_1k_2...k_N$ of length $N$, and prepares a key packet as 
$Key_{packet}= Pre || K || Post$, where $Pre$ and $Post$ are two fixed binary sequences that are concatenated with the key to mark the beginning and end of a key packet.    

\noindent \textit{\textbf{Step 2:}} The physician places the smartphone on the patient's body so that its 
flashlight is directed at the light sensor of the IMD (IMDs commonly have a 
fixed location and can be easily detected by the physician).  

\begin{figure}[ht]
\centering
\includegraphics[trim = 35mm 106mm 85mm 45mm ,clip, width=240pt,height=55pt]{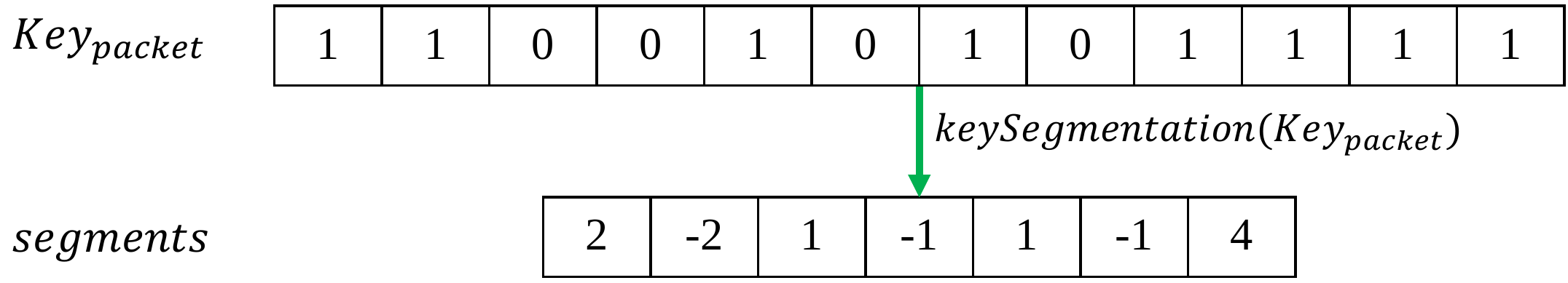}
\caption{$keySegmentation$ outputs $segments[]$ given $Key_{packet}$.} 
\label{fig:example}
\end{figure}

\noindent \textit{\textbf{Step 3:}} The external device uses on-off keying (OOK)
modulation to transmit $Key_{packet}$: the flashlight is turned on (off)
for a fixed period of time ($T_{step}$) to transmit bit ``1'' (``0'').
\textit{Algorithm 1} shows a simplified pseudo-code for this step. It
first computes $T_{step}=\frac{1000}{R}$ $ms$, where $R$ is the
transmission rate given by the user. Then, it calls the $keySegmentation$ 
procedure, which divides $Key_{packet}$ into smaller segments such that each 
segment only consists of all ``1''s or all ``0''s. The $keySegmentation$ 
procedure outputs an array of integer numbers ($segments[]$) so that: (i) the 
absolute value of each element in the array represents the length (the number 
of bits) of each of the above-mentioned segments, and (ii) the sign of the 
element shows whether bits of the segment are all ``1''s or all ``0''s, i.e., 
if all bits in the $i$th segment are ``1'', $segments[i]>0$, otherwise, 
$segments[i]<0$ (see Fig.~\ref{fig:example} for an example). Finally, 
\textit{Algorithm 1} turns on/off the flashlight with respect to the absolute 
values of the elements of $segments[]$ and $T_{step}$, i.e., 
$Abs(segment)*T_{step}$ determines how long the flashlight must be kept on/off. 

\noindent {\em Algorithm 1: flashlightControl procedure}

\noindent\makebox[\linewidth]{\rule{15.2cm}{0.1pt}}

\noindent Given: The key packet ($Key_{packet}$) and transmission rate ($R$)\\
\noindent\makebox[\linewidth]{\rule{15.2cm}{0.5pt}}
\begin{center}
\vspace{-1cm}
\noindent\makebox[\linewidth]{\rule{15.2cm}{0.5pt}}
\end{center}
{\small
1. $T_{step} \leftarrow 1000/R $ \\
2. $segments[] \leftarrow keySegmentation(Key_{packet})$\\
3. $For$ $each$ $segment$ $in$ $segments[]$\\
4. \indent $If (segment>0)$\\
5. \indent \indent$turnTheLightOn(Abs(segment)*T_{step})$\\
6.\indent $else$\\
7. \indent\indent $turnTheLightOff(Abs(segment)*T_{step})$\\
8. \indent $end$\\
9. $end$\\
}
\noindent\makebox[\linewidth]{\rule{15.2cm}{0.5pt}}

\begin{center}
\vspace{-1.05cm}
\noindent\makebox[\linewidth]{\rule{15.2cm}{0.5pt}}
\end{center}

\noindent \textit{\textbf{Step 4:}} The IMD demodulates the received visible light and recovers 
$Key_{packet}$. Then, it extracts $K$ from $Key_{packet}$ by removing $Pre$ 
and $Post$. Thereafter, it encrypts a fixed pre-defined confirmation message 
$M_{confirm}$ using $K$ and transmits this message $C=ENC(M_{confirm}, K)$ to the 
smartphone.  

\noindent \textit{\textbf{Step 5:}} The smartphone checks if it can successfully decrypt the received 
message $C$ using $K$, i.e., if $DEC(C, K)=M_{confirm}$. If the message 
can be successfully decrypted, the smartphone knows that the IMD received the 
key $K$ correctly, and then subsequent RF data transmissions are encrypted 
using key $K$.

In addition to key exchange, the above-mentioned protocol (the first three 
steps) can be used to transmit data/commands from the smartphone to the IMD 
without using the RF module. For example, a predefined stream of bits can be 
reserved for the shutdown command, i.e., a command that entirely disables the 
device, and sent using this protocol when needed. Note that the IMD cannot 
provide any feedback via OpSecure since the channel is unidirectional. However, modern IMDs commonly have an embedded beeping component that warn the patient in different scenarios, e.g., when the RF module is activated \cite{Zero-Power} or when the device's battery level is low \cite{BEEP}. Such a component can be also used to provide feedback when the IMD receives a predefined message over 
OpSecure, e.g., the beeping component can generate three beeps when the IMD 
receives the shutdown command via OpSecure.

\section{The prototype implementation and body model}
\label{EXP}
In this section, we first describe the wireless-enabled IMD that we 
implemented and the smartphone application that we developed. Then, we 
describe the human body model, which we used to evaluate the prototype 
implementation.

\subsection{Prototype implementation}
\begin{figure}[ht]
\centering
\includegraphics[trim =62mm 13mm 30mm 14mm ,clip, width=230pt,height=210pt]{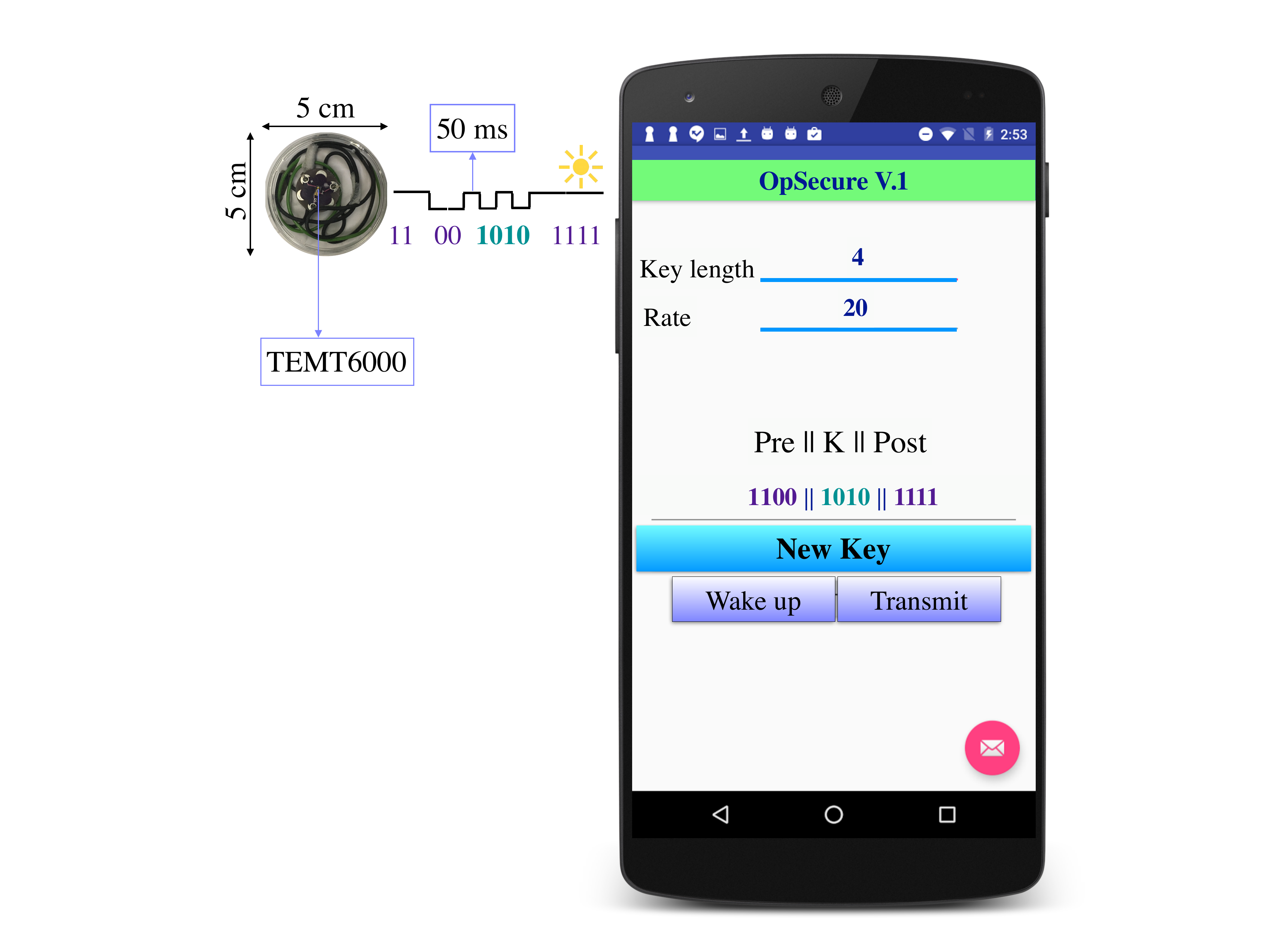}
\caption{The smartphone generates a 4-bit key and transmits the key over OpSecure. The application allows the user to control both the key length ($N$) and transmission rate ($R$).} 
\label{fig:IMP}
\end{figure}

As mentioned earlier, OpSecure establishes a unidirectional communication 
channel between the IMD and the external device. We 
implemented a wireless-enabled IMD prototype based on ATmega168V \cite{ATMEL} 
(a low-power microprocessor from Atmel), TEMT6000 \cite{TEMT6000} (an ambient 
light sensor from Vishay Semiconductors), and RFD77101 \cite{BLE_RF} (a 
Bluetooth Low Energy module from Simblee). The prototype does not offer any 
health monitoring/therapeutic operations. Indeed, it only implements the two 
proposed protocols. TEMT6000 enables OpSecure by receiving visible light 
generated by the user's smartphone, and RFD77101 provides the bidirectional RF communication that can be secured using a symmetric encryption mechanism for which the key can be exchanged over OpSecure. 
We also developed an Android application that can be used to either wake the 
IMD up or generate and transmit a random key to the IMD using the smartphone 
flashlight. The application allows the user to set the key length ($N$) and 
transmission rate ($R$). Fig.~\ref{fig:IMP} illustrates the application 
and the prototype. It also demonstrates how the application turns the 
flashlight on/off to transmit the key. For the key exchange example shown in 
Fig.~\ref{fig:IMP}, the transmission rate and key length are set to 
20 $b/s$ and 4 $b$ (in practice, the $N$ used would be much higher, e.g., 64 $b$), 
respectively. Thus, the smartphone needs 
$T_{step}=\frac{1000}{4}=50$ $ms$ for transmitting a single bit. In 
this implementation, the application uses two 4-bit sequences (``1100'' and 
``1111'') to mark the beginning and end of the key.

We evaluated our protocols using three different smartphones: Nexus 5s, Nexus 
6, and MotoX. 

\subsection{The bacon-beef body model}
\begin{figure}[ht]
\centering
\includegraphics[trim = 19mm 130mm 230mm 60mm, clip, width=200pt,height=130pt]{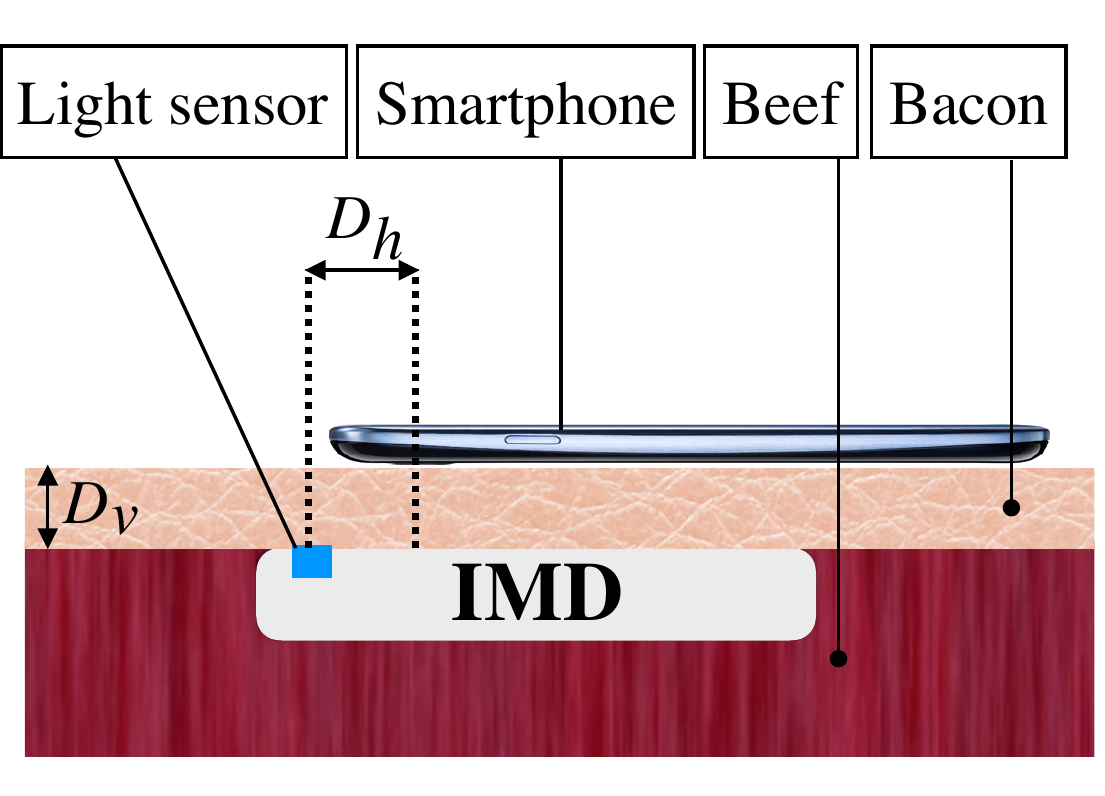}
\caption{Experimental setup: The smartphone is placed on top of the bacon layer above a transparent plastic sealing.} 
\label{fig:EXP}
\end{figure}

The bacon-beef model for the human body has been previously used in several 
research studies \cite{SECURE_VIBE,Zero-Power,GOL_MIT}. Although this model may 
not fully represent all the characteristics of the human body, it has been 
accepted as a valid testing methodology by researchers working on IMDs 
\cite{BEEF_4} due to the difficulties associated with more realistic 
experiments, e.g., laws that permit and control the use of animals for 
scientific experimentation. The bacon-beef body model consists of a 
thin layer of bacon on a thick layer of $85\%$ lean ground beef 
(Fig.~\ref{fig:EXP}). In our experiments, the IMD prototype is placed between 
the bacon and the ground beef, which reflects the typical placement of ICDs 
\cite{Zero-Power}. The smartphone is placed on top of the bacon layer above a 
transparent plastic sealing.

\section{Evaluation of the proposed protocols}
\label{EXAM}
In this section, we present evaluation results for the prototype 
implementation. In particular, we evaluate the transmission range (how far the 
smartphone can be placed from the IMD and have the visible light still reach 
it), wakeup/exchange time (the time needed by the wakeup/exchange 
protocol), protocol overheads (size and energy), and their security.  

\subsection{Transmission range}
\label{RANGE}
We evaluated the prototype using the bacon-beef model for the human body. We 
varied both the vertical distance and horizontal distance between the IMD and 
the smartphone (shown as $D_v$ and $D_h$ in Fig.~\ref{fig:EXP}, respectively) 
to evaulate the vertical and horizontal transmission range (maximum $D_v$ and 
$D_h$ at which the visible light can still reach the IMD). We found that both 
maximum $D_v$ and $D_h$ are independent of the key length and transmission 
rate. They mainly depend on the maximum light intensity that the flashlight 
has to offer. For the three smartphones we used in our experiments (Nexus 5s, 
Nexus 6, and MotoX), the maximum light intensity generated by their 
flashlights is almost the same. Indeed, for all three smartphones, the 
maximum vertical (horizontal) transmission range was about 2 $cm$ (1.5 $cm$). 
Thus, if the physician places the smartphone on the patient's body and keeps 
the smartphone within 1.5 $cm$ of the IMD's light sensor ($D_h < 1.5$ $cm$), the 
visible light can easily reach a depth of 1 $cm$ (the typical $D_v$ for IMDs 
such as ICDs \cite{Zero-Power}). The IMD location is fixed and easily 
recognizable by inspecting the patient's skin under which the IMD is 
implanted. Therefore, ensuring $D_h < 1.5$ $cm$ would be straightforward for 
a physician. 

\subsection{Transmission quality}
\label{QUAL}
We transmitted 100 different keys from each of the three smartphones to the 
IMD, with each smartphone placed within the horizontal transmission range of 
OpSecure ($D_h < 1.5$ $cm$), and with the IMD under a 2 $cm$ layer of bacon 
(2 $cm$ is the vertical transmission range). We found that all keys were 
transmitted over OpSecure without any error. Therefore, the bit error rate 
(the number of received bits that have been altered due to noise, 
interference, distortion, etc.) was zero in all these transmissions. In order 
to evaluate the effect of ambient noise (e.g., other light sources in the 
environment such as sunlight or a car's headlight) on transmission quality, 
we placed a powerful (3000-lumen) light source at a close distance (1 $m$) from 
the IMD. We noticed that the intensity of the visible light received by the 
IMD remained almost the same in the presence of the external light source. 
Indeed, the external light source did not negatively impact the quality of 
transmission at all.

\subsection{Wakeup/exchange time}
Next, we evaluated the wakeup time (the time that the wakeup protocol takes to 
detect the presence of the external device and turn on the RF module) and the 
exchange time (how long the key exchange protocol takes to exchange the 
encryption key).

\noindent \textit{\textbf{Wakeup time}}: As mentioned in Section \ref{PROPOSED}, the 
wakeup protocol periodically places the light sensor in the full operating 
mode, in which the sensor samples the light intensity, to check if the 
smartphone flashlight is present. The wakeup time depends on two parameters: 
(i) how long the light sensor is in the full operating mode ($T_{operation}$), 
and (ii) how long the light sensor remains in the standby mode 
($T_{standby}$) in which the sensor is disabled. $T_{operation}$ and 
$T_{standby}$ should be set with regard to the maximum tolerable wakeup time 
and energy consumption of the wakeup protocol. For example, if we set 
$T_{standby}=1.8$ $s$ and $T_{operation}=0.2$ $s$, the IMD turns on the light sensor 
for 0.2 $s$ and then disables it for 1.8 $s$. In this case, the worst-case wake 
up time will be $T_{standby}+T_{operation}=2$ $s$. As described later in 
Section \ref{ENERGY}, the worst-case wakeup time can be traded off against 
energy consumption by varying either $T_{standby}$ or $T_{operation}$.

\noindent \textit{\textbf{Exchange time}}: The exchange time can be readily calculated 
as $T_{EX}=N/R$, where $N$ and $R$ are the key length and transmission rate, 
respectively. $N$ depends on the encryption mechanism and is commonly 64 $b$ 
or 128 $b$. The transmission rate generally depends on two parameters: (i) the 
blinking frequency of the light source, i.e., how fast the light source can 
be turned on and off, and (ii) how fast the light sensor can sample the 
visible light. In our experiments, the maximum blinking frequency offered by 
the smartphones was within a 20-30 $Hz$ range, and the light sensor was able 
to sample visible light with a sampling rate of a few hundred $Hz$ (a sampling 
rate of 60 $Hz$ is sufficient to recover the key when the blinking frequency 
is 30 $Hz$). Therefore, the maximum blinking frequency of the smartphone 
flashlight limited the transmission rate. In fact, the maximum transmission 
rate was within the range of 20 $b/s$ (for MotoX) to 30 $b/s$ (for Nexus 6). As a result, the minimum time needed for exchanging a packet, that includes a key of length 64 $b$ (128 $b$) and both $Pre$ and $Post$, was within the range of 2.4 $s$ to 3.6 $s$ (4.5 $s$ to 6.8 $s$). 

Note that different smartphones may offer different maximum transmission 
rates. However, the IMD does not need to know the transmission rate $R$ 
beforehand since $R$ can be computed based on the binary sequence $Pre$, which 
is known to the IMD. In our prototype implementation, where the first two bits 
of $Pre$ are always ``11'' (as mentioned in Section \ref{EXP}, $Pre$=``1100''), 
$R$ can be computed as follows: $R=\frac{1000}{T_{step}}$ $ms$, where 
$\frac{1000}{T_{step}}$ $ms$ is half of the duration of the time frame in which 
the IMD observes the $Pre$ sequence.

\subsection{Size and energy overheads}
\label{ENERGY}
Next, we examine the size and energy overheads that the proposed protocols 
add to the IMD. 

Light sensors commonly consist of a phototransistor in series with a small 
resistor that converts received light to a voltage. Light sensors
typically also have an analog-to-digital converter (ADC) that converts 
this voltage to a digital number. Therefore, a light sensor consists of simple 
circuitry that can be implemented in a very small area. In order to save more 
area on the chip, manufacturers may also use an ADC already
incorporated into the IMD and just add a phototransistor/resistor. In both 
cases, the size overhead is negligible. 

The energy overhead is the additional energy consumed by the light sensor, 
which is added to the IMD to enable transmission over OpSecure. The energy 
consumption of a light sensor (even one with a built-in ADC) is typically very 
small, and thus results in negligible energy overhead on the IMD. We
investigate the energy overheads for each protocol using a realistic example 
next.

Consider a typical ICD with a 1.5-$Ah$ battery and 90-month lifetime (it 
consumes about 23.14 $\mu A$ current on an average). We can either use a light 
sensor with a built-in ADC such as MAX44007 \cite{ATMEL} or a light sensor 
without an ADC such as TEMT6000 \cite{TEMT6000} (used in our prototype). Next, 
we discuss the energy overheads of wakeup and key exchange protocols for the 
ICD in both cases. 

\noindent \textit{\textbf{Wakeup protocol}}: We configure the IMD so that the light 
sensor is in the full operating mode for $T_{operation}=0.2$ $s$ after 
being in the standby mode for $T_{standby}=1.8$ $s$. Thus, the light sensor only 
spends $10\%$ of the time in the full operating mode. MAX44007 drains 0.65 $\mu A$ (100 $pA$) from the battery in the full operating (standby) mode, 
thus draining 65.09 $nA$ on an average. In this case, the energy overhead of 
the wakeup protocol is less than $0.3\%$ of the total energy consumption. If 
we use TEMT6000, the phototransistor in series with the resistor and the 
built-in ADC, when operating in the full operating mode, drain a few $nA$ 
\cite{TEMT6000} and tens of $nA$ \cite{CNIA} from the battery, respectively. Therefore, their energy overheads are negligible in comparison to the total 
energy consumption of the ICD. Reducing
$\frac{T_{operation}}{T_{operation}+T_{standby}}$ makes the energy overhead 
even smaller. 

\noindent \textit{\textbf{Key exchange protocol}}: After waking up the RF module, the 
physician can use the smartphone to initiate the key exchange procedure in 
which the IMD configures the light sensor to sense in the full operating mode 
for a few seconds. However, key exchange is a very rare event for two reasons. 
First, a key that is exchanged once can typically be used for a long period of 
time unless the user suspects that the key is compromised. Second, the 
communication between the IMD and the external device is very sporadic 
(e.g., the number of transmissions varies from a few times per day to a few 
times per year). Thus, even if the external device transmits a new key for 
each communication session, the timeframe in which the light sensor operates 
in the full operating mode to exchange the key is negligible in comparison to 
the device lifetime. Consequently, the key exchange protocol adds almost-zero 
overhead to the IMD.

\subsection{Security analysis}
In this section, we evaluate the feasibility of two types of security attacks 
on OpSecure.

\noindent \textit{\textbf{Eavesdropping}}: We investigated the possibility of both 
near-IMD and remote eavesdropping attacks. 

We first placed the smartphone on the chest of a human subject and 
placed a light sensor close to the smartphone to measure the light intensity 
on the body surface at varying distances from the smartphone flashlight. As 
expected, the visible light attenuated very fast and the light sensor
was not able to detect the light from the flashlight as the distance between 
them became greater than 2 $cm$. Thus, an eavesdropping device to pick up the 
light and extract the key would need to be placed on the body surface within 
2 $cm$ of the IMD, which is not likely to be feasible since the patient can 
easily detect such a device.

We next investigated the feasibility of launching remote eavesdropping 
(without an on-body sensor). We noticed that the smartphone flashlight creates 
a red circular area on the user's chest when it is on. We investigated
if an attacker may be able to use a camera to capture a video from the user's 
chest and process the video to extract the key. In order to examine the 
feasibility of such an attack, we asked a subject to hold the smartphone over 
his chest and use his right hand to cover the smartphone. Then, we placed a 
12-megapixel camera 
at a distance of 1 $m$ (a reasonable distance for remote eavesdropping) from 
the user's chest, and captured two sets of videos in a dark room: four videos 
when the smartphone flashlight was on and four when the flashlight was off. 
The videos were captured in the dark room to simulate the worst-case scenario 
since in a dark environment the effect of ambient light sources is minimized 
and, as a result, the red spot created by the flashlight becomes more visible. 
We extracted 1000 frames from each video. For each video in the first
set, we compared all of its frames to the ones of all videos in the second set 
using the concept of RGB Euclidean distance \cite{RGB_E}. This is a metric 
that represents the color difference between two frames. For each video in the 
first set, the RGB Euclidean distances between its frames and the frames of 
other videos in the first set were similar to distances between its frames and 
the frames of videos in the second set, i.e., the videos in the first set were 
not distinguishable from the videos in the second set. This indicates that the 
attacker cannot detect the red spot created by the smartphone flashlight when 
the smartphone is placed on the user's chest and the user covers the 
smartphone by his hand. Thus, in this scenario, the attacker cannot distinguish 
bit ``1'' from ``0'' when the key exchange protocol is sending the key.

\noindent \textit{\textbf{Key injection}}: As mentioned earlier, the horizontal 
transmission range is about 1.5 $cm$ and the physician should keep the 
smartphone within this range to transfer the key to the IMD. Due to this 
proximity requirement, the attacker cannot place an unauthorized smartphone on 
the patient's body within the horizontal transmission range without raising 
suspicion. Moreover, as mentioned in Section \ref{QUAL}, the presence of 
an external powerful light source, which is not attached to the body, does 
not affect the intensity of the light received by the in-body light sensor. 
This significantly limits the attacker's ability to inject data (such as his 
encryption key) into OpSecure.

To sum up, the attacker can neither attach a device to the patient's body 
without raising suspicions nor remotely attack the device. 
\subsection{Summary of evaluations}
Table \ref{table:SummaryOFOp} summarizes the results of our analyses.
\begin{table}[ht] 
\caption{Summary of evaluations} 
\centering 
\begin{tabular}{|l| l|} 
\hline\hline 
Horizontal transmission range ($D_h$) & $1.5$ $cm$ \\ [0.5ex]
\hline 
Vertical transmission range ($D_v$) & $2$ $cm$ \\ [0.5ex]
\hline 
Bit error rate & 0\\ [0.5ex]
\hline 
Wakeup time & a few seconds\\ [0.5ex]
\hline 
Key exchange time & 2.4 $s$ to 3.6 $s$ (key of length 64 $b$) \\ [0.5ex]
    & 4.5 $s$ to 6.8 $s$ (key of length 128 $b$)\\ [0.5ex]
\hline 
Size overhead & negligible \\ [0.5ex]
\hline 
Energy overhead & less than $0.3\%$ \\ [0.5ex]
\hline 
Security guarantees & prevents eavesdropping and key injection \\ [0.5ex]
\hline 
\end{tabular} 
\label{table:SummaryOFOp}
\end{table}
\section{Chapter summary}
\label{CONC}
In this chapter, we described why attack-resilient wakeup and secure key exchange protocols are essential for establishing a secure RF-based communication link between the 
IMD and the external device. We discussed the shortcomings of 
previously-proposed protocols. We presented OpSecure, an optical secure 
communication channel between an IMD and an external device, e.g., smartphone, 
that enables an intrinsically short-range, user-perceptible one-way data 
transmission (from the external device to the IMD). Based on OpSecure, we 
proposed a wakeup and a key exchange protocol. In order to evaluate the 
proposed protocols, we implemented an IMD prototype and developed an Android 
application that can be used to wake up the IMD and transmit the encryption 
key from the smartphone to the IMD. We evaluated our prototype implementation 
using a human body model. The experimental results demonstrated that 
OpSecure can be used to implement both wakeup and key exchange protocols 
for IMDs with minimal size and energy overheads.

%% file: ch-physio/chapter-physio.tex
\chapter{Physiological Information Leakage \label{ch:physio}}

Information security has become an important concern in healthcare systems,
owing to the increasing prevalence of medical devices and the growing use of
wearable and mobile computing platforms for health and lifestyle monitoring.
Previous work in the area of health information security has largely focused
on attacks on the wireless communication channel of medical devices, or on
health data stored in online databases.

In this chapter, we pursue an entirely different angle to health information
security, motivated by the insight that the human body itself is a rich source
(acoustic, visual, and electromagnetic) of data.
We propose a new class of information security attacks that exploit {\em
physiological information leakage}, i.e., various forms of information that
naturally leak from the human body, to compromise privacy. As an example,
we demonstrate attacks that exploit acoustic leakage from the heart and lungs.

The medical devices deployed within or on our bodies also add to natural 
sources of physiological information leakage, thereby increasing opportunities 
for attackers. Unlike previous attacks on medical devices, which target the 
wireless communication to/from them, we propose privacy attacks that exploit 
information leaked by the very operation of these devices. As an example, we 
demonstrate how the acoustic leakage from an insulin pump can reveal important 
information about its operation, such as the duration and dosage of insulin 
injection. Moreover, we show how an adversary can estimate blood pressure 
(BP) by capturing and processing the electromagnetic radiation of an ambulatory 
BP monitoring device \cite{PHYSIO}. 

\section{Introduction}
Implantable and wearable medical devices (IWMDs) promise to transform 
healthcare, by enabling diagnosis, monitoring, and therapy for a wide range of 
medical conditions and by facilitating improved and healthier 
lifestyles. Rapid advances in electronic devices are revolutionizing the 
capabilities of IWMDs \cite{SUG_2}. New generations 
of IWMDs feature increased functional complexity, programmability, and 
wireless connectivity to body-area networks (BANs). 
Standardized communication protocols, such as Bluetooth \cite{BlueTooth} and 
ZigBee \cite{ZIGBEE}, are opening up new opportunities for providing low-power 
and reliable communication to IWMDs. These features facilitate convenient 
collection of medical data and personalized tuning of therapy through 
communication between different IWMDs and an external device (e.g., smartphone 
or clinical diagnostic equipment).

Advances in IWMDs have, unfortunately, also greatly increased the possibility 
of security attacks against them. Many recent research efforts have addressed 
the possibility of exploiting the wireless communication of IWMDs to 
compromise patients' privacy, or to send malicious commands that can cause 
unintended behavior. For example, Halperin et al.~showed that the 
unencrypted wireless channel of a pacemaker can be exploited to compromise the 
confidentiality of data or to send unauthorized commands that cause the 
pacemaker to deliver therapy even when it was not needed \cite{Zero-Power}. 
Subsequently, a successful attack on an insulin pump, exploiting  the wireless 
channel between the device and remote controller, was shown in 
\cite{PrincetonInsulin}. By reverse-engineering the customized radio 
communication and interpreting the unencrypted packets sent from a remote 
controller to an insulin pump, the attacker can launch radio 
attacks to inject insulin into the patient's body beyond the dosage regimen. 
Finally, attacks that drain the battery of IWMDs by sending packets that fail 
authentication have also been proposed \cite{Zero-Power}.

In this chapter, we demonstrate that medical privacy concerns extend far beyond 
the wireless communication to/from IWMDs. We make two main contributions. 
First, we describe the possibility of privacy attacks that target 
{\em physiological information leakage}, i.e., signals that continuously 
emanate from the human body due to the normal functioning of its organs. 
These attacks are a concern even when there is no medical device present, and 
hence have a much wider scope.

As our second contribution, we target IWMDs. We propose several novel attacks 
on privacy by leveraging information leaked from them during their normal 
operation. We demonstrate attacks 
on two medical devices based on acoustic and electromagnetic (EM) leakage from 
them.  Moreover, we investigate a novel metadata-based attack 
that extracts critical health-related information by monitoring the 
communication channel, although the data may be completely encrypted. We note 
that the proposed attacks are applicable even when medical devices have no 
wireless communication, or when the wireless communication is encrypted, 
unlike previous attacks that compromise unencrypted wireless 
channels \cite{PrincetonInsulin,Zero-Power}. 

The rest of the chapter is organized as follows. Section \ref{PHYS_THREAT} describes the 
threat model. Section \ref{PHYS_LEAK} discusses the sources and various types of 
physiological information leakage. Section \ref{PHYS_CAPTURE} describes how we capture different types of leaked physiological signals. Section \ref{PHYS_PAT} presents our bevy of proposed privacy attacks. Section \ref{PHYS_COUNT} suggests some countermeasures against the attacks, 
and Section \ref{PHYS_CONC} concludes the chapter.

\section{Threat model}
\label{PHYS_THREAT}
In this section, we first describe potential adversaries. Then, we describe 
potential risks that may arise from loss of privacy.
\subsection{Adversary}
We consider an adversary to be any potentially untrusted person who has a 
short-term physical proximity to the patient. The proposed attacks, while not impossible, may be difficult to deploy in a secure location such as the patient's home or a medical facility such as a hospital. However, none of our attacks require access to specialized medical equipment such as the ones used in hospitals. We also assume that long-term physical access to the 
patient or monitoring of the patient, e.g., using a camera that continuously 
monitors the subject's activities, is not feasible. In our attack
scenarios, the adversary gains the required physical access to the patient in 
any public location. Crowded places, such as train stations, bus 
stops, and shopping malls, may provide opportunities for the adversary to 
come closer to the subject, while hiding the required equipment. A potential 
adversary might be an employer who intends to discriminate against a 
chronically-ill patient, a criminal group seeking 
to sell valuable medical information to the highest bidder
\cite{STIGMA}, a private investigator who has been hired to spy on
the subject, or a political operative who wants to expose the medical 
condition of the subject for political advantage.

\subsection{Potential risks}
The patient's physiological signals may be exploited in various
ways.  We describe some of the consequences of such information leakage next.

\noindent \textbf{\textit{1. Job/insurance loss}}: Revelation of medical conditions 
may negatively impact a person's employment prospects or make it more difficult for him to obtain 
insurance. Leakage of this sensitive information from the human body or IWMDs, 
such as the presence of a serious illness, level of the illness,
exposure of a condition that may carry social stigma, and exposure of physical, 
emotional or mental conditions would naturally raise serious privacy concerns.    

\noindent \textbf{\textit{2. Unauthorized interviews}}: An unauthorized interviewer may be 
able to combine lie detection (also called deception detection) questioning 
methods with the privacy attack techniques proposed in this chapter to ascertain 
the truth or falsehood of responses given by the subject, without his consent. 
Several researchers have investigated variations in vital health signals, such 
as the respiratory rate and heart rate, in the presence of acute emotional 
stress (e.g., when the person is lying) or a mental problem 
\cite{STRESS1,STRESS2,STRESS3}. For instance, Sung et al.~have demonstrated 
changes in the heart and respiratory rates in live poker game 
scenarios \cite{POKER_METRICS}.

\noindent \textbf{\textit{3. Indirect consequences}}: Although disclosure of medical 
information using the proposed privacy attacks might not be directly lethal, 
unlike attacks on the integrity of the medical
device \cite{PrincetonInsulin, Zero-Power}, it may lead to a subsequent tailored integrity attack. For instance, as described later, extracting medical device 
information, model, type, and configuration from the device using EM leakage
may provide enough information to an adversary to design a more effective 
integrity attack using the extracted parameters.
Moreover, detection of usage of certain medical devices by adversaries may endanger the safety of the patient, e.g., if the device is very
expensive and attracts theft, or embarrass the subject if the medical condition carries a social 
stigma \cite{STIGMA}.

\section{Information leakage}
\label{PHYS_LEAK}
In this section, we discuss the possible sources of information leakage, followed by brief descriptions of different 
types of signal leakages addressed in this chapter.

\subsection{Leakage sources}
We consider two sources of information leakage: (i) human body 
and (ii) IWMDs. Each source continuously leaks information through different 
types of signals.

Several organs in our body generate biomedical signals. Some of these 
signals can be remotely captured and analyzed. For example, our lungs generate 
an acoustic wave called {\em respiration sound}, which can be captured by a 
microphone.

In addition to body organs, IWMDs may also reveal critical health information 
under normal operation even when not using any wireless communication to 
transmit data.  For example, the electrical motor of an insulin pump 
generates an acoustic signal when injecting insulin. As described later
in Section \ref{PHYS_PAT}, performing simple signal processing on this 
acoustic signal can reveal the prescribed insulin dose.

\subsection{Leakage types}
\begin{figure}[h]
\centering
\includegraphics[trim = 70mm 50mm 70mm 55mm ,clip, width=250pt,height=200pt]{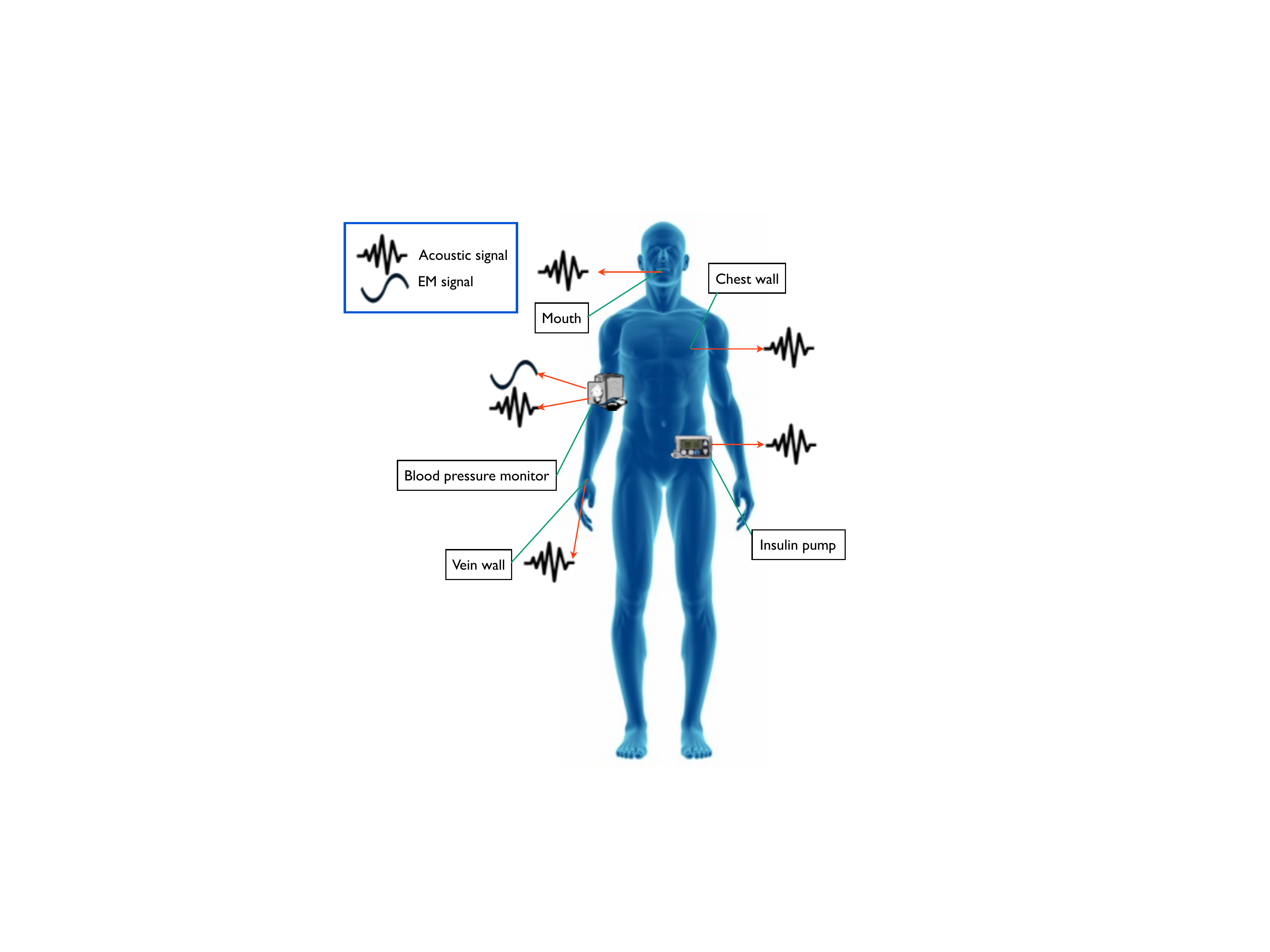}
\caption{Sources of leakage and different types of signals that are continuously leaking from the human body and IWMDs.} 
\label{fig:sourceType}
\end{figure}

In general, leaked physiological signals can be divided into two types: 
(i) acoustic and (ii) EM signals.  Fig.~\ref{fig:sourceType} demonstrates the 
sources of leakage, as well as the different types of signals that we consider 
in this chapter. Body organs, such as heart and lungs, produce an acoustic signal 
that can be captured remotely and analyzed.  IWMDs, such as an insulin pump 
or BP monitor, may also generate acoustic and EM signals during their normal 
operation even if they are not transmitting any data. The following subsections describe these signals in detail.
\subsubsection{Body-related information}
The human body consists of several continuously-operating organs. 
Various acoustic and EM signals are generated as a result. The majority of 
these signals are too weak to be captured without physical contact or may be 
absorbed by other organs before emanating from the body. For example, 
electrical activities originating from nerves carry real-time information 
about health status. The two commonly-used methods for measuring these signals 
are electroencephalography (EEG) and electrocardiography (ECG). The amplitudes 
of EEG and ECG signals vary from tens of microvolts to few millivolts. 
The frequencies of most of these signals are below 40 $Hz$ \cite{EEGECG}. Another 
example is the acoustic signal generated by blood circulating through internal 
organs. However, this is absorbed by the surrounding muscles and tissues. 

If an EM or acoustic signal generated by an organ emanates from the human body, 
it may be captured and analyzed to reveal health-related information. For 
example, one such signal is the respiration sound that is generated by 
chest vibration and airflow through the mouth. Next, 
we discuss different types of signals that 
might leak from the human body during normal operation.

\noindent \textbf{\textit{Acoustic signals emanating from the human body}}: Some of the body organs generate acoustic signals during their normal operation. In this chapter, we examine the feasibility of capturing such 
naturally-occurring acoustic signals from a distance to reveal confidential 
health information of a person. Specifically, we show how capturing acoustic 
signals generated by two organs, heart and lungs, can reveal critical 
information.

As discussed later in Section \ref{PHYS_PAT}, a simple signal 
processing algorithm enables us to count the number of peaks in 
the raw heart sound signal and thus compute the heart rate. The heart
rate may be an indicator of several critical illnesses \cite{CIRIT}
or a sudden emotional stress. For example, when a person lies, his heart rate 
gets elevated above the normal \cite{LieTruth}. Therefore, if an adversary can 
monitor the heart rate remotely, he may even be able to assess whether the 
person is telling the truth.

Respiratory sounds also reveal valuable information about the health 
condition of an individual. The process of recording respiratory sounds and 
analyzing them is referred to as computerized respiratory sound analysis 
\cite{CORSA}; it provides crucial information on respiratory dysfunction, and 
changes in the respiratory characteristics (e.g., duration, timing). 

\noindent \textbf{\textit{EM signals emanating from the human body}}: The human body continuously emits infrared radiation that carries health information. These raw data can be captured and processed by an 
attacker at a distance.

The use of thermal images has increased dramatically in the medical 
applications during the last decade. Thermal imaging cameras highlight warm 
objects against cooler backgrounds. As a result, the human body is easily 
visible in the environment. Moreover, some physiological variations in the 
human body can also be detected with thermal imaging techniques employed in 
medical diagnostic procedures. Several research projects on thermal 
imaging have been discussed in the medical literature. Using these methods 
\cite{Medical1, Medical2, Medical3}, an eavesdropper can easily reveal the 
health status of a person. For example, in \cite{Medical1}, Arora et al.~showed 
the effectiveness of detecting breast cancer using digital infrared thermal 
imaging. The possibility of fever detection using thermal imaging 
techniques is described in \cite{Medical2}.

\subsubsection{IWMD-related information}
As mentioned earlier, IWMDs are used for monitoring and therapeutic 
purposes. An IWMD may leak health-related data or metadata that compromise the 
patient's privacy. We describe next how IWMDs can leak 
information through acoustic and EM signals.

\noindent \textbf{\textit{Acoustic signals emanating from IWMDs}}: First, we describe acoustic leakage from IWMDs. Acoustic waves propagate through a transmission medium using adiabatic compression and decompression. These waves are generated by a source. The source vibrates the medium, leading 
to propagation of vibrations from the source.

Electronic devices with microelectromechanical parts generate unintentional 
acoustic signals during normal operation.  Some recent research efforts have 
demonstrated the feasibility of revealing critical information by interpreting 
acoustic emanations from peripheral computer devices. For example, researchers 
have shown that acoustic emanations from matrix printers carry substantial 
information about the printed text \cite{print}. Moreover, Zhuang et al.~have 
demonstrated the feasibility of recovering keystrokes typed on a keyboard from 
a sound recording of the user typing \cite{keyboard}. 

In this chapter, we demonstrate how acoustic signals generated by
an IWMD (e.g., an insulin pump) may carry significant information about the 
patient's health status and the functioning of the medical device.

\noindent \textbf{\textit{EM signals emanating from IWMDs}}: We divide the EM radiations into two classes: (i) unintentionally-generated and (ii) intentionally-generated. 
Generally, an electronic equipment may emit unintentional EM signals that can 
be used as side-channel information, allowing eavesdroppers to reconstruct 
processed data at a distance \cite{EM}. This has been a concern in the design 
of military hardware for over half a century \cite{TEMP}. IWMDs can also 
unintentionally generate EM signals while performing their regular tasks. 
These signals may reveal critical information about the status of the medical 
device and patient's health condition. In this chapter, we demonstrate how an 
insulin pump can leak information about its function by emitting unintentional 
EM radiations. Medical devices may also use EM signals intentionally to 
wirelessly transmit medical data. Eavesdropping 
on unencrypted wireless communication has been addressed in several research 
articles \cite{PrincetonInsulin, Zero-Power}. Moreover, potential information 
leakage from fully-encrypted data packets has been discussed in computer 
science literature \cite{CSL1,CSL2,CSL3,CSL4}. In this chapter, we focus on the 
metadata that leaks through wireless communication even when the packets are 
fully encrypted. 

\section{Leaked signals and capture methods}
\label{PHYS_CAPTURE}
Before discussing various attacks on the privacy of
medical data, we need to describe how we capture different types
of signals that naturally leak from the human body.  Table \ref{table:PHYSIO_SUM_TABLE} summarizes the
sources of physiological information leakage, the types of leakage, and
a short description of each form of leakage, which we have used to
extract valuable medical information. Then, we briefly describe different 
capture methods that we have utilized in our experiments.


\begin{table*}[t]
  \centering
  \caption{Sources of leakage, types of leakage, and descriptions}
  \begin{tabular}{|l|l|l|}
  \hline
  	Source & Type & Description\\ [0.5ex]
    \hline
    \hline
    Human Body & Acoustic & The vibrations generated by a human organ \\
    & & due to the normal functioning of organs\\
    \hline
    \multirow{2}{*}{IWMDs} & Acoustic & The sound made by electrical components \\ 
    & & during the normal operation of the device   \\
    \hhline{~--}         &  EM (unintentional)&  The unintentionally-generated EM emanat- \\
    & & ion due to the normal operation of the circ-\\
    & & uitry inside the device\\
	\hhline{~--}         &  EM (wireless)& The EM radiation that is intentionally gen-\\
	& & erated by the device for communication\\
    \hline
  \end{tabular}
  \label{table:PHYSIO_SUM_TABLE}
\end{table*}


\subsection{Capturing acoustic signals emanating from body organs}
Here, we describe how we can capture the acoustic signals leaked during the 
normal functioning of lungs and heart. We describe two methods for remotely 
capturing small vibrations or weak acoustic signals generated by these organs. 


\subsubsection{Displacement-based laser microphone}
\begin{figure}[h]
\centering
\includegraphics[trim = 80mm 80mm 80mm 55mm ,clip, width=250pt,height=180pt]{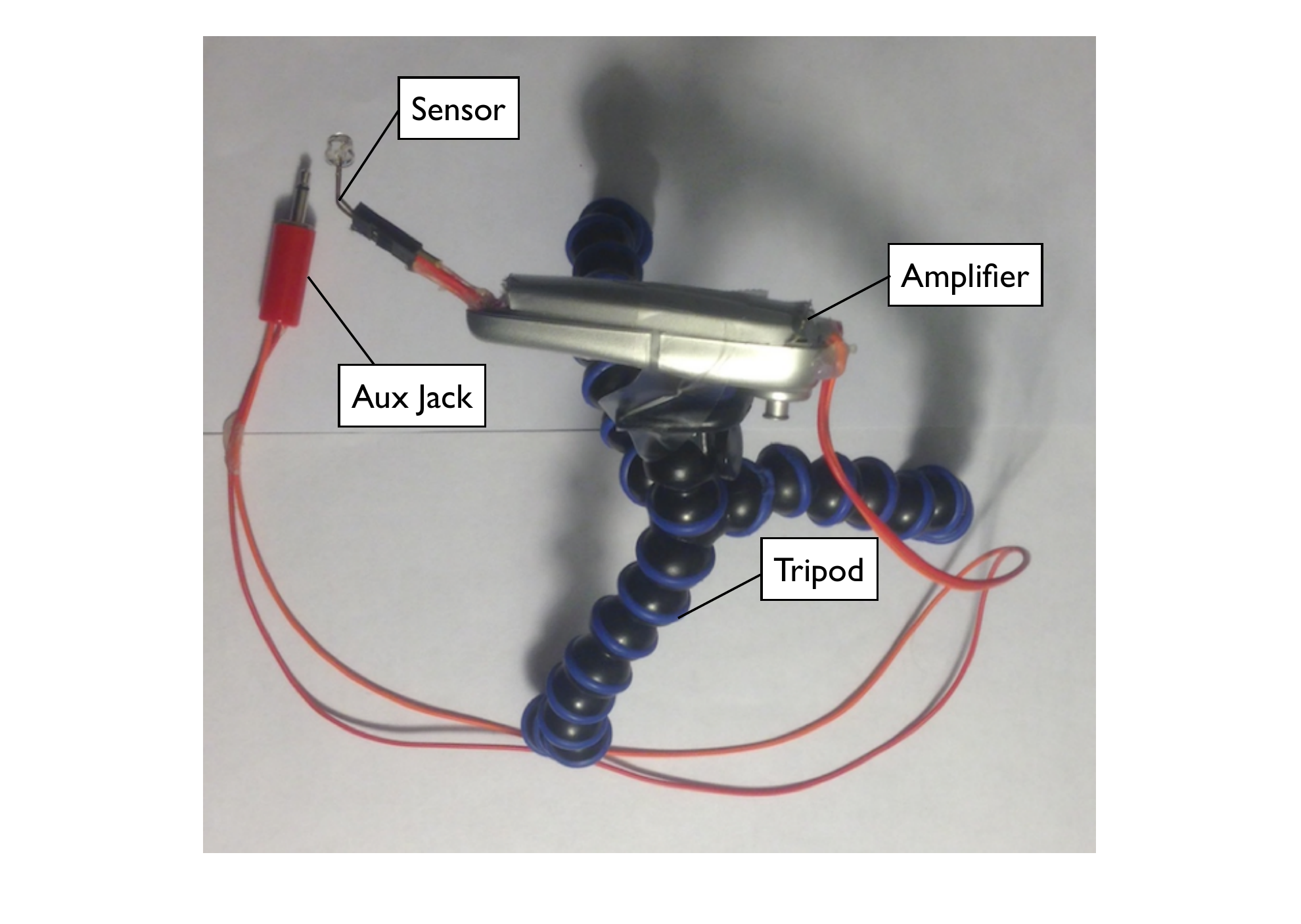}
\caption{The displacement-based laser microphone.} 
\label{fig:experimental}
\end{figure}

\begin{figure}[h]
\centering
\includegraphics[trim = 100mm 80mm 90mm 85mm ,clip, width=250pt,height=160pt]{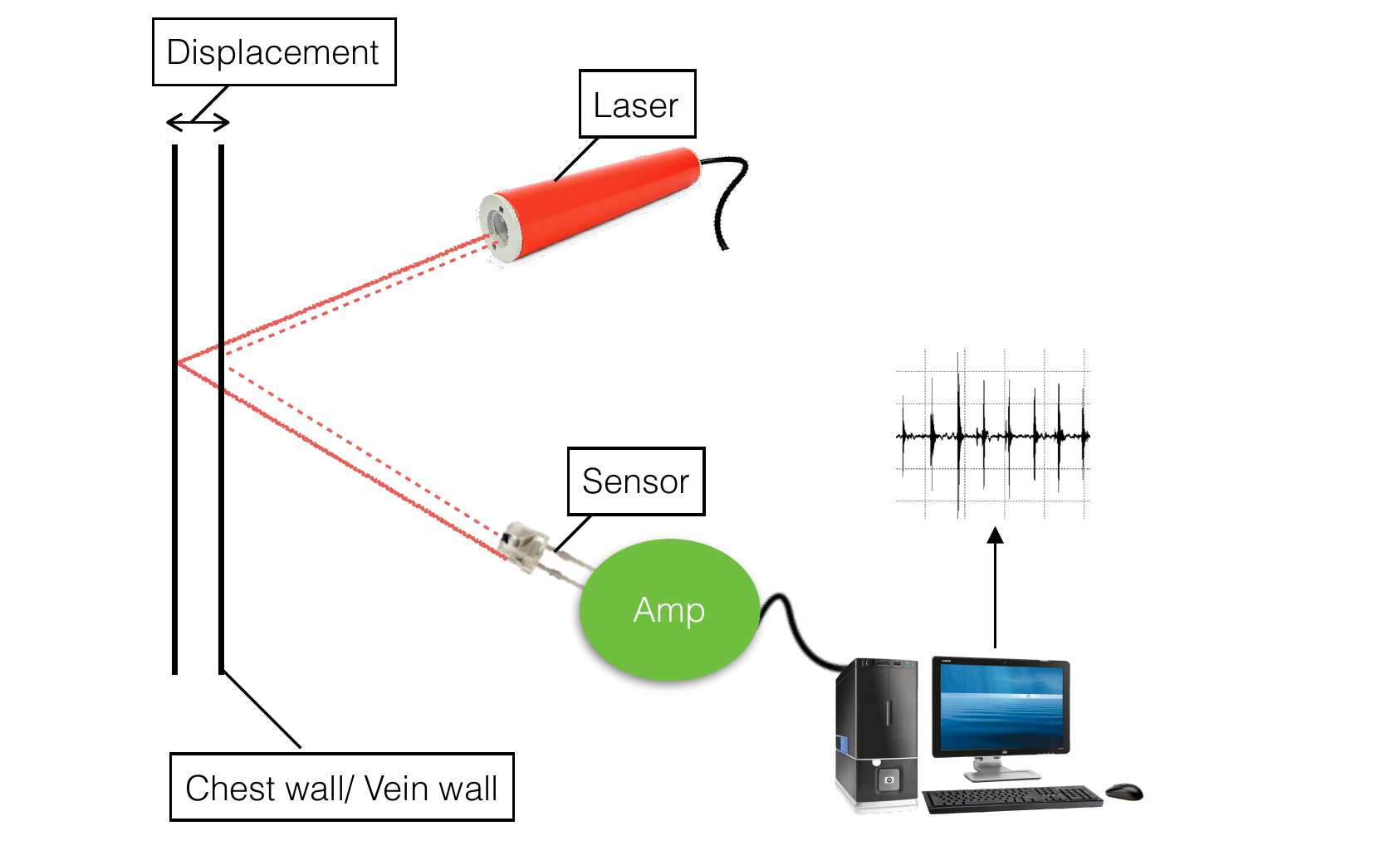}
\caption{Schematic for displacement-based laser microphone: The laser beam 
forms a small incident angle with the surface. The fraction of light beam 
received by the light sensor depends on the vibration of the surface.} 
\label{fig:laser}
\end{figure}

We have designed and built a displacement-based laser microphone that uses a laser beam to detect sound vibrations from a distance (Fig.~\ref{fig:experimental}). Laser microphones were invented to eavesdrop on a conversation with a minimal chance 
of exposure.  Although they have been used for surveillance purposes for a 
long time \cite{LaserSUR}, for the first time, we employ these microphones in 
the context of a privacy attack on patients' medical data. We have built an 
inexpensive laser microphone to detect vibrations emanating from the human 
heart and lungs. This device is based on detecting the varying amounts of 
reflected laser beam received by a single ambient light sensor. As illustrated 
in Fig.~\ref{fig:laser}, the laser beam forms a small incident angle with the 
surface. Surface vibration along the normal vector causes displacement of the 
reflected beam, and as a result, the amount of laser signal reaching the 
receiver varies for different displacements. 

\subsubsection{Parabolic microphone} 
The second capture method we propose is based on a parabolic 
microphone (KJB-Det \cite{KJB}) to capture weak acoustic signals generated by 
the lungs. It uses a parabolic reflector to collect and focus sound waves onto 
a receiver.  It amplifies the acoustic signal by concentrating all of the 
sonic energy at the focal point, thus increasing the signal-to-noise ratio 
(SNR). KJB-Det comes with parabolic dish that has a diameter of 50 $cm$. In 
addition, electronic amplifiers used in KJB-Det can increase the overall level 
of both the noise and acoustic signal, without degrading the SNR.

\subsection{Capturing acoustic signal generated by IWMDs}
In order to capture the acoustic signal leaked from IWMDs, several sound-recording equipments can be used, ranging from simple microphones to sophisticated parabolic microphones. In this chapter, we describe two different privacy attacks based on acoustic signals leaked from an insulin pump (electrical motor's sound and acoustic alarm generated by its safety system) and one privacy attack based on the acoustic leakage from an ambulatory BP monitoring device (acoustic signals 
generated by the various components of the device when it is measuring 
BP). In the privacy attack based on the signal generated by insulin pump motor, we capture and amplify the sound of the electrical motor using a parabolic microphone (KJB-Det). In other attacks, we use a simple microphone to record the required acoustic signal. 

\subsection{Capturing unintentional EM signals}
We use an oscilloscope 
to detect the unintentionally-generated EM signals. The EM side-channel 
information that we capture is available during the normal operation of 
the medical device even when the device is not transmitting any data 
(e.g., using a USB cable or wireless communication). We capture the raw EM 
signal directly from the antenna that is connected to the oscilloscope, 
instead of a filtered and demodulated signal with limited bandwidth. We use 
an antenna (75 Ohms VHA 9103 Dipol Balun) to improve the SNR for signals in 
the 25 $MHz$ to 500 $MHz$ frequency band. Moreover, 
we check if these EM signals can be captured using a small 
portable antenna, such as a simple loop of copper wire with 50 $cm$
circumference.

EM signals may remain undetected using standard techniques. Spectral analyzers 
need static carrier signals of significant amplitude. The demodulation process 
may eliminate the interesting components of unintentionally-emitted EM signals. 
In addition, the scanning process of wide-band receivers may take a lot of 
time\cite{keyboardW}.

\subsection{Capturing the metadata of wireless communication}
In order to monitor fully-encrypted wireless communication and extract the 
metadata from the communication channel, we first need to find the frequency 
band of the transmission. If the model and type of the device are known, the 
frequency range can be extracted from manufacturer's documentation. 

In general, an IWMD should make its existence and type unknown to unauthorized 
parties. If a device reveals its existence, its type should still remain 
hidden to unauthorized persons for many different reasons. For 
example, the device might be extremely expensive. More importantly, knowing 
the specific model of a device may provide critical information to potential 
adversaries. As we elaborate later, if the type, characteristics, and settings 
of an IWMD are known, designing a tailored attack becomes much easier. 
A tailored attack is a smart attack based on the specific features and 
configurations of a known device.  Therefore, we assume that the model and 
type of the IWMD are not known to the attacker. 

A fast approach for detecting the frequency band of a wireless transmission is 
through an oscilloscope that uses a loop of wire as an antenna. The 
eavesdropper can scan different frequency ranges when the communication 
channel is active and guess the frequency range. In addition, the frequency 
band of communication for an unknown IWMD can usually be obtained by scanning 
some specific bands based on the fact that FDA regulations impose specific 
limits on the frequency bands of medical devices. The majority of medical 
devices communicate at 450 $MHz$, 600 $MHz$, 900 $MHz$, 1.4 $GHz$, and 2.4 $GHz$.

After finding the frequency band of transmission, the encrypted packets can be 
captured using one or multiple universal software radio peripherals
(USRPs) \cite{USRP}. We demonstrate how examining the frequency band of the 
channel and characteristics of the packets can reveal critical health 
information.

\section{Proposed privacy attacks}
\label{PHYS_PAT}
\begin{figure*}[t]
\centering
\includegraphics[trim = 25mm 135mm 25mm 25mm ,clip, width=450pt,height=120pt]{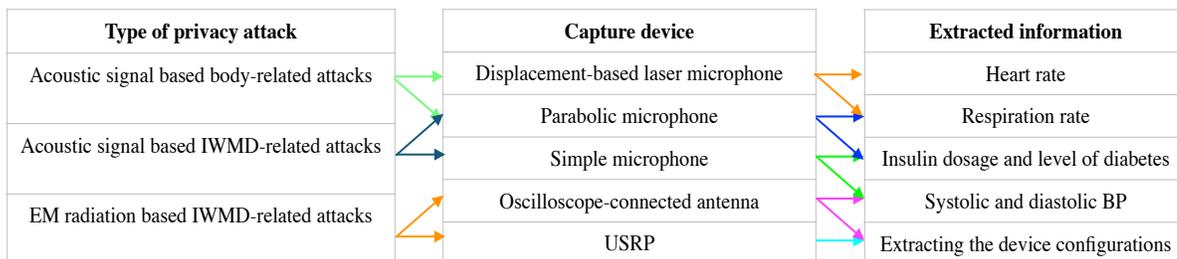}
\caption{
Different types of privacy attacks, capture methods, and the private information that each type of attack can extract from different leaked signals
} 
\label{fig:SUMF}
\end{figure*}

In this section, we propose several privacy attacks to demonstrate how
processing physiological leakage from the human body and IWMDs can reveal
various medical conditions and provide valuable information about the
device (e.g., device type, model, and manufacturer). Fig.~\ref{fig:SUMF}
depicts different types of privacy attacks, capture methods that we use
to record the leaked signals, and the private information, which each
type of attacks can extract from different physiological leaked signals.
We demonstrate next several privacy attacks for each type.

\subsection{Acoustic signal based body-related attacks}
Here, we first describe a method to estimate the heart and respiration 
rates from the captured acoustic signal. Then, we discuss the parameters that affect the accuracy and detection range of the attack when each of the two capture methods (parabolic microphone and laser-displacement microphone) is used.

For obtaining the heart and respiration rates, we use a
simple algorithm to find the local maxima.  In order to reduce the effect of 
noise, the algorithm ensures that the distance between two consecutive peaks 
is more than the value of a parameter called $distanceThreshold$. The maximum 
possible human heart rate (200 pulses per minute) and respiration rate 
(80 breaths per minute) are used to define $distanceThreshold$. Thus, 
$distanceThreshold$ is set to 300 $ms$ and 750 $ms$ for the heart and respiration 
rates, respectively. We perform 20 trials to estimate the heart 
and respiratory rates based on the leaked acoustic signals. For each
trial, we compare our estimates with the actual
values. We discuss how accurately our algorithm can estimate the heart and 
respiratory rates. We discuss the accuracy of our algorithm for each capture 
method (parabolic microphone and laser-displacement microphone).

\noindent \textbf{\textit{Laser-displacement microphone}}: we use the laser-displacement microphone for capturing acoustic signals from both the lungs and heart.  The sound quality obtained by this 
microphone depends on two factors: (a) reflection fraction, which is the 
fraction of the incident beam that is reflected by the surface, and (b) the 
displacement of the received beam. The first parameter depends on the nature 
of the surface. For example, the human skin absorbs a large fraction of the 
incident beam; therefore, the sensor should be placed close to the skin to 
receive the beam. However, the displacement of the received beam on the 
sensor decreases as the sensor gets nearer the reflecting surface. We were 
able to accurately extract the respiration rate from 5 $cm$ away. If the person 
wears a metallic/reflecting necklace, we can point the incident beam towards 
the necklace instead, which is a better reflector than the human skin. We were 
able to accurately extract the respiration rate from 6 $m$ away when the person 
wore a flat steel necklace. We also used a displacement-based laser microphone 
to detect the heart rate. In the absence of an attached reflector surface, the 
acoustic signal was used by the laser microphone to detect the heart rate with 
over $95\%$ accuracy at a distance of 5 $cm$. At greater distances, the amount 
of received beam reduces drastically and the accuracy drops. 

\noindent \textbf{\textit{Parabolic microphone}}: The audio gain of a parabolic microphone increases 
as the frequency increases.  The gain of an ideal dish with a diameter of 
50 $cm$ and a perfect parabolic shape and focus is characterized by a curve 
starting from 0 $dB$ at 200 $Hz$. In order to enhance the amplification of our 
parabolic microphone, we replaced its dish with a larger dish with a diameter 
of 1 $m$ that provides a 6 $dB$ amplification at 200 $Hz$. At lower frequencies, 
the most important parameters are dish size and the quality of the microphone. 
The modified parabolic microphone was able to detect the respiration rate at 
a distance of 5 $m$. However, the parabolic microphone was unable to detect the 
heart rate.

\subsection{Acoustic signal based IWMD-related attacks}
Here, we describe the privacy attacks that exploit acoustic information 
leakage from IWMDs. Acoustic signals generated by an IWMD can provide valuable 
information to an unauthorized party.  Each IWMD consists of several 
components. Some of these components (e.g., electrical motors and relays) can 
unintentionally produce a capturable sound during normal operation. An 
unintentionally-generated acoustic signal can be used as a side-channel 
information to reveal the status of the medical device and the patient's 
condition. In addition to this class of acoustic signals, some IWMDs 
intentionally produce acoustic signals to notify the users of conditions 
that require immediate attention. Many medical devices have alarm systems; 
among them are insulin pumps, pulse oximetry devices, and BP monitors. 
These alarms offer necessary warnings to inform patients of changes in 
their health condition. They usually provide sophisticated mechanisms for 
safety checks. These alarms make the patient aware of an unusual situation. 
Generally, the audible frequency range for a human is between 20 $Hz$ and 
20 $kHz$. Frequency ranges of 2 $kHz$ to 4 $kHz$ are most easily heard. For this 
reason, most alarms emit sound in this frequency range. Different acoustic 
warnings and alarms generated by IWMDs may also provide valuable information 
to an adversary.

We discuss next different attacks based on unintentionally- and  intentionally-generated acoustic signals generated by IWMDs.

\subsubsection{Extracting injected dosage and estimating level of diabetes}
\begin{figure}[h]
\centering
\includegraphics[trim = 50mm 80mm 140mm 100mm ,clip, width=250pt,height=130pt]{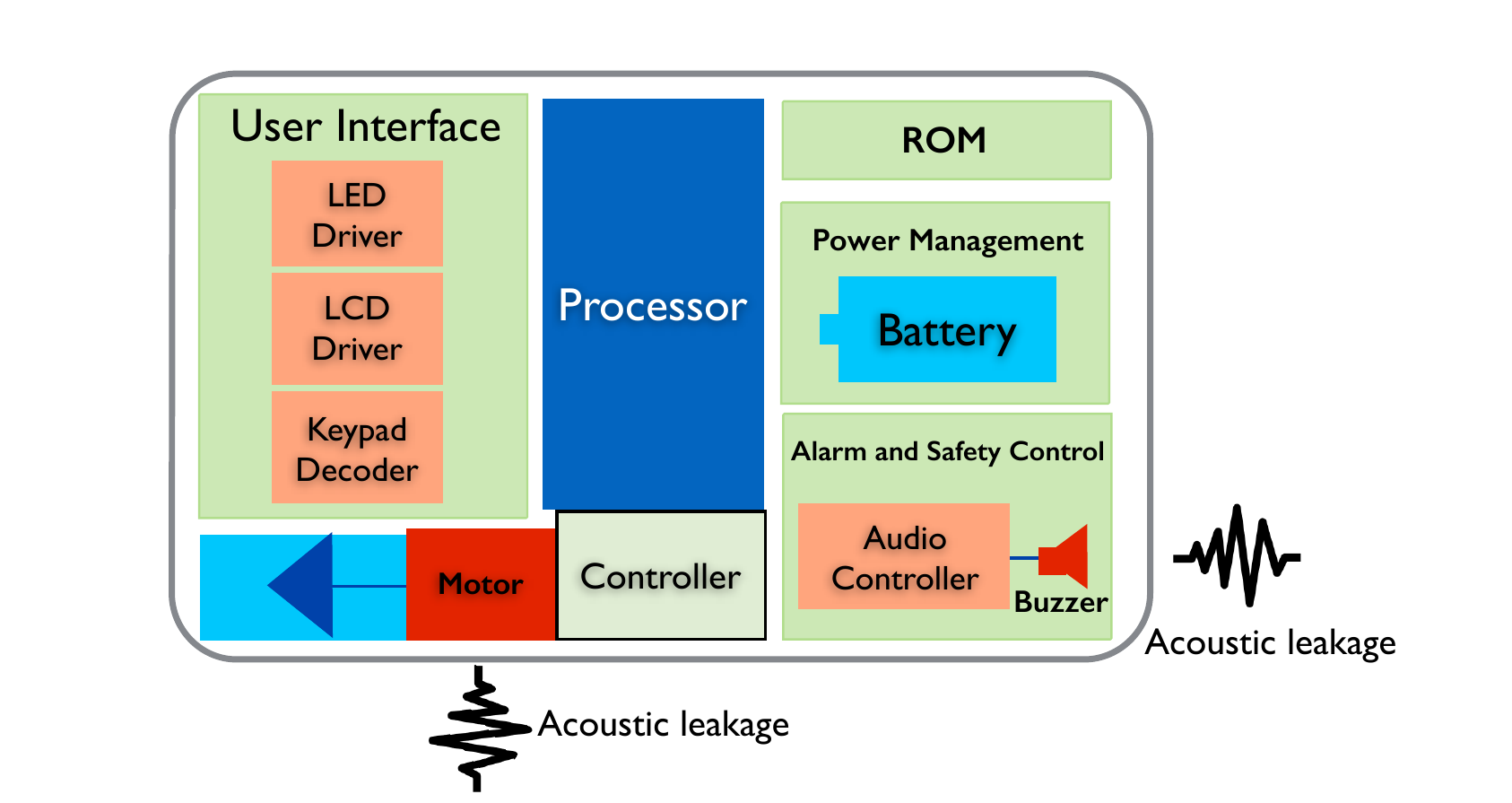}
\caption{A schematic view of an insulin pump. The components marked in red (motor and buzzer) generate the acoustic signals that can be interpreted to reveal the medical data.} 
\label{fig:InsulinPump}
\end{figure}

Here, we propose privacy attacks based on the acoustic signals leaked from an insulin delivery system (Fig.~\ref{fig:InsulinPump}). In this medical device, the electrical motor unintentionally generates acoustic signals and the speaker intentionally produces different alarms as 
reminders for calibration and high/low glucose, and as predictive high/low 
glucose alerts. The components marked in red (motor and buzzer) generate the 
acoustic signals that we can interpret to reveal the medical data. 

We present two attacks on an insulin pump using both unintentionally- and 
intentionally-generated acoustic signals. First, we demonstrate how capturing 
and interpreting the unintentional acoustic signal generated by its electrical 
motor can reveal the prescribed dosage, and hence the level of diabetes. 
Second, we use the acoustic signals generated by its safety system to remotely 
examine the status of the device and extract the prescribed dosage.

\noindent \textbf{\textit{Attack 1: Extracting information from motor sounds of an insulin pump}}: We show below how processing the weak acoustic signal generated by the electrical motor of an insulin pump can reveal the exact amount of injected insulin, and as a result, provide an estimation of initial blood sugar, and 
level of diabetes. As mentioned earlier, we used a parabolic microphone
to capture the acoustic signal in this case.  We propose two signal processing 
algorithms for this attack.
\begin{figure}[ht]
\centering
\includegraphics[trim = 117mm 75mm 95mm 105mm ,clip, width=250pt,height=150pt]{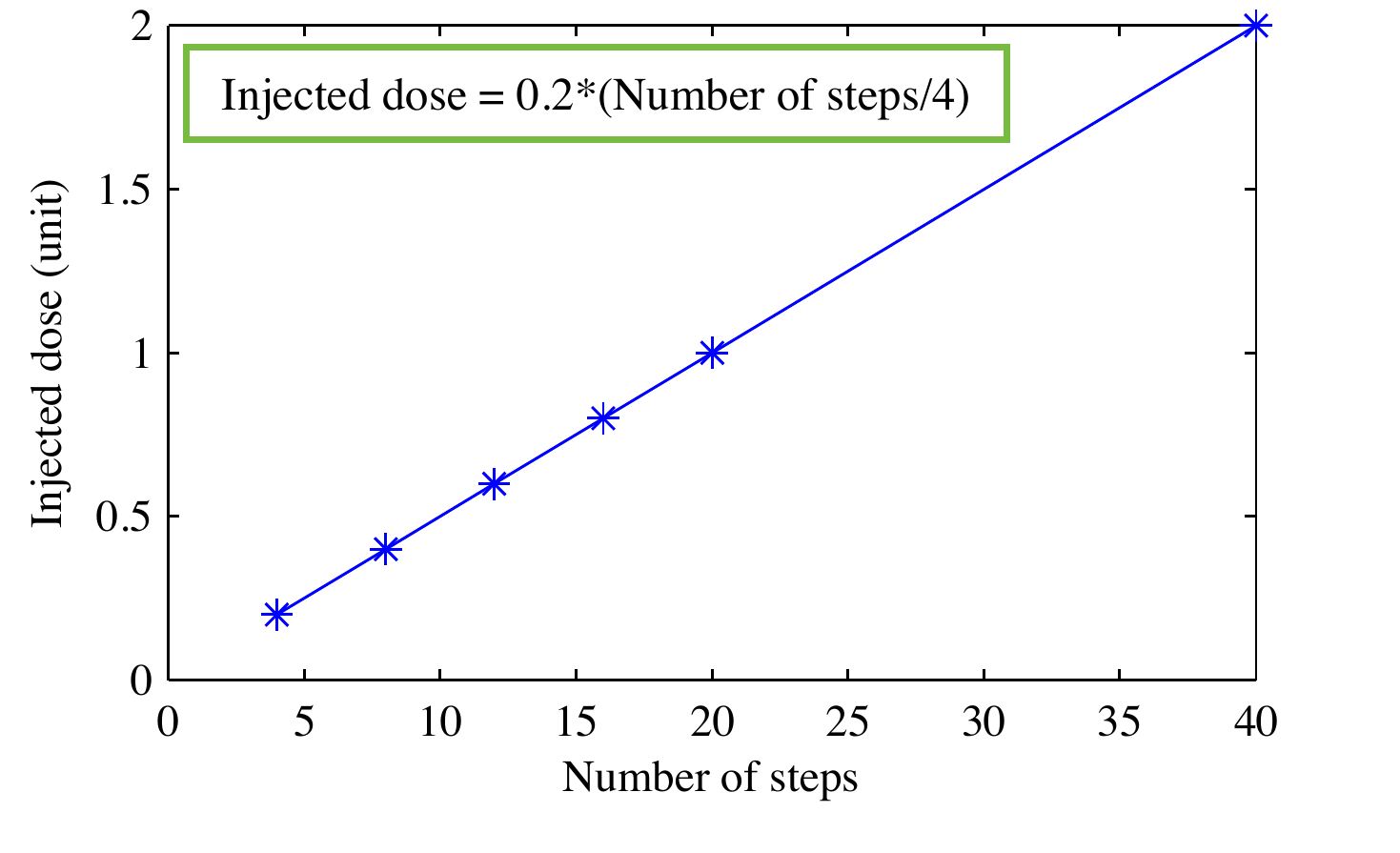}
\caption{Dose of injected insulin vs. the number of rotation steps of the electrical motor.}
\label{fig:StepDose}
\end{figure}

\begin{figure}[ht]
\centering
\includegraphics[trim = 115mm 40mm 75mm 107mm ,clip, width=250pt,height=150pt]{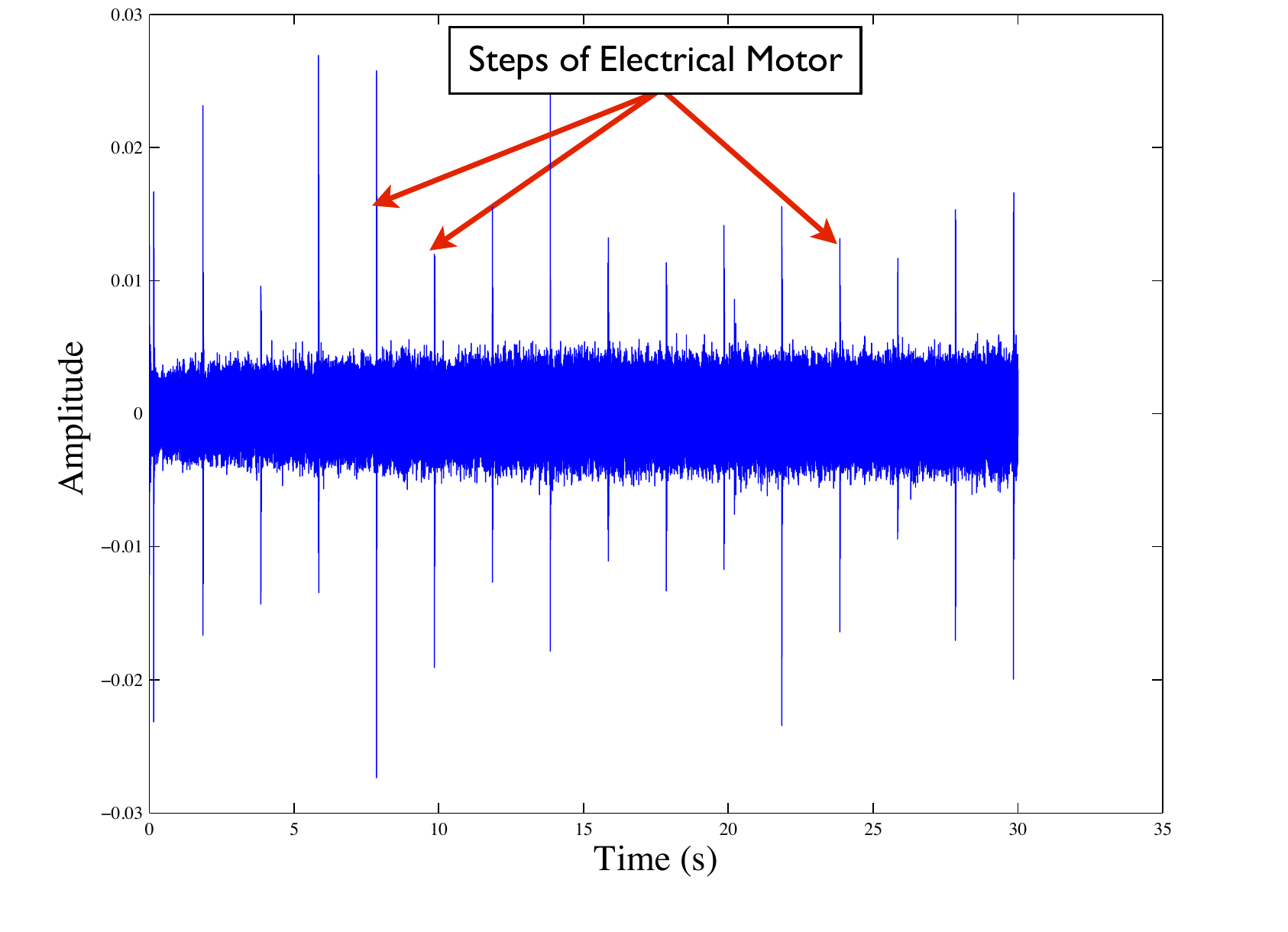}
\caption{Acoustic signal generated by the electrical motor of an insulin pump while injecting 0.8 unit of insulin.} 
\label{fig:MotorSound}
\end{figure}
In an insulin pump, a step motor is used in the injection procedure. Our 
experiments demonstrate that there is a linear relationship between the number 
of rotation steps of the electrical motor and the amount of injected insulin 
(Fig.~\ref{fig:StepDose}).  Fig.~\ref{fig:MotorSound} illustrates the acoustic 
signal generated by the electrical motor while injecting 0.8 unit of insulin. 
Each peak corresponds to one step of the motor. The first processing algorithm 
finds the number of peaks. Thus, for this case, the total number of steps is 
calculated as 16, thereby inferring that 0.8 unit of insulin was injected.
{\em Algorithm 1} shows the pseudo-code of the algorithm. It calculates 
the exact amount of injected insulin based on the number of motor steps.  Its 
four inputs are: (i) $acousticSignal$, which is the acoustic signal of the 
electrical motor sampled at 22 $kHz$, (ii) $distance$, which indicates the 
minimum acceptable distance between two consecutive peaks (steps), 
(iii) $threshold$, which is the minimum acceptable amplitude of a peak, 
and (iv) $widthThreshold$, which is the width of a step in the absence of 
environmental noise. We obtain the number of steps 
from the number of peaks in $acousticSignal$ using subroutine $stepCount$. 
Then, using another subroutine $stepWidth$, we calculate the width of each 
step that is defined as the time when the peak and its neighboring points 
are greater than $widthThreshold$. After finding the number of peaks and the 
width of each peak, we estimate the number of steps that might 
be corrupted by comparing the width of each peak to $widthThreshold$. If 
the width is more than $widthThreshold$, it is likely to contain noise in the 
area around the peak.

{ 
\noindent {\em Algorithm 1: Calculating the insulin dosage from the acoustic signal}. \\
\noindent\makebox[\linewidth]{\rule{15.1cm}{0.5pt}}
{\footnotesize
\noindent Given: $acousticSignal$, $distance$, $threshold$, and 
$widthThreshold$\\
\noindent \makebox[\linewidth]{\rule{15.1cm}{0.5pt}}
\noindent 1. $steps \leftarrow stepCount(acousticSignal, threshold, distance)$\\
2. $widths \leftarrow stepWidth(acousticSignal, threshold, distance)$\\
3. $for$ $each$ $width$ $in$ $widths$\\
4. \hspace{0.5cm} $if (width >widthThreshold)$\\
5. \hspace{1cm}  $noisy \leftarrow noisy + (width/widthThreshold)$\\
6. \hspace{0.5cm}  $end$ \\
7. $end$ \\
8. $if (noisy  > 3)$\\
9. \hspace{0.5cm} $Print$ `` $Warning$: $Inaccurate$ ''\\
10.\hspace{0.5cm} $return$ $-1$ \\
11. $else$\\ 
12. \hspace{0.5cm} $dosage \leftarrow \left \lceil{steps/4}\right \rceil * 0.2$\\
13. \hspace{0.5cm} $Print$ $dosage$\\
14. \hspace{0.5cm} $return$ $0$\\
15. $end$\\}
\noindent\makebox[\linewidth]{\rule{15.1cm}{0.5pt}}

This algorithm is able to automatically detect the number 
of peaks in $acousticSignal$ that are corrupted.  If the number of corrupted 
locations in $acousticSignal$ is more than three, there will not be enough 
information in $acousticSignal$ to reveal the exact insulin dose. Therefore, 
the attacker should discard that waveform, and try again later when background 
noise is less powerful.

In order to evaluate and compare our acoustic signal based algorithms, we 
constructed a test set consisting of 20 acoustic signals generated by the 
insulin pump when injecting four different doses of 0.2, 0.4, 0.6, and 0.8 
unit of insulin (five injections for each dose). We captured the first 10 
acoustic signals in a silent office (low-noise environment). We 
captured the other acoustic signals in the presence of background noise 
generated by a conversation. The algorithm could extract the injected dose 
exactly for the first 10 cases. In the presence of the conversation, the 
algorithm correctly detected the corrupted signal in four cases, and extracted 
the exact injected dose in the other six cases.

In addition to counting the number of steps, we can calculate the duration of 
an injection. Calculating the duration is more robust against noise.
Using our second algorithm, we show how an adversary can use an estimation of 
the injection duration to find the exact amount of injected insulin without 
knowing the exact number of steps.

The amount of injected insulin is quantized to a multiple of 0.2 unit of
insulin. As a result, the injection duration is quantized and is a multiple of 
7 seconds. For example, the injection of 0.2 and 0.4 unit of insulin takes 
about 7 and 14 seconds, respectively. Fig.~\ref{fig:DurDose} shows the amount 
of injected insulin with respect to injection duration. It shows there is an 
almost-linear relationship between the amount of injected insulin and 
injection duration. Therefore, if the attacker can only estimate the injection 
duration by calculating the time during which the sound of the electrical 
motor is present, he can find the exact amount of injected insulin even when a 
large fraction of the acoustic signal is dominated by background noise and 
counting the total number of steps is not feasible (Fig.~\ref{fig:OverPower}). 
Using the test set described earlier, our duration-based algorithm was able to 
extract the exact amount of insulin in 18 of the 20 cases (10 under low-noise 
signals and 8 under noisy signals). Similar to the previous method, this 
algorithm was also able to automatically detect the situations in which the 
presence of noise affects the computed results.

In summary, capturing and processing the acoustic signal generated by the 
electrical motor of an insulin pump may reveal the injected dosage, and as a 
result, reveal the medical condition of the patient. The medical literature 
suggests that one unit of insulin is required per 50 mg/dl above 120 mg/dl of 
blood sugar\cite{dosage}. Therefore, after measuring the insulin dosage, we 
can also estimate the level of blood sugar before injection.

\begin{figure}[h]
\centering
\includegraphics[trim = 113mm 65mm 95mm 110mm ,clip, width=250pt,height=150pt]{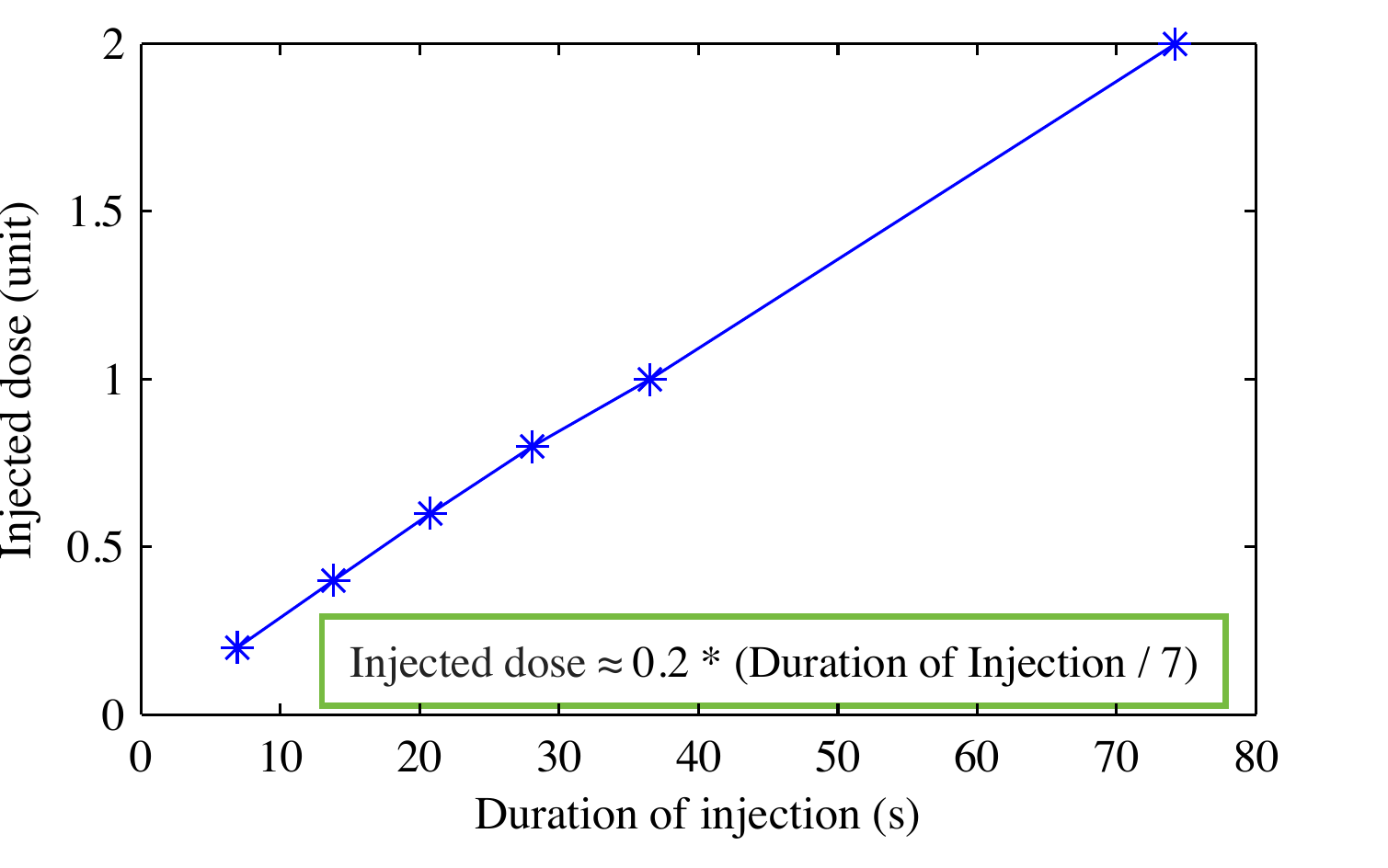}
\caption{Dose of injected insulin vs. injection duration.}
\label{fig:DurDose}
\end{figure}

\begin{figure}[h]
\centering
\includegraphics[trim = 20mm 8mm 30mm 40mm ,clip, width=250pt,height=150pt]{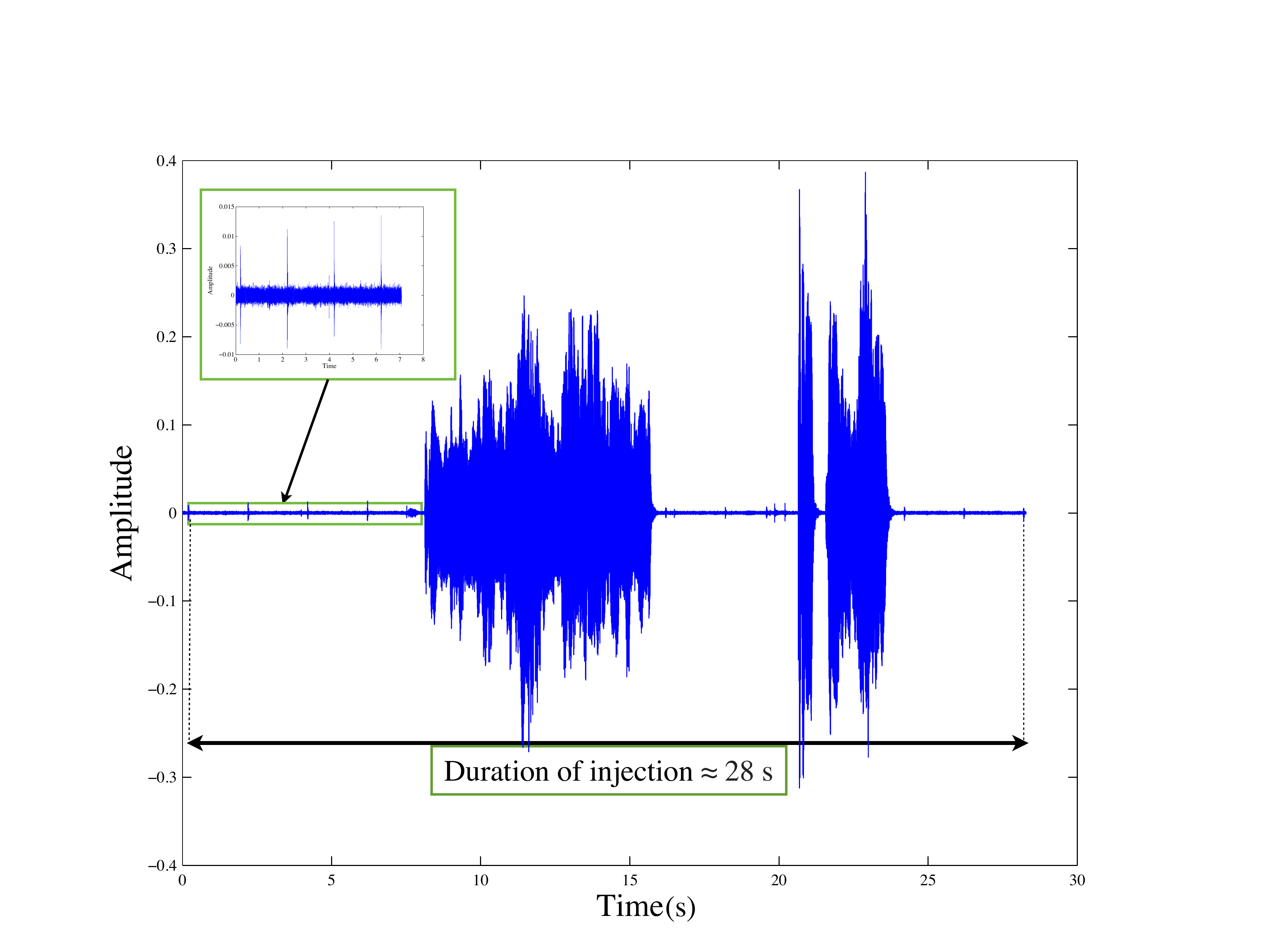}
\caption{Acoustic signal generated by the electrical motor of an insulin pump when 0.8 unit of insulin is injected. For a 
large fraction of time, the acoustic signal is dominated by background noise, and counting the number of rotation
steps is not feasible.} 
\label{fig:OverPower}
\end{figure}

\noindent \textbf{\textit{Attack 2: Eavesdropping on alarms of an insulin pump}}:
We describe below how the safety system of an insulin pump, which intentionally 
generates acoustic signals to inform patients, can unintentionally leak 
critical information about the health condition of a patient. As mentioned 
earlier, alarms are intended to alert patients of special events. The 
controller unit of the insulin pump (Fig.~\ref{fig:InsulinPump}) is 
responsible for handling alerts and alarms, and the speaker generates audible 
signals in various situations, including blockage, low/high sugar level, 
initialization, and end of an injection.

Each injection procedure has four different phases: (i) initialization, 
(ii) confirmation, (iii) injection, and (iv) end of injection. 
Fig.~\ref{fig:Alarm} shows the acoustic signal generated by the alarm system 
of an insulin pump when a user tries to inject 0.8 unit of insulin. The four 
phases of the injection procedure are demonstrated in this figure. 
After the patient sends the injection command, the beginning of the 
initialization phase is reported by a single beep sound.  Then, the user sets 
the dosage by repeatedly pressing a button. In the confirmation phase, multiple 
beeps are generated based on 
the desired dosage. In this phase, one beep is generated by the safety system 
for every increment of 0.2 unit in insulin dose. However, the 
frequency of beeps in this phase is 2$\times$ higher than that in the 
initialization phase. Then, the injection begins and finally the end of 
injection is reported to the patient by a single beep. 

We used three methods to find the exact amount of injected insulin by 
interpreting the acoustic signal: (i) initialization-based method that counts 
the number of peaks (beeps) in the initialization phase, (ii)
confirmation-based method that counts the number of peaks in the confirmation 
phase, and (iii) duration-based method that calculates injection duration. 
Although the first two methods are straightforward, the third method is
more accurate, especially in noisy environments. Similar to \textit{Attack 1}, by extracting the quantized values of injection 
duration and dose, the exact prescribed dosage can be calculated, even if the 
attacker cannot count the number of beeps, but only estimate the injection 
duration. The injected dosage can be directly computed based on the 
almost-linear relationship between injection duration and injected dosage 
(Fig.~\ref{fig:DurDose}). 

In order to evaluate the accuracy of each algorithm, we constructed a similar 
test set to the one we used earlier. We captured the acoustic signal from 10 $m$ away. The raw signal was amplified using a cheap amplifier before 
processing.  All three methods could accurately extract the injected dose in 
the low-noise environment. Table \ref{table:PHYSIO_compare} shows the number of 
correct and incorrect calculations and accuracy of each method in the noisy 
environment. The duration-based method showed the best accuracy, where 
accuracy is defined as the percentage of correctly-calculated doses.  

\begin{table}[ht] 
\caption{Accuracy of the three methods for eavesdropping on the alarm system of an insulin pump} 
\centering 
\begin{tabular}{c c c c} 
\hline\hline 
Method & Correct & Incorrect & Accuracy ($\%$) \\ [0.5ex]
\hline 
Initialization-based & 6 & 4 & 60\\[1ex] 
Confirmation-based  & 7 & 3 & 70 \\[1ex]
Duration-based & 10 & 0 & 100 \\[1ex]
\hline 
\end{tabular} 
\label{table:PHYSIO_compare} 
\end{table} 

In addition to compromising  health-related information of a patient, the 
status of the medical device, such as blockage and low-battery state, can also 
be directly extracted by capturing and analyzing the alarms generated by the 
insulin pump.

\begin{figure}[h]
\centering
\includegraphics[trim = 70mm 60mm 70mm 40mm ,clip, width=250pt,height=150pt]{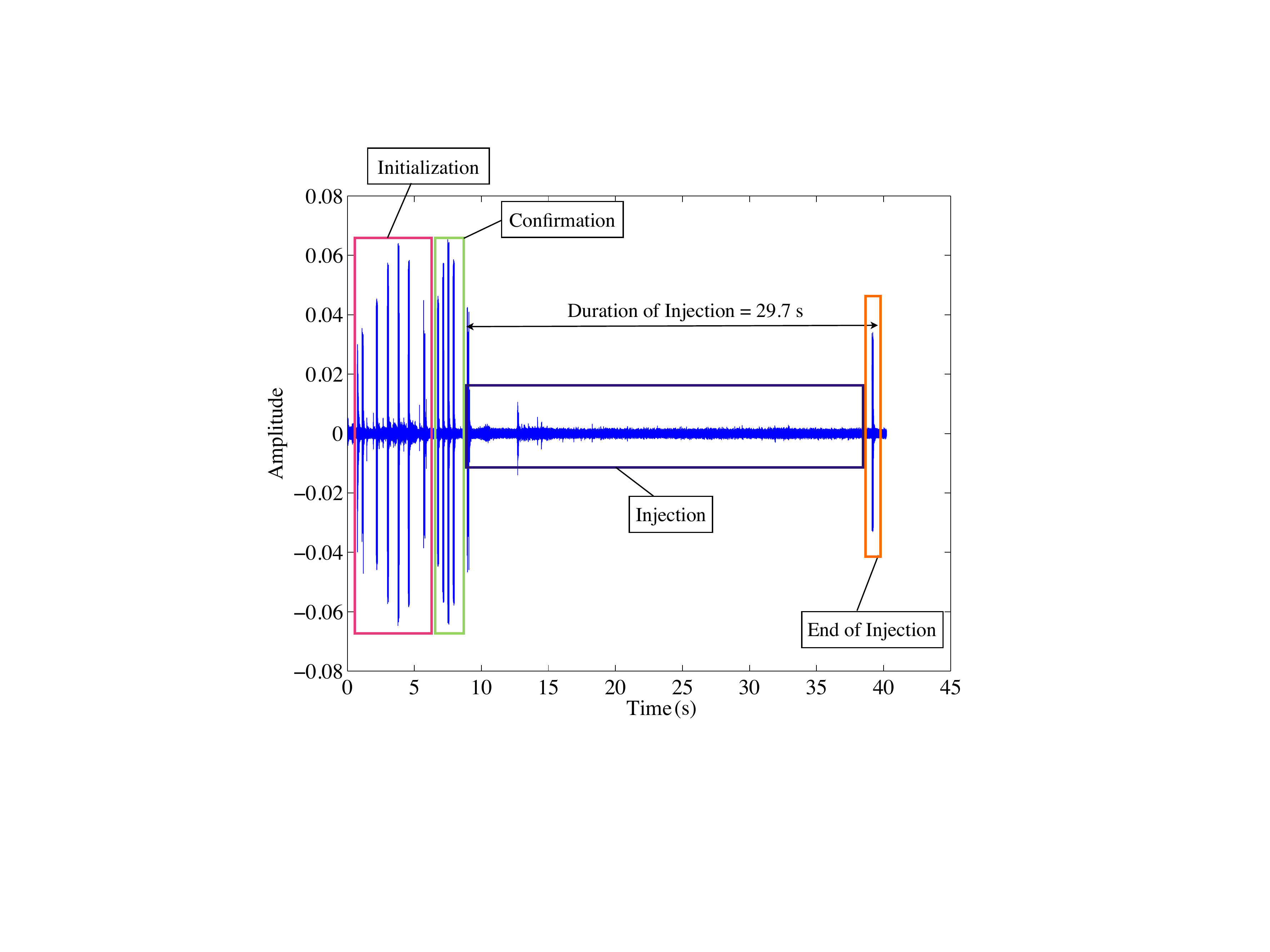}
\caption{Acoustic signal generated by the safety system of an insulin pump when the user tries to inject 0.8 unit of insulin.} 
\label{fig:Alarm}
\end{figure}

\subsubsection{Estimating BP by interpreting leaked acoustic signal of an ambulatory BP monitoring device}
In this section, we target an ambulatory BP monitoring device. 
Fig.~\ref{fig:BP} shows a block diagram of such a device. The components shown 
in red are the major sources of acoustic leakage.  We interpret the sound 
generated by its electrical pump to estimate the patient's BP, which 
is the pressure generated by circulating blood upon the walls of blood veins. 
A simple microphone suffices to capture the acoustic signals from a
distance of up to 10 $m$ in this case.
BP is commonly represented by two numbers (systolic and diastolic), and is 
measured in millimeters of mercury ($mm Hg$).  Non-invasive ambulatory BP 
monitoring is being increasingly used to continuously monitor patients' BP. 
A digital BP monitor has a cuff and digital pressure sensor. When a 
user inserts his arm in the cuff, it is automatically inflated by an electric 
motor. The digital monitor determines the BP and heart rate by measuring the 
small oscillations when the pressure is slowly released from the cuff. Common 
BP monitoring devices use a simple algorithm to derive an upper 
bound on systolic BP. They inflate the cuff to reach the upper bound in every 
measurement.  However, in order to ensure patient's comfort, some new BP 
devices often use a technology known as fuzzy logic, which anticipates 
systolic BP to prevent over-inflation. In these devices, the highest pressure 
in the cuff is approximately 10 $mm Hg$ to 15 $mm Hg$ more than the actual 
systolic pressure. In this chapter, we have targeted a commercially available 
BP monitoring system. We choose not to disclose its brand and model number.  
We discuss how the sound generated by the electrical pump can 
provide enough information for an eavesdropper to accurately estimate the BP 
(both systolic and diastolic).

\begin{figure}[h]
\centering
\includegraphics[trim = 70mm 60mm 70mm 100mm ,clip, width=250pt,height=130pt]{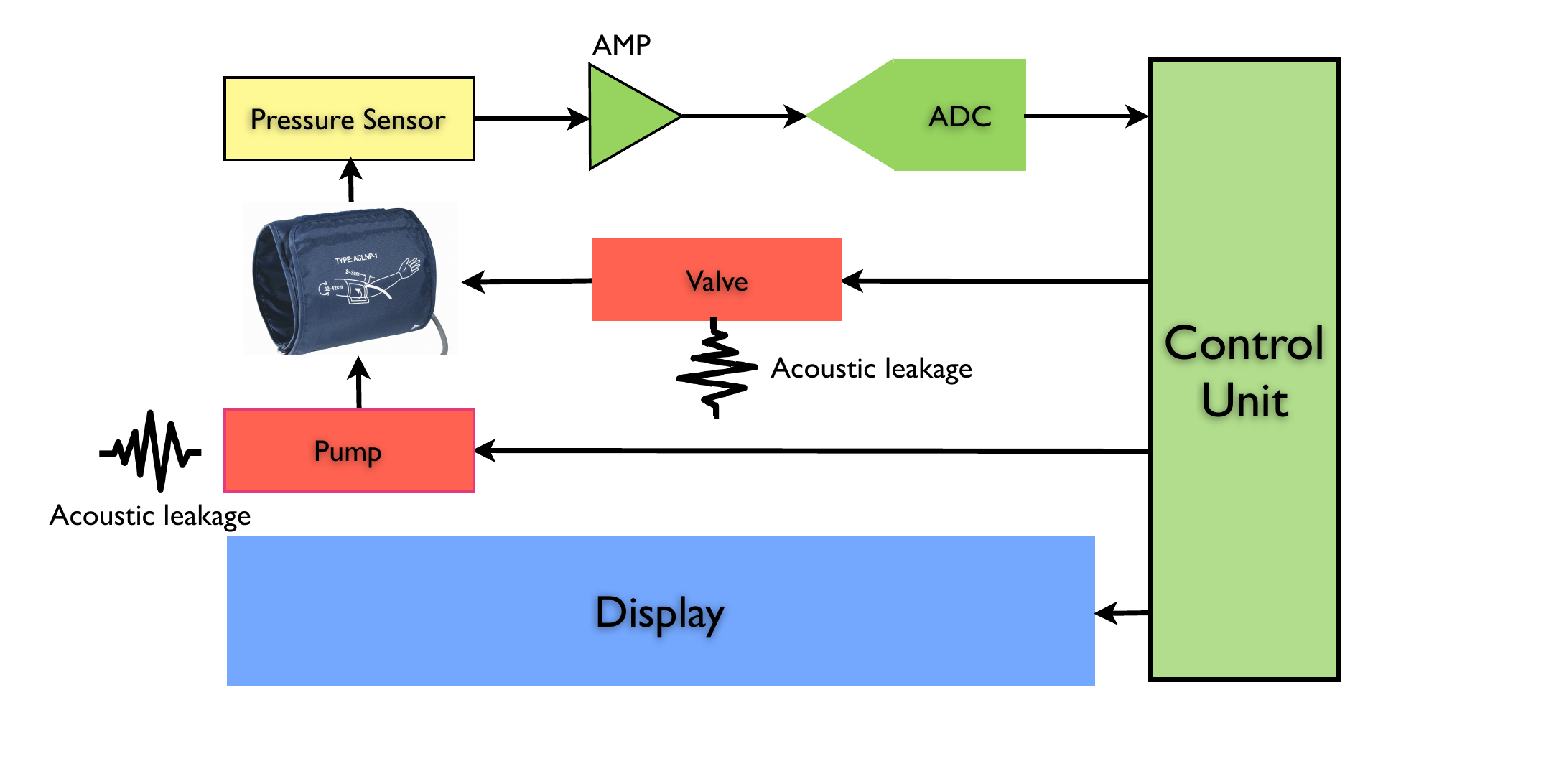}
\caption{Block diagram of an ambulatory BP monitoring device. The components shown 
in red are the major sources of acoustic leakage.} 
\label{fig:BP}
\end{figure}

Our experiments demonstrate that each measurement consists of three 
consecutive phases: (i) inflation phase in which the cuff pressure increases 
to reach its upper bound value, (ii) step-wise deflation phase in 
which the monitoring device opens an air valve to slowly decrease the cuff 
pressure and measure the BP, and (iii) restart phase. Fig.~\ref{fig:BPwave} 
shows the acoustic signal generated during the measurement.

\begin{figure}[h]
\centering
\includegraphics[trim = 130mm 65mm 90mm 100mm ,clip, width=250pt,height=150pt]{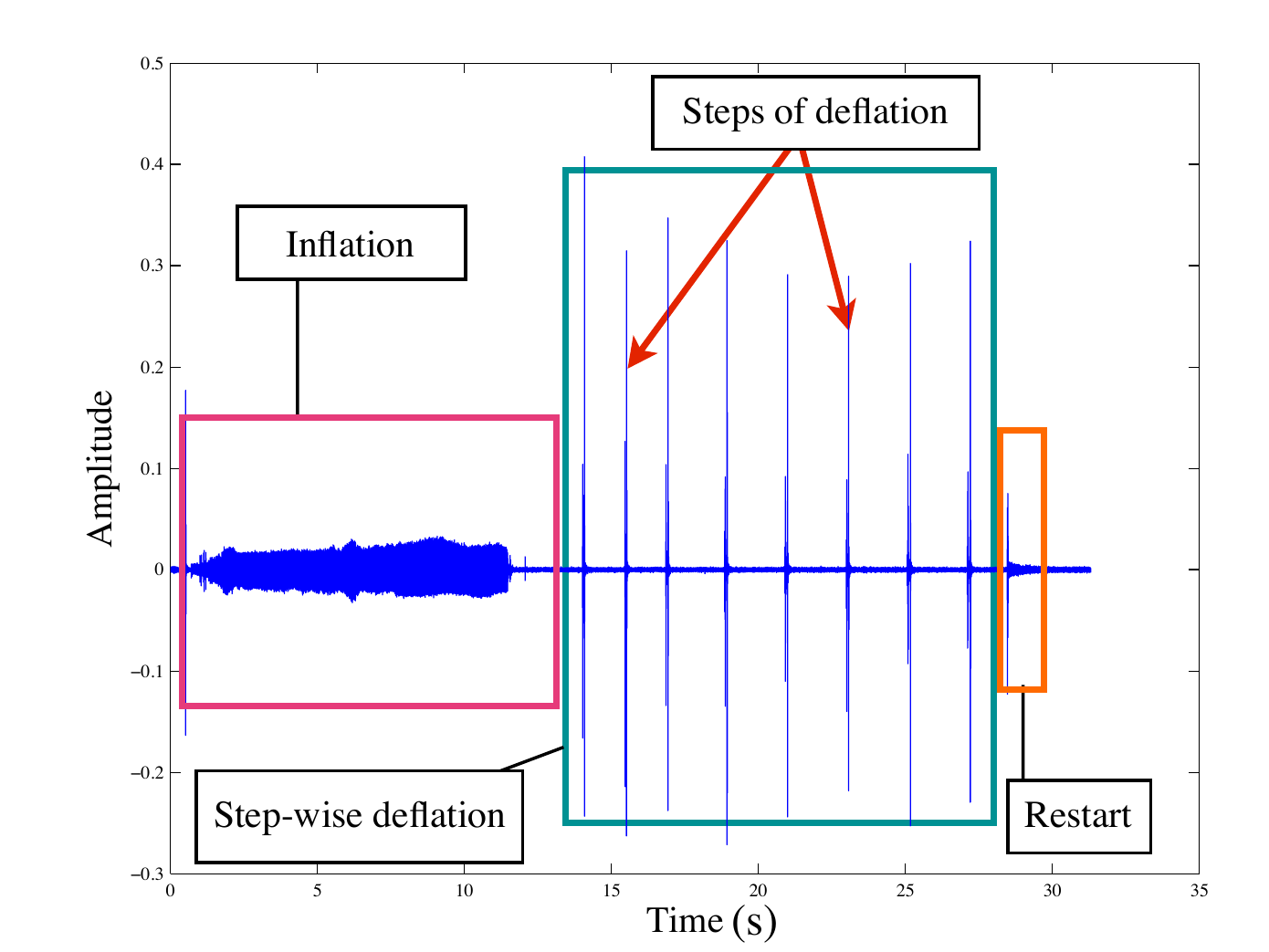}
\caption{Acoustic signal generated by the ambulatory BP monitoring device. 
Three phases of measurement are shown.} 
\label{fig:BPwave}
\end{figure}

In the BP monitoring device used in our experiments, the cuff pressure 
decreases about 9 $mm Hg$ for each step of deflation in the second phase of 
measurement. In addition, we found that the systolic BP was detected after 
three or four steps in the step-wise deflation phase, which suggests that 
the systolic BP should be in the range of $(P_h-27)$ $mm Hg$ to $(P_h-36)$ $mm Hg$, 
where $P_h$ is the maximum cuff pressure in the inflation phase. Moreover, 
based on our experimental results, the diastolic pressure is usually detected 
in the range of $P_l$ $mm Hg$ to $(P_l+9)$ $mm Hg$, where $P_l$ is the minimum cuff 
pressure during the step-wise deflation phase before the device enters the 
restart phase. In order to examine the accuracy of the above claim, we used 
25 BP measurements. The systolic BP was in the range of $(P_h-27)$ $mm Hg$ to 
$(P_h-36)$ $mm Hg$ for 21 out of 25 measurements. Moreover, for 23 out of 
25 measurements, the diastolic BP was in the range of $P_l$ $mm Hg$ to $(P_l + 9)$ $mm Hg$. Therefore, if we can develop a method to detect $P_h$ and $P_l$, the 
systolic and diastolic pressure can be estimated as $(P_h-27+P_h-36)/2$ $mm Hg$ and $(P_l+P_l+9)/2$ $mm Hg$, respectively.

We describe next how we can use the acoustic signal generated by the 
electrical pump to extract $P_h$ and $P_l$ and thus estimate the BP. The cuff 
pressure reaches its maximum value at the end of the inflation phase. In order 
to find the maximum value for an arbitrary measurement, we construct a 
look-up table that maps the maximum pressure ($P_h$) to the duration of 
inflation ($T_h$), where $P_h$ varies from 100 $mm Hg$ to 180 $mm Hg$. For each 
measurement, we first calculate $T_h$ by finding the part of 
the acoustic signal in which the pumping sound is present. Then, we use the 
look-up table to find $P_h$.  Thereafter, we count the number of steps before 
deflation and compute $P_l=P_h - numberOfSteps * 9$.  Then, we calculate the range of systolic and 
diastolic pressures, and report the middle points of these ranges as their 
estimate. Our experimental results show that for 19 out of 25 arbitrary 
measurements, this algorithm calculates both systolic and diastolic 
pressures with maximum absolute error less than 8\% (4.5\% average
absolute error for systolic pressure and 7.0\% for diastolic pressure), where 
error is defined as the 
difference between the estimated and actual values divided by the actual 
value. The main reason for the six failed cases was re-inflation. 
Re-inflation occurs when the patient suddenly changes his arm position during 
the step-wise deflation phase.  In this case, the monitoring device increases 
the cuff pressure again. It is easy to modify the method to detect the 
situation in which re-inflation occurs.  {\em Algorithm 2} gives the 
pseudo-code for the improved version of our algorithm. We define a function, 
called $infCount$, which finds the number of inflation phases separated by 
deflation steps. This improved algorithm automatically detects whether the 
algorithm is unable to calculate the BP accurately. 
\subsection{EM radiation based IWMD-related attacks}
We target two classes of EM radiations: (i) unintentional EM radiations that 
are signals generated by different components of an IWMD (e.g., processor, 
controller), and (ii) intentional EM radiations that are encrypted wireless 
communications that transmit medical data. We discuss two EM radiation 
based attacks using each class of EM radiations, namely from the pump in a 
BP monitor, and based on the metadata of wireless communications of an insulin 
pump.

\subsubsection{Estimating BP from unintentional EM radiations}
Here, we discuss an attack based on capturing and analyzing the EM radiation 
that is unintentionally generated by the BP monitoring device. 

Using the EM signals captured from the BP monitoring device, we were able to 
estimate the patient's BP. EM radiations reveal the activity of the electrical 
pump in the different phases of a measurement (inflation, step-wise deflation, 
and restart). The duration of the inflation phase can be revealed by
calculating the time when the electrical motor produces the EM radiations, and 
as a result, the systolic BP can be extracted by using the method discussed 
earlier. Moreover, by 
monitoring the activity of the device in the deflation phase, the
number of deflation steps could be detected. Estimating the BP using EM 
signals was as accurate as when it was estimated from acoustic signals. However, this method can be easily used in a crowded environment, where the acoustic signal may be dominated by background noise.  

{
\noindent {\em Algorithm 2: Estimating systolic and diastolic BP}\\
\noindent\makebox[\linewidth]{\rule{15.1cm}{0.5pt}}
{\footnotesize
\noindent Given: $acousticSignal, table$ where $table: T_h \rightarrow P_h$\\
\noindent\makebox[\linewidth]{\rule{15.1cm}{0.5pt}}
\indent 1. $infNumber \leftarrow  infCount(acousticSignal)$\\
\indent 2. 		   $if (InfNumber > 1)$ \\
\indent 3.\hspace{0.5cm} $Print$ `` $Warning$: $Inaccurate$ ''\\
\indent 4.\hspace{0.5cm} $return -1$\\ 
\indent 5. $end$\\
\indent 6. $T_h \leftarrow  calculateTimeOfInflation(acousticSignal)$\\
\indent 7. $P_h \leftarrow lookUp(T_h, table)$\\
\indent 8. $Steps \leftarrow CountPeaks (acousticSignal)$\\
\indent 9. $upperSystolic \leftarrow P_h - 27$\\
\indent 10. $lowerSystolic \leftarrow P_h - 36$\\
\indent 11. $P_l \leftarrow P_h - numberOfSteps * 9$\\
\indent 12. $ upperDiastolic  \leftarrow P_l$\\
\indent 13. $lowerDiastolic \leftarrow P_l + 9 $\\
\indent 14. $systolic \leftarrow  \frac{upperSystolic+lowerSystolic}{2}$\\
\indent 15. $diastolic \leftarrow \frac{upperDiastolic+lowerDiastolic}{2}$\\
\indent 16. $Print$ $diastolic$, $systolic$\\
\indent 17. $return$ $0$\\}
\noindent\makebox[\linewidth]{\rule{15.1cm}{0.5pt}} 

The activity of the electrical pump in the inflation phase was completely detectable from 
15 $cm$ away when we used the VHA antenna. Moreover, when we replaced the VHA 
antenna with a wire of 50 $cm$ circumference, we were able to detect the 
activity from 10 $cm$ away. The deflation steps were detectable using the VHA 
antenna and wire from 10 $cm$ and 5 $cm$ away, respectively.

\subsubsection{Extracting insulin dosage regimen from the wireless communication metadata of the insulin pump} 
Here, we describe how capturing and processing the metadata leaked from the 
communication channel of an insulin pump can reveal critical medical 
information, including the injected dose of insulin, number of injections, 
and level of diabetes.

Different manufacturers have different priorities and considerations. 
As a result, the 
metadata of the communication channel of one device are different from those 
of others. The metadata-based attack that we discuss next consists of two 
main steps: (i) the eavesdropper first extracts the metadata from the 
communication channel to reveal valuable information about the type and model 
of the IWMD, and (ii) when the device type is known, the attacker designs a 
tailored attack that specifically targets the known device.  We discuss six 
classes of metadata leaked from the communication channel that can be used 
to find valuable information about the device: (i) frequency of communication, 
(ii) time between two consecutive transmissions, (iii) communication protocol, 
(iv) packet size, (v) detection range, and (vi) modulation protocol. However, 
in most cases, a subset of these classes can uniquely identify the model 
and type of the device. 

We describe an example of metadata-based privacy attack using the insulin pump delivery system. For the 
insulin pump we used in this research, the frequency of communication 
(around 900 $MHz$), time between two packets (5 minutes), and modulation 
protocol (on-off keying) would be conclusive enough for an adversary to 
uniquely identify the insulin pump and its manufacturer. In addition, the 
detection range (20 $m$) and packet size (80 $b$) match the
information given in the documentation of the device.

We describe a tailored attack against an insulin pump. We assume 
all communications are fully encrypted.  In the first step of the 
metadata-based attack, we find the model and type of the insulin pump. This 
specific model comes with a remote control. The remote control is a device 
that controls and programs the insulin pump and allows the user to deliver 
a discrete bolus dose or stop/resume insulin delivery. Each button on the 
remote control sends a specific command to the insulin pump. The size of 
remote control is usually small to ensure patient's convenience, and as a 
result, there are only a few buttons on the remote control. Different 
sequences of buttons on the insulin pump are to be pressed in different 
situations. For the remote control that we used in our experiment, the 
patient should use at least three button presses to start the injection: 
(i) the first button tells the device to initialize the injection, (ii) the 
second button is used to set the dosage of injection, and (iii) the third 
button confirms the injection.  The patient can press the second button 
multiple times to increase the dosage. In this scenario, interpreting the 
number of consecutive packets can uniquely reveal the occurrence of the 
injection, and the insulin dosage. In order to verify this claim, we 
designed an experimental setup using the insulin delivery system, its remote 
control, and a USRP. We performed 10 trials in which the amount of injected 
insulin varies from 0.2 to 2 units of insulin. All trials confirmed that the 
amount of injected insulin can be easily computed as: $dosage = (N-2)*0.2$ 
$unit$, where $N$ refers to the number of transmitted packets. For example, 
if seven packets are captured by the USRP in this case, the first and last 
packets would represent initialization and confirmation.  The other five 
packets can be assumed to be sent to increase the amount of injected
insulin. Therefore, $dosage = (7-2)*0.2=1.0$ $unit$.

To sum up, the number of injections can be extracted by 
counting the number of transmissions that have more than three packets. 
This approach is completely accurate when the distance between the USRP and 
the remote control is less than 6 $m$.

\section{Possible countermeasures}
\label{PHYS_COUNT}
In this section, we briefly discuss some possible countermeasures to protect 
the patient against the privacy attacks described in this chapter. We hope 
these initial suggestions would spur further research on countermeasures 
against such attacks. We discuss different countermeasures for each source 
of leaked signals (human body and IWMDs). Hiding information that leaks from 
the body is difficult because there are many local sources of leakage, e.g., 
lungs, heart, and skin. We can hide some of this information using cloth as 
a shield. However, since it is typically not possible to cover the whole body, 
medical information may at least leak from the face. For example, the EM 
radiation from the face leaks enough information to detect if a person 
has fever. 

As mentioned earlier, many components inside medical devices may
generate acoustic or EM signals: the motherboard, communication cables,
processor, and actuators. Potential solutions for eliminating the
leakage of  compromising information from IWMDs can be divided into
three main categories: strength reduction, information reduction
\cite{CSL0,COUNT1}, and noise addition techniques
\cite{CSL1,COUNT1,COUNT2}. We describe each category next.

\subsubsection{Signal strength reduction techniques} 
The main goal of these 
techniques is to reduce the strength of unintentionally-leaked signals such 
that their detection range reduces drastically. These methods include circuit 
redesign, using low-power components, utilizing shielding, and designing 
isolating containers (i.e., containers that are usually made of a metal mesh 
and can block acoustic/EM signals of certain frequencies).  However, 
incorporating these techniques will increase IWMD price and thus may not be 
desirable from a cost perspective.

\subsubsection{Information reduction techniques}
These techniques focus on minimizing 
the amount of information leakage. One obvious solution for reducing 
information leakage from IWMDs is to use standard communication protocols and 
implementations. If different manufacturers adopt such standards (e.g., they 
use the same frequency band, modulation protocol, and packet 
size), unintentionally-leaked signals from various devices would be similar 
to each other and carry minimum discriminatory information. However, 
enforcing the use of a single standard communication protocol and a set
of design requirements is difficult due to the fact that different medical 
companies have different priorities. For example, many manufacturers prefer to 
use standard communication protocols at 2.4 $GHz$ (e.g., Bluetooth and ZigBee) 
in their medical devices for connection to smart phones, whereas other 
manufacturers mainly use a customized communication protocol to minimize 
energy or increase the communication range. 
\subsubsection{Noise addition techniques} Such techniques can be implemented in
either hardware or software. In software-based defense schemes, IWMDs and 
their remote controllers can be programmed to produce superfluous 
packets to significantly degrade the accuracy of attacks based on counting 
the number of packets or any other statistical analysis. For example, our 
EM-based attack against the insulin pump can be mitigated as follows. As 
mentioned earlier, the variation in the number of packets can reveal the exact 
amount of injected insulin. Our experimental results demonstrate that the 
total number of packets that might be sent by the remote control to the 
insulin pump for each injection process is between 3 to 52. Thus, if the 
remote control is programmed such that it produces some fake packets for each 
injection step and always sends 52 packets, then counting the number of 
packets will not provide any meaningful information. However, for the majority 
of patients, adding fewer fake packets may be sufficient to drastically reduce 
the accuracy of the attack. Unfortunately, producing fake packets may 
drastically increase the energy consumption of the device, and shorten its 
battery lifetime significantly. Hardware-based countermeasures can also 
be useful in protecting the user from the privacy attacks based on 
physiological information leakage. The local sources of leakage 
(e.g., motherboard, wires, and display board) can be identified during the 
design or manufacturing process. Afterwards, extra components may be added 
to generate intended acoustic/EM noise in a specific frequency range to 
overpower the valuable signal during the normal operation of the device. 
However, the noise generator should not corrupt intentionally-generated 
signals (e.g., alarms from the insulin delivery system). Although noise 
addition techniques can provide promising countermeasures to protect the 
patient against privacy attacks, they have two main drawbacks. First, they 
increase the cost of manufacture. Second, adding a noise generator may increase 
the energy consumption of the device and thus reduce its battery lifetime.

\section{Chapter summary}
\label{PHYS_CONC}
In this chapter, we discussed two sources, namely the human body and IWMDs, that 
continuously leak health information under normal operation. We targeted two 
types of signals for each source: acoustic and EM.  We then described a 
variety of attacks on the privacy of health data by capturing and processing 
unintentionally-generated leaked signals. Moreover, we discussed the 
feasibility of using intentionally-generated acoustic signals (as a 
side-channel information) and EM signals (as a carrier of metadata) to 
compromise the patient's health privacy. Finally, we suggested some 
countermeasures.

%% file: ch-DISASTER/chapter-DISASTER.tex
\chapter{DISASTER: Dedicated Intelligent Security Attacks on Sensor-triggered Emergency Responses \label{ch:DISASTER}}

In this chapter, we introduce a new class of attacks against cyber-physical systems (CPSs),
called \textit{dedicated intelligent security attacks against sensor-triggered 
emergency responses (DISASTER)}. DISASTER targets safety mechanisms deployed 
in automation/monitoring CPSs and exploits design flaws and security weaknesses 
of such mechanisms to trigger emergency responses even in the absence of a 
real emergency. Launching DISASTER can lead to serious consequences for three 
main reasons. First, almost all CPSs offer specific emergency responses and, as a result, are potentially susceptible to such attacks. Second, DISASTER can be easily designed to target a large number of CPSs, e.g., the anti-theft 
systems of all buildings in a residential community. Third, the widespread 
deployment of insecure sensors in already-in-use safety mechanisms along 
with the endless variety of CPS-based applications magnifies the impact of 
launching DISASTER.

In addition to introducing DISASTER, we describe the serious 
consequences of such attacks. We demonstrate the feasibility of launching 
DISASTER against the two most widely-used CPSs: residential and industrial 
automation/monitoring systems. Moreover, we suggest several countermeasures 
that can potentially prevent DISASTER and discuss their advantages and drawbacks \cite{DISASTER_PAPER}.

\section{Introduction}
CPSs offer a transformative approach to automation 
and monitoring through integration of processing, networking, and control. This 
combination and active collaboration of computational elements, e.g., powerful 
base stations, and small embedded devices, e.g., sensors, enable CPSs to 
reliably and efficiently control physical entities.

Recent and ongoing advances in microelectronics, networking, and computer 
science have resulted in significant CPS growth. Such systems
facilitate automation and monitoring in various application
domains, e.g., smart manufacturing lines, smart homes, smart cities, smart 
grids, and smart vehicles. Moreover, with the emergence of the 
Internet-of-Things (IoT) paradigm and IoT-enabled CPSs in the last decade, 
it has become clear that the economic and societal potential of such systems 
is far beyond what may have been imagined. Thus, major investments have been made worldwide to design and develop CPSs.

As a side effect of the rapid development and pervasive use of CPSs, the 
number of potential threats and possible attacks against the security of such 
systems is increasing drastically, while, unfortunately, their security needs 
are not yet well-recognized \cite{SURVEY0,SURVEY2,SURVEY3}.

An essential component of a majority of CPSs is a safety mechanism, which aims 
to minimize harm to users' well-being or damage to equipment upon the 
detection of risks, hazards, or unplanned events. The security of safety 
mechanisms is an emerging research topic that is attracting increasing 
attention in academic, industrial as well as governmental research. A few 
real-world attacks and recent research efforts have demonstrated that 
generic classes of security attacks, e.g., computer worms, man-in-the-middle 
attacks, and denial of service (DoS), which have been extensively 
studied in the network/computer security domains, can be modified to 
\textit{disable} the safety mechanisms of CPSs. For example, in 2003, the SQL 
Slammer worm infected the Davis-Besse nuclear power plant in Ohio, USA, 
and disabled the plant's safety parameter display system and plant process 
computer for several hours \cite{DISABLE0}. Stuxnet \cite{ST2,ST1}, 
a real-world high-impact man-in-the-middle attack, was launched against safety 
mechanisms employed in thousands of industrial CPSs in 2012. Stuxnet faked 
industrial process control sensor signals so that safety mechanisms of 
infected systems were disabled, and as a result, their emergency responses 
were not activated even in the presence of a real emergency. Furthermore, Lamb 
developed a DoS attack against residential intrusion detection systems in 
which the attacker continuously jams the communication channel between 
motion sensors and the base station to suppress the system's alarm that
is supposed to be triggered in the presence of an intruder \cite{INSEC}.

In this chapter, we present and extensively describe a new class of attacks against CPSs, refereed to as \textit{DISASTER}. As opposed to the previously-proposed 
attacks that mainly aim to \textbf{disable} emergency responses of safety 
mechanisms in the presence of an emergency situation, DISASTER attempts to 
\textbf{trigger} the system's emergency responses in the absence of a 
real emergency. 

Our key contributions can be summarized as follows: 
\begin{enumerate}
\item We introduce DISASTER and discuss potential attackers who may be 
motivated to launch such security attacks.
\item We discuss the impact of DISASTER by describing the consequences of 
launching such attacks. 
\item We examine common design flaws and security weaknesses of safety 
mechanisms and their components, which may be exploited by an attacker 
to launch DISASTER.
\item We demonstrate the feasibility of launching DISASTER in realistic 
scenarios, e.g., residential and industrial automation/monitoring systems.
\item We suggest several countermeasures to proactively address DISASTER, and 
discuss their advantages and drawbacks.
\end{enumerate}

The remainder of the chapter is organized as follows. Section \ref{Threat} describes the 
threat model. Section \ref{COMP_WEAK} describes the typical architecture of CPSs that we 
consider in this chapter. Then, it discusses different components, 
design flaws, and security weaknesses of their safety mechanisms. 
Section \ref{CONSEQ} describes potential consequences of launching DISASTER. 
Section \ref{Syst} demonstrates how it is feasible to launch the proposed attacks 
against two real CPSs. Section \ref{COUNTER} suggests several countermeasures to 
prevent DISASTER and describes why proactive countermeasures might not 
always be able to provide sufficient protection against the proposed 
attacks. Finally, Section \ref{DISASTER_CONC} concludes the chapter. 

\section{Threat model}
\label{Threat}
In this section, we first describe what enables DISASTER and makes CPSs 
susceptible to such attacks, and discuss why launching DISASTER can be disastrous in real-word scenarios. Second, we discuss who the potential attackers may be who exploit vulnerabilities of CPSs to launch the proposed 
security attacks, and what their motivations may be. 

\subsection{Problem definition}
As described later in Section \ref{COMP_WEAK}, in a typical CPS, a
centralized processing unit (commonly referred to as base station)
obtains a description of the environment based on the data that it
collects from different sensors, and it processes the sensory data along
with user inputs to control physical objects. Given direct interactions
of such a system with both the environment and users, safety and
security are two fundamental requirements of CPSs. Safety mechanisms
employed in CPSs protect users from undesirable outcomes, risks, hazards, or 
unplanned events that may result in death,
injury, illness, or other harm to individual's well-being, damage to equipment 
or harm to organizations, while security protocols are focused on protecting 
the system from intentional attacks \cite{SSCONF}.

Although safety and security seem to share very similar goals at first glance, 
a close examination of various safety and security requirements demonstrates 
that ensuring both safety and security of CPSs is not always possible due to 
the existence of unavoidable safety-security conflicts 
\cite{CONFLICT_0,CONFLICT_1}. When ensuring both safety and security is not 
feasible, safety is typically given preference and safety mechanisms willingly 
sacrifice security of the system to ensure users' safety. For example, 
modern vehicles commonly support an automatic door unlocking mechanism 
\cite{PATENT_COL}, which opens the vehicle's doors upon the detection of a 
collision. This safety mechanism ensures passengers' safety by completely 
disabling the car's security system after detecting an accident. 

The unavoidable safety-security conflicts along with different design flaws 
and security weaknesses of components, e.g., sensors and base stations,
used in safety mechanisms facilitate
DISASTER. In DISASTER, the attacker exploits such conflicts/weaknesses
to \textit{fool the under-attack safety mechanism into falsely labeling
a normal situation as an emergency in an attempt to activate emergency
responses when they are not needed}. As discussed later in Section
\ref{CONSEQ}, activating emergency responses in the absence of an
emergency can lead to catastrophic situations, ranging from system shutdown 
to life-threatening conditions.

Launching DISASTER can have severe negative consequences in real-world 
scenarios for three reasons. First, since the majority of CPSs offer emergency 
responses, they are susceptible to such attacks. Second, as demonstrated later 
in Section \ref{Attacks}, DISASTER can be implemented to simultaneously target 
a large number of CPSs, e.g., the anti-theft systems of all buildings in a 
residential community. Third, the widespread use of vulnerable sensors along 
with the endless variety of CPS-based applications magnifies negative 
consequences of launching DISASTER.

Despite the fact that safety mechanisms are designed to control hazards and emergency situations and minimize their associated risks, a careless design of a safety mechanism endangers \textbf{both} users' safety and the system's security.

\subsection{Potential attackers}
Next, we discuss potential attackers who may target CPSs, and what 
their motivations might be. 

As discussed earlier, DISASTER is widely applicable since CPSs used for 
automation/monitoring are in widespread use in our everyday lives. Such 
systems may manage a huge amount of information and be used for many services, 
ranging from industrial management to residential monitoring. This has made 
such CPSs targets of interest for a multitude of attackers, including, but not 
limited to, cyberthieves, hacktivists, occasional hackers, and
cyberterrorists. Unfortunately, as described later in Section \ref{Attacks}, 
an attacker with very 
limited resources, e.g., a very cheap radio transmitter such as HackRF 
\cite{HACKRF}, can easily launch powerful large-scale attacks against CPSs.

As extensively discussed later in Section \ref{CONSEQ}, the attackers might launch DISASTER to access restricted areas, cause economic 
damage to companies or individuals, trigger life-threatening operations, 
or halt automation/monitoring processes. Moreover, they might try to 
make CPS use so inconvenient to the user that he is forced to 
shut down the whole system. 

\section{Typical components and weaknesses of safety mechanisms}
\label{COMP_WEAK}
In this section, we first describe the typical architecture of CPSs that 
we consider in this chapter, and discuss different components of their safety 
mechanisms, and two main types of emergency responses that they provide. 
Second, we discuss design flaws and security weaknesses, which are commonly 
present in widely-used safety mechanisms.   

\subsection{Typical CPS architecture}
\begin{figure}[ht]
\centering
\includegraphics [width=250pt,height=180pt]{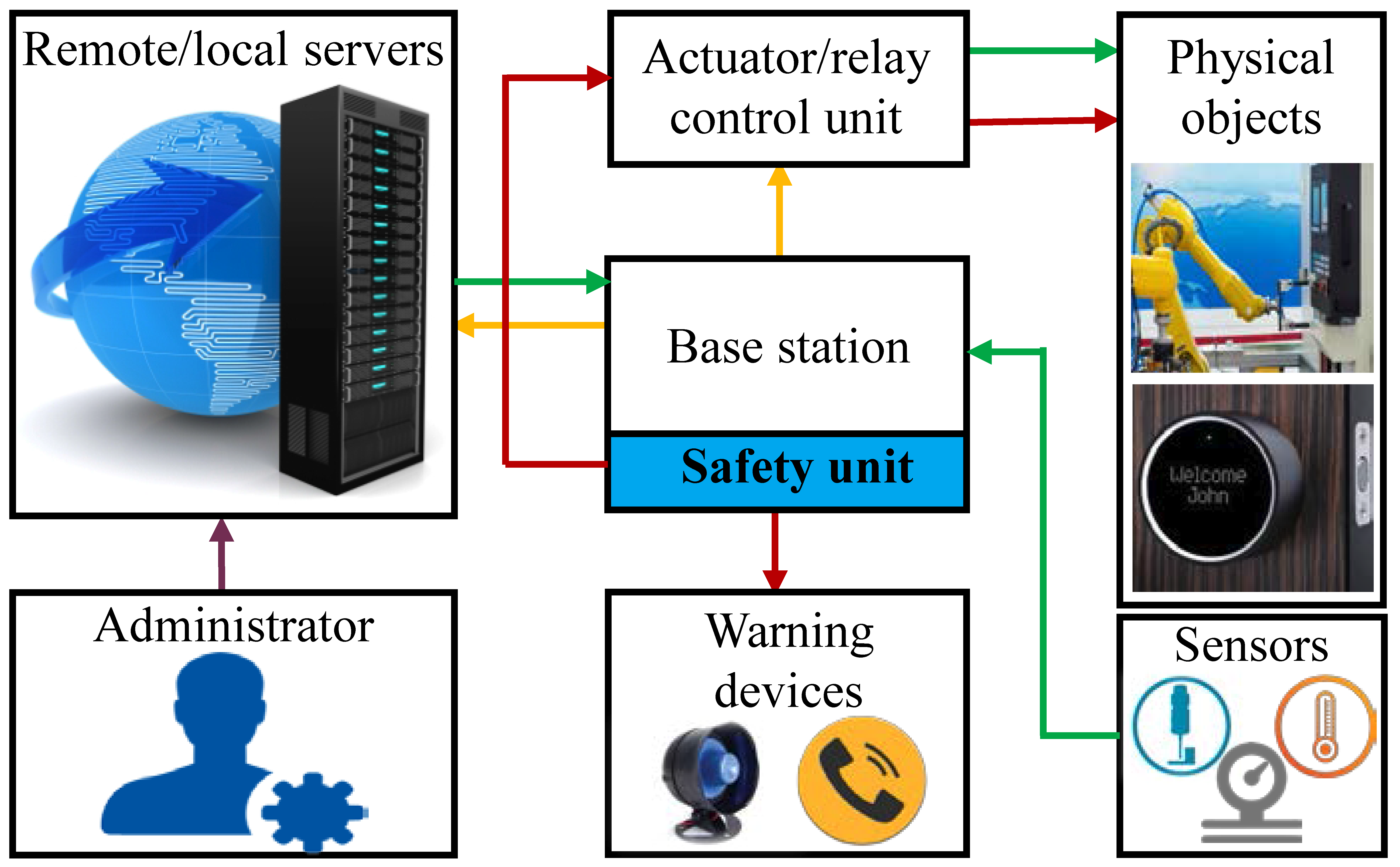}
\caption{Common architecture of CPS. Upon the detection of an emergency,
the safety unit directly controls the physical objects or warns the users 
by activating the passive components.}
\label{fig:ARCH_1}
\end{figure}

Fig.~\ref{fig:ARCH_1} illustrates a common CPS architecture that 
includes safety mechanisms. A typical CPS consists of: (i) a base 
station, which collects and processes environment-related data and controls 
other components, (ii) wireless sensors that continuously collect data and 
transmit them to the base station, and (iii) physical objects that are 
controlled by the base station. State-of-the-art CPSs may also allow the 
user to remotely control, configure, or access the system over the 
Internet. The base station gathers data from different sources, 
e.g., sensors, cloud servers, and user inputs, and processes them to 
control different physical objects. Furthermore, a majority of modern 
CPSs have a safety mechanism, which typically needs two extra components: a 
safety unit and warning devices, e.g., speakers. The safety unit is usually 
integrated into the base station. When it detects an emergency, e.g., a 
fire or an accident, it activates warning devices or overrides control 
signals of physical objects to minimize safety risks associated with the 
situation. In each application domain, certain conditions and states 
are defined as emergency situations in which safety risks are present 
and should be actively and aggressively addressed.

Although emergency responses vary significantly from one application to 
another, there are two main types of emergency responses: active and passive. 
Next, we briefly describe each.

\noindent \textit{\textbf{1. Active response}}: Such a response actively attempts to minimize the safety risks by controlling various actuators and components in the system. When the human operator is unable to provide a sufficiently fast response, it is required that the CPS offer an active response to minimize damages associated with the emergency situation. Typically, when a CPS provides an active response, it initiates a specific emergency procedure or halts the system's normal operation. For example, consider an insulin delivery system that continuously monitors blood glucose and injects insulin into the 
patient's body in order to regulate the blood glucose level. If the system 
detects a life-threatening low level of blood sugar, it immediately
halts the injection procedure to ensure patient safety.

\noindent \textit{\textbf{2. Passive response}}: A passive emergency response is provided by the system to warn human operators, proximate people, e.g., residents of a building, or emergency departments, e.g., fire department, about the need to 
take immediate action. Unlike an active response, it does not directly control physical entities. For example, consider a simple fire detection system.  Upon detecting a fire, the system provides a passive response that activates various notification appliances, e.g., flashing lights and electromechanical horns.

\subsection{Common design flaws and security weaknesses}
\label{WEAK}
Due to the inevitable complexities of CPSs, heterogeneity of entities that form 
them, and limitation of on-sensor resources (e.g., small amount of on-sensor 
storage), designing completely-secure safety mechanisms and finding perfect 
safety design strategies are very daunting. As a result, it is easy to find 
several already-in-market products that are vulnerable to DISASTER. Next, we 
discuss three design flaws and security weaknesses of safety mechanisms 
that may enable an attacker to launch such attacks. 

\subsubsection{Using insecure sensors} 
As mentioned earlier, wireless sensors continuously collect and transmit 
data that are needed for detecting emergency situations. A fundamental 
consideration in the design of wireless sensors is choosing the communication 
protocol. Many manufacturers have decided to design and use customized 
wireless protocols in an attempt to minimize expenses and maximize 
battery lifetime. We examined 70 already-in-market sensors from 
10 different manufacturers (we reviewed documentations of 62 sensors, and 
as discussed later in Section \ref{Attacks}, we closely inspected 
and attacked eight sensors under realistic scenarios) that are widely used 
in safety mechanisms. We noticed that the communication protocol of 
each sensor has at least one of the following security weaknesses.   

\noindent \textit{\textbf{1. Lack of data encryption or obfuscation}}:
Unfortunately, a great number of customized wireless communication protocols 
used in already-in-market sensors support neither strong encryption 
nor obfuscation mechanisms and are, hence, susceptible to various forms 
of security attacks. In particular, due to the specific requirements of 
safety systems (e.g., fast response time and monitorability) and common 
limitations of sensing platforms (e.g., limited on-sensor energy, small 
amount of available memory, and limited computation power), the majority 
of sensors utilized in safety mechanisms transmit unencrypted non-obfuscated 
packets over the communication channel. As a result, 
an attacker can reverse-engineer the system's communication protocol 
and generate illegitimate packets (packets not created by legitimate sensors) 
using his equipment.

\noindent \textit{\textbf{2. Lack of timestamp}}: A
great number of sensors do not include a timestamp (i.e., a sequence 
of encrypted information identifying when the transmission occurred) in 
their packets. As a result, the system's base station is unable to 
distinguish new legitimate packets generated by the sensor from old 
seemingly legitimate ones, which are recorded and retransmitted by an attacker.

\noindent \textit{\textbf{3. Using default passwords}}:
Setting non-random default passwords at the manufacturing/installation time 
is a very common mistake that can lead to severe security 
attacks against the system even when strong encryption mechanisms are 
utilized to secure the communication channel. It is very common for system 
administrators or user to forget to change the system's default password 
at installation time. A recent article provided a list of more than 
73,000 cameras in 256 countries that use standard communication protocols and 
a strong encryption mechanism, yet are not immune to security attacks 
because they use a default password for encrypting their communications 
\cite{CAMERA}.

\noindent \textit{\textbf{4. Using short sensor identifiers}}:
In order to distinguish different sensors from each other and enhance the 
security of the communication protocol, most sensors include 
their identifier (also referred to as the pin code or identification number) 
in all packets they transmit over the communication channel. The 
base station uses the sensor's identification code to ensure that the 
incoming packet comes from one of the already-registered sensors, which 
are known to the system. However, as demonstrated later in 
Section \ref{Attacks}, several in-market sensors from
well-known manufacturers use very short sensor identifiers (4-8 bits).
As a result, they are susceptible to brute-force attacks (i.e., attacks 
consisting of systematically checking all possible sensor identifiers 
until the correct one is found).

\subsubsection{Offering inessential sacrifices} 
As mentioned earlier, upon the detection of an emergency situation, 
automation/monitoring CPSs willingly sacrifice some of their security 
mechanisms to ensure users' safety. As an example, fire evacuation systems 
open all doors to enable firefighters to access different rooms and allow 
the occupants to safely leave the building. Although this evacuation 
mechanism seems essential to ensuring occupant safety, it might enable 
an attacker to access restricted areas by triggering an emergency response. 
This example demonstrates that designers should take both safety and 
security considerations into account, when designing emergency responses 
of a CPS.

\subsubsection{Relying on a single sensor type}
In order to provide a reactive emergency response when required, 
the automation/monitoring CPS must be able to correctly distinguish 
abnormal situations from normal ones. In fact, the most important steps 
in minimizing safety risks is detecting emergency situations. Therefore, 
before providing any response, the CPS needs to collect sufficient sensory 
data to obtain a clear description of its environment. If insufficient 
information is given to the system, it might fail to correctly recognize 
emergency situations. Unfortunately, in order to minimize costs, the 
majority of the already-in-use automation/monitoring CPSs only process a 
single environmental attribute.  For example, consider a fire evacuation 
system that only relies on smoke detection sensors. Such a system provides 
an emergency response when at least one of the sensors detects the 
existence of smoke. Thus, an attacker can easily trigger the emergency 
response by only targeting a single vulnerable smoke detection sensor. 

\section{Potential consequences of launching DISASTER}
\label{CONSEQ}
As mentioned earlier, CPSs are in widespread use and handle sensitive tasks in 
various application domains. Hence, launching tailored attacks, like DISASTER, 
that are applicable to various forms of automation/monitoring CPSs, 
can lead to serious consequences. Such consequences depend on the type of 
emergency response activated by the attack. Generally, the negative impact of 
triggering an active response is more significant than the impact of triggering 
a passive response due to the fact that the former can actively control 
various critical operations and even bypass human operators' decisions. Next, 
we describe possible consequences associated with launching DISASTER. 

\subsection{Life-threatening conditions}
Triggering the emergency response of a CPS that handles critical
operations, e.g., medical or industrial automation tasks, can lead to serious 
life-threatening conditions. This can range from conditions affecting an 
individual to those affecting a large number of people. For example, consider 
an insulin delivery system that is equipped with a safety mechanism, which 
monitors the blood glucose and immediately stops the injection procedure when 
it detects the patient has hypoglycemia, i.e., a life-threatening low level of 
blood glucose. Triggering the active emergency response of the insulin delivery 
system immediately shuts down the device. An attacker might be able to trigger 
such an active response even when the blood glucose level is normal/high 
to halt the delivery system and cause hyperglycemia, i.e., a life-threatening 
high level of blood glucose.
 
\subsection{Economic collateral damage} Economic damages refer to monetary 
losses, including, but not limited to, loss of property, machinery, equipment, 
and business opportunities, costs of repair or replacement, the economic value 
of domestic services, and increased medical expenses. Almost all emergency 
responses have associated costs, even when they are triggered by
a real emergency situation. For example, consider a vehicular CPS that is 
designed to inflate the vehicle's airbags in a collision to provide 
protection in an accident. In a minor collision that results in the 
deployment of airbags, the whole dashboard panel, steering wheel, and all 
airbags have to be replaced. In such cases, the active emergency response 
provided by the CPS is quite costly. Therefore, if an attacker can activate 
such an emergency response, he will be able to cause collateral economic 
damage.

The cost associated with triggering emergency responses varies
significantly from one application domain to another. For example, the 
economic collateral damage that results from triggering an emergency response 
of a vehicular CPS is much less than that of an industrial CPS that controls a 
manufacturing line.

\subsection{Overriding access control mechanisms}
In the presence of an emergency situation, a CPS might also be able to command 
the access control systems that control which users are authorized to access 
different restricted areas. Such a control is important for two reasons. First, 
the CPS can facilitate the evacuation procedure in the case of an emergency. 
Second, the system can lock down particular areas to prevent an intruder from 
escaping. For example, a residential CPS, which is able to control door locks, 
may open the main entrance to ensure that firefighters can easily access 
restricted areas and residents can safely evacuate the building. However, if 
an attacker can trigger this safety mechanism in the absence of an emergency 
situation, he might be able to bypass physical security mechanisms and access 
restricted areas.   

\subsection{Unintended ignorance}
As mentioned earlier, CPSs provide both passive and active responses to 
minimize damage associated with an emergency situation. A majority of 
emergency responses could be extremely annoying to the proximate people if 
they are activated in the absence of an emergency situation. For example, 
upon the detection of an emergency situation, a great number of CPSs activate 
notification appliances, e.g., electromechanical horn and speaker, that 
generate a high-pitched noise to inform the nearby people about the need to 
take immediate action. If an attacker launches DISASTER that activates 
notification appliances several times in a short time frame, 
the system administrator might be convinced that the system is faulty and turn 
off the emergency responses. This might lead to serious safety/security risks 
for the duration the emergency responses remain off.

\section{Launching DISASTER}
\label{Syst}
In this section, we demonstrate the feasibility of launching DISASTER in 
realistic scenarios, e.g., residential and industrial automation/monitoring 
systems. As mentioned in Section \ref{WEAK}, communication 
protocols utilized in wireless sensors commonly have various security 
weaknesses. Next, we first briefly describe two well-known types of attacks 
that exploit security weaknesses of communication protocols to create and 
transmit illegitimate packets. Second, we demonstrate how an attacker can 
tailor these generic forms of attacks to trigger the emergency responses of 
safety mechanisms and endanger both user safety and system security.

\subsection{Creating and transmitting illegitimate packets}
\label{APPR}
Here, we briefly describe two attacks against sensors that enable the attacker 
to send illegitimate packets to the base station. In both attacks, we use 
GNURadio \cite{GNU}, a development toolkit that can be used along with an 
external radio frequency (RF) hardware, e.g., HackRF\cite{HACKRF}, to 
implement various software-defined transmitters/receivers that are implemented 
as software programs to control external RF devices.

\noindent \textit{\textbf{Attack 1: Retransmitting recorded packets}}: In this approach, an attacker aims to record data packets and 
retransmit them to the base station without processing or modifying their 
contents. To do so, the attacker first builds an RF receiver that listens to 
the communication channel between sensors and the base station and records 
the transmitted packets. Then, he uses an RF transmitter, which can 
retransmit previously-recorded packets on the same communication channel. 
If the frequency of the communication channel is known to the attacker, he 
can implement the above-mentioned receiver/transmitter using the 
built-in libraries of GNURadio. 

The frequency of the communication channel can be extracted from documents 
submitted to Federal Communications Commission (FCC). FCC is an independent 
U.S. government agency that tests all wireless products sold in the U.S. and 
provides a public database, which includes test reports and documentations 
of the products \cite{FCC}. In order to find the frequency of the communication 
channel used by a sensor, the attacker only needs to find the sensor's 
documentations by searching its FCC code (i.e., an identification code that 
specifies the sensor's manufacturer and type) in FCC's public database. FCC 
codes are commonly written on the sensor's cover. 

\noindent \textit{\textbf{Attack 2: Reverse engineering the communication protocol}}:
\indent In this approach, the attacker records several data packets and 
processes them to explore how the communication protocol sends digital data 
over the communication channel. We describe next how an attacker can
reverse-engineer the communication protocol of an arbitrary sensor using 
HackRF and GNURadio.  

\begin{figure}[ht]
\centering
\includegraphics[width=240pt,height=230pt] {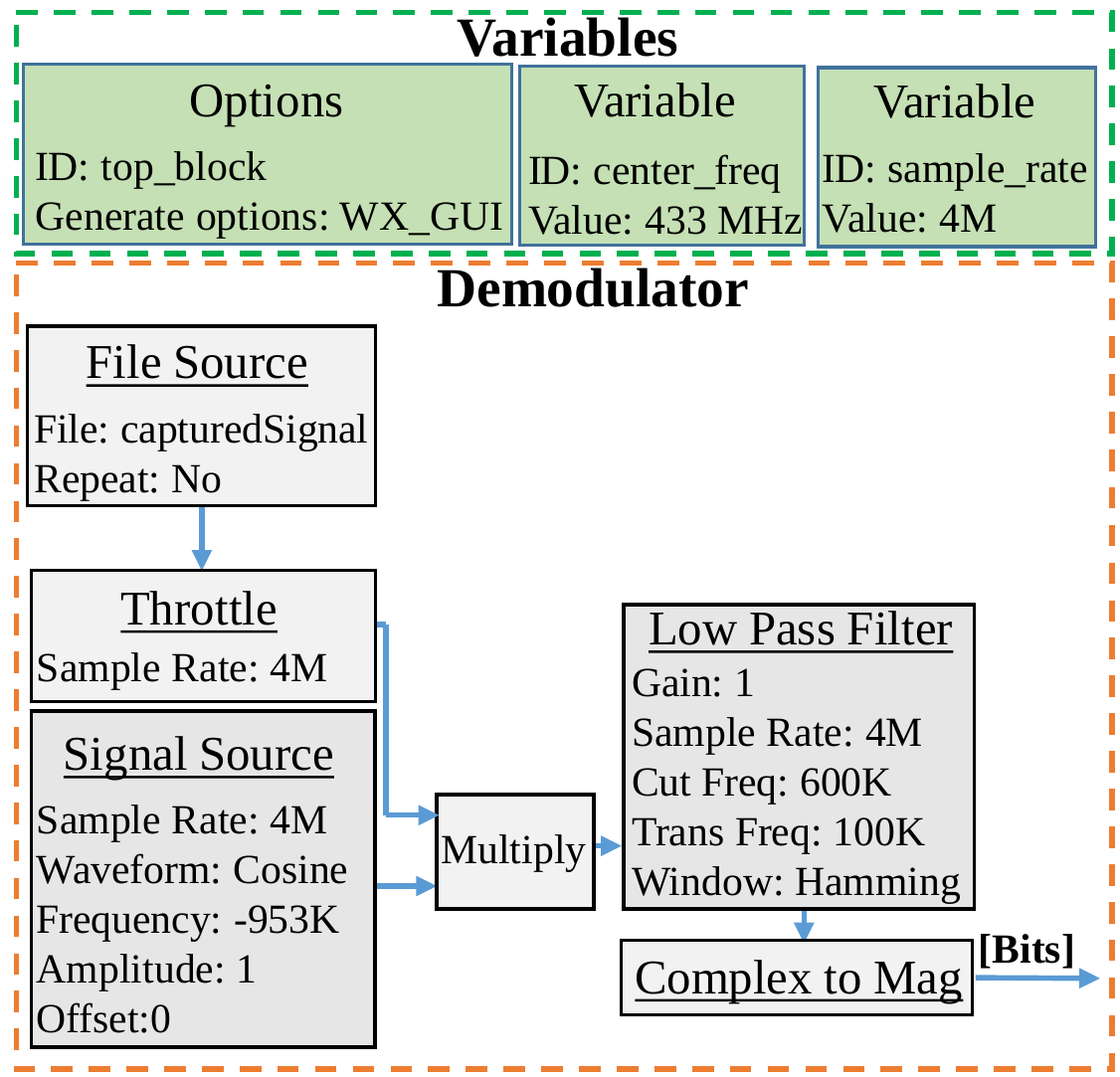}
\caption{The implementation of an OOK demodulator in GNURadio}
\label{fig:trans}
\end{figure}

\begin{enumerate}
\item The attacker first obtains the communication frequency of the sensor 
and records several packets from the sensor using the method discussed in 
the previous approach.
\item Then, the attacker finds the modulation type of the communication 
protocol. The two most commonly-used communication protocols used in wireless 
sensors are on-off keying (OOK) and binary frequency-shift keying (BFSK). 
OOK is the simplest form of amplitude-shift keying modulation in which 
digital data are presented as the presence/absence of a carrier wave, and 
BFSK is the simplest form of frequency-shift keying (FSK) that uses a pair 
of discrete frequencies to transmit binary digital data. The modulation 
type of a communication protocol can be easily detected by examining the 
Fourier transform of a packet received by HackRF. A single peak (two discrete 
peaks) in the Fourier transform represents an OOK (BFSK) modulation. 
\item After finding the modulation type of a communication protocol, the 
attacker implements a software-defined demodulator in GNURadio that extracts 
the transmitted digital data from the recorded analog signal. 
Fig.~\ref{fig:trans} demonstrates the implementation of an OOK demodulator 
in GNURadio. 
\item If the sensor does not support any encryption mechanism, the attacker can 
easily examine the digital data to determine what each bit represents, and 
how he can generate seemingly legitimate packets with arbitrary content.
\end{enumerate}
This approach provides two fundamental advantages over the previous approach. 
First, the attacker does not need to capture any packets from the actual 
sensors that he targets. In fact, he can conduct several experiments in a 
test environment, e.g., a laboratory, to discover potential vulnerabilities 
in the system. Second, the attacker can improve the quality of the 
transmitted signal and increase its signal-to-noise ratio in an attempt 
to launch an attack against the system from a large distance.

\subsection{DISASTER case studies}
\label{Attacks}
In this section, we demonstrate the feasibility of launching DISASTER against 
the two most widely-used automation/monitoring CPSs: residential and industrial.

\subsubsection{Case I: Residential automation/monitoring CPSs} We first briefly 
discuss what a typical residential CPS does, what types of sensors are 
commonly used in such a system, and what emergency responses it offers in 
the presence of an emergency. Second, we demonstrate how an attacker can use 
the two well-known generic types of attacks discussed in Section \ref{APPR} 
to trigger the emergency responses of residential CPSs in real-world scenarios.

\noindent \textit{\textbf{Residential CPSs, their services and emergency responses}}: A residential CPS is mainly designed to offer physical security 
mechanisms, enhance residents' convenience, and minimize energy consumption. 
It can also offer increased quality of life to the residents who need special 
assistance, e.g., the elderly and disabled people. It processes the data 
collected by various sensors to control lighting, heating, and cooling, 
and to monitor/command the security locks of gates and doors. In addition, 
the base station continuously collects and processes real-time 
environment-related data gathered by sensors to detect emergency situations. 
Table \ref{table:sensorsR} includes different sensors that are commonly used 
in residential CPSs, a short description of each sensor, and services that 
rely on each sensor. 

\begin{table*}[t]
\caption{Different sensors used in a typical residential CPS, their
descriptions, and services} 
\centering 
\begin{tabular}{|l |l| l|} 
\hline\hline 
Sensor type & Description & Service\\ [0.5ex]
\hline
Humidity sensor & measures moisture content&
Heating/cooling\\ [0.5ex]
Temperature sensor & measures the current temperature &
Heating/cooling \\ [0.5ex]
Light sensor & measures the luminance & Lighting\\
[0.5ex]
Motion sensor & detects the presence of a person & Anti-theft mechanism\\ [0.5ex]
Door sensor & checks if a door has been opened &
Anti-theft mechanism\\ [0.5ex]
Smoke detector & detects the presence of
smoke/fire & Fire detection\\ [0.5ex]
\hline 
\end{tabular}
\label{table:sensorsR}
\end{table*}

State-of-the-art residential CPSs are able to detect two common emergency 
situations: fire and ongoing burglary, and provide three typical emergency 
responses: two passive and one active. Next, we elaborate on these responses 
and discuss the negative consequences of activating each response.

\noindent \textit{Passive response I: Activating warning devices:} In the 
presence of an emergency situation, notification appliances, e.g., flashing 
lights, electromechanical horns, or speakers, are activated to warn the 
proximate people about the need to take immediate action. A majority of 
notification appliances generate a high-pitched sound to attract the attention 
of those nearby. The generated sound could be extremely annoying if it is 
activated in the absence of an emergency situation. Hence, if a potential 
attacker can trigger this emergency response several times in the absence of 
an emergency situation, residents might be convinced that the system is faulty 
and turn off the emergency response. This might lead to serious 
safety/security risks and concerns. For example, a burglar might try to 
trigger the anti-theft alarm several times in a short period of time, e.g., in 
an hour, in the hope of convincing the user to turn off the monitoring system.

\noindent \textit{Passive response II: Informing police/fire department:} 
Requesting immediate help from the police/fire department, when a real 
threat is not present, puts firefighters, police officers, as well as the 
public at risk by needlessly placing heavy, expensive equipment on the streets 
while wasting fuel and causing traffic jams. Moreover, if an attacker can 
initiate a help request several times in a short period of time, he might be 
able to persuade firefighters, police officers, and occupants to believe that 
when an alarm goes off it is likely a false alarm. As demonstrated later, 
DISASTER can be launched from a large distance (e.g., over  100 $m$ 
from the base station). Therefore, an attacker might be 
able to design a large-scale attack (e.g., he can launch DISASTER using 
HackRF, while driving in a residential community, to trigger the alarm 
systems of all houses in the community) to impose significant additional 
cost on both residents and the responsible governmental department, and 
convince the residents to turn off their security/safety alarms.

\noindent \textit{Active response: Controlling door locks:} As mentioned 
earlier, a residential automation/monitoring system may be able to 
automatically control the locks upon the detection of a fire or 
burglary. In the presence of a fire, it opens the main doors/entrances to 
ensure that firefighters can enter the affected areas and residents can safely 
evacuate the building. Moreover, in the presence of an ongoing burglary, it 
locks the main entrances to ensure that the thief is not able to leave the 
crime scene until police officers arrive. Although this emergency response is 
offered to minimize potential safety risks, triggering this response by an 
attacker in the absence of an emergency situation could lead to
serious security issues. For example, if the attacker triggers the fire 
evacuation procedure, he will be able to bypass the physical security 
mechanism of the building by unlocking main entrances. Similarly, the attacker 
might be able to confine the residents inside the house by initiating the 
anti-theft lock-down procedure.

\noindent \textit{\textbf{Demonstration of DISASTER against residential CPSs}}: In order to examine the feasibility of launching DISASTER against 
residential CPSs, we developed two experimental scenarios using the approaches 
described in Section \ref{Syst}. In both scenarios, we targeted three 
types of sensors (highlighted in red in Table \ref{table:sensorsR}). The 
residential CPS processes the data gathered by these sensors to detect 
emergency situations (fire or burglary). In our experimental setup, we 
closely inspected six already-in-market sensors (two motion detectors, three 
smoke detectors, one door sensor) made by well-known manufacturers that
cater to the home automation industry. Since these sensors are deployed 
in numerous already-in-use systems, we choose not to disclose their brand 
and model number in this chapter.

Next, we describe how the two previously-mentioned generic attacks can be 
used to launch DISASTER and activate emergency responses of the system. 

\begin{figure}[ht]
\centering
\includegraphics [trim = 85mm 45mm 90mm 45mm ,clip, width=270pt,height=180pt]{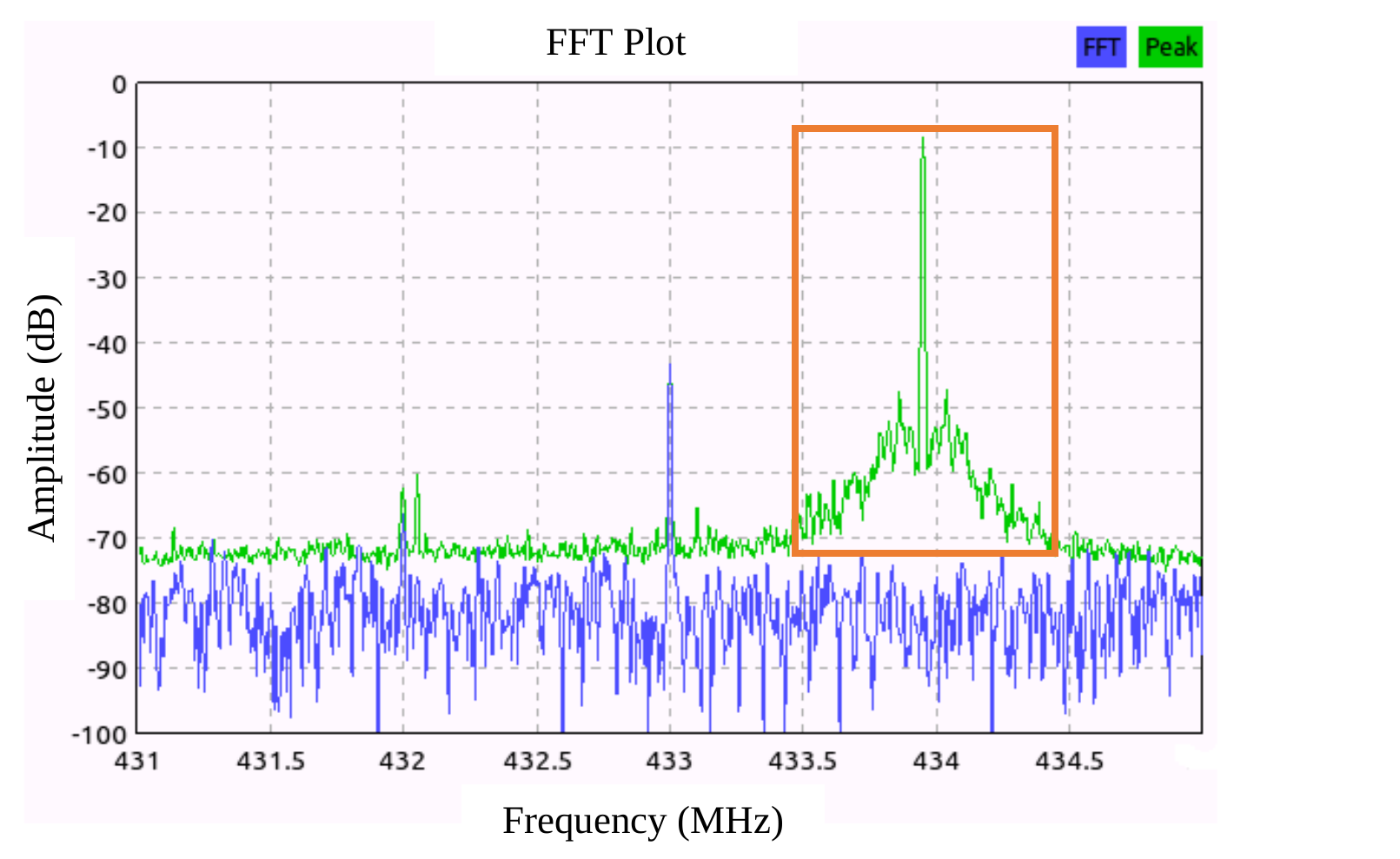}
\caption{The door sensor generates a packet as soon as it detects the door is open. The spike in the fast Fourier transform of the analog signal shows a single transmission using OOK modulation.}
\label{fig:doorp}
\end{figure}

\noindent \noindent \textit{Experimental scenario 1: Retransmitting packets:} 
In our experimental setup, we captured and retransmitted 20 packets 
from each sensor (120 packets in all) using the software-defined 
transmitter/receiver described in Section \ref{APPR}. Fig.~\ref{fig:doorp} 
demonstrates a packet generated by the door sensor and captured by HackRF. The 
base station of all the under-experiment sensors accepted previously-recorded 
packets. This indicates that the packets generated by these sensors 
include neither a timestamp nor a sequence number. Thus, an attacker can record 
a packet from each of these sensors and retransmit it to the base station of 
the CPS that utilizes the sensor in an attempt to trigger emergency responses.  

In this experimental scenario, we placed HackRF at different distances from 
each under-experiment sensor to find the maximum recording distance from which 
the attacker can record a packet using HackRF. Moreover, we increased the 
distance between HackRF and the base station to find the maximum 
retransmission distance for each sensor from which a previously-captured 
packet can be received and accepted by the sensor's base station. 
Table \ref{table:resultsDD} summarizes results of this experiment for 
different sensors. 

\begin{table}[ht] 
\caption{Maximum recording distance and maximum retransmission distance for each sensor in Experimental scenario 1} 
\centering 
\begin{tabular}{l c c} 
\hline\hline 
Sensor & Maximum recording & Maximum retransmission\\ 
& distance (m) &  distance (m) \\ [0.5ex]
\hline 
Motion sensor I & 58 & 54\\ [0.5ex]
Motion sensor II & 110 & 105 \\ [0.5ex]
Smoke detector I & 67 & 50\\ [0.5ex]
Smoke detector II & 52 & 55\\ [0.5ex]
Smoke detector III & 54 & 48\\ [0.5ex]
Door sensor & 56 & 54 \\ [0.5ex]

\hline 
\end{tabular} 
\label{table:resultsDD}
\end{table}
 
Sensors that enable the anti-theft mechanism (motion and door sensors) 
transmit data very frequently even when the mechanism is completely disabled. 
The motion sensor transmits a packet when it detects a moving object, and the 
door sensor transmits a packet when it detects an open door. For such sensors, 
an attacker can simply capture a packet when the mechanism is disabled, e.g., 
when the residents are inside, and retransmit the packet when it is enabled. 
Unlike motion/door sensors, smoke sensors rarely transmit a packet to the 
base station since their event-driven transmission protocol only transmits 
a packet to the base station 
when an actual threat, e.g., a fire, is present. Thus, capturing and 
retransmitting a packet that is generated by smoke sensors are difficult and 
potentially very time-consuming for the attacker. In the second experimental 
scenario, we discuss how the attacker can reverse-engineer communication 
protocols deployed in smoke detectors to easily launch DISASTER against the 
system.

\noindent \noindent \textit{Experimental scenario 2: Reverse engineering:} As 
mentioned in Section \ref{Syst}, in this approach, the attacker 
records and demodulates transmitted packets in a test environment, e.g., a 
laboratory, to examine how a sensor transfers digital data over the 
communication channel.

\begin{figure}[ht]
\centering
\includegraphics [width=250pt,height=180pt]{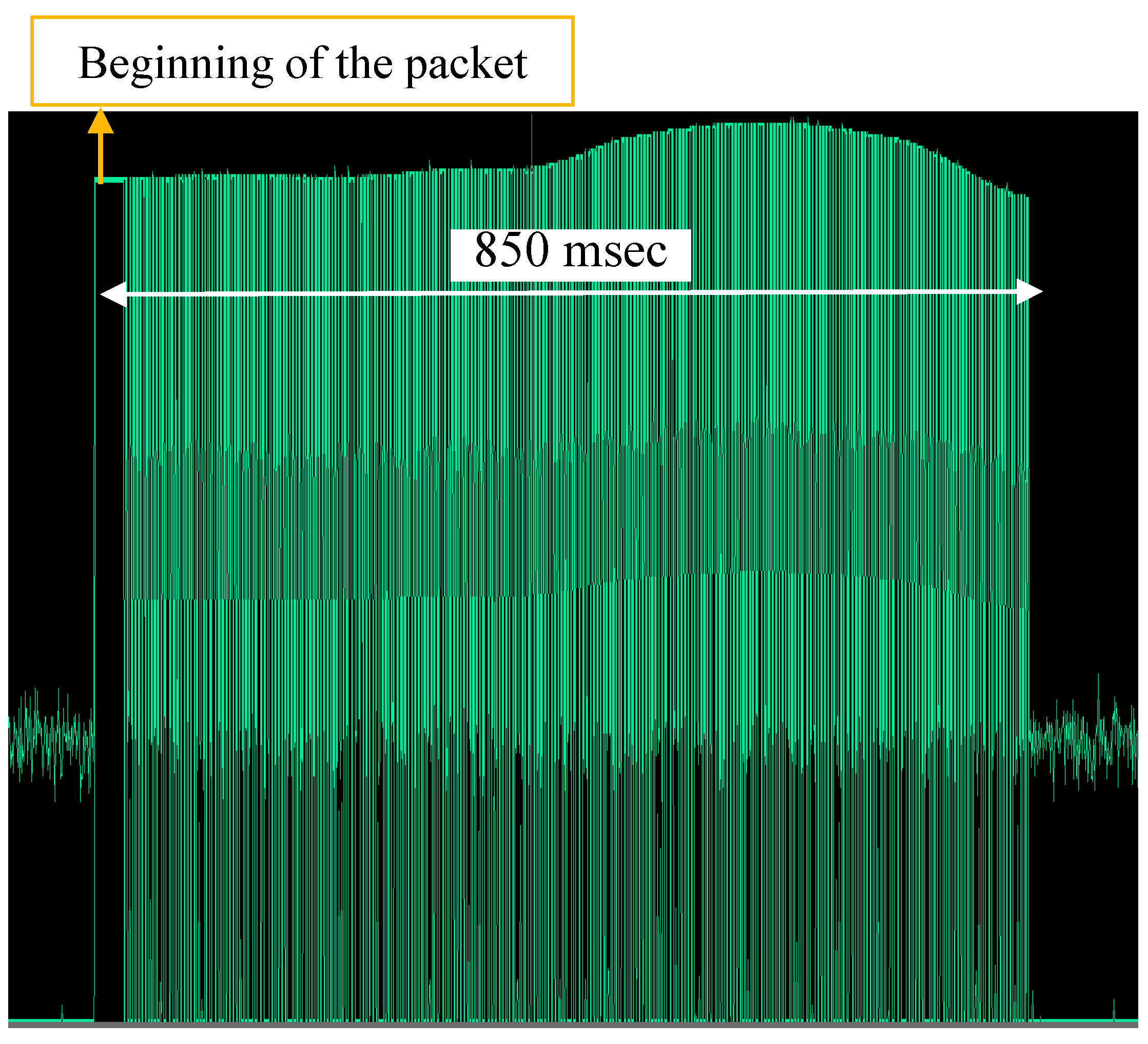}
\caption{The bitstream transmitted by the door sensor to the base station of 
the residential CPS. The door sensor repeatedly transmits a single static 
packet, which includes its 4-bit pin number, to its base station.}
\label{fig:DOORB}
\end{figure}

We examined the communication protocol used in the six under-experiment 
sensors. The examination revealed that all sensors share a common 
security weakness: in order to provide a cost-effective solution, the 
manufacturers used very simple non-standard transmission protocols that do not 
provide any cryptographic mechanism. Indeed, the packets transmitted from these 
sensors to their base stations are neither cryptographically protected nor 
completely obfuscated. Fig.~\ref{fig:DOORB} demonstrates the bitstream 
transmitted by a door sensor to the base station of a residential CPS. The 
analog signal (Fig. \ref{fig:doorp}) captured by HackRF is demodulated using 
the OOK demodulator (Fig.~\ref{fig:trans}). This sensor repeatedly (20 times) 
transmits a single static packet (that does not change over time), which 
includes its 4-bit pin number (a very short sensor identification code), to 
its base station. 

We were able to completely reverse-engineer the six communication protocols 
used in these sensors and determine that all of them generate static packets, 
which include fixed pin numbers. In other words, the only unknown field in the 
bitstream of an arbitrary packet generated by the sensors was the device's pin 
number. Table \ref{table:sensorSpec} specifies the communication frequency, 
modulation type, and pin length of each under-experiment sensor.

\begin{table}[ht] 
\caption{Communication frequency, modulation type, and pin length of each 
residential sensor} 
\centering 
\begin{tabular}{l c c c} 
\hline\hline
Sensor & Communication & Modulation & Pin length\\ 
& frequency (MHz) & & (bits)\\ [0.5ex]
\hline 
Motion sensor I & 433 & OOK & 4\\ [0.5ex]
Motion sensor II & 433 & OOK & 8\\ [0.5ex]
Smoke detector I & 920 & OOK & 8\\ [0.5ex]
Smoke detector II & 920 & OOK & 8\\ [0.5ex]
Smoke detector III & 433 & OOK & 8 \\ [0.5ex]
Door sensor & 433 & OOK & 4 \\ [0.5ex]
\hline 
\end{tabular} 
\label{table:sensorSpec}
\end{table}

After completely reverse engineering the protocol, we implemented a brute-force 
attack using different possible values of the pin numbers of the sensors. We 
were able to find the actual pin numbers of all sensors in less than five 
seconds. Then, we placed HackRF at different 
distances from the base station of the under-experiment sensors and used the 
maximum transmission power of HackRF to determine the maximum transmission 
distance from which this attack is possible. Table \ref{table:resultsR2} 
summarizes the results of this experiment for different sensors. 
   
\begin{table}[ht] 
\caption{Maximum transmission distance for each residential sensor in Experimental scenario 2} 
\centering 
\begin{tabular}{l c} 
\hline\hline 
Sensor & Maximum transmission distance (m)\\ [0.5ex]
\hline 
Motion sensor I & 75 \\ [0.5ex]
Motion sensor II & 85 \\ [0.5ex]
Smoke detector I & 75\\ [0.5ex]
Smoke detector II & 80\\ [0.5ex]
Smoke detector III & 80 \\ [0.5ex]
Door sensor & 75 \\ [0.5ex]
\hline 
\end{tabular} 
\label{table:resultsR2}
\end{table}
 
\subsubsection{Case II: Industrial automation/monitoring CPS} We first briefly 
discuss different services and emergency responses offered by a typical 
industrial CPS. Second, we demonstrate how an attacker can trigger the 
emergency responses of industrial CPSs.

\noindent \textit{\textbf{Industrial CPSs, their services and emergency responses}}: A typical industrial automation/monitoring CPS offers various automatic 
mechanisms to operate the equipment, e.g., machinery and boilers, with minimal 
human intervention, and several approaches that enable remote monitoring of 
the industrial environment. Generally, industrial automation/monitoring CPSs 
deal primarily with the automation of manufacturing, quality control, and 
material-handling processes. In addition, almost all modern industrial 
automation/monitoring CPSs continuously monitor the environment to detect 
emergency situations. These situations need to be aggressively addressed due 
to the fact they can be catastrophic in a large industrial setting. 

State-of-the-art industrial CPSs are able to detect a variety of emergency 
situations, e.g., a tank overflow, system failure, or a fire. Upon the 
detection of an emergency situation, they commonly provide four emergency
responses, including two passive responses and two active responses. The two 
passive responses are similar to the ones provided by residential CPSs.
Hence, we discuss the two active responses that are commonly offered by 
industrial CPSs and discuss the negative consequences of activating each 
response.

\noindent \textit{Active response I: Halting normal operation:} A halting procedure is initiated to shut down a part, e.g., a plant or control unit, of the industrial setting or the whole production line when 
necessary. Upon the detection of an emergency situation, the centralized base 
station responds by placing the controllable elements, e.g., valves and
pumps, into a safe state. For example, a halting procedure controls valves to 
stop the flow of a hazardous fluid or external gases upon the detection of a 
dangerous event. This provides protection against possible harm to people, 
equipment or the environment. Launching DISASTER against an industrial 
CPS, which activates the halting procedure, 
may lead to two consequences: production loss and profit penalty. A halting 
procedure shuts down specific units or the entire facility. This can lead to 
a significant production loss in chemical industries, e.g., gasoline-centric refinery, where shutting down a unit may stop chemical reactions from 
completing. Moreover, several time-consuming safety checks need to be done 
before restarting the normal operation. Thus, the facility might need to be 
shut down for a substantial amount of time.  This could cause a significant 
impact on profits. For example, an average-sized U.S. Gulf Coast oil refinery 
loses 68,000 dollars a day for a downed unit \cite{FAIL}. 

\noindent \textit{Active response II: Initiating a damage control mechanism}: Damage control mechanisms include any prudent action aimed at preventing/reducing any expected damage to the industrial setting, stabilize 
the situation caused by the damage or alleviate the effects of damage. The 
main purpose of damage control is to offer a way to return the production line 
to its normal operation with minimal loss of property or life. A common damage 
control mechanism in an industrial environment is automatic fire suppression, 
which employs a combination of dry chemicals and wet agents to extinguish a 
fire. It applies an extinguishing agent to a three-dimensional enclosed space 
in order to achieve a concentration of the agent that is sufficient to 
suppress the fire. A fire suppression system that primarily 
injects gases into enclosed spaces presents a risk of suffocation. Numerous 
incidents have been documented where individuals in such spaces have been 
killed by carbon dioxide agent release \cite{SHE1,SHE2}. Moreover, the 
positive pressure caused by these gases may be sufficient to break windows and 
even walls and destroy the surrounding equipment. Thus, launching DISASTER 
against an industrial CPS that triggers its fire suppression mechanism may 
lead to serious consequences, ranging from severe damage to the equipment to 
life-threatening conditions.

\begin{figure}[ht]
\centering
\includegraphics [trim = 40mm 16mm 0mm 10mm ,clip, width=270pt,height=170pt]{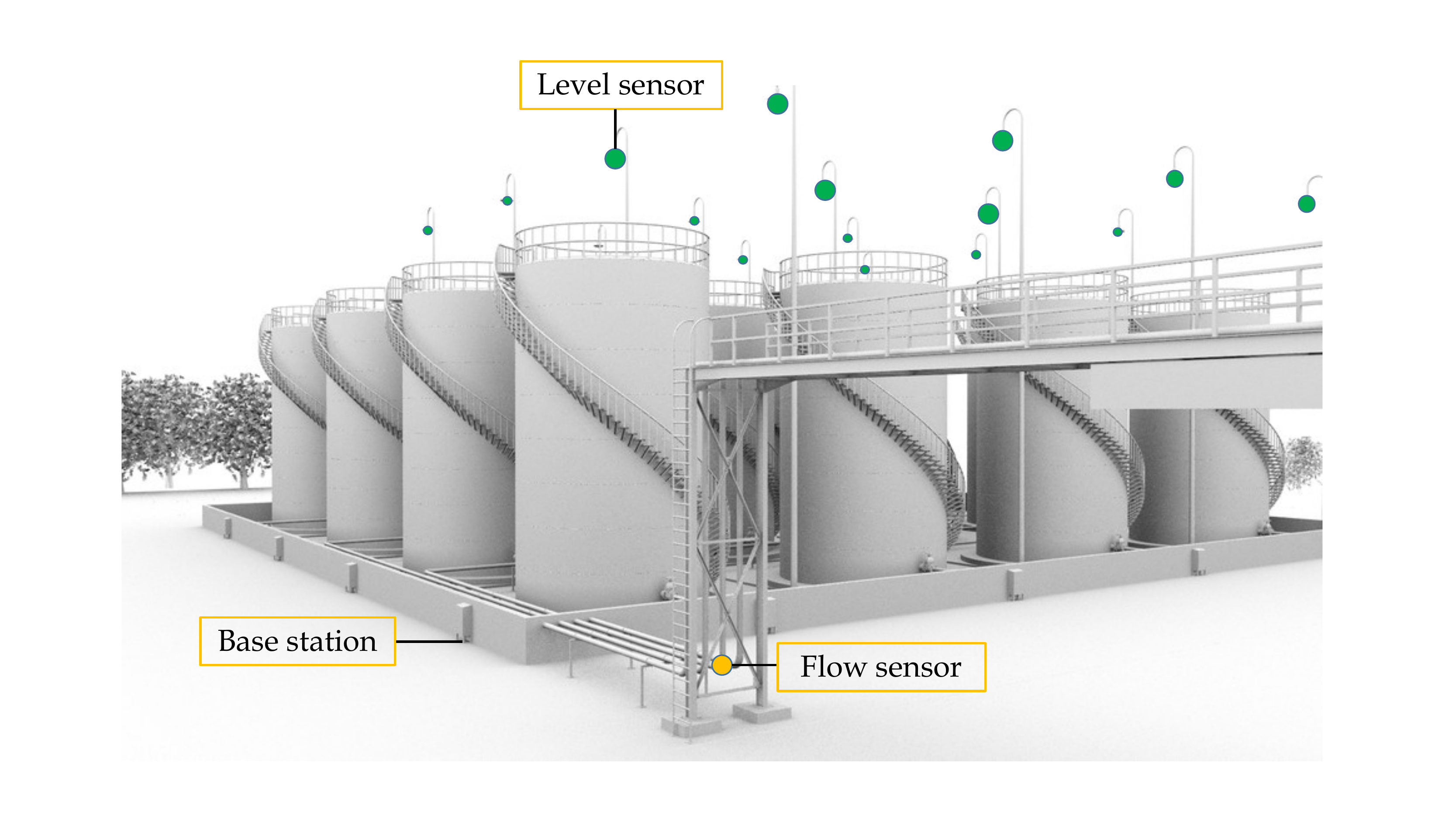}
\caption{A simple industrial automation/monitoring CPS}
\label{fig:INDT}
\end{figure}

\noindent \textit{\textbf{Demonstration of DISASTER against industrial CPSs}}: In order to investigate the feasibility of launching DISASTER against 
industrial CPSs, we targeted two industrial systems that use level sensors to 
monitor liquid level changes in storage tanks (Fig.~\ref{fig:INDT}). Level 
monitoring-based safety mechanisms are commonly used in various industrial
environments, e.g., the oil industry, to detect an emergency situation that is 
called tank overflow. 

To activate the emergency responses of such a system, the attacker can 
transmit an illegitimate packet to the base station that indicates that the 
storage tank is full. Using each of the two approaches described in 
Section \ref{Syst}, an attacker can generate such packets. 
However, capturing and retransmitting level sensors' regular data packets 
(i.e., the packets that are periodically transmitted to the base station 
to report the level of liquid) cannot activate the system's emergency 
responses. In fact, the attacker needs to record a packet in the presence of 
a real emergency situation, which is extremely rare in real-world industrial 
environments, and use it later. Thus, the first approach may not be 
practical for launching DISASTER against industrial systems described above. 

\begin{figure}[ht]
\centering
\includegraphics [width=250pt,height=210pt]{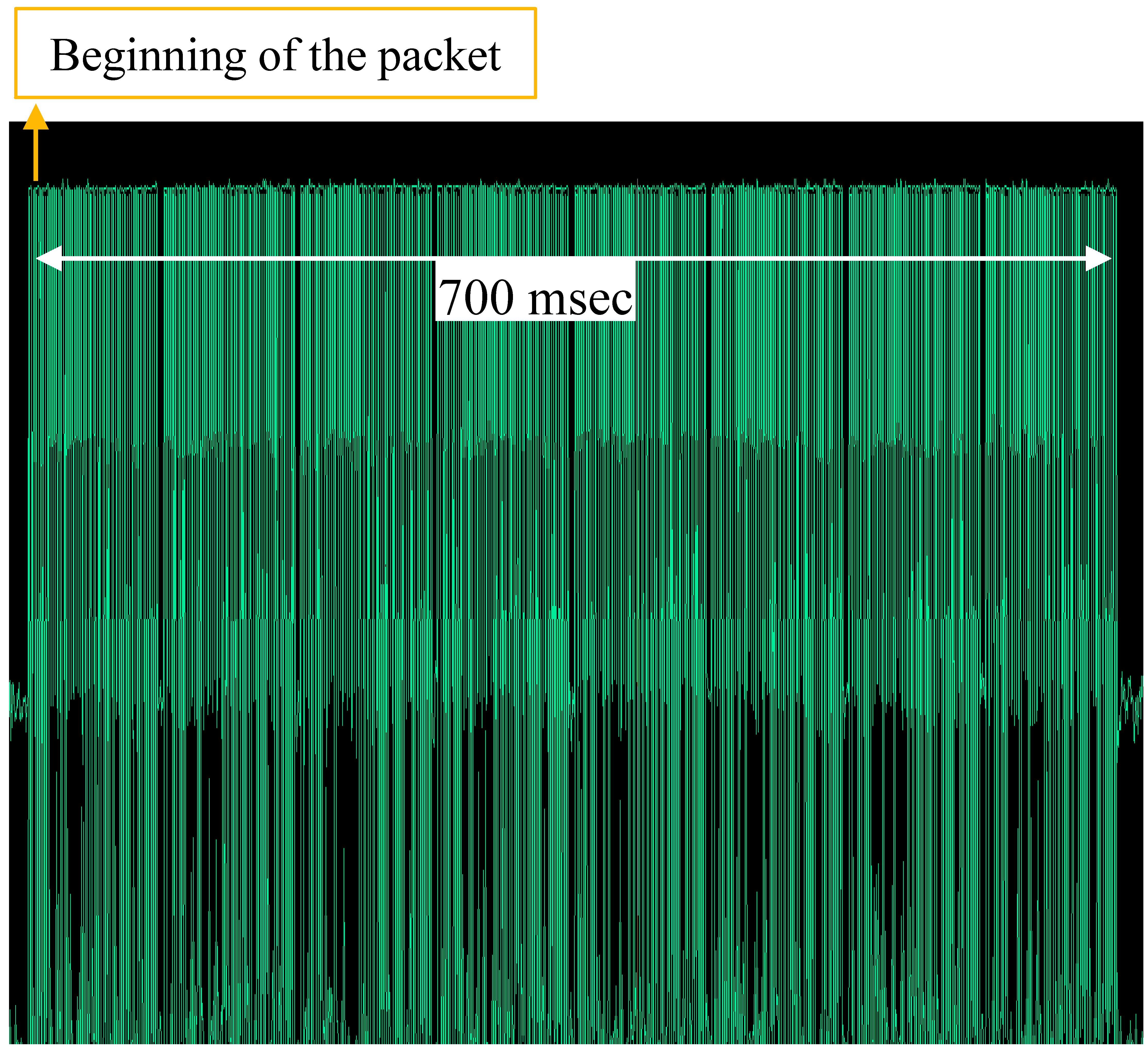}
\caption{The bitstream transmitted by one of the level sensors to the base station of the industrial CPS.}
\label{fig:LTB}
\end{figure}

A close examination of two commonly-used industrial level monitoring-based 
CPSs revealed that none of their sensors uses a secure transmission protocol. 
Indeed, communications between the sensors and their corresponding base 
stations are not cryptographically-protected. Therefore, we were able to 
completely reverse-engineer the communication protocols used in these sensors. 
To do this, we captured and demodulated 40 data packets (80 packets in all) 
generated by each sensor. Fig.~\ref{fig:LTB} demonstrates the bitstream 
transmitted by one of the level sensors.

We found that one of the level sensors simply transmits an unecrypted packet 
that includes a 10-bit pin number and a data field that represents the liquid 
level in the tank. In order to trigger the emergency response of the CPS that 
uses this sensor, we transmitted 1024 packets with different pin numbers. We 
were able to trigger the emergency alarm in less than 10 seconds for this 
sensor. 

We observed that the other sensor transmits unencrypted packets that only 
contain data and a 1-byte sequence number. Therefore, in order to trigger the 
emergency response of the CPS that uses this sensor, an attack can be 
implemented as follows. The attacker first captures a packet from the sensor 
and extracts the sequence number. Then he creates and transmits a packet in 
which the data field is set to its maximum value and the sequence number field 
is set to the sequence number extracted from the captured packet plus one. We 
implemented this attack and were able to successfully trigger the emergency 
responses of the system.

Table \ref{table:sensorSpecL} specifies communication frequency, modulation 
type, and pin length for each level sensor.

\begin{table}[ht] 
\caption{Communication frequency, modulation type, and pin length for each level sensor} 
\centering 
\begin{tabular}{l c c c c} 
\hline\hline 
Sensor & Communication & Modulation & Pin length & Sequence\\ 
& frequency (MHz) & & (bits) &  number\\ [0.5ex]
\hline 
LS I & 433 & OOK & 10 & No\\ [0.5ex]
LS II & 920 & OOK & 0 (no pin) & Yes (1 byte)\\ [0.5ex]
\hline 
\end{tabular} 
\label{table:sensorSpecL}
\end{table}

Moreover, we placed HackRF at different distances from the base stations of 
the two sensors to find the maximum transmission distance. 
Table \ref{table:resultsRR2} summarizes results of this experiment. In 
real-world industrial CPSs, where signal repeaters (i.e., electronic devices 
that receive signals and retransmit them at a higher power) are in 
widespread use to support long-range communications, sensors may be located 
several miles away from the base station. Unfortunately, the attacker can 
also exploit these repeaters to extend the attack range up to tens of miles, 
e.g., the attacker can place a HackRF hundred meters away from a repeater 
that is located several miles away from the under-attack base station of
the CPS.

\begin{table}[ht] 
\caption{Maximum transmission distance for each industrial level sensor 
examined in Experimental scenario 2} 
\centering 
\begin{tabular}{l c} 
\hline\hline
Sensor & Maximum transmission distance (m)\\ [0.5ex]
\hline 
LS I & 70 \\ [0.5ex]
LS II & 250 \\ [0.5ex]
\hline 
\end{tabular} 
\label{table:resultsRR2}
\end{table}

\section{Suggested countermeasures}
\label{COUNTER}
In this section, we first suggest three approaches to mitigate the consequences 
of launching DISASTER, and for each approach, we briefly describe its 
limitations and disadvantages. Second, we discuss why preventing DISASTER may 
not always be feasible due to the existence of unpredictable situations.

\subsection{Proactive countermeasures}
Next, we describe three proactive approaches, which can be 
deployed in the design and verification phases of manufacturing, to 
prevent DISASTER. 

\subsubsection{Utilizing cryptographic mechanisms}
As demonstrated in Section \ref{Attacks}, using simple customized 
communication protocols to provide short-range communication between sensors 
and the centralized base station can lead to serious security issues and enable 
an attacker to reverse-engineer the protocol. The main weakness of the 
majority of non-standard communication protocols is that they do not offer 
any cryptographic mechanisms to ensure confidentiality and integrity of data 
transmitted by sensors. Utilizing standard communication protocols, which 
provide strong encryption mechanisms to ensure confidentiality and integrity, 
or adding encryption mechanisms to customized communication protocols can 
significantly limit the ability of an attacker to launch a security attack 
against the system.  Bluetooth\cite{BLE} and ZigBee\cite{COMPARISON} are instances 
of standardized communication protocols that offer lightweight encryption 
mechanisms (e.g., encrypted timestamp and data encryption), yet provide a 
long battery lifetime. Despite the advantages of encryption 
mechanisms, utilizing strong encryption in the sensors used in CPS safety 
mechanisms may not be feasible in the current state of technology due to 
several domain-specific limitations of CPSs (e.g., lack of enough sources 
of randomness and limited energy/computation resources) and special 
requirements of safety mechanisms (e.g., low cost and short response time).

In addition to encryption, obfuscation (i.e., a procedure applied to data 
to intentionally make them hard to understand without knowing the procedure 
that was applied) can make reverse engineering of the protocol 
harder. However, obfuscation cannot truly secure the system since a skilled 
attacker may eventually be able to reverse-engineer the obfuscation procedure. 
In fact, obfuscation can only delay (not prevent) reverse engineering of the 
communication protocol. 

\subsubsection{Security/safety-oriented verification}
As mentioned earlier, common design flaws and security weaknesses of safety 
mechanisms along with ignorance of common security-safety 
conflicts/trade-offs can endanger both security and safety of the system. 
There is a great body of literature on different types of design verification 
approaches and several commercialized verification approaches that 
manufacturers can use to detect and address common design flaws before 
introducing their product into the market.  Such verification approaches have 
traditionally been used to ensure that a product, service, or system works 
correctly, meets specific requirements, and fulfills its intended purposes. 
Unfortunately, traditional verification mechanisms do not typically target 
a comprehensive set of security and safety requirements. Recently, a 
few verification approaches \cite{VERF1,VERF2, NKJ1,NKJ2} have begun to 
take various security and safety considerations into account. Moreover, 
a few proposals (e.g., \cite{CONFLICT_1}) have offered theoretical approaches to 
detect different safety-security conflicts/trade-offs in CPSs. 

Utilizing such newly-proposed approaches can enable designers 
and manufacturers to predict or detect design flaws, security weaknesses, 
and safety-security conflicts before releasing a product to the market. 
However, relying on such methods cannot completely prevent DISASTER due 
to: (i) the existence of serious disagreements between some security and 
safety requirements that forces the manufacturer to consider some 
requirements and ignore others, (ii) inefficiency and imperfections of 
verification mechanisms, and (iii) unpredictability of threats against CPSs.

\subsubsection{Designing multimodal systems}
In order to plan and execute an appropriate emergency response, a CPS must 
reason about its surroundings and obtain a precise description of the 
environment. Such a system can gather and process data from its surroundings 
using various types of sensors. Generally, a CPS that uses multiple types of 
sensors (referred to as a multimodal system) can obtain more information 
about the environment than a CPS that only relies on one type (referred to 
as a unimodal system). As a result, multimodal CPSs provide two 
fundamental advantages over unimodal systems. First, they offer a higher 
accuracy in detecting emergency situations due to the fact that they 
can obtain more information about the environment and utilize sensor fusion 
methods (that combine various sensory data to improve the resolution 
and accuracy of specific sensor data) to achieve an accurate description 
of the environment. Second, multimodal CPSs are typically more difficult 
to attack since the attacker has to simultaneously target several sensor 
types to launch DISASTER. For example, a fire detection mechanism that relies 
on both temperature and smoke detection sensors is clearly more accurate and 
secure in comparison to a system that only uses smoke detection sensors.

However, multimodal CPSs generally have two disadvantages over unimodal 
systems: (i) multimodal systems are more complex and expensive, 
(ii) since the multimodal systems need to process more information, 
they are slower in detecting emergency situations. Moreover, if all 
sensors utilized in a multimodal CPS have common design flaws and 
weaknesses (e.g., all sensors use the same communication protocol that 
support neither obfuscation nor encryption), launching DISASTER against the 
multimodal system may not be significantly harder than launching an
attack against a unimodal system.

\subsection{Unpredictable situations}
Although the three previously-mentioned approaches can significantly limit the 
ability of a potential attacker, providing a comprehensive solution for 
eliminating the security risk associated with DISASTER is hard for two reasons. 
First, DISASTER might be feasible as a result of the existence of a situation 
that is not predictable at design time. Second, modeling human errors, e.g., 
pitfalls in the installation procedure, which might make a system susceptible 
to the proposed attacks, is very difficult. 
For instance, suppose a company sends certified installers to install a 
residential automation system that provides a fire evacuation mechanism, which 
is able to unlock the doors upon the detection of a fire. A month later, 
another company installs an air conditioner unit that is accessible from 
outside the building.  In this scenario, an attacker might be able to inject 
smoke into the residence using the routing paths of the air conditioner and 
trigger the fire evacuation mechanism even when there is no fire. In fact, the 
air conditioner provides the means to a potential attacker to have physical 
access to the home automation system and trigger its emergency responses. 

\section{Chapter summary}
\label{DISASTER_CONC}
In this chapter, we introduced DISASTER, which exploits design flaws 
and security weaknesses of safety mechanisms deployed in CPSs along with 
safety-security conflicts to trigger the system's emergency responses even 
in the absence of a real emergency situation.  This can lead to serious 
consequences, ranging from economic collateral damage to life-threatening 
conditions. 

We examined several already-in-use sensors and listed common design flaws and 
security weaknesses of safety mechanisms. We discussed the various impacts of 
DISASTER and described potential consequences of such attacks. We also 
demonstrated the feasibility of launching DISASTER in realistic scenarios, 
e.g., residential and industrial automation/monitoring CPSs. Finally, we 
suggested several countermeasures against the proposed attacks, and 
discussed how unpredictable situations may give rise to significant 
security problems in presumably secure CPSs.  

%% file: ch-CABA/chapter-CABA.tex
\chapter{CABA: Continuous Authentication Based on BioAura \label{ch:CABA}}
Most computer systems authenticate users only once at the time of initial login,
which can lead to security concerns. Continuous authentication has been 
explored as an approach for alleviating such concerns. Previous methods for 
continuous authentication primarily use biometrics, e.g., fingerprint and face 
recognition, or behaviometrics, e.g., key stroke patterns. We describe CABA, 
a novel continuous authentication system that is inspired by and leverages the 
emergence of sensors for pervasive and continuous health monitoring. CABA 
authenticates users based on their BioAura, an ensemble of biomedical signal streams that can be collected 
continuously and non-invasively using wearable medical
devices. While each such signal may not be highly discriminative by itself, we
demonstrate that a collection of such signals, along with robust machine 
learning, can provide high accuracy levels. We demonstrate the feasibility of 
CABA through analysis of traces from the MIMIC-II dataset \cite{MIMIC}. We propose various 
applications of CABA, and describe how it can be extended to user 
identification and adaptive access control authorization. Finally, we discuss 
possible attacks on the proposed scheme and suggest corresponding 
countermeasures \cite{CABA}.

\section{Introduction}
\label{intro}
Authentication refers to the process of verifying a user
based on certain credentials, before granting access to a secure system, 
resource, or area \cite{AUTH_KEYS}. Traditionally, authentication is only performed when the user initially 
interacts with the system \cite{MULTI1}. In these scenarios, the user faces a 
knowledge-based authentication challenge, e.g., a password inquiry, and the 
user is authenticated only if he offers the correct answer, e.g., the 
password. 

Although one-time authentication has been the dominant authentication
mechanism for decades \cite{QUEST}, several issues spanning user inconvenience to security flaws have been investigated by researchers 
\cite{FLAW1,FLAW5}. For example, the user has to 
focus on several authentication steps when he tries to unlock a smartphone
based on a password/pattern-based authentication method. This 
may lead to safety risks, e.g., distraction when the user is driving. A serious security flaw of one-time 
authentication is its inability to detect intruders after initial 
authentication has been performed. For example, an unauthorized user can 
access private resources of the authorized user if the latter leaves his
authenticated device unattended, or forgets to log out \cite{F1}.

The above concerns have led to the investigation of continuous authentication 
mechanisms. Such mechanisms monitor the user's interactions with the device even after initial login to ensure that the initially-authenticated 
user is still the one using the device. Initial efforts in this direction were
based on simple security policies that lock the user's device after a period of 
inactivity, and ask the user to re-enter the password. However, such schemes 
may be annoying to users while they still expose a window of vulnerability, leaving much room for 
improvement \cite{LOCK}. Thus, a rapidly-growing body of literature on the 
usage of biometrics, i.e., strongly-reliable biological traits such 
as facial features, and behaviometrics, i.e., measurable behavior such as frequency 
of keystrokes, for continuous authentication has emerged in the last decade 
\cite{F1,SU3}.

Recently, wearable medical sensors (WMSs), which measure biomedical signals, 
e.g., heart rate, blood pressure, and body temperature, have drawn a lot of 
attention from researchers and begun to be adopted in practice \cite{WMSA1,WMSA5}. A recent report by Business Insider \cite{BUS} 
claims that 33 million wearable health monitoring devices were sold in 2015. 
It forecasts that this number will reach 148 million by 2019, and continue to grow rapidly thereafter. We suggest that, since such biomedical signals will be collected anyway for health
monitoring purposes, they can also be used to aid authentication. The use of continuously-collected biomedical
data for user verification and identification seems promising for three 
reasons. First, if the biomedical signals are collected by WMSs for medical purposes, using them for authentication does not require any extra device 
that is not already on the body. Second, this information is collected
transparently to the user, i.e., with minimal user involvement. Third, unlike 
traditional biometrics/behaviometrics, e.g., face features and keystroke
patterns, information that may frequently 
become unavailable, the stream of biomedical signals collected by WMSs 
is always available when the person is wearing WMSs.

In this chapter, we present CABA, a transparent continuous authentication 
system based on an ensemble of biomedical signal streams (Biostreams in short) that we 
call \textit{BioAura} \footnote{Aura is traditionally defined as the energy field around a person. Analogously, we use the term BioAura to define the biological field around a person, manifested as a set of Biostreams.}. A Biostream is a sequence of
biomedical signal samples that are continuously gathered by a WMS for medical 
diagnosis and therapeutic purposes. The most 
important difference between a Biostream and a biometric trait is that a 
Biostream alone does not have enough discriminatory power to distinguish 
individuals. Thus, an authentication decision based on a single Biostream, 
e.g., body temperature or blood pressure, is unlikely to be sufficiently discriminative. However, when multiple Biostreams are combined into a BioAura, 
it leads to a powerful continuous authentication scheme. \\

Our key contributions can be summarized as follows: 
\begin{enumerate}
\item We suggest a list of design requirements for any continuous authentication system.
\item In order to analyze the discriminatory power of BioAura, we propose a continuous authentication system based on BioAura (called CABA) and investigate it from both accuracy and scalability perspectives. 
\item We suggest an adaptive authorization scheme and describe how it can be used to alleviate user inconveniences associated with the use of continuous authentication systems that might falsely reject the user. 
\item We describe various possible attacks against the proposed continuous authentication system along with several countermeasures to prevent such attacks.
\end{enumerate}

The rest of the chapter is organized as follows. Section \ref{OCTAGON} describes the 
requirements that should be targeted in the design of continuous
authentication systems and discusses how CABA addresses such
requirements. Section \ref{PROPOSED_BIO} describes BioAura and the Biostreams that form the proposed BioAura. Section \ref{SCOPE_CABA} discusses the scope of CABA applications. Section \ref{PROT} describes the CABA prototype and our
experimental setup. Section \ref{ACC} investigates CABA from both accuracy and
scalability perspectives. Section \ref{CABA_IDENT} describes how CABA can support
identification, i.e., the process of recognizing a user without knowing his 
user ID. Section \ref{CABA_ADAPT} presents an adaptive authorization scheme that can be used along with CABA to enhance user convenience. Section \ref{CABA_THREAT} discusses possible attacks against the proposed authentication system and describes different countermeasures against each attack. Section \ref{COMPARE} discusses related work and compares CABA with previously-proposed continuous authentication systems. Section \ref{DISC} briefly 
describes possible privacy concerns surrounding the use of biomedical
signals, how CABA can be used as a stand-alone one-time authentication system, 
and the effects of temporal conditions on authentication results. Finally, 
Section \ref{CABA_CONC} concludes the chapter. 

\section{Desirable authentication requirements}
\label{OCTAGON}
In this section, we first describe the desirable requirements that every continuous authentication system must satisfy. Then, we discuss how CABA addresses such requirements.
\subsection{Design-octagon}
Even though several continuous authentication systems have been
proposed in the past, they have not been evaluated against a comprehensive list of 
design requirements. A few studies, e.g., \cite{DESIGN2,F2,COST_1}, consider 
a small set of requirements, e.g., cost and accuracy. However, there is no standard list of design requirements that a continuous authentication system must satisfy. We suggest such a list below (that also includes some of the desiderata of typical WMS-based systems, as shown in Fig. \ref{fig:DESIGN}). We call it the \textit{Design-octagon}
since it comprises eight design requirements (Fig.~\ref{fig:requirements}):

\begin{figure}[h]
\centering
\includegraphics[trim = 79mm 37mm 80mm 55mm ,clip, width=200pt,height=180pt]{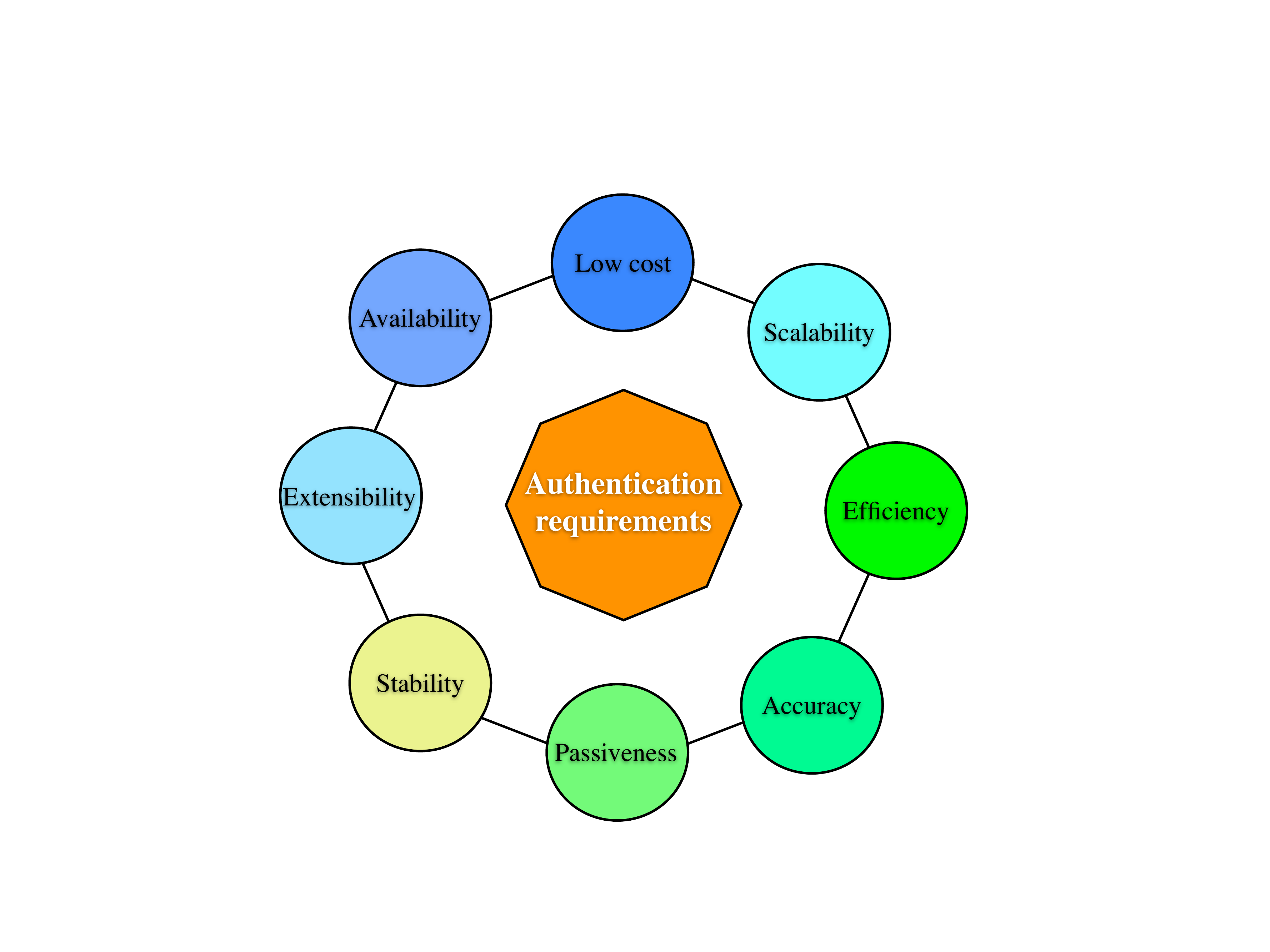}
\caption{Design-octagon: Desiderata for a continuous authentication 
system.} 
\label{fig:requirements}
\end{figure}

\noindent \textit{\textbf{1. Passiveness}}: A user-friendly system must not require frequent user
involvement \cite{PASSIVE_1}. For example, if the authentication system asks the user
to re-enter his credentials often, it may be quite annoying to the user \cite{F2}.\\
\noindent \textit{\textbf{2. Availability}}: The system should provide a reliable authentication 
system at all time instances \cite{F2}. Lack of continuous availability is a 
significant drawback of several previously-proposed continuous authentication 
systems -- they may often fail due to a lack of sufficient information \cite{SU1}. 
For instance, a keyboard-based system may unintentionally reject a legitimate user when he is watching a movie and not using the keyboard.\\ 
\noindent \textit{\textbf{3. High accuracy}}: Undoubtedly, the most important requirement of every 
authentication system is high accuracy. The system should be able to 
confidently and accurately distinguish legitimate users from 
impostors, and reject impostors' requests.\\
\noindent \textit{\textbf{4. Scalability}}: The system should be able to handle a growing amount of 
work when the number of users increases \cite{QUEST}. In particular, its time and space 
complexity should increase modestly with an increase in the number of users \cite{TSC_1}. \\
\noindent \textit{\textbf{5. Efficiency}}: A short response time, i.e., the time 
required to capture a test sample, process it, and provide a decision, is 
very desirable for two reasons. First, it is desirable for the system to quickly authenticate 
a legitimate user and reject an impostor to ensure user convenience \cite{QUEST}. Second, security may also suffer if there is an appreciable delay. For example, if authorization takes five 
minutes, an impostor may be able to control the system and access 
restricted resources in that five-minute timeframe, while the system is still 
processing.\\
\noindent \textit{\textbf{6. Low cost}}: Cost is an important factor in authentication systems used in 
low-security environments, e.g., in personal computers \cite{F2,PERM_1}. In such environments, the cost of adding or modifying the authentication system should ideally be negligible. Thus,
systems that do not need extra peripherals, such as retina scanners, would be 
generally preferred. However, for highly-secure environments, e.g., military 
bases, expensive authentication systems could be deployed \cite{COST_1}.\\
\noindent \textit{\textbf{7. Stability}}: Any trait that is recorded for processing for authentication 
purposes must ideally have only slight changes or maintain its pattern over a 
certain time period \cite{PERM_1,PERM_2}. \\
\noindent \textit{\textbf{8. Extensibility}}: The authentication system should be able to function on a 
wide variety of devices regardless of underlying hardware. Ideally, the system 
should not require dedicated hardware. One of the advantages of password-based authentication is that it can be easily extended to protect a large number of systems, devices, and resources with minimal system modification \cite{QUEST}. 

\subsection{Addressing desirable requirements}
In this subsection, we describe how CABA ensures all of the requirements discussed above.\\
\noindent \textit{\textbf{1. Passiveness}}: In CABA, passiveness is addressed through the use of WMSs. These are small and compact sensors that are specifically designed to
take the passiveness requirement into account, since continuous health
monitoring needs to minimize user involvement. Thus, passiveness is not 
only a very desirable requirement for continuous authentication, but also a significant consideration in designing WMSs. Unfortunately, major 
biometrics-based systems, e.g., fingerprint-based, do not provide a high level of passiveness.\\
\noindent \textit{\textbf{2. Availability}}: The use of WMSs as capture devices also ensures a continuous stream of information since this is also required for continuous health monitoring. However, neither biometrics nor behaviometrics guarantees availability. For example, keyboard/mouse-based continuous authentication systems fail when the user stops using the dedicated peripherals.\\
\noindent \textit{\textbf{3. Accuracy}}: The accuracy of CABA is extensively
investigated in 
Section \ref{ACC} in various experimental scenarios. Section \ref{COMPARE} 
demonstrates that the accuracy of CABA, which is based on an ensemble of 
weakly discriminative Biostreams, is comparable to previously-proposed 
systems, which are based on strong biometrics. \\
\noindent \textit{\textbf{4. Scalability}}: The scalability of CABA is investigated in 
Section \ref{ACC} based on two scalability metrics (time complexity and space 
complexity). Our analysis shows that an increase in the number of users can be 
easily handled in this system.\\
\noindent \textit{\textbf{5. Efficiency}}: Authentication can be done in a few
milliseconds. For each authentication attempt, the user can immediately
provide the required data since the data are already collected using WMSs. 
The efficiency of the system is described in more detail in Section \ref{ACC}.\\
\noindent \textit{\textbf{6. Low cost}}: As discussed later in Section \ref{PROPOSED_BIO}, 
the proposed BioAura consists of Biostreams that are collected for continuous health monitoring. If the user is already using a continuous health monitoring system, CABA can offer continuous authentication with minimal cost.\\
\noindent \textit{\textbf{7. Stability}}: Our investigations of different Biostreams and 
their high authentication accuracy over different timeframes demonstrate that 
the collected Biostreams maintain their pattern over time. Therefore, they can 
be used as authentication traits.\\
\noindent \textit{\textbf{8. Extensibility}}: By decoupling the collection of Biostreams from the authenticating device, CABA can be implemented in any general-purpose computing device with sufficient memory capacity and computation power. Unfortunately, neither biometrics- nor behaviometrics-based systems provide significant extensibility. For example, the nature of keyboard/mouse-based authentication schemes inherently limits their applications, i.e., they can only be used for implementing an authentication mechanism in a system that has a keyboard or a mouse.

\section{BioAura}
\label{PROPOSED_BIO}
In this section, we first briefly describe how Biostreams can be collected 
using WMSs. Then, we discuss which Biostreams constitute the BioAura.

\begin{figure}[h]
\centering
\includegraphics[trim = 0mm 40mm 140mm 42mm ,clip, width=250pt,height=220pt]{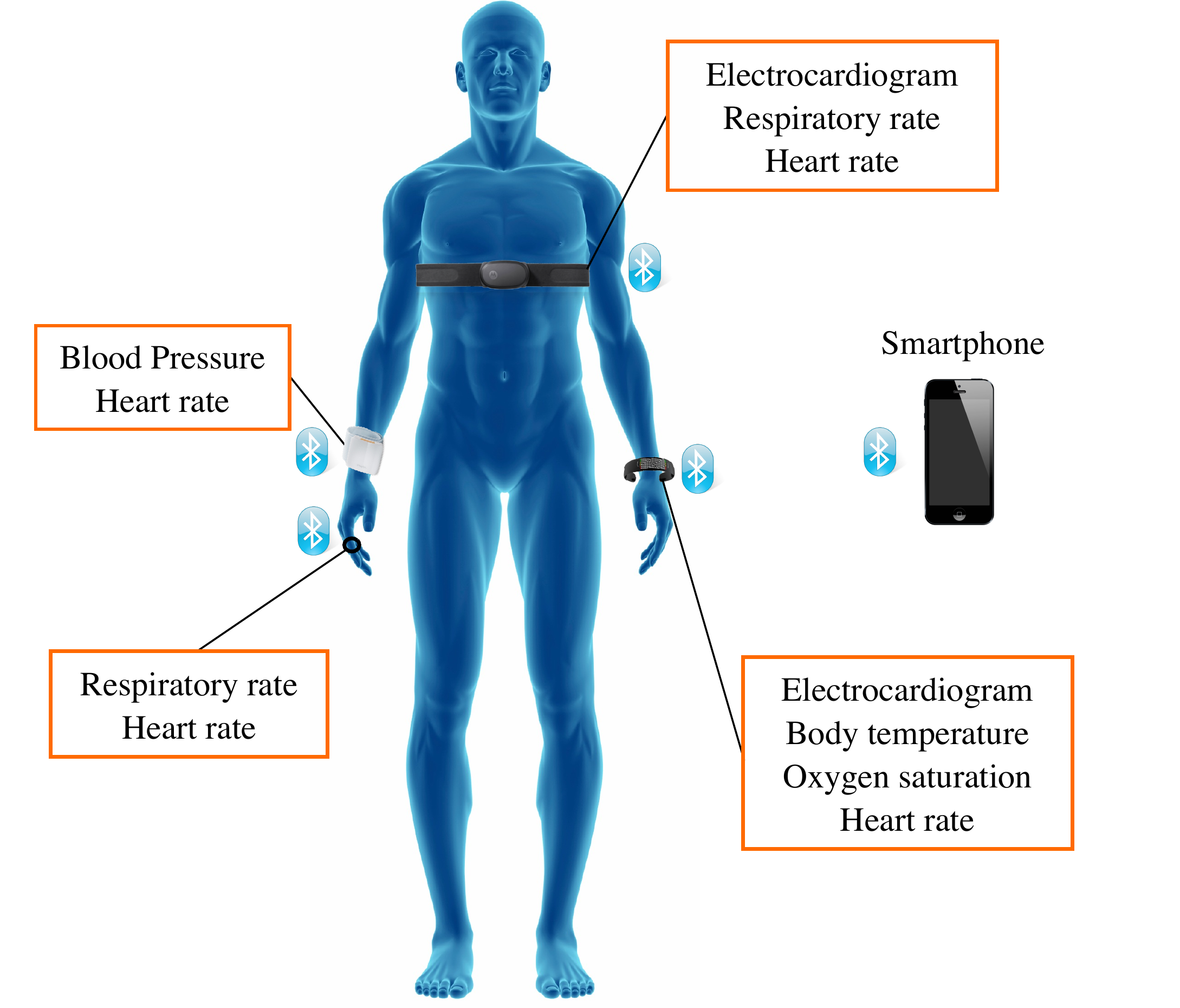}
\caption{A continuous health monitoring system consisting of several small 
lightweight WMSs that transmit biomedical data to the smartphone.} 
\label{fig:basestation}
\end{figure}

As mentioned earlier, BioAura is an ensemble of Biostreams, which are 
gathered by WMSs for medical diagnosis and continuous health monitoring. The 
most widely-used scheme for continuous health monitoring consists of two 
classes of components: (i) WMSs and (ii) a base station \cite{CNIA}. All
WMSs transmit their data to the base station either for further
processing or long-term storage. In recent years, smartphones have
become the dominant base station since they are powerful and ubiquitous,
and their energy resources are less limited relative to WMSs
\cite{SMARTPHONE_1,CNIA}. Fig.~\ref{fig:basestation} illustrates a
simple continuous health monitoring system that consists of several
small lightweight WMSs, which transmit their biomedical data to the smartphone over a Bluetooth communication link (similar to the left side of the personal health care system illustrated in Fig. \ref{fig:WSN}}). 

Smartphones can perform simple preprocessing to extract values of some 
important features from the data, and transmit those values. In CABA, 
the smartphone first executes a very simple feature extraction function that 
computes the average value of the samples in each Biostream over the last 
one-minute timeframe of data. Then, it only transmits a feature vector that 
contains these average values. 

As mentioned earlier in Section \ref{intro}, with the expected widespread use of WMSs, CABA can be used to provide a continuous authentication system as a side benefit of continuous health monitoring systems. Our proposed BioAura consists of Biostreams that are essential for routine continuous health monitoring, and their collection needs minimum user involvement. Such Biostreams are expected to be included in long-term continuous health monitoring systems. 

Table \ref{table:BIOAURA} shows the most widely-used Biostreams, their abbreviations/notations used in the medical literature, and their units \cite{CNIA,CNIA_2}. In this chapter, we exclude the first three ones from the proposed BioAura, and include the other nine. Next, we discuss why the three Biostreams are excluded.\\
\noindent \textit{\textbf{1. Electroencephalogram (EEG)}}: EEG is excluded from BioAura 
because it cannot be conveniently captured. The current method for capturing 
EEG requires the user to wear a cap. Moreover, its capture devices cannot be miniaturized further because electrodes need to form a minimum diameter to be noise-robust \cite{EEG_CAP}.     \\
\noindent \textit{\textbf{2. Electrocardiogram (ECG)}}: Even performing a
low-complexity feature extraction on one minute of ECG signals requires
at least $400 \times$ more operations than performing a simple
statistical feature extraction, e.g., averaging, on the respiratory rate
values \cite{ECGC}. This would place a significant computational and energy demand on battery-powered devices such as smartphones and wearables. If
we try to avoid the preprocessing, i.e., feature extraction, on the smartphone and just transmit the ECG signals to the authentication system, this would also entail significant energy consumption since ECG waveforms contain at least 200 $samples/s$ \cite{SAM1,CNIA}. \\ 
\noindent \textit{\textbf{3. Blood glucose (BG)}}: BG is excluded because currently
the devices that measure BG are invasive, i.e., they require a sample of the user's blood.\\
\indent Although we have currently used nine Biostreams to form the BioAura in the 
prototype implementation, CABA need not necessarily be limited to these
nine.  As other compact WMSs become available in the future, they could also be 
made part of the BioAura.
\begin{table}[ht] 
\caption{Biostreams, their abbreviations/notations, and units} 
\centering 
\begin{tabular}{l c c} 
\hline\hline 
Biostream & Abbreviations/Notations& Unit\\ [0.5ex]
\hline 
\cellcolor{red!25} Electroencephalogram & EEG & $\mu V$ \\[1ex]
\cellcolor{red!25} Electrocardiogram & ECG & $\mu V$ \\[1ex]
\cellcolor{red!25} Blood glucose & BG &  $mg/dL$ \\[1ex]
\cellcolor{green!25}Arterial systolic blood pressure & ABPSYS & $mmHg$ \\[1ex] 
\cellcolor{green!25}Arterial diastolic blood pressure & ABPDIAS & $mmHg$ \\[1ex]
\cellcolor{green!25}Arterial average blood pressure & ABPMEAN & $mmHg$ \\[1ex]
\cellcolor{green!25}Heart rate & HR & $1/min$ \\[1ex]
\cellcolor{green!25}Pulmonary systolic artery pressure & PAPSYS & $mmHg$ \\[1ex]
\cellcolor{green!25}Pulmonary diastolic artery pressure & PAPDIAS & $mmHg$ \\[1ex]
\cellcolor{green!25}Body temperature & T & $Celsius$ \\[1ex]
\cellcolor{green!25}Oxygen saturation & SPO2 & $\%$ \\[1ex]
\cellcolor{green!25}Respiratory rate & RESP & $1/min$ \\[1ex]
\hline 
\end{tabular} 
\label{table:BIOAURA} 
\end{table}
\section{Scope of applications}
\label{SCOPE_CABA}
In this section, we describe the possible applications of CABA. The concept of 
continuous authentication based on BioAura can be used to protect 
(i) personal computing devices and servers, (ii) software applications, and 
(iii) restricted physical spaces. Next, we conceptually describe how CABA can 
be used to protect each domain. 

Computing devices, e.g., personal computers, laptops, tablets, and cell 
phones, or servers can employ two different approaches to utilize CABA: (i) they can use their own computing resources
to implement a stand-alone version of CABA, or (ii) they can simply use decisions made
by a version of the scheme implemented on a trusted server. We investigate 
both approaches.

\begin{figure}[h]
\centering
\includegraphics[trim = 70mm 44mm 65mm 10mm ,clip, width=250pt,height=240pt]{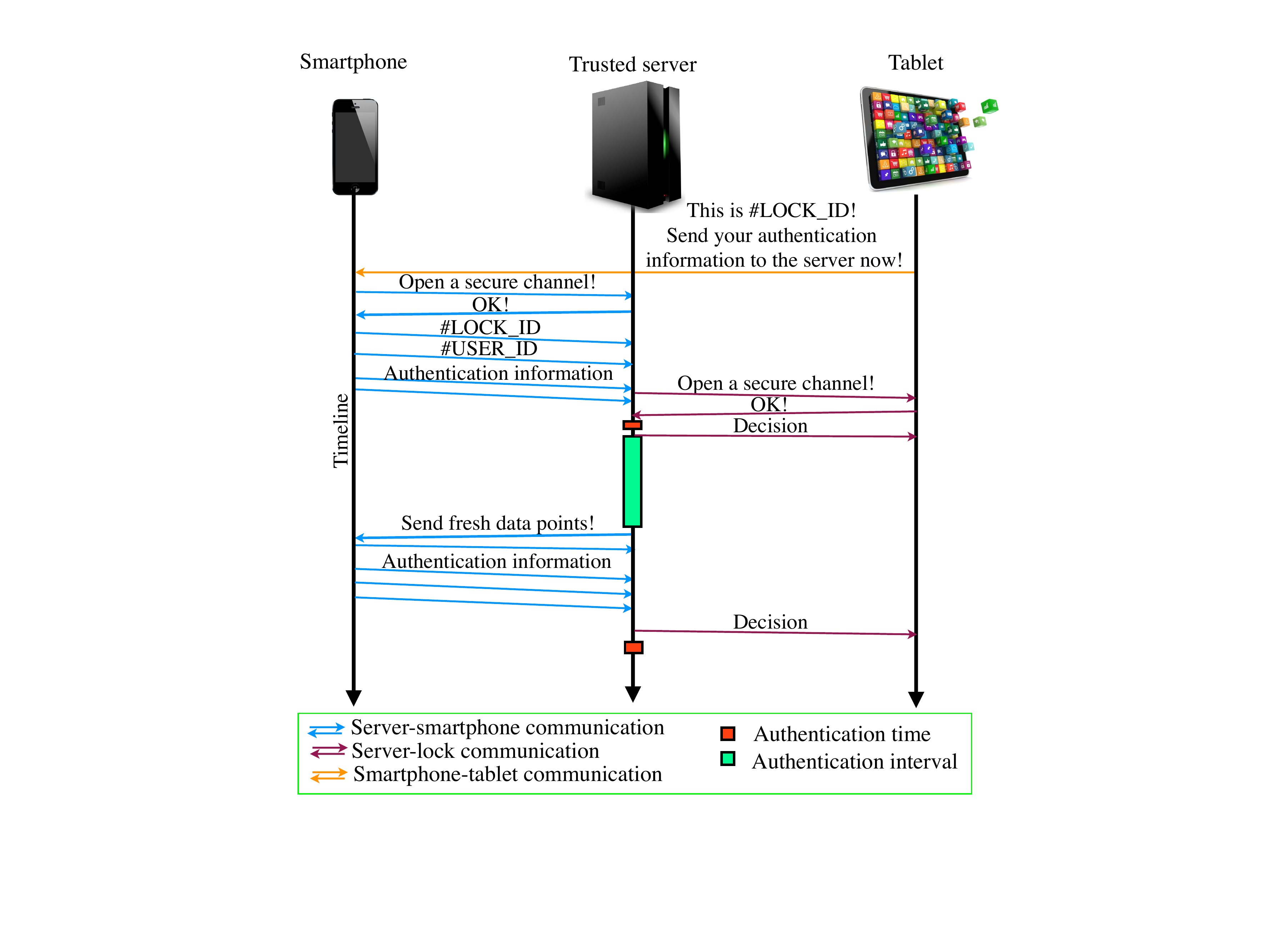}
\caption{The tablet wants to authenticate the user. The vertical arrows 
depict the timeline.} 
\label{fig:tablet}
\end{figure}

\noindent \textit{\textbf{Example 1}}: Suppose a tablet wants to authenticate
its user. The tablet may be unable to dedicate its
limited memory/energy resources to support the
whole authentication process. In such a scenario,
it can use decisions made by a trusted server running
CABA. Fig. \ref{fig:tablet} illustrates this scenario.
When the user tries to unlock the tablet, it informs
the user's smartphone. The smartphone asks
the trusted server to open a secure communication
channel. The smartphone then sends the information required for specifying the device that needs to be
unlocked, e.g., the tablet ID, along with the information that needs to
be processed to authenticate the user,
e.g., the user ID and a preprocessed frame of data points from his
BioAura, to the trusted server. The
server then authenticates the user and sends this decision to the tablet. After initial login, the trusted server
demands fresh data points at certain intervals.

\begin{figure}[h]
\centering
\includegraphics[trim = 75mm 65mm 70mm 19mm ,clip, width=250pt,height=215pt]{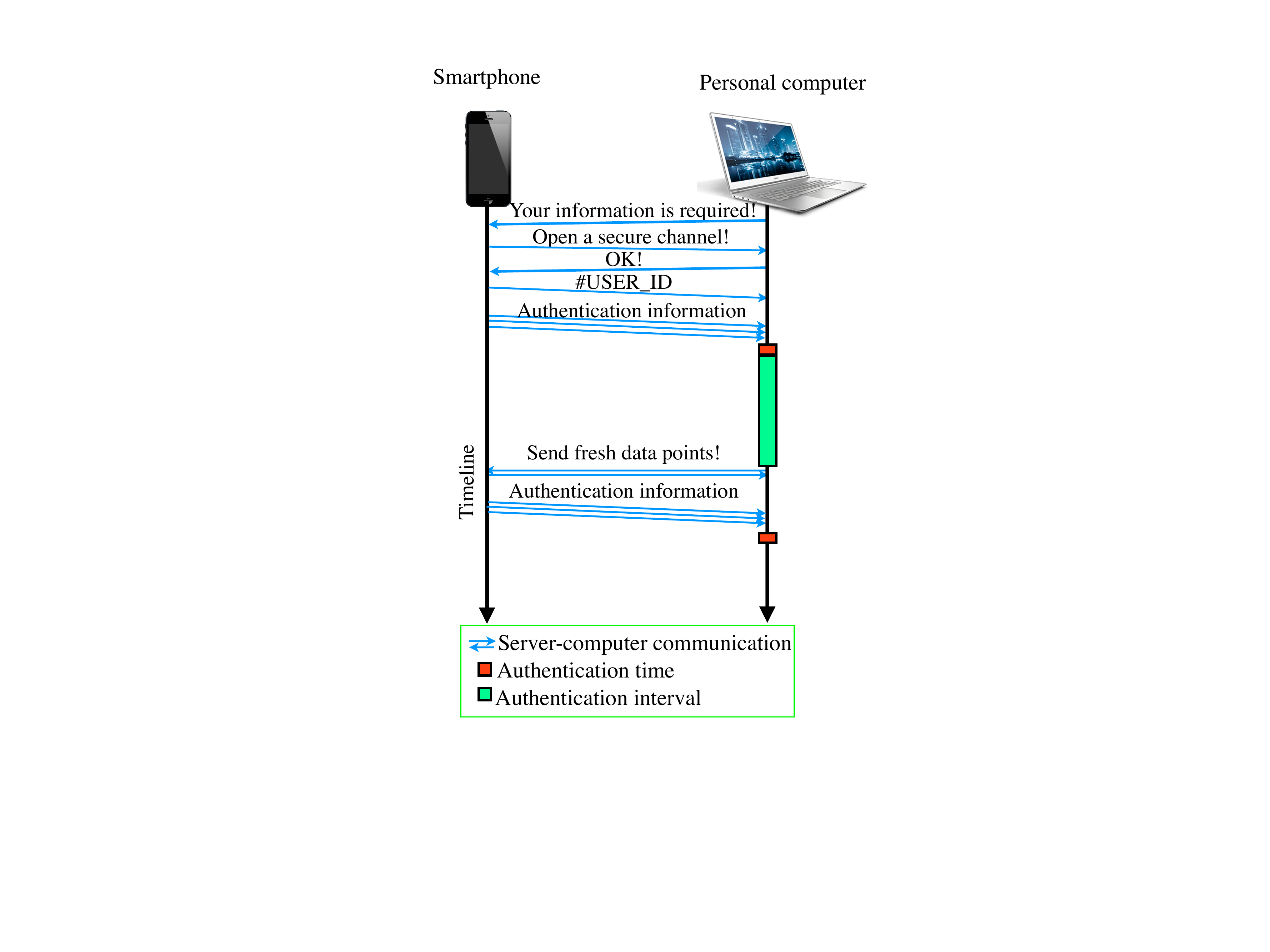}
\caption{The laptop wants to authenticate the user before allowing the user to 
utilize its resources or software applications.} 
\label{fig:laptop}
\end{figure}

\noindent \textit{\textbf{Example 2}} Suppose the user wants to login to his personal computer. In this case, the computer has enough
computational power and energy capacity to implement a stand-alone version of CABA. This case is
similar to the one in Example 1, except that there is no need for a trusted server (Fig. \ref{fig:laptop}).

Similarly, CABA has the potential to provide continuous authentication
for applications that need strong authentication, e.g., e-commerce 
applications. Its authentication decisions can be made on the same device that runs the application or on a powerful trusted server and then transmitted to the device that runs the application.\\
\noindent \textit{\textbf{Example 3}}: Consider an online banking application that
is installed on the user's smartphone. When the user opens the
application to access his bank account, the smartphone opens a secure
communication channel with the trusted server. Then, the smartphone
sends the required information for specifying the application, e.g., the 
application ID, along with information needed for authenticating the user. The rest of the protocol is the same as before.

Finally, CABA can be used to control access to restricted physical
spaces, e.g., buildings. Typically, the electronic device that controls
the entrance, e.g., a smart lock, would not have enough computation power to use a stand-alone version of CABA. Hence, in such cases, the scheme can be implemented on a trusted server, and decisions then transmitted to the device. This case is similar to the one depicted in Fig. \ref{fig:tablet}, with the tablet replaced by the lock.
\section{Implementation and experimental setup}
\label{PROT}
In this section, we first describe our implementation of CABA. We then discuss
our experimental setup and different metrics that we used to investigate 
the proposed system.
\subsection {Prototype implementation}
Similar to other authentication systems, CABA has two operating phases:
(i) enrollment phase in which CABA builds machine learning-based models
for each user, given the training data, and (ii) user authentication phase in which the system continuously authenticates the user. The description of the 
two phases is presented next.

\subsubsection {Enrollment phase}
In the enrollment phase, the authentication system is given a training
dataset. The system builds the model using a supervised learning
approach, i.e., a machine learning approach in which the model is built based 
on labeled training data points. 

Generally, the amount of information needed to build a model varies from one 
application to another. As we elaborate later in Section \ref{ACC}, we 
evaluated the number of training data points needed to investigate how much 
information should be sent to the authentication system to build a reliable 
and accurate model. Each data point in the training set is
nine-dimensional and consists of the average values of successive measurements of a Biostream 
over a one-minute timeframe. The value of each dimension is represented using 
half-precision floating-point format that requires two bytes of storage. 
Therefore, if the smartphone needs to transmit data points extracted 
over a one-hour period, it only needs to send 1080 bytes of data to the 
authentication system over this period.  

In order to maintain reliability, CABA should train a new model based on fresh 
biomedical data obtained at certain intervals. In other words, CABA should 
update the model regularly to ensure that the model maintains accuracy and 
can distinguish legitimate users from impostors. The frequency of model 
update, i.e., how frequently CABA should repeat the enrollment phase, depends on 
several factors, such as required accuracy and learning time. As we show later in Section \ref{ACC}, our experimental results indicate that when CABA re-trains the model every four hours, it achieves the best accuracy and the learning time is only a few minutes. Learning can be done transparently to the user.  In other words, CABA can re-train the model while the user 
continues to be authenticated. For example, suppose the enrollment phase takes 
five minutes each time and is repeated every four hours, i.e., each model 
is used for four hours. CABA can start re-training to generate a new model after 
3 hours and 55 minutes, and be ready with it after four hours have elapsed.

\subsubsection {User authentication phase}
In this phase, the system makes decisions using the already-trained model. 
In a continuous authentication scenario, the system should verify the user's 
identity at certain intervals. The frequency of authentication depends on 
several factors, such as the required level of security and the amount of 
information required for one authentication. In our prototype implementation, CABA re-authenticates the user every minute based on a given 
nine-dimensional data point $Y$ that contains the average values of the 
chosen Biostreams over a specified time interval. When the user 
approaches the authentication system and requests authentication, the 
smartphone performs a simple computation on the already-gathered Biostreams 
and provides $Y$. Therefore, unlike most previously-proposed continuous 
authentication systems, e.g., keyboard/mouse-based systems, that require 
the user to wait while they collect authentication information, CABA 
obtains the required information almost \textit{instantaneously} because the 
information has already been gathered and stored on the smartphone for the 
purpose of health monitoring. 

\begin{figure}[h]
\centering
\includegraphics[trim = 70mm 83mm 80mm 62.5mm ,clip, width=250pt,height=145pt] {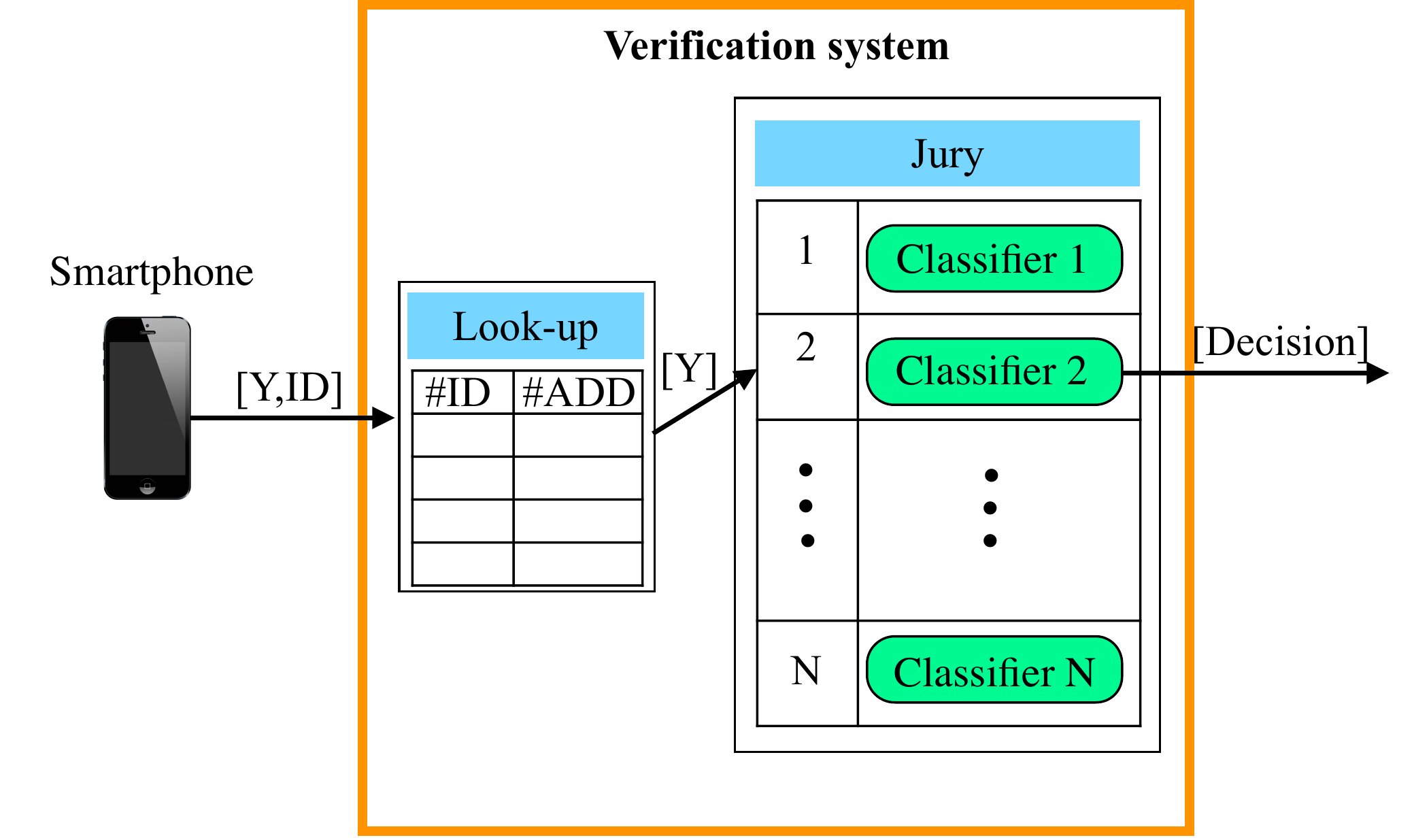}
\caption{User authentication phase: The user's smartphone provides $Y$ and 
the user ID, and CABA outputs the decision.}
\label{fig:VER}
\end{figure}

Fig. \ref{fig:VER} illustrates how CABA works when the user requests 
authentication. In a single verification attempt: 
\begin{enumerate}
\item The smartphone preprocesses one minute of Biostreams collected from the 
user's BioAura. Then, it transmits the preprocessed information ($Y$) along 
with user ID to the authentication system. 
\item The Look-up stage sends $Y$ to the appropriate classifier in 
the Jury stage based on the given user ID.
\item The dedicated classifier processes $Y$ and outputs a binary decision 
(accept or reject).
\end{enumerate}

Next, we provide a detailed description of Look-up and Jury stages.\\
\noindent \textit{\textbf{1. Look-up stage}}: This stage forwards the nine-dimensional vector $Y$ provided by the smartphone to the appropriate classifier based on the given user ID. In order to provide a fast search mechanism to find the appropriate classifier, this stage can be implemented using a hash table that associates user IDs with pointers to the classifiers.\\
\noindent \textit{\textbf{2. Jury stage}}: The Jury stage consists of $N$ binary classifiers, where $N$ is the number of people who need to be authenticated. The $i$-th classifier is trained to only accept the data point $Y$ that is extracted by the $i$-th user's smartphone from his BioAura. The training set of the $i$-th classifier consists of the $i$-th user's data points labeled as ``accept" and others' data points labeled as ``reject". 

We have used two well-known binary classification methods to build our model: Support Vector Machine (SVM) \cite{SVM} and Adaptive Boosting 
(AdaBoost) \cite{ADA1}. Next, we briefly describe each method.

\begin{itemize}
\item SVM: The basic concept in an SVM is to find a hyperplane that separates 
the $n$-dimensional data into two classes. However, since the data points in 
the dataset are not usually linearly separable, SVMs introduce the concept
of kernel trick that projects the points into a higher-dimensional space, 
where they are linearly separable. When no prior knowledge about the
dataset is available, SVMs usually demonstrate promising results and
generalize well. A great number of previous research studies that
perform authentication using machine learning methods only consider SVM
with a linear kernel \cite{SU2} or radial basis function (RBF) kernel 
\cite{BE_1}. We decided to investigate both.

\item AdaBoost: Although SVM has been commonly used in
previously-proposed continuous authentication systems in a variety of
scenarios, we decided to include AdaBoost as well. The idea behind
AdaBoost is to build a highly accurate classifier by combining many weak
classifiers that always perform a little bit better than random
guessing on every distribution over the training set \cite{ADA1}. Since
biomedical signals are individually slightly discriminative, they lead to weak 
classifiers, which can be collectively turned into a strong classifier using 
AdaBoost. Choosing appropriate types of weak classifiers is a significant 
consideration in AdaBoost. The most commonly used weak learning methods for 
implementing AdaBoost-based classifiers are decision stumps (also called 
one-node tree) and decision trees. 
\end{itemize}

\subsection{Experimental setup and metrics}
Here, we first describe the parameters and dataset used in
our experiments. Then, we discuss the accuracy and scalability metrics used 
to investigate the proposed authentication system.

\subsubsection{Experimental parameters and dataset}
Next, we discuss the parameters used in our experimental setup and 
describe the dataset. \\
\noindent \textit{\textbf{Parameters}}: We need the following five parameters 
in our experiments.
\begin{itemize}
\item Dataset length ($L$): This is the duration of Biostreams measurements, i.e., the number of hours of information we have 
for each person in our data set. In our experiments, we used 14 hours of
data for each individual, i.e., $L=14$ $h$. 
\item Dataset dimension ($n$): This is the number of Biostreams that form the 
BioAura of an individual. In our setup, we have included nine Biostreams for 
each person, i.e., $n=9$.  
\item Dataset size ($N$): This is the number of people in the dataset. 
 
\item Training window size ($TRW$): This represents the duration of the signal measurements (expressed in hours) for each individual that we used for training the model in the enrollment phase. For example, if $TRW$ is 1 hour, it means we have included 60 data 
points in our training set, where each point is a nine-dimensional vector 
consisting of the average values of successive measurements of the nine 
Biostreams over a one-minute timeframe. We vary TRW in our experiments to study its impact on the model's accuracy.

\item Testing window size ($TEW$): This represents the duration of 
signal measurements (expressed in hours) for each individual for
investigating the accuracy of the trained model in the user authentication 
phase. We vary the value of TEW in our
experiments to investigate its impact on the performance of CABA.

\end{itemize}

\noindent \textit{\textbf{Dataset}}: In order to investigate the accuracy of CABA, we 
used a freely available multi-parameter dataset, called MIMIC-II \cite{MIMIC}. 
MIMIC-II was gathered in a controlled environment in which each user
remains almost stationary during data collection. It has been extensively used 
in the medical and biomedical fields. It consists of several anonymized 
high-resolution vital sign trends, waveforms, and sampled biomedical signals 
for many individuals. For our experiments, we could only find 37 users ($N=37$) in MIMIC-II for whom the data: (i) include the nine targeted Biostreams, and (ii) are available over several hours (at least 14 hours) with minimal missing values (we excluded a user for whom the data were not available for more than two consecutive hours). Biostreams were sampled using patient monitors (Component Monitoring System Intellivue MP-70 and Philips Healthcare) at the sampling rate of 125 Hz \cite{MIMIC}.

\subsubsection {Accuracy metrics}
Next, we describe five metrics that we used for analyzing the accuracy 
of the proposed authentication system. The first three are traditionally used 
for examining authentication systems. We define two more to investigate
the accuracy in the context of continuous authentication. 

\begin{itemize}
\item False acceptance rate ($FAR$): This is the ratio of falsely accepted 
unauthorized users to the total number of invalid requests made by impostors 
trying to access the system. In the context of continuous authentication, we 
use the notation $FAR_{t=TEW}$ to denote FAR under $TEW$.  A lower FAR is 
preferred in cases in which security is very important \cite{SU_1_2}. 

\item False rejection rate ($FRR$): This refers to the ratio of falsely 
rejected requests to the total number of valid requests made by legitimate 
users trying to access the system. We use the notation $FRR_{t=TEW}$ 
to denote FRR under $TEW$. A lower FRR is preferred for user convenience 
\cite{SU_1_2}.

\item Equal error rate ($EER$): This is the point where $FAR$ equals $FRR$. 
Reporting only $FRR$ or $FAR$ does not provide the complete picture because 
there is a trade-off between the two since we can make one of them low by 
letting the other one become high. Therefore, we use $EER$ (instead of 
$FRR$ or $FAR$) for reporting CABA's accuracy. As before, we use the 
notation $EER_{t=TEW}$ to denote EER under $TEW$.

\item False acceptance worst-case interval ($FAW$): The output of the 
authentication system in a time period $T$ is a sequence of accept/reject 
decisions. As an example, Fig.~\ref{fig:FAW} shows two possible output 
sequences over a ten-minute authentication timeframe when an impostor is 
trying to get authenticated.  In both sequences, the number of falsely 
accepted requests is the same. However, in a continuous authentication 
scenario, the second sequence would be considered worse since the impostor 
can use the system over a four-minute timeframe without being detected, 
whereas in the first case the impostor can only use the system over a 
one-minute timeframe. We define $FAW$ as the longest time interval (expressed 
in minutes for CABA) over which an impostor can be falsely accepted as a 
legitimate user. In the example of Fig.~\ref{fig:FAW}, 
$FAW$ is one minute and four minutes for the first and second cases, 
respectively.  

\begin{figure}[h]
\centering
\includegraphics[trim = 10mm 145mm 17mm 68mm ,clip, width=250pt,height=40pt] {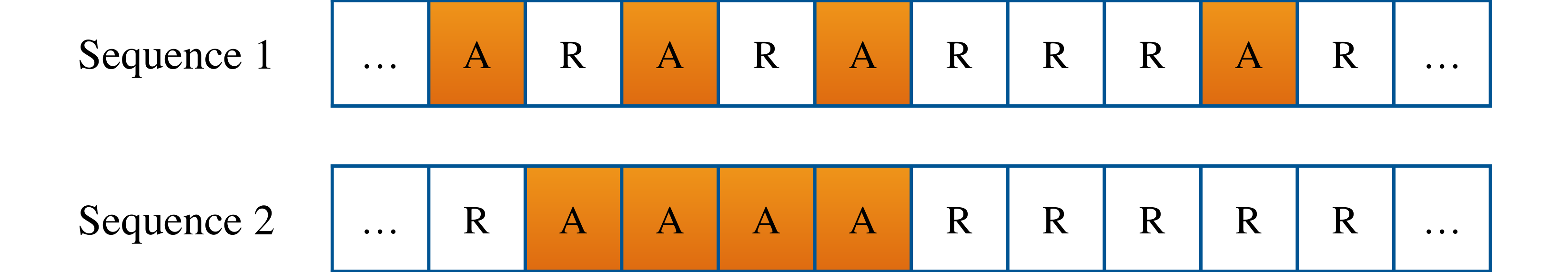}
\caption{Two possible output sequences over a ten-minute authentication
timeframe. A (R) refers to an accept (reject) decision.}
\label{fig:FAW}
\end{figure}

\item False rejection worst-case interval ($FRW$): Analogously to $FAW$, we 
define $FRW$ as the longest time interval (expressed in minutes) over
which a legitimate user might be falsely rejected and marked as an impostor.
\end{itemize}

\subsubsection{Scalability metrics}
As mentioned in Section \ref{OCTAGON}, the time and space complexity of
the authentication system should increase modestly with an increase in
the number of users. In order to investigate the scalability of the proposed method, we express the time and space complexities of CABA using the well-known $O$ notation, as a function of $N$ (number of the people in the dataset).

\section{Authentication results}
\label{ACC}
In this section, we investigate CABA from both the accuracy and scalability
perspectives. 
\subsection{Authentication accuracy}
In order to investigate the accuracy of the authentication system, we 
implemented a prototype of CABA in MATLAB.
    
The accuracy of a model is generally investigated using a set of data points 
that is different from the set used in constructing the model. Thus, in order 
to train and test a model, the dataset can be divided into two parts: training 
and test sets. The classical $K$-fold cross-validation is not suitable 
for estimating the performance of a system that processes a time series, i.e., 
a sequence of data points consisting of successive measurements, because 
potential local dependencies across observations in a time series define a 
structure in the data that will be ignored by cross-validation \cite{CVN1}. 
Thus, in this work, instead of using traditional cross-validation, we 
designed several experimental scenarios for evaluating the accuracy of the 
authentication system. We describe these scenarios next. 
\begin{enumerate}
\item \textit{Baseline:} In the baseline scenario, we break the available 
dataset into two equal parts, i.e., $TEW=TRW=7$ $h$. We use the first half of 
the dataset (the first seven hours) of each individual to train the
model and the second half to test it. We use all the Biostreams, 
i.e., $n=9$, to train and test our system. We use two classification
methods: SVM and AdaBoost. In the case of SVM, we use two kernels (linear and 
RBF). In the case of AdaBoost, we consider decision 
stumps (one-node tree) and decision trees with 5, 10, 15, and 20 nodes as 
weak classifiers. We run 40 iterations for all Adaboost-based classifiers 
since we determined experimentally that the training error becomes zero within these many iterations and testing error becomes minimum. The value of 
$EER_{t=7h}$ is reported in Table \ref{table:BASE_VER_EER} for all classifiers. 
AdaBoost with a tree size of 15, i.e., with 15 nodes in the tree, has the 
minimum value of $EER_{t=7h}$. Increasing tree size usually improves the 
accuracy of Adaboost-based classifiers. However, using larger trees leads to 
more complex models, which are more susceptible to overfitting \cite{OVF}. 
This can be seen when we move from a tree size of 15 to 20.

\begin{table}[ht]
  \centering
  \caption{Classifiers and their $EER_{t=7h}$}
  \begin{tabular}{|l|l|c|}
  \hline
  	Type of classifier & Specification & $EER_{t=7h}$ ($\%$)\\ [0.5ex]
    \hline
    \multirow{2}{*}{SVM} & Kernel = Linear & 3.0 \\
    \hhline{~--}         & Kernel = RBF & 2.6  \\
    \hline
    \hline
    \multirow{2}{*}{AdaBoost} & Tree size = 1  & 3.1 \\
    \hhline{~--}         &  Tree size = 5 &  2.9 \\
    \hhline{~--}         &  Tree size = 10 &  2.9 \\
    \hhline{~--}         &  Tree size = 15 &  2.4 \\
    \hhline{~--}         &  Tree size = 20 &  2.5 \\
    \hline
  \end{tabular}
  \label{table:BASE_VER_EER}
\end{table}

Table \ref{table:BFAW} summarizes $FAW$ and $FRW$ for all classification 
schemes.  Consider RBF SVM as an example.  Its $FAW$ is 4 minutes, 
which suggests that, in the worst case, an impostor can deceive the 
authentication system for a 4-minute timeframe. Its $FRW$ is 3 minutes,
which suggests that, in the worst case, a legitimate user is falsely 
rejected for a stretch of 3 minutes.

\begin{table}[h!]
  \centering  
  \caption{Classifiers and their $FAW$ and $FRW$}
  \begin{tabular}{|l|l|c|c|}
  \hline
  	Type of classifier & Specification & $FAW$ ($min$) & $FRW$ ($min$) \\ [0.5ex]
    \hline
    \multirow{2}{*}{SVM} & Kernel = Linear & 4 & 3 \\
    \hhline{~---}         & Kernel = RBF & 4 & 3 \\
    \hline
    \hline
    \multirow{2}{*}{AdaBoost} & Tree size = 1 & 5 & 3 \\
    \hhline{~---}         & Tree size = 5 & 4 & 3 \\
    \hhline{~---}         & Tree size = 10 & 4 & 3 \\
    \hhline{~---}         & Tree size = 15 & 4 & 3 \\
    \hhline{~---}         & Tree size = 20 & 4 & 4 \\
    \hline
  \end{tabular}
  \label{table:BFAW}
\end{table}

\item \textit{Biased $FAR_t/FRR_t$:} Even though it is easier to compare
authentication methods based on their $EER_t$, we may want to minimize 
$FAR_t$ in highly-secure environments in order to ensure that an impostor 
is not authorized or minimize $FRR_t$ to enhance user convenience. A low 
$FAR_t$ indicates a high security level and a low $FRR_t$ ensures user 
convenience. In this experimental scenario, we use the same parameters that 
are used in the baseline. However, false acceptance and false rejection are 
penalized differently. We consider two cases: (i) try to make $FAR_t$ close to 
zero ($FAR_{t=7h} < 0.1 \% $) and measure $FRR_t$, and (ii) try to make 
$FRR_t$ close to zero ($FRR_{t=7h} < 0.1 \% $) and measure $FAR_t$. 
Tables \ref{table:B2} and \ref{table:B1} summarize the results for these two 
cases. Based on Table \ref{table:B2}, CABA can be seen to ensure that impostors 
are not accepted, but at the cost of an increase in $FRR$. Based on 
Table \ref{table:B1}, CABA can be seen to not negatively
impact user convenience, i.e., not falsely reject the user, while rejecting 
impostors in more than $90\%$ of the cases. 

\begin{table}[h!]
  \centering  
  \caption{Classifiers and their $FRR$ ($FAR \approx 0$)}
  \begin{tabular}{|l|l|c|}
  \hline
  	Type of classifier & Specification & $FRR$($\%$)\\ [0.5ex]
    \hline
    \multirow{2}{*}{SVM} & Kernel = Linear & 10.2  \\
    \hhline{~--}         & Kernel = RBF & 9.6  \\
    \hline
    \hline
    \multirow{2}{*}{AdaBoost} & Tree size = 1 & 10.0  \\
    \hhline{~--}         & Tree size = 5 & 9.7 \\
    \hhline{~--}         & Tree size = 10 & 8.7  \\
    \hhline{~--}         & Tree size = 15 & 8.4  \\
    \hhline{~--}         & Tree size = 20 & 8.9  \\
    \hline
  \end{tabular}
  \label{table:B2}
\end{table}
\begin{table}[h!]
  \centering
  \caption{Classifiers and their $FAR$ ($FRR \approx 0$)}
  \begin{tabular}{|l|l|c|}
  \hline
        Type of classifier & Specification & $FAR$($\%$)\\ [0.5ex]
    \hline
    \multirow{2}{*}{SVM} & Kernel = Linear & 8.9  \\
    \hhline{~--}         & Kernel = RBF & 7.6  \\
    \hline
    \hline
    \multirow{2}{*}{AdaBoost} & Tree size = 1 & 10.7  \\
    \hhline{~--}         & Tree size = 5 & 9.2  \\
    \hhline{~--}         & Tree size = 10 & 7.8  \\
    \hhline{~--}         & Tree size = 15 & 7.6  \\
    \hhline{~--}         & Tree size = 20 & 8.2  \\
    \hline
  \end{tabular}
  \label{table:B1}
\end{table}

\item \textit{Variable window size:} As mentioned earlier, we set the 
training and testing window sizes to 7 $h$ in the baseline. Here, we change  
the size of the training and testing windows such that $TRW= 2, 3, \cdots , 12h$ and $TEW + TRW=14$ $h$. Fig.~\ref{fig:EERToTRW} shows the average $EER_{t}$ 
for different classifiers with respect to $TRW$. For all classifiers, as we 
increase $TRW$ from 2 $h$ to 6 $h$, $EER_t$ decreases drastically. Then it 
remains almost constant until $TRW$ reaches $11$ $h$. Above this $TRW$, $EER$ 
starts increasing for two possible reasons. First, the model may become 
overfitted. Second, the number of test points may be inadequate.

\begin{figure}[h]
\centering
\includegraphics[trim = 30mm 81mm 30mm 82mm ,clip, width=250pt,height=160pt]{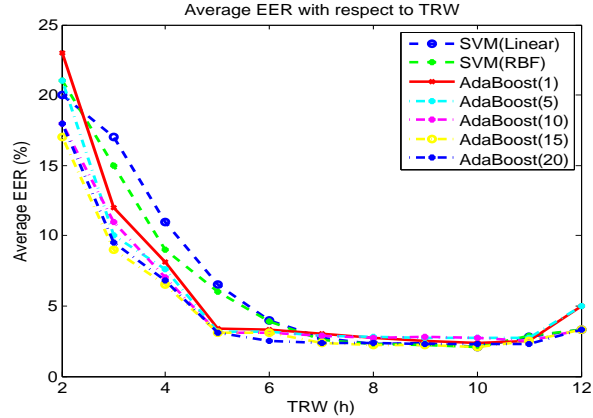}
\caption{Average $EER_{t}$ for different classifiers with respect to $TRW$.} 
\label{fig:EERToTRW}
\end{figure}

\item \textit{Moving training window:} In this scenario, the training window 
moves behind the testing window (Fig.~\ref{fig:Moving}). We consider 
$TEW=TRW=1, 2, \cdots, 7$ $h$. Our experimental results demonstrate that this 
verification scheme provides the best result for $TEW=TRW= 4$ $h$, for
which the average $EER_t$ is $1.9\%$ and the classification method is 
AdaBoost with a tree size of 15 nodes. This suggests that we can achieve the 
best accuracy for $TRW=4$ $h$, under the assumption that the trained model is 
valid for the next four hours. 

\begin{figure}[h]
\centering
\includegraphics[trim = 70mm 115mm 70mm 105mm ,clip, width=250pt,height=50pt]{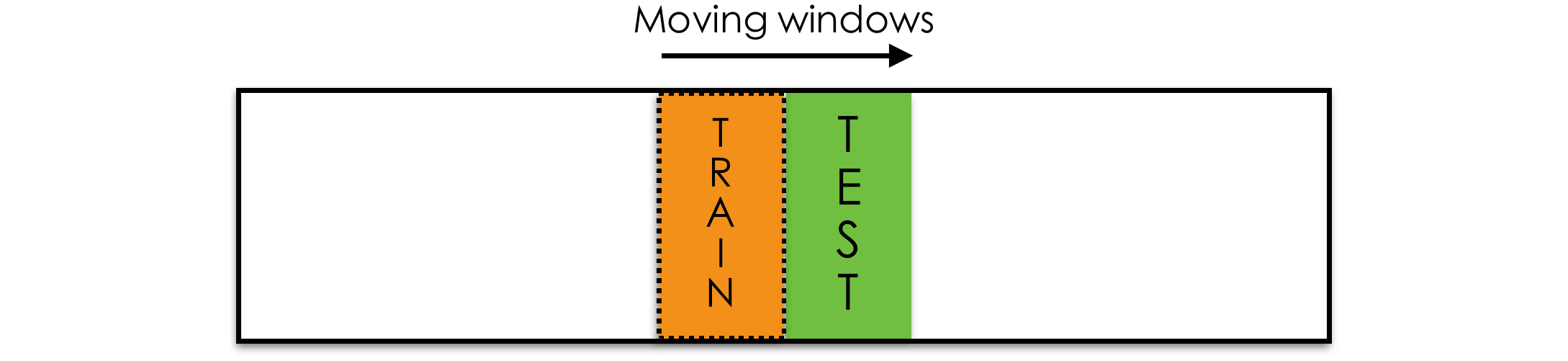}
\caption{Moving training window.} 
\label{fig:Moving}
\end{figure}

\item \textit{Reducing the number of Biostreams:}  We also investigate
the impact of dropping a Biostream.  Traditionally, feature reduction is
used to remove redundant or irrelevant features from the data set before 
commencing on the training process in order to decrease unnecessary 
computational cost. However, in our scenario, the main purpose of feature 
reduction is to investigate how each feature affects accuracy. If CABA can 
provide an acceptable accuracy with fewer features, fewer WMSs would be 
required. We dropped one feature at a time and computed $EER_{t=7h}$ of the 
system. All other configurations are kept the same as in the baseline. 
Fig.~\ref{fig:DROP} illustrates how $EER_{t=7h}$ changes for each of the
seven classifiers used in our experiments (two SVM classifiers with
different kernel types and five AdaBoost classifiers with different tree
sizes) when we drop different Biostreams. The green bar shows the
baseline scenario in which no feature is dropped. We can see that dropping 
the respiratory rate (temperature) has maximum (minimum) negative impact on 
authentication accuracy. Thus, the most and least important Biostreams are 
respiratory rate and body temperature, respectively.

\begin{figure}[h]
\centering
\includegraphics[trim = 10mm 61mm 12mm 22mm ,clip, width=250pt,height=160pt]{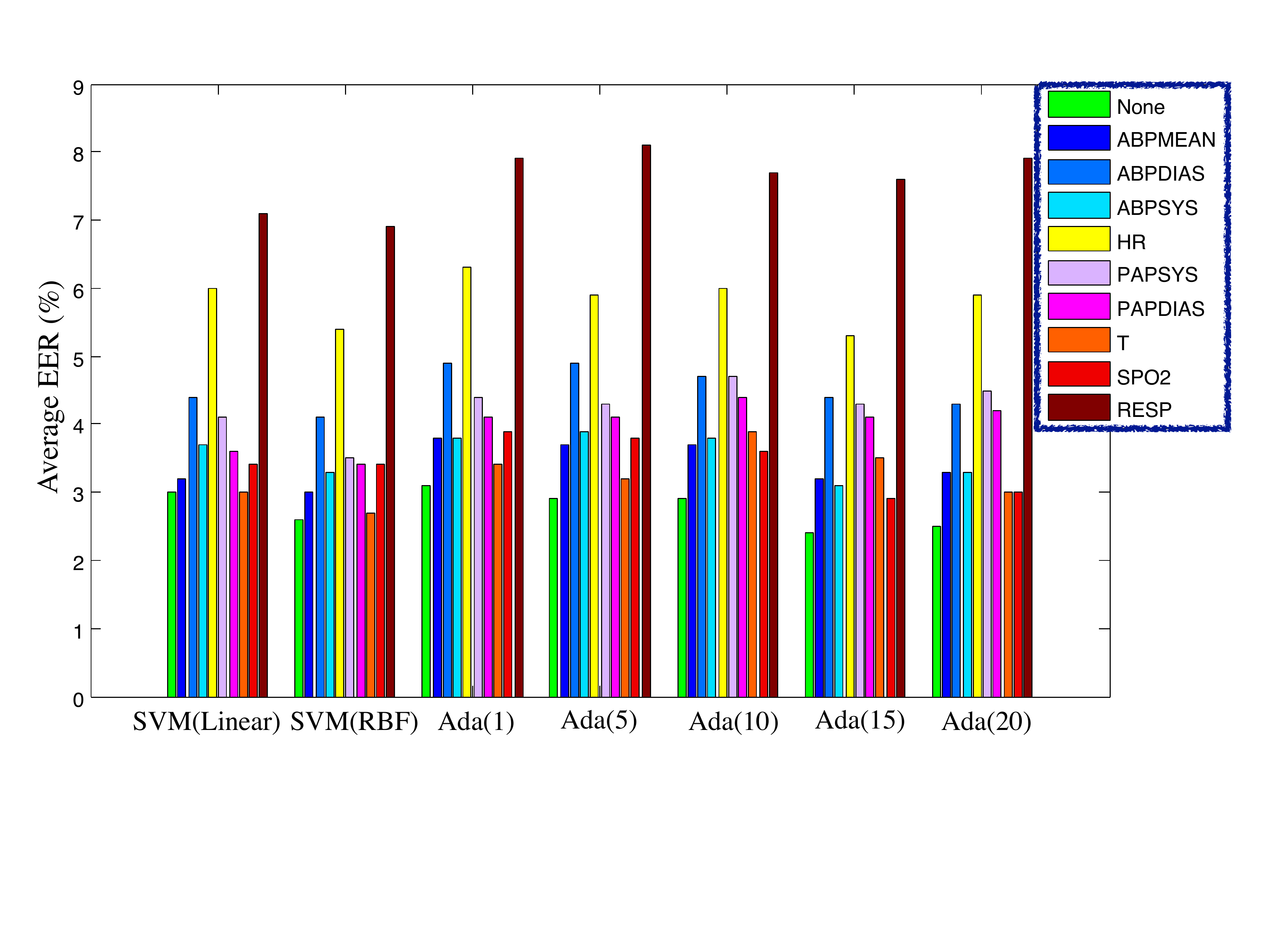}
\caption{$EER_{t=7h}$ for different classifiers when Biostreams are dropped one at a time. The green bar depicts the baseline scenario in which no feature is dropped. The abbreviations/notations provided in Table \ref{table:BIOAURA} are used to label other bars.} 
\label{fig:DROP}
\end{figure}

\end{enumerate}

\subsection{CABA scalability}
We discuss below the worst-case time and space complexities of CABA. 

\subsubsection{Time complexity}
As discussed earlier, CABA can be implemented in such a manner that the time required by the enrollment phase is hidden from the user's perspective. Hence, we focus on the time complexity of the user authentication process. We found that the required time for processing an authentication 
request for $N=37$ was on the order of a few milliseconds for all
classification methods, when CABA was implemented on a MacBook Pro
(2.3 $GHz$ Intel Core i7 processor with 8 $GB$ memory). This suggests that 
CABA can re-authenticate the user very quickly. 

When a person requests authentication by providing his ID and feature vector 
$Y$, the Look-up stage forwards $Y$ to one and only one classifier in 
the Jury stage based on the given user ID. Then, the classifier's decision is 
the final decision of the authentication system. Hence, in order to analyze 
the time complexity of a single user authentication process, we need to consider 
the time complexity of the Look-up stage, and one classifier in the Jury 
stage, as follows:

\begin{itemize}
\item Look-up stage: If the Look-up stage is implemented using a hash table 
that associates user IDs with pointers to classifiers, then its search 
operation (finding the location of the classifier associated with the user ID) 
can be performed in $O(1)$ time.

\item  One classifier in the Jury stage: The time complexity of the
classifier varies from one classification algorithm to another. The time
complexities of AdaBoost classifiers and the SVM classifier with a
linear kernel do not depend on $N$, i.e., they have time complexity of
$O(1)$. The time complexity of SVM with an RBF kernel is $O(n_{SV})$,
where $n_{SV}$ is the number of support vectors. Theoretically, $n_{SV}$
grows linearly with a linear increase in $N$. Thus, the SVM classifier
with an RBF kernel has a time complexity of $O(N)$.
\end{itemize}

Hence, the overall time complexity of user authentication is $O(1)$ for
AdaBoost classifiers and the SVM classifier with a linear kernel, and
$O(N)$ for the SVM classifier with an RBF kernel.

\subsubsection{Space complexity}
We first investigate how much memory is required for our prototype implementation of CABA. Then, we discuss how the amount of memory required to store the two stages (Look-up and Jury) increases with $N$. 

The amount of memory required for storing the Look-up stage in our prototype, 
where $N=37$, was less than 1 $kB$. The amount of memory required for storing a 
single classifier in the Jury stage varies from tens of bytes (for SVM with a 
linear kernel) to a few $kB$ (for AdaBoost with a tree size of 20). Therefore, 
the total amount of memory allocated to the authentication system is less than 1 $MB$.

We investigate the space complexity as a function of $N$.

\begin{itemize}
\item Look-up stage: If the Look-up stage is implemented using a hash
table, its space complexity is $O(N)$. 

\item  Jury stage: The space complexity of a single classifier in the
Jury stage depends on the type of classifier. The space complexities of
AdaBoost classifiers and the SVM classifier with a linear kernel do not
depend on $N$, i.e., they have space complexity of $O(1)$. However, the
space complexity of the SVM with an RBF kernel is $O(N)$. Since the number of classifiers in the Jury stage
increases linearly with $N$, its space complexity is $O(N)$ (when an
AdaBoost classifier or SVM classifier with a linear kernel is used) or
$O(N^2)$ (when the SVM classifier with the RBF kernel is employed).
\end{itemize}


\section{Using BioAura for identification}
\label{CABA_IDENT}
The majority of continuous authentication systems only support 
continuous verification in which the user provides a user ID and the system 
checks if the user is the person he purports to be. In this section, we 
describe how CABA can be slightly modified to also identify the user from a
database of users by processing feature vector $Y$ provided by the smartphone. 
An identification scenario consists of four steps. The first three steps are 
similar to the ones discussed in Section \ref{PROT} for continuous 
authentication. In the fourth step, CABA processes the decisions of all 
classifiers in the Jury stage to indicate that the user is not in the 
database, or conclude that he is, in which case it returns his user ID. 
This step can be implemented in different ways. In our implementation, 
CABA processes all outputs of the Jury stage and outputs the user ID if there 
is only one classifier whose output is an accept decision. Otherwise, it 
indicates no match. Our experimental results demonstrate that this scheme provides the best result, for which the identification rate is $96.1\%$, with the AdaBoost classification method with a tree size of 15 nodes. Identification rate is a commonly used metric for this scenario
\cite{SU_1_2}. It is defined as the percentage of attempts correctly identified to the total number of attempts made.

\section{Real-time adaptive authorization}
\label{CABA_ADAPT}
In this section, we first define the concept of authorization. Then, we 
propose a real-time adaptive authorization (RAA) scheme, which uses the 
decisions from CABA to provide an extremely flexible access control model. 
The RAA concept is not limited to CABA. It provides an adjustable access 
control model for any authorization system that authorizes the user based on 
decisions of a continuous authentication system.

Authorization is defined as the process of establishing if the user, who has been already authenticated, should be allowed access to a resource, system, or 
area \cite{AUTHOR}.

Traditional authorization schemes grant a specific access level to the 
authenticated user based on his user ID. However, the fact that continuous 
authentication systems have a non-zero $FRR$ implies that such a simple 
scheme may unintentionally block a legitimate access when the authentication 
system fails to recognize a valid user for a short period of time. Consider 
a scenario in which a continuous authentication system is used to protect a 
personal laptop from unauthorized users. The authentication system first 
authenticates the user. Then, the authorization scheme specifies the user's 
access level based on the user ID. However, the laptop may log out the user 
when the authentication scheme falsely rejects him. RAA schemes can be
used to alleviate user inconvenience caused by false reject decisions. 
They continuously adjust the user's access level based on the last decision 
of the authentication system. Next, we propose an RAA scheme that can be used with a continuous authentication system.

A trust level-based RAA adaptively changes the user's access level based on a 
parameter called trust level (TRL). TRL is a recently-suggested parameter 
that represents how much we trust a user based on previous decisions of the 
continuous authentication system \cite{MET1}. TRL has a value between 0 
and 100, where a higher number indicates a higher level of trust. The 
initial value of TRL is 100 when the user is authenticated and authorized for 
the first time. The value of TRL is continuously updated using a trust update 
procedure after each user authentication. A simple trust update procedure 
may be to just increase (decrease) the TRL by a constant step after each 
accept (reject) decision. \textit{Algorithm 1} shows the 
pseudo-code for such an approach.  We need to set two parameters: 
$W_{Accept}$ and $W_{Reject}$. The values of $W_{Accept}$ and $W_{Reject}$ 
should be chosen such that the TRL value becomes 0 as soon as we detect the 
presence of an impostor and becomes 100 when we confidently verify that the 
user is legitimate. Consider AdaBoost classification with a tree size of 15 
nodes that yields $FRW =3$. This indicates that the authentication system may 
falsely reject three consecutive requests of a legitimate user in the 
worst case. Therefore, if the RAA scheme gets at least four consecutive reject 
decisions from the authentication system, it becomes confident that the user 
is an impostor ($TRL=0$). Hence, we can set $W_{Reject}$ for this
classifier as follows: $W_{Reject}=\frac{-100}{FRW+1}=\frac{-100}{4}=-25$. 
$FAW=4$ for the above-mentioned classification method, which indicates that 
in the worst case, an impostor may be falsely accepted as a legitimate user 
in four successive trials. Therefore, if the authentication system outputs 
five consecutive accept decisions, TRL should become 100. Thus, we can set 
$W_{Accept}$ as follows: $W_{Accept}=\frac{+100}{FAW+1}=\frac{+100}{5}=+20$.

We can set different threshold values for different applications. We set
the threshold value to 100 for accessing email and financial accounts to
ensure that the user can access such accounts only when the system is
confident that the user is legitimate. However, for less sensitive
applications, e.g., simple web surfing, a lower level of trust might be sufficient. Using CABA in conjunction with RAA can enhance user convenience, while ensuring high security for critical applications. 

\noindent {\em Algorithm 1: trustUpdate procedure}

\noindent\makebox[\linewidth]{\rule{15.1cm}{0.1pt}}
{\footnotesize
\noindent Given: The latest $decision$ of the authentication system and current TRL value\\
\noindent\makebox[\linewidth]{\rule{15.1cm}{0.5pt}}
1. $TRL \leftarrow TRL+F_{Update} $, where\\
\begin{equation*}
F_{Update} (decision)= \begin{cases}
  W_{Accept}, & \text{if } decision = Accept, \\
  W_{Reject}, & \text{otherwise}.
\end{cases}
\end{equation*}
\noindent 2. \textit{If} $(TRL > 100)$\\
3. \indent \indent \indent $TRL \leftarrow 100$\\
4. $end$\\
5. \textit{If} $(TRL < 0)$\\ 
6. \indent \indent \indent $TRL \leftarrow 0$\\
8.    $end$\\
9. \textit{Return} $TRL$ \\
}
\noindent\makebox[\linewidth]{\rule{15.1cm}{0.5pt}}

\section {Potential threats and countermeasures}
\label{CABA_THREAT}
Next, we describe possible attacks/threats against CABA that can be exploited by attackers to bypass CABA. For each attack, we also suggest possible countermeasures. \\
\noindent \textbf{1. Eavesdropping:} This is defined as the act of covertly 
listening to confidential conversations of others \cite{IOT_SURVEY}, which, in our context, can 
be done by intercepting the communication between two devices using an 
appropriate equipment, e.g., HackRF \cite{HACKRF}. Eavesdropping can occur when unencrypted information is transmitted over an untrusted channel. \\ 
\textbf{Countermeasures:} The most effective and well-known defense against eavesdropping is encryption. For example, the transmitted message can be 
encrypted using Advanced Encryption Standard \cite{AESBOOK}. However, implementing a strong encryption in WMSs may not be possible in the current state of the technology since they have limited energy and memory capacity. 
Fortunately, eavesdropping does not pose a direct threat to the authentication 
system.  In other words, it is possible to design the authentication system assuming that eavesdropping does occur on the communication between the WMSs and the smartphone. In this case, CABA would require that the data be sent 
from a smartphone that is previously registered in the system to ensure that 
the attacker is not able to capture the biomedical information and send the 
captured information to CABA using another smartphone. The smartphone can send 
its unique ID over a secure communication link to CABA before transmitting the 
biomedical information.\\
\noindent \textbf{2. Phishing:} This is an attack that attempts to fool the user 
into submitting his confidential or private information, e.g., username, 
password, email address, and phone number, to an untrusted server or 
device \cite{PHIS}. 
For example, the attacker might 
attempt to fool the user's smartphone by sending a counterfeit request that 
asks the smartphone to send its authentication-related information to the 
attacker's server.\\
\textbf{Countermeasures:} The most effective way to address phishing attacks is to use 
a digital certificate, i.e., an electronic document that allows a device to 
exchange information securely using the public key infrastructure
\cite{CERT}. The certificate carries information about the key and its owner. 
In CABA, the server's digital certificate can be examined by the smartphone to ensure that the server is trusted.\\
\noindent \textbf{3. Replay attack:} In a replay attack, an attacker records the data, 
packets, and user's credentials, which are transmitted between two devices, 
e.g., a WMS and the smartphone, and exploits them for a malicious purpose. In 
a replay attack against the authentication system, the attacker attempts to 
impersonate a legitimate user in order to bypass the authentication procedure 
and gain full access to the protected device, application, or area. Unlike the 
attacks based on eavesdropping, in a replay attack, the attacker does not need 
to interpret the packets. In fact, he can even record encrypted packets
and retransmit them in order to bypass the
system.\\\textbf{Countermeasures:} An encrypted timestamp, i.e., a sequence of encrypted 
information identifying when the transmission occurred, can be utilized 
to enable the authentication system to check that the packets were not 
previously recorded. Moreover, the packet should include a field that contains 
the encrypted information, e.g., a hashed device ID, which can be used in the 
authentication system to uniquely specify the sender of the packets and check 
if the sender is known and trusted.  \\
\noindent \textbf{4. Poisoning attack:} In a poisoning attack, the attacker changes the final 
learning model by adding precisely-selected invalid data points to the 
training dataset \cite{PODEF}. In CABA, the attacker might threaten the 
integrity of the machine learning algorithm by using an untrusted WMS that 
aims to add malicious data points to the training set.\\
\textbf{Countermeasures:} We describe two types of countermeasures against 
poisoning attacks. \\
\indent \textit{1. Outlier detection:} One of the common goals of defenses against poisoning 
attacks is to reduce the effect of invalid data points on the final result. In a machine learning method, such invalid data points are considered 
outliers in the training dataset. Several countermeasures against poisoning 
attacks have been discussed in \cite{MEHRAN}.\\
\indent \textit{2. Digitally-signed biomedical information:} A digital signature can be
used to check the authenticity of the information. It is a 
mathematical method for demonstrating the authenticity of a transmitted 
message. Thus, it provides the means to the recipient to check if the message 
is created by a legitimate sender. The WMSs and the smartphone can digitally 
sign the biomedical information before transmitting it.   

\section{Comparison between CABA and previously-proposed systems}
\label{COMPARE}
In this section, we first describe why previously-proposed authentication 
systems based on biomedical signals (EEG and ECG) may not be
well-suited to continuous authentication. Then, we compare CABA to three 
promising biometrics-/behaviometrics-based continuous authentication systems. 

The use of EEG \cite{EEG_4} and ECG
\cite{ECG_3,ECG_4} signals, as biomedical traits with high
discriminatory power for user authentication, has received widespread
attention in recent years. Although such authentication systems
show promising results, they do not provide a convenient method
for long-term continuous user authentication for two reasons.
First, they commonly need long measurement times and impose a heavy
computational load on the system \cite{EEG_PRO}. Second, due to the
size/position requirements of the electrodes that enable EEG/ECG
acquisition \cite{EEG_CAP,ECG_3}, these systems can mainly be used 
for one-time user authentication (or short-term
continuous authentication) systems. For example, the user needs to wear a 
large cap to collect the data for EEG-based authentication \cite{EEG_4}, 
which is not convenient for long-term continuous authentication. 

As mentioned in Section \ref{intro}, several biometrics-/behaviometrics-based continuous authentication systems have been proposed. Among them, facial recognition systems \cite{FACE_1,FACE_2,F1,F2}, which use 
facial features (as biometrics), and keyboard-/mouse-based authentication 
systems \cite{KM_1,K_2,K_3,AUTH,SU1, SU3}, which rely on 
keystroke/mouse dynamics (as behaviometrics), are the most promising.

Facial recognition systems make use of low-cost cameras that are commonly 
built into most laptops.  They are accurate when the user
looks straight at the webcam. However, their performance is significantly
affected by illumination, pose, expression or changes in the image
acquisition method \cite{F2}. Moreover, the user's facial images
is unavailable when the user turns his head or does not look at the 
camera.  Such systems are also not useful for tablets and smartphones since 
the user typically does not face a built-in camera in these cases. 

Previous keyboard-/mouse-based authentication systems report promising results 
and provide user authentication in a convenient manner. However, 
they have four drawbacks that limit their
applicability: (i) their performance is easily impacted by environmental
variables, such as changes in software environments, input devices,
task, or interaction modes \cite{SU3,SU_1_2}, (ii) they can only be employed
when system has a keyboard/mouse, (iii) the data often become
unavailable, e.g., when the user is watching a movie on his computer, 
and (iv) keyboard-based systems need active involvement of the user for long 
sessions, e.g., several minutes \cite{SU1}, to guarantee acceptable accuracy, 
and mouse dynamics based systems have still not reached an acceptable 
accuracy levels \cite{SU3}.    

Unlike most continuous authentication systems that support personal computers 
and laptops, CABA can be used to protect personal computers, servers, software 
applications, and restricted physical spaces. Moreover, WMSs ensure a continuous 
data stream. This enables the user to freely move and change his posture while 
being authenticated. In addition, unlike previous systems, CABA can be 
implemented on any general-purpose computing unit
with sufficient memory capacity and computation power. 

\section{Discussion}
\label{DISC}
Here, we address three items not yet explained in detail. First, we discuss an 
important privacy concern surrounding the use of biomedical signals. 
Second, we describe how CABA can also be used as a stand-alone one-time 
authentication system. Third, we discuss the impact of temporal conditions on 
authentication results.

\subsection{Health information leakage}
An important privacy concern associated with the use of biomedical signals 
is the possibility of health information leakage. For example, an
adversary might extract disease-specific information from such signals, e.g., certain heart rate ranges may be correlated with 
cardiovascular disease \cite{HR_1}. Exposure of a serious illness or a 
condition that carries social stigma would naturally raise serious privacy 
concerns \cite{PHYSIO}. However, since CABA does not rely on high-precision 
measurements (it only processes the average values of Biostreams over 
specific time frames), the amount of health-related information potentially 
leaked by CABA is less than leaked by EEG/ECG-based approaches that rely on 
high-quality EEG/ECG signals. Similar concerns have been discussed in previous 
research efforts for both biometrics and behaviometrics, and usually addressed 
by suggesting legislation \cite{PRIVACY_2}.

\subsection{One-time authentication based on BioAura}
CABA can also be used as a stand-alone one-time authentication system.
We discuss several such scenarios next.

\begin{itemize}
\item Battery-powered devices: Incurring overheads of continuous 
authentication on a battery-powered device may drain 
its battery quickly, and lead to user inconvenience. 
\item Low-security environment: Continuous authentication may not be required 
in a low-security environment, e.g., a common room in an apartment.
\item Intentionally-shared resources: A user might want to intentionally 
authorize a group of users to access some specific locations or resources. 
For example, consider a user who uses a smart lock, which grants 
access to him when he approaches the door of his house. He may want to open 
the door for his guests and leave the house.
\end{itemize}
Generally, a continuous authentication system that has high accuracy and 
a short response time may be able to provide stand-alone one-time 
authentication or complement a traditional authentication system (whose 
decision is only considered at the time of initial login). As discussed in 
Section \ref{ACC}, CABA provides an accurate decision within a few 
milliseconds and, hence, is also useful for one-time authentication.

\subsection{The impact of temporal conditions}
The negative impact of temporal conditions, e.g., emotional/physical conditions 
and changes in posture, gesture or facial expressions, on
widely-used biometrics-/behaviometrics-based systems have been discussed
earlier \cite{SU_1_0}. Similarly, some biomedical signals may change 
significantly due to a change in physical activity. This may negatively 
impact authentication accuracy. For example, when the user suddenly starts 
running, his blood pressure, heart rate, and respiration rate increase. 
Therefore, if the authentication system has only been trained using data 
collected when the user was at rest, it might fail to authenticate the user 
after he finishes running. A solution would be to design a state-aware system 
that takes different emotional states and physical activities into account. 
Algorithms exist for recognizing emotional states \cite{EMOTION} and the type \cite{ACT1} and intensity \cite{ALINIA} of physical activities using WMSs. Such algorithms can be 
used in conjunction with CABA. 

\section{Chapter summary}
\label{CABA_CONC}
In this chapter, we proposed CABA, a novel user-transparent system for 
continuous authentication based on information that is already gathered by 
WMSs for diagnostic and therapeutic purposes. We described a prototype implementation of CABA and 
comprehensively investigated its accuracy and scalability. We also described 
how CABA can be used to support user identification. We then presented
an RAA scheme that uses the decisions from CABA to enable flexible access 
control. We compared CABA to previously-proposed continuous authentication systems (biometrics- and behaviometrics-based), and highlighted its 
advantages. We discussed several attacks against the proposed authentication 
system along with their countermeasures. Finally, we briefly described an 
privacy concerns surrounding the use of biomedical signals, how CABA can 
also be used for one-time authentication, and impact of temporal conditions on 
authentication.

%% file: ch-conclusion/chapter-conclusion.tex
\chapter{Conclusion\label{ch:conclusion}}

\input{ch-conclusion/conclusion} 
\input{ch-conclusion/futurework} 

%% file: ch-conclusion/conclusion.tex
The emergence of the Internet of Things (IoT) paradigm provides an opportunity to transform isolated devices into communicating things. As a side effect of rapid advances in the development of IoT-enabled systems, the number of potential security/privacy attacks against such systems, \textit{in particular wearable medical sensor (WMS)-based systems}, has grown exponentially. Therefore, common security threats and privacy concerns need to be studied and addressed in depth. This thesis attempted to explore and address different security/privacy-related issues associated with IoT-enabled systems with a special focus on WMS-based systems.
 
\section{Thesis summary}
In Chapter \ref{ch:intro}, we discussed different IoT reference models and the scope of IoT applications. We described what security means in the scope of IoT and who the attackers that target the IoT might be, and what motivations they might have. Furthermore, we discussed different WMS-based systems along with their applications, components, and design requirements.  

In Chapter \ref{ch:relatedwork}, we described related work and provided background for several key concepts used in this thesis. We summarized different attacks and threats on the edge-side layer of IoT and described possible countermeasures against them.  We discussed several research directions that are closely related to the domain of WMSs and described how previous research studies have  facilitated the design and development of WMS-based systems.

In Chapter \ref{ch:hmonitoring}, we described a secure energy-efficient system for long-term continuous health monitoring. We constructed a wireless body area network (WBAN) using eight biomedical sensors. Furthermore, we presented different processing and transmission schemes and evaluated our schemes using the WBAN. We also examined the storage requirements for long-term analysis and storage. Among the four schemes we evaluated (including the baseline scheme), we showed that the compressive sensing (CS)-based scheme provides the most computational energy savings because it needs to process much fewer signal samples. For low-sample-rate sensors, we can achieve significant energy savings by simply accumulating the raw data before transmitting them to the base station. In addition, the CS-based scheme also allows us to reduce the storage requirements significantly. The results indicate that long-term continuous health monitoring is indeed feasible from both energy and storage points of view.

In Chapter \ref{ch:OpSecure}, we presented OpSecure, an optical secure 
communication channel between an implantable medical device (IMD) and an external device, e.g., smartphone, that enables an intrinsically short-range, user-perceptible one-way data 
transmission (from the external device to the IMD). Based on OpSecure, we 
proposed a wakeup and a key exchange protocol. In order to evaluate the 
proposed protocols, we implemented an IMD prototype, that supports both protocols, and developed an Android application, which can be used to wake up the IMD and transmit the shared key from the smartphone to the IMD. We evaluated our prototype implementation using a human body model. Our experimental results indicate that OpSecure can be used to implement both wakeup and key exchange protocols for IMDs with minimal size and energy overheads. 

In Chapter \ref{ch:physio}, we discussed two sources of information leakage, namely the human body and implantable and wearable medical devices (IWMDs), which continuously leak health information. We described two types of signals for each source: acoustic and electromagnetic. We presented a variety of attacks on the privacy of health data by capturing and processing unintentionally-generated leaked signals and also discussed the feasibility of using intentionally-generated acoustic/electromagnetic signals to compromise the patient's health privacy. 

In Chapter \ref{ch:DISASTER}, we discussed Dedicated Intelligent Security Attacks on Sensor-triggered Emergency Responses (DISASTER). It exploits design flaws 
and security weaknesses of safety mechanisms utilized in cyber-physical systems (CPSs) to trigger the system's emergency responses even when a real emergency situation is not present. We demonstrated that DISASTER can lead to serious consequences, ranging from economic collateral damage to life-threatening 
conditions. We examined several already-in-use sensors and summarized common design flaws and security weaknesses of safety mechanisms. We demonstrated the possibility of launching DISASTER in realistic scenarios and suggested several countermeasures against the proposed attacks.

In Chapter \ref{ch:CABA}, we proposed a continuous authentication system based on biological aura (CABA). We described a prototype implementation of CABA and examined its accuracy and scalability. We discussed how CABA can be used to support user identification. We presented a real-time adaptive authorization scheme, which uses decisions provided by CABA to enable flexible access control. We compared CABA to previously-proposed continuous authentication systems and highlighted its advantages. We discussed several attacks against the proposed authentication system along with their countermeasures. We also described privacy concerns surrounding the use of biomedical signals, how CABA can also be used for one-time authentication, and impact of temporal conditions on authentication.

%% file: ch-conclusion/futurework.tex
\section{Future directions}
Next, we describe several avenues along which the proposed schemes and techniques we presented can be further explored.
\subsubsection{Targeting weak links of IoT-enabeled systems}
The majority of IoT-based services rely on compact battery-powered devices with limited storage and computation resources. Due to the special characteristics of these devices 
and cost factors considered important by manufacturers, several 
\textit{already-in-market devices} do not support highly-secure cryptographic 
protocols. This has led to the emergence of an enormous number of weak links 
in the network/system that can be exploited by an attacker to target other 
presumably-secure entities in the network. A few research efforts \cite{WE1, 
WE2} have recently demonstrated the possibility of targeting weak edge nodes 
to extract the home user's WiFi password. Chapman \cite{WE1} has demonstrated how Internet-connected light bulbs can reveal the user's WiFi password to the attacker. In \cite{WE2}, a similar attack is discussed, which extracts the WiFi password by targeting the user's smart lock. The endless variety of IoT applications magnifies the impact of these weak edge nodes. DISASTER, which we described in Chapter \ref{ch:DISASTER}, also targets insecure sensors in home automation and industrial CPSs. Similar attacks can be proposed and investigated in other application domains.  

\subsubsection{Unexpected uses of collected data}
The widespread use of ubiquitous computing enabled by IoT technologies
has led to the pervasive deployment of Internet-connected sensors in
modern day living. In recent years, a few research efforts have
attempted to shed light on unexpected uses of different types of
environment-/user-related data collected by Internet-connected sensors
\cite{S1,S2,S3,S4}. For example, McKenna et al. have provided a list of
privacy-sensitive information, e.g., number of residents, personal habits, and 
daily routines, that can be inferred from smart homes' electricity load data collected by smart meters \cite{S1}. Despite the existence of previous research efforts, the extent of private information that can be inferred from presumably non-critical data is neither well-known nor well-understood. 

\subsubsection{Promising authentication solutions}
As we described in Chapter \ref{ch:CABA}, password-based authentication has several issues spanning user inconvenience to security flaws. For example, a serious security flaw of one-time password-based 
authentication is its inability to detect unauthorized users after initial authentication. Such issues have boosted the design and development of continuous authentication mechanisms, e.g., face recognition and keystroke-based authentication systems. In Chapter \ref{ch:CABA}, we presented a novel continuous authentication system based on biomedical signals collected by WMSs and described why the proposed system has an advantage over previously-proposed systems. Similar systems can be designed using another set of biomedical signals, which can potentially address a variety of issues associated with the use of one-time password-based systems. Moreover, researchers may want to investigate the effect of temporal changes in the user's biomedical signals on authentication results.